\documentclass[a4paper,11pt,twoside,openright]{ThesisStyle}
\usepackage{graphicx}
\usepackage{caption}
\usepackage{tocloft}
\usepackage{amsmath}
\usepackage{amssymb}
\usepackage{alphalph}
\makeatletter
\newalphalph{\fnsymbolwrap}[wrap]{\@fnsymbol}{}
\makeatother

\usepackage{upgreek}
\usepackage{gensymb}
\usepackage{subfig}
\usepackage{rotating}
\usepackage{fancyhdr}
\usepackage{geometry}
\usepackage{setspace}
\usepackage[normalem]{ulem}
\usepackage{url}
\usepackage{color}
\definecolor{light-gray}{gray}{0.9}
\definecolor{background-color}{gray}{0.5}
\definecolor{dark-yellow}{cmyk}{0.0, 0.0, 0.8, 0.4}
\usepackage{makeidx}
\makeindex
\usepackage{titlesec}
\usepackage{breakurl}
\usepackage[breaklinks,plainpages=false,pageanchor=false]{hyperref}
\usepackage[all]{hypcap}

\hypersetup{
    bookmarks=true,         
    unicode=false,          
    pdftoolbar=true,        
    pdfmenubar=true,        
    pdffitwindow=false,     
    pdfstartview={FitH},    
    pdftitle={Numerical Estimation Schemes for Quantum Tomography},    
    pdfauthor={Teo Yong Siah},     
    pdfsubject={Quantum Information},   
    pdfcreator={Teo Yong Siah},   
    pdfproducer={Teo Yong Siah}, 
    pdfnewwindow=true,      
    colorlinks=true,       
    linkcolor=blue,          
    citecolor=red,        
    filecolor=magenta,      
    urlcolor=cyan           
}

\titleformat{\section}[frame]
{\normalfont}
{\filright
\footnotesize
\enspace SECTION \thesection\enspace}
{8pt}
{\Large\bfseries\filcenter}

\fancypagestyle{plain}{ %
\fancyhf{} 
}

\newcommand{\HRule}{\rule{\linewidth}{0.1mm}}
\newcommand{\HRulecolor}{\color{dark-yellow}{\rule{\linewidth}{0.8mm}}\color{black}}
\newcommand{\ket}[1]{\left|{#1}\right>}
\newcommand{\bra}[1]{\left<{#1}\right|}
\newcommand{\sket}[1]{\big|{#1}\big>\!\!\big>}
\newcommand{\sbra}[1]{\big<\!\!\big<{#1}\big|}
\newcommand{\superinner}[2]{\big<\!\!\big<{#1}\big|{#2}\big>\!\!\big>}

\newcommand{\tr}[1]{\textnormal{tr}{\left\{#1\right\}}}
\newcommand{\D}{\mathrm{d}}
\newcommand{\I}{\mathrm{i}}
\newcommand{\transpose}[1]{{#1}^\textsc{\,t}}

\newcommand{\E}{\mathrm{e}}

\newcommand{\ID}{D_\text{i}}
\newcommand{\OD}{D_\text{o}}

\newcommand{\sitem}{\addtocounter{enumi}{-1}\item}
\newcommand{\ssitem}{\addtocounter{enumi}{-2}\item}

\numberwithin{equation}{section}

\def\listofsymbols{
\begin{tabbing}
$\overleftrightarrow{1}$~~~~~~~~~~~~~~~~~~~~~~\=\parbox{3.9in}{unit dyadic \dotfill \pageref{symbol:1dyadic}}\\\\
\addsymbol 1_\mathcal{H}: {identity operator in $\mathcal{H}$}{symbol:identityh}
\addsymbol 1_\mathcal{K}: {identity operator in $\mathcal{K}$}{symbol:identityk}
\addsymbol \lVert O\rVert_2: {2-norm of an operator $O$}{symbol:2norm}
\addsymbol \mathfrak{A}_k: {aperture operator}{symbol:ak}
\addsymbol \alpha: {a complex number given by $x+\I p$ for real $x$ and $p$}{symbol:alpha}
\addsymbol A: {annihilation operator}{symbol:ann}
\addsymbol \mathcal{A}: {auxiliary complex operator for parameterizing $\rho$ and $E$}{symbol:auxa}
\addsymbol a_k(x): {aperture function}{symbol:akx}
\addsymbol \overline{\quad\vphantom{M}}: {equivalent to $\sum_\mathfrak{D}$}{symbol:avgsym}
\addsymbol C: {qubit Clifford unitary operator}{symbol:cliffordqubit}
\addsymbol \mathfrak{C}(\rho_\text{true},\hat\rho): {cost functional of $\hat\rho$ for $\rho_\text{true}$}{symbol:costfunc}
\addsymbol \overline{\mathfrak{C}}(\hat\rho): {average cost functional of $\hat\rho$}{symbol:costfuncavg}
\addsymbol \mathcal{C}_\text{H-S}(\rho_\text{true}-\hat\rho): {covariance between $\rho_\text{true}$ and $\hat\rho$}{symbol:covariance}
\addsymbol \overleftrightarrow{\mathcal{C}}\left(\vec{t},\vec{r}\right): {covariance dyadic between $\vec{t}$ and $\vec{r}$}{symbol:covariancedyadic}
\addsymbol \overleftrightarrow{\mathfrak{C}}_\text{ML}: {ML covariance dyadic evaluated at $\vec{t}=\vec{t}_\text{ML}$}{symbol:covariancedyadicml}
\addsymbol \mathcal{C}(E,\rho): {cross entropy}{symbol:crossent}
\addsymbol \mathfrak{D}: {measurement data}{symbol:data}
\addsymbol \sum_\mathfrak{D}: {average over all possible $\mathfrak{D}$}{symbol:dataavg}
\addsymbol \frac{\partial}{\partial\vec{r}}: {gradient operator with respect to $\vec{r}$}{symbol:deedeer}
\addsymbol \Delta: {operator variance of a convex set of $\hat E_\text{ML}$s (QPT)}{symbol:delta}
\addsymbol \mathcal{D}: {Lagrange functional}{symbol:dfunctional}
\addsymbol D: {dimension of the Hilbert space}{symbol:dim}
\addsymbol \mathcal{D}(\alpha): {displacement operator for a given $\alpha$}{symbol:dispop}
\addsymbol \mathcal{D}_\text{tr}: {trace-class distance}{symbol:distancetraceclass}
\addsymbol (\D\rho): {prior}{symbol:drho}
\addsymbol D_\text{sub}: {dimension of the truncated Hilbert space}{symbol:dsub}
\addsymbol (\D\tau_\textsc{d}): {integration measure for the $D$-dimensional Hilbert space}{symbol:dtau}
\addsymbol \overleftrightarrow{\vphantom{M}}: {symbol for dyadic}{symbol:dyadicsym}
\addsymbol E: {Choi-Jami{\'o}\l kowski operator for a quantum process}{symbol:e}
\addsymbol \hat E_\text{ML}: {ML process estimator}{symbol:eml}
\addsymbol \overline{E}_\text{ML}: {operator centroid of a convex set of $\hat E_\text{ML}$s}{symbol:emlbar}
\addsymbol \hat E_\text{MLME}: {MLME process estimator}{symbol:emlme}
\addsymbol E_\text{prior}: {prior information about $E_\text{true}$}{symbol:eprior}
\addsymbol \epsilon,\epsilon_k: {small positive step size in an iterative algorithm}{symbol:epsilon}
\addsymbol \varepsilon: {precision for terminating an iterative algorithm}{symbol:epsilonprec}
\addsymbol \eta: {overall detection efficiency}{symbol:eta}
\addsymbol E_\text{true}: {Choi-Jami{\'o}\l kowski operator for the true process}{symbol:etrue}
\addsymbol F: {Fisher's information dyadic}{symbol:fisherdyadic}
\addsymbol \mathcal{F}: {frame superoperator}{symbol:framesuperop}
\addsymbol f_j: {measured frequency of outcome $\Pi_j$}{symbol:freq}
\addsymbol \Gamma_j: {$D$-dimensional trace-orthonormal basis operators}{symbol:gammaj}
\addsymbol \gamma_{k+1}: {Polak-Ribi{\` e}re criterion evaluated in the $k$th step}{symbol:gammakp1}
\addsymbol G\,(G'): {sum of all the POM outcomes}{symbol:gop}
\addsymbol \mathfrak{G},\mathfrak{H}: {positive operators that parametrize a quadratic form}{symbol:ghop}
\addsymbol \hat{\,\,}: {symbol for an estimator}{symbol:hatsym}
\addsymbol H_n(x): {degree-$n$ Hermite polynomial in $x$}{symbol:hermite}
\addsymbol \mathcal{H},\mathcal{H'}: {Hilbert space of the input states (QPT)}{symbol:hhilbert}
\addsymbol \vec{h}_k\,(\vec{H}_k): {conjugate direction vector operators for the $k$th step}{symbol:hk}
\addsymbol \mathfrak{h}_k(x): {impulse response function}{symbol:hkx}
\addsymbol \textsc{h},\textsc{v}: {horizontal and vertical polarizations of photons}{symbol:hpolvpol}
\addsymbol \mathcal{I}: {identity superoperator}{symbol:identitysuperop}
\addsymbol \mathcal{I}(\lambda;\rho): {information functional of $\rho$}{symbol:infofunctional}
\addsymbol \mathcal{I}(\lambda;E): {information functional of $E$}{symbol:infofunctionale}
\addsymbol I_k(x_j): {intensity of a beam, at the $j$th pixel, on the focal plane of the $k$th microlens aperture}{symbol:ik}
\addsymbol I^\text{prop}_k(x): {intensity of a beam, at position $x$, after propagating from the $k$th microlens aperture}{symbol:ipropk}
\addsymbol J_\mu(y): {Bessel function in $y$ of order $\mu$}{symbol:jmuy}
\addsymbol \ket{\,\,\,},\bra{\,\,\,}: {kets and bras respectively}{symbol:ket}
\addsymbol \mathcal{K}: {Hilbert space of the output states (QPT)}{symbol:khilbert}
\addsymbol K_m: {Kraus operators}{symbol:kraus}
\addsymbol L: {total number of input states (QPT)}{symbol:l}
\addsymbol \lambda_j,\lambda,\mu: {Lagrange multipliers}{symbol:lagmult}
\addsymbol \Lambda: {Lagrange operator}{symbol:lagop}
\addsymbol L^\nu_n(y): {degree-$n$ associated Laguerre polynomials in $y$ of order $\nu$}{symbol:laguerre}
\addsymbol \text{LG}_l: {Laguerre-Gaussian mode of order $l$}{symbol:lgl}
\addsymbol \mathcal{L}(\mathfrak{D};\rho),\,\mathcal{L}(\{n_j\};\rho): {likelihood functional of $\rho$ (perfect measurements)}{symbol:likelihood}
\addsymbol \mathcal{L}(\{n_{lm}\};E): {likelihood functional of $E$ (perfect measurements)}{symbol:likelihoode}
\addsymbol \mathcal{L}'(\{n_{lm}\};E): {likelihood functional of $E$ (imperfect measurements)}{symbol:likelihoodeimp}
\addsymbol \log\tilde{\mathcal{L}}(\{\nu_{lm}\};E,\rho): {projected log-likelihood functional (QPT)}{symbol:likelihoodeproj}
\addsymbol \mathcal{L}'(\{n_j\};\rho): {likelihood functional of $\rho$ (imperfect measurements)}{symbol:likelihoodimp}
\addsymbol \mathcal{L}_\text{H}(\{n_j\};\rho): {hedged likelihood functional of $\rho$}{symbol:likelihoodhedged}
\addsymbol L_z: {orbital angular momentum operator in the $z$ direction}{symbol:lz}
\addsymbol M: {total number of POM outcomes (QPT)}{symbol:m}
\addsymbol \mathcal{M}: {completely-positive map}{symbol:mapcp}
\addsymbol M_\eta: {efficiency matrix}{symbol:meta}
\addsymbol M_\text{obs}: {observable matrix}{symbol:matrixobs}
\addsymbol \mathfrak{M}: {Gram matrix}{symbol:mgram}
\addsymbol \hat E_\text{MPL},\hat\rho_\text{MPL}: {MPL process and state estimators (QPT)}{symbol:mplestimators}
\addsymbol \ket{n}: {Fock states}{symbol:nstate}
\addsymbol n_j: {number of occurrences of outcome $\Pi_j$}{symbol:nj}
\addsymbol n_{lm}: {number of occurrences of outcome $\Pi_m$ with $\rho^{(l)}_\text{i}$ (QPT)}{symbol:nlm}
\addsymbol n'_l: {defined as $\sum_{m}n_{lm}=N$}{symbol:nlmprime}
\addsymbol \tilde{n}_l: {true number of copies of $\rho^{(l)}_\text{i}$ (QPT)}{symbol:ntildel}
\addsymbol N_\text{ports}: {number of output ports in TMD detection}{symbol:nports}
\addsymbol N: {measured total number of copies of quantum systems}{symbol:numcopies}
\addsymbol N_\text{true}: {true total number of copies}{symbol:numcopiestrue}
\addsymbol n_{>0}: {number of positive eigenvalues of $\mathfrak{M}$}{symbol:numlindep}
\addsymbol \mathcal{P}: {parity operator}{symbol:parity}
\addsymbol P(\alpha): {Glauber-Sudarshan $P$ function of $\alpha$}{symbol:pglauber}
\addsymbol P: {momentum quadrature}{symbol:pmom}
\addsymbol \vec{\partial}: {two-component gradient operator}{symbol:partialvec}
\addsymbol \Pi_j: {outcomes of a probability operator measurement (POM)}{symbol:pij}
\addsymbol \Pi'_j: {outcomes of an imperfect POM}{symbol:pijprime}
\addsymbol \Pi^\text{wit}_j: {outcomes of a witness basis}{symbol:pijwit}
\addsymbol \Pi_k(x_j): {outcomes describing the SH detections}{symbol:pikxj}
\addsymbol p_{lm}: {probability of an outcome $\Pi_m$ with $\rho^{(l)}_\text{i}$ (QPT)}{symbol:plm}
\addsymbol p'_l: {defined as $\sum_mp_{lm}=1/L$}{symbol:plmprime}
\addsymbol p_j: {probability of an outcome $\Pi_j$}{symbol:prob}
\addsymbol p(\mathfrak{D}\cap\rho): {probability of having $\mathfrak{D}$ and $\rho$ simultaneously}{symbol:proband}
\addsymbol p(\mathfrak{D}|\rho): {conditional probability of obtaining $\mathfrak{D}$ given $\rho$}{symbol:probcond}
\addsymbol \hat p_j: {estimated probabilities}{symbol:probest}
\addsymbol p^\text{true}_j: {true probabilities}{symbol:probtrue}
\addsymbol \ket{\,\,\,}_\text{prod}: {product ket}{symbol:prodket}
\addsymbol \pi(\rho): {prior probability distribution of $\rho$}{symbol:priorprob}
\addsymbol \tilde p_{lm},\tilde\nu_m,\tilde p_m: {projected quantites (QPT)}{symbol:projquantities}
\addsymbol \psi(x): {complex amplitude of a light beam, at position $x$, from the source}{symbol:psix}
\addsymbol \psi^{(k)}_\text{prop}(x): {complex amplitude of a propagated light beam, at position $x$, from the $k$th aperture}{symbol:psipropx}
\addsymbol \psi'_k(x): {complex amplitude of a light beam, at position $x$ on the focal plane of the $k$th aperture}{symbol:psixprime}
\addsymbol \mathcal{Q}(\alpha): {Husimi $\mathcal{Q}$ function for a given $\alpha$}{symbol:qfunc}
\addsymbol \mathcal{R}(x,p,\tau),\mathcal{R}(\alpha,\tau): {function of $\alpha$ and $\tau$, with $0\leq\tau\leq 1$, for computing $\tilde\tau$}{symbol:rfunc}
\addsymbol \rho: {state or statistical operator}{symbol:rho}
\addsymbol \rho_{2\text{ph}}: {two-photon state}{symbol:rho2ph}
\addsymbol \rho_\text{coh}: {coherence operator}{symbol:rhocoh}
\addsymbol \hat\rho^\text{MLME}_\text{coh}: {MLME estimator for $\rho^\text{true}_\text{coh}$}{symbol:rhocohmlme}
\addsymbol \rho^\text{true}_\text{coh}: {true coherence operator describing a light beam}{symbol:rhocohtrue}
\addsymbol \rho_\text{i},\ID: {input state and its dimension (QPT)}{symbol:rhoin}
\addsymbol \rho^{(l)}_\text{i}: {$l$th input state (QPT)}{symbol:rhoinl}
\addsymbol \rho_\text{o},\OD: {output state and its dimension (QPT)}{symbol:rhoout}
\addsymbol \rho^{(l)}_\text{o}: {$l$th output state (QPT)}{symbol:rhooutl}
\addsymbol \hat\rho_\text{B},\hat\rho_\text{B}(\mathfrak{H}): {Bayesian estimator, with and without bias $\mathfrak{H}$}{symbol:rhobayes}
\addsymbol \rho_\text{ent}: {entangled state}{symbol:rhoent}
\addsymbol \hat\rho: {state estimator}{symbol:rhoest}
\addsymbol \hat\rho_{\text{I},\lambda}: {state estimator that maximizes $\mathcal{I}(\lambda;\rho)$ for fixed $\lambda$}{symbol:rhoestinfo}
\addsymbol \hat{\rho}_\text{HML}: {HML state estimator}{symbol:rhohml}
\addsymbol \hat\rho_\text{ME}: {ME state estimator}{symbol:rhome}
\addsymbol \hat\rho_\text{ML}: {ML state estimator}{symbol:rhoml}
\addsymbol \hat\rho_\text{MLME}: {MLME state estimator}{symbol:rhomlme}
\addsymbol \rho_\text{sep}: {separable state}{symbol:rhosep}
\addsymbol \rho_\text{ss}: {stationary state of a laser}{symbol:rhoss}
\addsymbol \rho_\text{true}: {true state obtained from measuring infinite copies}{symbol:rhotrue}
\addsymbol R: {operator defined as $\sum_{j}f_j\Pi_j/p_j$}{symbol:rmatrix}
\addsymbol r^\text{true}_j: {coefficients of $\rho_\text{true}$ expressed in terms of $\Gamma_j$s}{symbol:rtrue}
\addsymbol \vec{r}_\text{true}: {column of $r^\text{true}_j$}{symbol:rtruevec}
\addsymbol S(E): {von Neumann entropy of $E$}{symbol:se}
\addsymbol \hat\sigma: {linear-inversion estimator}{symbol:sigmahat}
\addsymbol \sket{\Psi},\sbra{\Psi}: {superkets and superbras repectively for the operator $\Psi$}{symbol:sket}
\addsymbol S(\{f_j\}|\{p_j\}): {relative entropy}{symbol:srel}
\addsymbol S(\rho)\,(S_\text{max}): {von Neumann entropy of $\rho$ (maximum value)}{symbol:srho}
\addsymbol \tilde\tau: {non-classicality depth}{symbol:taudepth}
\addsymbol \Theta_j: {dual operator of the outcome $\Pi_j$}{symbol:thetaj}
\addsymbol \textsc{t}_l: {partial transpose on the $l$th subsystem}{symbol:tl}
\addsymbol T_j: {transmission probabilities}{symbol:transmission}
\addsymbol \vec{t}: {column of coefficients for $\hat\rho$ expressed in terms of $\Gamma_j$s}{symbol:tvec}
\addsymbol \vec{t}_\text{ML}: {column of coefficients for $\hat\rho_\text{ML}$ expressed in terms of $\Gamma_j$s}{symbol:tmlvec}
\addsymbol \mathfrak{U}_k: {unitary response operator for the $k$th microlens}{symbol:uk}
\addsymbol U_1^\text{wp},U_2^\text{wp}: {unitary transformations effected by wave plates}{symbol:unitarywp}
\addsymbol \mathcal{W}_{00}: {Wigner functional at the phase space origin}{symbol:w00}
\addsymbol \mathcal{W}(x,p),\mathcal{W}(\alpha): {Wigner function in phase space}{symbol:wigner}
\addsymbol \mathfrak{W}: {entanglement witness}{symbol:witness}
\addsymbol \xi: {parameter that modifies $\gamma_{k+1}$}{symbol:xi}
\addsymbol X: {position quadrature}{symbol:xpos}
\addsymbol \ket{x_\vartheta}: {eigenkets of $X\cos\vartheta+P\sin\vartheta$}{symbol:xthetastate}
\addsymbol \vec{z}_k\,\left(\vec{Z},\vec{Z}_k\right): {direction vector (operator version)}{symbol:zvec}
\end{tabbing}  \clearpage}
\def\addsymbol #1: #2#3{$\displaystyle{#1}$ \> \parbox{3.9in}{#2 \dotfill \pageref{#3}}\\\\}
\def\addsymbolWO #1: #2{$#1$ \> \parbox{3.9in}{#2 \dotfill *}\\}

\makeatletter
\def\idxquad{\hskip 10\p@}
\makeatother

\begin{document}
\thispagestyle{empty}
\begin{titlepage}
\begin{center}
\noindent {\large \textbf{NATIONAL UNIVERSITY OF SINGAPORE}} \\
\vspace*{0.6cm}
\vspace*{0.5cm}
\noindent \large \textbf{DOCTORAL\ \ THESIS} \\
\vspace*{0.3cm}
\noindent {submitted in partial fulfilment of the requirements for the degree of} \\
\vspace*{0.3cm}
\noindent \large \textbf{Doctor of Philosophy in Science} \\
\vspace*{0.4cm}
\noindent \large Teo Yong Siah \\
\vspace*{0.8cm}
\HRule \\[0.4cm]
\noindent {\large \textbf{Numerical Estimation Schemes for Quantum Tomography}} \\
\HRule
\vspace*{0.8cm}
\noindent \large Thesis Advisor: Berthold-Georg \textsc{ENGLERT} \\
\vspace*{0.5cm}
\vfill
\noindent {\textbf{Department of Physics/}} \\
\noindent \textbf{Centre for Quantum Technologies/} \\
\noindent \textbf{NUS Graduate School for Integrative Sciences and Engineering} \\
\vspace*{2.0cm}
{\large 2012/2013}
\end{center}
\end{titlepage}
\pagecolor{white}
\thispagestyle{empty}
\begin{titlepage}
\singlespacing
\begin{center}
\begin{flushleft}
\noindent {\Huge {\textbf{\sf Numerical Estimation Schemes for}}}
\end{flushleft}
\begin{flushleft}
\noindent {\Huge {\textbf{\sf Quantum Tomography}}}
\end{flushleft}
\HRulecolor\\
\vspace*{0.5cm}
\begin{flushleft}
\Large \emph{\sf A survey of novel numerical techniques for quantum estimation}
\end{flushleft}
\doublespacing
\vspace*{3cm}
\hfill\Large \sf Teo Yong Siah
\vspace*{2cm}
\includegraphics[width=0.9\textwidth]{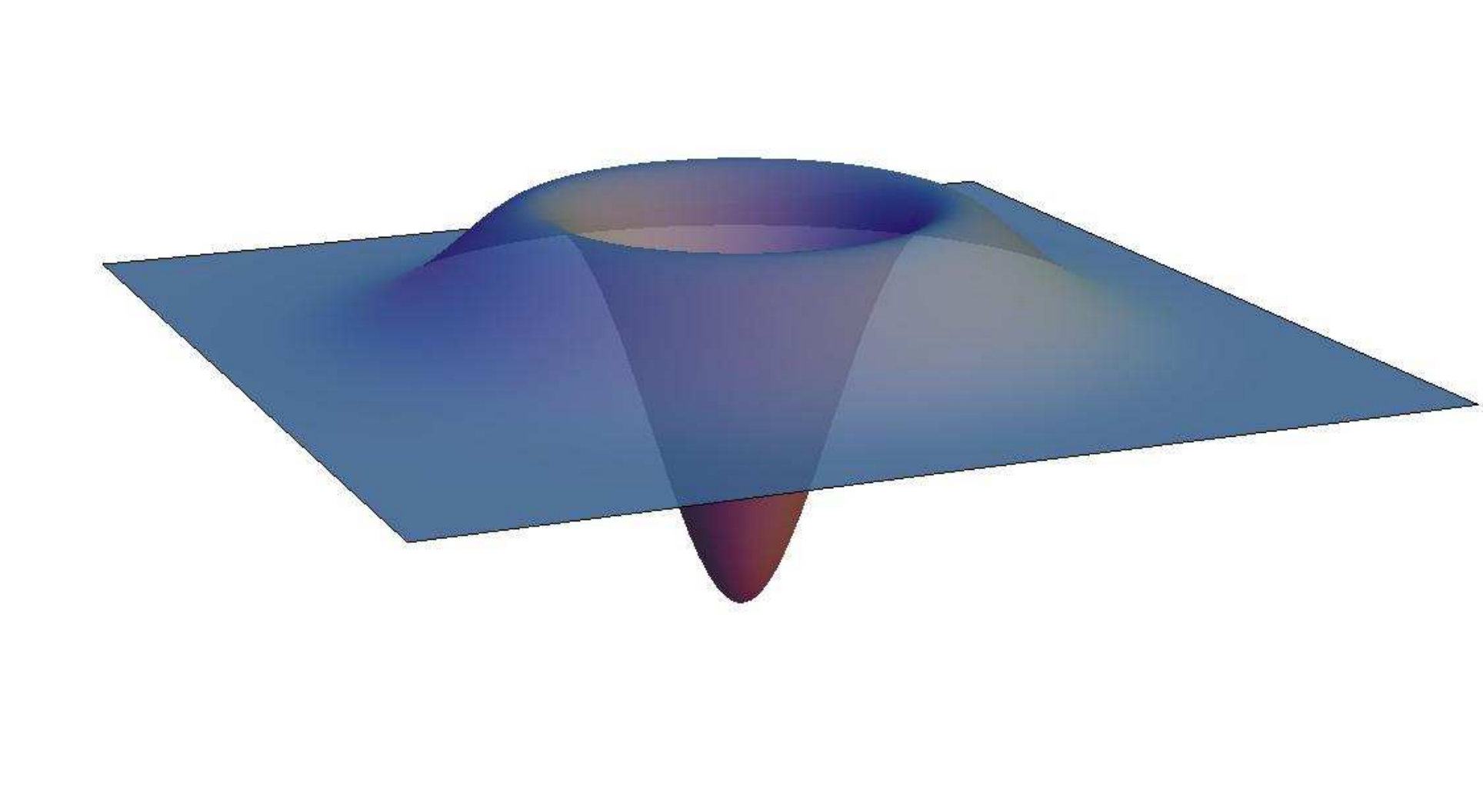}\\
\vfill
\HRulecolor\\
\noindent {\large \textbf{\sf National University of Singapore}\hfill \large 2012} \\
\end{center}
\end{titlepage}
\thispagestyle{empty}
\sloppy

\titlepage

\thispagestyle{plain}
\pagenumbering{roman}
\cleardoublepage
\phantomsection 
\addcontentsline{toc}{chapter}{Acknowledgements}
\chapter*{Acknowledgements}
The author would like to express his gratitude to his Ph.D. thesis supervisor Prof. Berthold-Georg~Englert, a Principal Investigator at the Centre for Quantum Technologies, National University of Singapore, for his patient guidance. The most part of this dissertation involves work that was done in collaboration with Prof.~Jaroslav~{\v R}eh{\'a}{\v c}ek and Prof.~Zden{\v e}k~Hradil from the Department of Optics at Palack{\'y} University in Olomouc, Czech Republic. The author would also like to thank Zhu~Huangjun, Thiang~Guo~Chuan and Ng~Hui~Khoon for the many insightful discussions. Finally, the author thanks the NUS Graduate School for Integrative Sciences and Engineering and the Centre for Quantum Technologies for their support.

\vfill
Y. S. Teo  

\cleardoublepage
\tableofcontents

\cleardoublepage
\phantomsection
\addcontentsline{toc}{chapter}{Summary}
\chapter*{Summary}
\markboth{\hfill Summary}{Summary \hfill}
One statistically meaningful technique to estimate the unknown quantum state based on a set of informationally complete measurement data is the maximum-likelihood method (ML). This technique yields a unique ML estimator for a given complete set of data. An iterative algorithm was proposed by Jaroslav {\v R}eh{\'a}{\v c}ek \emph{et al.} to search for a positive estimator that maximizes the likelihood functional. We first show that this algorithm coincides with the steepest-ascent technique and develop a new algorithm based on the conjugate-gradient method that can be more efficient than the steepest-ascent version. We inspect the performance of this new algorithm with Monte Carlo numerical simulations.\\\\
In general, however, the measurement data obtained from complex quantum systems are informationally incomplete and, as a rule, do not yield a unique state estimator. We establish an estimation scheme where both the likelihood and the von Neumann entropy functionals are maximized in order to systematically select the most-likely estimator with the largest entropy, that is, the least-bias maximum-likelihood and maximum-entropy estimator (MLME), consistent with a given set of measurement data. This is equivalent to the joint consideration of our partial knowledge and of our ignorance about the source to reconstruct its identity. The MLME technique is then applied to both experimental and simulation data.\\\\
Next, we take a look at a recent proposal by R. Blume-Kohout ---~the hedged maximum-likelihood method~--- for quantum state estimation and derive an iterative scheme (HML) to look for the estimator that maximizes the hedged likelihood functional. We then report some interesting features of these HML estimators in the context of informationally incomplete measurements and compare them with the MLME estimators using numerical simulations.\\\\
Entanglement detection via witness measurements is a useful technique to check if an unknown quantum state is an entangled one. The MLME algorithm can also be used to increase the efficiency of entanglement detection, using the data obtained from measuring sets of witness bases. This is better than the conventional witness measurement strategy in which only the expectation value of each witness is estimated and used to infer the existence of entanglement in the unknown quantum state. In our proposed strategies, all information from the collected data is used to detect entanglement and when this fails, state estimation can be performed to estimate the unknown state. Adaptive strategies to measure these witness bases will also be presented.\\\\
Finally, we also propose a similar algorithm, as in quantum state estimation, for incomplete quantum process estimation based on the combined principles of maximum-likelihood and maximum-entropy, to yield a unique estimator for an unknown quantum process when one has a set of informationally incomplete data. We apply this iterative algorithm adaptively to various situations in order to minimize the amount of measurement resources required to estimate the unknown quantum process with incomplete data. 

\singlespacing
\cleardoublepage
\phantomsection
\addcontentsline{toc}{chapter}{List of Tables}
\listoftables

\cleardoublepage
\phantomsection
\addcontentsline{toc}{chapter}{List of Figures}
\listoffigures

\singlespacing
\cleardoublepage
\phantomsection
\addcontentsline{toc}{chapter}{List of Symbols}
\chapter*{List of Symbols}
\markboth{\hfill List of Symbols}{List of Symbols \hfill}
\begin{tabbing}
$\overleftrightarrow{1}$~~~~~~~~~~~~~~~~~~~~~~\=\parbox{3.9in}{unit dyadic \dotfill \pageref{symbol:1dyadic}}\\\\
\addsymbol 1_\mathcal{H}: {identity operator in $\mathcal{H}$}{symbol:identityh}
\addsymbol 1_\mathcal{K}: {identity operator in $\mathcal{K}$}{symbol:identityk}
\addsymbol \lVert O\rVert_2: {2-norm of an operator $O$}{symbol:2norm}
\addsymbol \mathfrak{A}_k: {aperture operator}{symbol:ak}
\addsymbol \alpha: {a complex number given by $x+\I p$ for real $x$ and $p$}{symbol:alpha}
\addsymbol A: {annihilation operator}{symbol:ann}
\addsymbol \mathcal{A}: {auxiliary complex operator for parameterizing $\rho$ and $E$}{symbol:auxa}
\addsymbol a_k(x): {aperture function}{symbol:akx}
\addsymbol \overline{\quad\vphantom{M}}: {equivalent to $\sum_\mathfrak{D}$}{symbol:avgsym}
\addsymbol C: {qubit Clifford unitary operator}{symbol:cliffordqubit}
\addsymbol \mathfrak{C}(\rho_\text{true},\hat\rho): {cost functional of $\hat\rho$ for $\rho_\text{true}$}{symbol:costfunc}
\addsymbol \overline{\mathfrak{C}}(\hat\rho): {average cost functional of $\hat\rho$}{symbol:costfuncavg}
\addsymbol \mathcal{C}_\text{H-S}(\rho_\text{true}-\hat\rho): {covariance between $\rho_\text{true}$ and $\hat\rho$}{symbol:covariance}
\addsymbol \overleftrightarrow{\mathcal{C}}\left(\vec{t},\vec{r}\right): {covariance dyadic between $\vec{t}$ and $\vec{r}$}{symbol:covariancedyadic}
\addsymbol \overleftrightarrow{\mathfrak{C}}_\text{ML}: {ML covariance dyadic evaluated at $\vec{t}=\vec{t}_\text{ML}$}{symbol:covariancedyadicml}
\addsymbol \mathcal{C}(E,\rho): {cross entropy}{symbol:crossent}
\addsymbol \mathfrak{D}: {measurement data}{symbol:data}
\addsymbol \sum_\mathfrak{D}: {average over all possible $\mathfrak{D}$}{symbol:dataavg}
\addsymbol \frac{\partial}{\partial\vec{r}}: {gradient operator with respect to $\vec{r}$}{symbol:deedeer}
\addsymbol \Delta: {operator variance of a convex set of $\hat E_\text{ML}$s (QPT)}{symbol:delta}
\addsymbol \mathcal{D}: {Lagrange functional}{symbol:dfunctional}
\addsymbol D: {dimension of the Hilbert space}{symbol:dim}
\addsymbol \mathcal{D}(\alpha): {displacement operator for a given $\alpha$}{symbol:dispop}
\addsymbol \mathcal{D}_\text{tr}: {trace-class distance}{symbol:distancetraceclass}
\addsymbol (\D\rho): {prior}{symbol:drho}
\addsymbol D_\text{sub}: {dimension of the truncated Hilbert space}{symbol:dsub}
\addsymbol (\D\tau_\textsc{d}): {integration measure for the $D$-dimensional Hilbert space}{symbol:dtau}
\addsymbol \overleftrightarrow{\vphantom{M}}: {symbol for dyadic}{symbol:dyadicsym}
\addsymbol E: {Choi-Jami{\'o}\l kowski operator for a quantum process}{symbol:e}
\addsymbol \hat E_\text{ML}: {ML process estimator}{symbol:eml}
\addsymbol \overline{E}_\text{ML}: {operator centroid of a convex set of $\hat E_\text{ML}$s}{symbol:emlbar}
\addsymbol \hat E_\text{MLME}: {MLME process estimator}{symbol:emlme}
\addsymbol E_\text{prior}: {prior information about $E_\text{true}$}{symbol:eprior}
\addsymbol \epsilon,\epsilon_k: {small positive step size in an iterative algorithm}{symbol:epsilon}
\addsymbol \varepsilon: {precision for terminating an iterative algorithm}{symbol:epsilonprec}
\addsymbol \eta: {overall detection efficiency}{symbol:eta}
\addsymbol E_\text{true}: {Choi-Jami{\'o}\l kowski operator for the true process}{symbol:etrue}
\addsymbol F: {Fisher's information dyadic}{symbol:fisherdyadic}
\addsymbol \mathcal{F}: {frame superoperator}{symbol:framesuperop}
\addsymbol f_j: {measured frequency of outcome $\Pi_j$}{symbol:freq}
\addsymbol \Gamma_j: {$D$-dimensional trace-orthonormal basis operators}{symbol:gammaj}
\addsymbol \gamma_{k+1}: {Polak-Ribi{\` e}re criterion evaluated in the $k$th step}{symbol:gammakp1}
\addsymbol G\,(G'): {sum of all the POM outcomes}{symbol:gop}
\addsymbol \mathfrak{G},\mathfrak{H}: {positive operators that parametrize a quadratic form}{symbol:ghop}
\addsymbol \hat{\,\,}: {symbol for an estimator}{symbol:hatsym}
\addsymbol H_n(x): {degree-$n$ Hermite polynomial in $x$}{symbol:hermite}
\addsymbol \mathcal{H},\mathcal{H'}: {Hilbert space of the input states (QPT)}{symbol:hhilbert}
\addsymbol \vec{h}_k\,(\vec{H}_k): {conjugate direction vector operators for the $k$th step}{symbol:hk}
\addsymbol \mathfrak{h}_k(x): {impulse response function}{symbol:hkx}
\addsymbol \textsc{h},\textsc{v}: {horizontal and vertical polarizations of photons}{symbol:hpolvpol}
\addsymbol \mathcal{I}: {identity superoperator}{symbol:identitysuperop}
\addsymbol \mathcal{I}(\lambda;\rho): {information functional of $\rho$}{symbol:infofunctional}
\addsymbol \mathcal{I}(\lambda;E): {information functional of $E$}{symbol:infofunctionale}
\addsymbol I_k(x_j): {intensity of a beam, at the $j$th pixel, on the focal plane of the $k$th microlens aperture}{symbol:ik}
\addsymbol I^\text{prop}_k(x): {intensity of a beam, at position $x$, after propagating from the $k$th microlens aperture}{symbol:ipropk}
\addsymbol J_\mu(y): {Bessel function in $y$ of order $\mu$}{symbol:jmuy}
\addsymbol \ket{\,\,\,},\bra{\,\,\,}: {kets and bras respectively}{symbol:ket}
\addsymbol \mathcal{K}: {Hilbert space of the output states (QPT)}{symbol:khilbert}
\addsymbol K_m: {Kraus operators}{symbol:kraus}
\addsymbol L: {total number of input states (QPT)}{symbol:l}
\addsymbol \lambda_j,\lambda,\mu: {Lagrange multipliers}{symbol:lagmult}
\addsymbol \Lambda: {Lagrange operator}{symbol:lagop}
\addsymbol L^\nu_n(y): {degree-$n$ associated Laguerre polynomials in $y$ of order $\nu$}{symbol:laguerre}
\addsymbol \text{LG}_l: {Laguerre-Gaussian mode of order $l$}{symbol:lgl}
\addsymbol \mathcal{L}(\mathfrak{D};\rho),\,\mathcal{L}(\{n_j\};\rho): {likelihood functional of $\rho$ (perfect measurements)}{symbol:likelihood}
\addsymbol \mathcal{L}(\{n_{lm}\};E): {likelihood functional of $E$ (perfect measurements)}{symbol:likelihoode}
\addsymbol \mathcal{L}'(\{n_{lm}\};E): {likelihood functional of $E$ (imperfect measurements)}{symbol:likelihoodeimp}
\addsymbol \log\tilde{\mathcal{L}}(\{\nu_{lm}\};E,\rho): {projected log-likelihood functional (QPT)}{symbol:likelihoodeproj}
\addsymbol \mathcal{L}'(\{n_j\};\rho): {likelihood functional of $\rho$ (imperfect measurements)}{symbol:likelihoodimp}
\addsymbol \mathcal{L}_\text{H}(\{n_j\};\rho): {hedged likelihood functional of $\rho$}{symbol:likelihoodhedged}
\addsymbol L_z: {orbital angular momentum operator in the $z$ direction}{symbol:lz}
\addsymbol M: {total number of POM outcomes (QPT)}{symbol:m}
\addsymbol \mathcal{M}: {completely-positive map}{symbol:mapcp}
\addsymbol M_\eta: {efficiency matrix}{symbol:meta}
\addsymbol M_\text{obs}: {observable matrix}{symbol:matrixobs}
\addsymbol \mathfrak{M}: {Gram matrix}{symbol:mgram}
\addsymbol \hat E_\text{MPL},\hat\rho_\text{MPL}: {MPL process and state estimators (QPT)}{symbol:mplestimators}
\addsymbol \ket{n}: {Fock states}{symbol:nstate}
\addsymbol n_j: {number of occurrences of outcome $\Pi_j$}{symbol:nj}
\addsymbol n_{lm}: {number of occurrences of outcome $\Pi_m$ with $\rho^{(l)}_\text{i}$ (QPT)}{symbol:nlm}
\addsymbol n'_l: {defined as $\sum_{m}n_{lm}=N$}{symbol:nlmprime}
\addsymbol \tilde{n}_l: {true number of copies of $\rho^{(l)}_\text{i}$ (QPT)}{symbol:ntildel}
\addsymbol N_\text{ports}: {number of output ports in TMD detection}{symbol:nports}
\addsymbol N: {measured total number of copies of quantum systems}{symbol:numcopies}
\addsymbol N_\text{true}: {true total number of copies}{symbol:numcopiestrue}
\addsymbol n_{>0}: {number of positive eigenvalues of $\mathfrak{M}$}{symbol:numlindep}
\addsymbol \mathcal{P}: {parity operator}{symbol:parity}
\addsymbol P(\alpha): {Glauber-Sudarshan $P$ function of $\alpha$}{symbol:pglauber}
\addsymbol P: {momentum quadrature}{symbol:pmom}
\addsymbol \vec{\partial}: {two-component gradient operator}{symbol:partialvec}
\addsymbol \Pi_j: {outcomes of a probability operator measurement (POM)}{symbol:pij}
\addsymbol \Pi'_j: {outcomes of an imperfect POM}{symbol:pijprime}
\addsymbol \Pi^\text{wit}_j: {outcomes of a witness basis}{symbol:pijwit}
\addsymbol \Pi_k(x_j): {outcomes describing the SH detections}{symbol:pikxj}
\addsymbol p_{lm}: {probability of an outcome $\Pi_m$ with $\rho^{(l)}_\text{i}$ (QPT)}{symbol:plm}
\addsymbol p'_l: {defined as $\sum_mp_{lm}=1/L$}{symbol:plmprime}
\addsymbol p_j: {probability of an outcome $\Pi_j$}{symbol:prob}
\addsymbol p(\mathfrak{D}\cap\rho): {probability of having $\mathfrak{D}$ and $\rho$ simultaneously}{symbol:proband}
\addsymbol p(\mathfrak{D}|\rho): {conditional probability of obtaining $\mathfrak{D}$ given $\rho$}{symbol:probcond}
\addsymbol \hat p_j: {estimated probabilities}{symbol:probest}
\addsymbol p^\text{true}_j: {true probabilities}{symbol:probtrue}
\addsymbol \ket{\,\,\,}_\text{prod}: {product ket}{symbol:prodket}
\addsymbol \pi(\rho): {prior probability distribution of $\rho$}{symbol:priorprob}
\addsymbol \tilde p_{lm},\tilde\nu_m,\tilde p_m: {projected quantites (QPT)}{symbol:projquantities}
\addsymbol \psi(x): {complex amplitude of a light beam, at position $x$, from the source}{symbol:psix}
\addsymbol \psi^{(k)}_\text{prop}(x): {complex amplitude of a propagated light beam, at position $x$, from the $k$th aperture}{symbol:psipropx}
\addsymbol \psi'_k(x): {complex amplitude of a light beam, at position $x$ on the focal plane of the $k$th aperture}{symbol:psixprime}
\addsymbol \mathcal{Q}(\alpha): {Husimi $\mathcal{Q}$ function for a given $\alpha$}{symbol:qfunc}
\addsymbol \mathcal{R}(x,p,\tau),\mathcal{R}(\alpha,\tau): {function of $\alpha$ and $\tau$, with $0\leq\tau\leq 1$, for computing $\tilde\tau$}{symbol:rfunc}
\addsymbol \rho: {state or statistical operator}{symbol:rho}
\addsymbol \rho_{2\text{ph}}: {two-photon state}{symbol:rho2ph}
\addsymbol \rho_\text{coh}: {coherence operator}{symbol:rhocoh}
\addsymbol \hat\rho^\text{MLME}_\text{coh}: {MLME estimator for $\rho^\text{true}_\text{coh}$}{symbol:rhocohmlme}
\addsymbol \rho^\text{true}_\text{coh}: {true coherence operator describing a light beam}{symbol:rhocohtrue}
\addsymbol \rho_\text{i},\ID: {input state and its dimension (QPT)}{symbol:rhoin}
\addsymbol \rho^{(l)}_\text{i}: {$l$th input state (QPT)}{symbol:rhoinl}
\addsymbol \rho_\text{o},\OD: {output state and its dimension (QPT)}{symbol:rhoout}
\addsymbol \rho^{(l)}_\text{o}: {$l$th output state (QPT)}{symbol:rhooutl}
\addsymbol \hat\rho_\text{B},\hat\rho_\text{B}(\mathfrak{H}): {Bayesian estimator, with and without bias $\mathfrak{H}$}{symbol:rhobayes}
\addsymbol \rho_\text{ent}: {entangled state}{symbol:rhoent}
\addsymbol \hat\rho: {state estimator}{symbol:rhoest}
\addsymbol \hat\rho_{\text{I},\lambda}: {state estimator that maximizes $\mathcal{I}(\lambda;\rho)$ for fixed $\lambda$}{symbol:rhoestinfo}
\addsymbol \hat{\rho}_\text{HML}: {HML state estimator}{symbol:rhohml}
\addsymbol \hat\rho_\text{ME}: {ME state estimator}{symbol:rhome}
\addsymbol \hat\rho_\text{ML}: {ML state estimator}{symbol:rhoml}
\addsymbol \hat\rho_\text{MLME}: {MLME state estimator}{symbol:rhomlme}
\addsymbol \rho_\text{sep}: {separable state}{symbol:rhosep}
\addsymbol \rho_\text{ss}: {stationary state of a laser}{symbol:rhoss}
\addsymbol \rho_\text{true}: {true state obtained from measuring infinite copies}{symbol:rhotrue}
\addsymbol R: {operator defined as $\sum_{j}f_j\Pi_j/p_j$}{symbol:rmatrix}
\addsymbol r^\text{true}_j: {coefficients of $\rho_\text{true}$ expressed in terms of $\Gamma_j$s}{symbol:rtrue}
\addsymbol \vec{r}_\text{true}: {column of $r^\text{true}_j$}{symbol:rtruevec}
\addsymbol S(E): {von Neumann entropy of $E$}{symbol:se}
\addsymbol \hat\sigma: {linear-inversion estimator}{symbol:sigmahat}
\addsymbol \sket{\Psi},\sbra{\Psi}: {superkets and superbras repectively for the operator $\Psi$}{symbol:sket}
\addsymbol S(\{f_j\}|\{p_j\}): {relative entropy}{symbol:srel}
\addsymbol S(\rho)\,(S_\text{max}): {von Neumann entropy of $\rho$ (maximum value)}{symbol:srho}
\addsymbol \tilde\tau: {non-classicality depth}{symbol:taudepth}
\addsymbol \Theta_j: {dual operator of the outcome $\Pi_j$}{symbol:thetaj}
\addsymbol \textsc{t}_l: {partial transpose on the $l$th subsystem}{symbol:tl}
\addsymbol T_j: {transmission probabilities}{symbol:transmission}
\addsymbol \vec{t}: {column of coefficients for $\hat\rho$ expressed in terms of $\Gamma_j$s}{symbol:tvec}
\addsymbol \vec{t}_\text{ML}: {column of coefficients for $\hat\rho_\text{ML}$ expressed in terms of $\Gamma_j$s}{symbol:tmlvec}
\addsymbol \mathfrak{U}_k: {unitary response operator for the $k$th microlens}{symbol:uk}
\addsymbol U_1^\text{wp},U_2^\text{wp}: {unitary transformations effected by wave plates}{symbol:unitarywp}
\addsymbol \mathcal{W}_{00}: {Wigner functional at the phase space origin}{symbol:w00}
\addsymbol \mathcal{W}(x,p),\mathcal{W}(\alpha): {Wigner function in phase space}{symbol:wigner}
\addsymbol \mathfrak{W}: {entanglement witness}{symbol:witness}
\addsymbol \xi: {parameter that modifies $\gamma_{k+1}$}{symbol:xi}
\addsymbol X: {position quadrature}{symbol:xpos}
\addsymbol \ket{x_\vartheta}: {eigenkets of $X\cos\vartheta+P\sin\vartheta$}{symbol:xthetastate}
\addsymbol \vec{z}_k\,\left(\vec{Z},\vec{Z}_k\right): {direction vector (operator version)}{symbol:zvec}
\end{tabbing}  \clearpage
\cleardoublepage

\doublespacing
\pagenumbering{arabic}
\cleardoublepage
\chapter{Quantum State Estimation}\index{quantum state estimation (tomography)}
\label{chap:qse}

\section{Introduction}

Quantum state preparation is the first important step for any protocol that makes use of quantum resources. Examples of such protocols are quantum state teleportation and quantum key distribution which require entangled quantum states. In order to verify the integrity of the quantum state prepared by the source, one carries out \emph{quantum state tomography} on the source. Measurements are performed on a collection of identical copies of quantum systems (electrons, photons, etc.) that are emitted from the source. Then, the quantum state of the source is inferred from the measurement data obtained from this collection. The measurements are generically described by a set of positive operators $\Pi_j$\label{symbol:pij} that compose a \emph{probability operator measurement} (POM). After that, the measurement data obtained are used to infer the quantum state of the source. Such a procedure of state inference, which shall be our main focus in this dissertation, is also known as \emph{quantum state estimation}.

The central idea of quantum state estimation is to attribute a well-defined objective \emph{true} state to each measured quantum system that is emitted from the source, making a connection with the frequentist's definition of classical estimation. An observer, after measuring a finite number of copies, will obtain a state estimator that is generally different from that obtained by another observer, after measuring his own copies in a different way. This is not surprising since the quantum state of the source directly reflects the amount of information an observer gains after measuring his copies \cite{quantum_definetti}. As the number of copies approaches infinity, different estimation procedures ultimately lead to the same \emph{true} quantum state of the source if the measurements completely characterize the source. However, such an idealized situation is never achievable in any laboratory setting, as one can only perform measurements on \emph{finite} copies of quantum systems. As a result, the state estimator obtained will be different from the true state and depends on the details of the estimation procedure. To make statistical predictions, the corresponding operator $\hat\rho$\label{symbol:rhoest}\label{symbol:hatsym} describing this estimator must be a \emph{statistical operator}, which is positive. This will ensure that the estimated probability $\hat p_j=\tr{\hat\rho\Pi_j}$\label{symbol:probest} for an outcome $\Pi_j$ of \emph{any} set of POM is positive. We shall denote all estimated quantities with a ``hat'' symbol.

The frequentist's\index{frequentist} notion of quantum state estimation, described above, is fundamentally different from the Bayesian\index{Bayesian} point of view \cite{qstateest,quantum_definetti}, in which there is no objective true state of the source to be characterized. Rather, the quantum state of a given source is treated purely as knowledge that is to be updated by the measurement data obtained from finite copies, subjected to some prior information about the distribution of statistical operators. In the latter viewpoint, the quantum state of the source is naturally regarded as a subjective reality that is based on the measurements performed by an observer, rather than a definite state that is associated to the source. Unfortunately, due to its technical difficulty, a feasible Bayesian estimation scheme for quantum states is presently undeveloped.

There are two popular methods for the frequentist's\index{frequentist} version of quantum state estimation: \emph{Bayesian state estimation}\footnote{Not to be confused with the Bayesian view of quantum estimation as discussed previously.} and \emph{maximum-likelihood estimation} (ML). The Bayesian\index{Bayesian} state estimation method \cite{bayesian1,bayesian2,bayesian3} constructs a state estimator from an integral average over all possible quantum states to estimate the unknown true state. The \emph{likelihood functional}, which yields the likelihood of obtaining a particular sequence of measurement detections given a quantum state, serves as a weight for the average. This approach includes all the neighboring states near the maximum of the likelihood functional as possible guesses for the unknown $\rho_\text{true}$\label{symbol:rhotrue}. These neighboring states are given especially significant weight when $N$\label{symbol:numcopies}, the measured total number of copies, is small, in which case the likelihood functional is only broadly peaked at the maximum. However, the integral average unavoidably depends on how one measures volumes in the state space, and there is no universal and unambiguous method for that. The ML method \cite{ml1,ml2,qstateest,rehacek1}, on the other hand, simply chooses the estimator as the statistical operator that maximizes the likelihood functional. For a sufficiently large number of copies, both methods give the same estimator since the likelihood functional peaks very strongly at the maximum.

When the measurement outcomes form an \emph{informationally complete} set, the measurement data obtained will contain maximal information about the source. Thus, a unique\index{unique estimator} state estimator can be inferred with ML. Unfortunately, in tomography experiments performed on complex quantum systems with many degrees of freedom, it is not possible to implement such an informationally complete set of measurement outcomes. As a result, some information about the source will be missing and its quantum state cannot be completely characterized. For example, if a source produces a mode of light that is described by an infinite-dimensional statistical operator $\rho_\text{true}$, then no matter how ingeniously a measurement scheme is designed to probe incoming photons prepared by this source, an infinite amount of information about the mode of light will always remain unknown. The ML estimator obtained from these informationally incomplete data is no longer unique\index{unique estimator} and there will in general be infinitely many other ML estimators that are consistent with the data.

The standard approach to this problem is to apply an \textit{ad hoc} truncation\index{state-space truncation} on the Hilbert space and perform the state reconstruction in a particular subspace. This results in a smaller number of unknown parameters that can then be \emph{uniquely} determined by the measurement scheme. Since the truncation is largely based on the observer's intuition about the expected result, that is the true state that describes an infinite number of copies of such quantum systems, this cannot be a truly objective method \cite{quantum_diag}. A more objective alternative is to consider the largest possible reconstruction subspace that is compatible with any existing \emph{prior} knowledge\index{prior information (knowledge)} about the source. For example, if an observer has prior knowledge about the range of the energy spectrum a given light source can have, he should consider the largest possible reconstruction subspace that contains quantum states describing the source in this range of energies. This inevitably introduces more unknown parameters that cannot be uniquely determined by the measurements and one should select the state estimator in this subspace that is least biased.

In Refs.~\cite{mlme} and \cite{mlme_comprehensive}, we reported an iterative algorithm (MLME) to estimate unknown quantum states from incomplete measurement data by maximizing the likelihood and von Neumann entropy functionals. The application of this algorithm was illustrated with simulations and experimental data and we concluded that, together with a more objective Hilbert space truncation, this approach can serve as a reliable and statistically meaningful quantum state estimation with incomplete data.

In this first chapter, we will discuss, at great lengths, the principles of quantum state estimation and establish some novel algorithms using various numerical methods. 

\section{Preliminaries of quantum state estimation}
\label{sec:qseprelim}

\subsection{Estimation theory}

At the heart of estimation theory\index{estimation theory} lies the principles of \emph{functional optimization} \cite{ml2}. Typically, an objective functional involving the \emph{cost functional} $\mathfrak{C}(\rho_\text{true},\hat\rho)$\label{symbol:costfunc}\index{cost functional} of an estimator $\hat\rho$ for the unknown quantum state $\rho_\text{true}$ of a source is minimized based on the measurement data $\mathfrak{D}$\label{symbol:data}. These measurement data are collected in an experiment carried out on the unknown source producing multiple copies $N$ of quantum systems, each prepared in the state $\rho_\text{true}$. The data collection is usually done with a probability operator measurement\index{probability operator measurement (POM)} (POM) such that $\sum_j\Pi_j=1$.

Since $\rho_\text{true}$ is always unknown, in order to obtain a generically reliable estimator, the objective functional to be minimized has to be independent of $\rho\equiv\rho_\text{true}$\label{symbol:rho}. There are many kinds of such objective functionals we can use. A typical kind of objective functional, which we will consider here as the main example, is one that accounts for all possible experimental data $\mathfrak{D}$ one can obtain in an experiment. This allows us to find the estimator that is, in this sense, a universally optimal estimator for the cost functional that is independent of the data. To this end, we introduce the average cost functional
\begin{equation}
\overline{\mathfrak{C}}(\hat\rho)=\sum_\mathfrak{D}\int(\D\tau_\textsc{d})\,p(\mathfrak{D}\cap\rho)\mathfrak{C}(\rho,\hat\rho)\,,
\end{equation}\label{symbol:costfuncavg}\label{symbol:avgsym}where $(\D\tau_\textsc{d})$\label{symbol:dtau} is a pre-chosen integration measure for the $D$-dimensional\label{symbol:dim} Hilbert space and $p(\mathfrak{D}\cap\rho)$\label{symbol:proband} is the probability of having $\mathfrak{D}$ and the state $\rho$ simultaneously. The summation notation $\sum_\mathfrak{D}$\label{symbol:dataavg} refers to an average over all possible $\mathfrak{D}$. The statistical identity $p(\mathfrak{D}\cap\rho)=p(\mathfrak{D}|\rho)\pi(\rho)$\label{symbol:probcond}\label{symbol:priorprob} separates $p(\mathfrak{D}\cap\rho)$ into a product of a conditional probability distribution and a \emph{prior} probability distribution $\pi(\rho)$\index{prior probability (quantum states)} of all possible states $\rho$. The conditional probability $p(\mathfrak{D}|\rho)$, which involves the data, is defined in terms of the \emph{likelihood functional}\index{likelihood functional} $\mathcal{L}(\mathfrak{D};\rho)$\label{symbol:likelihood} inasmuch as
\begin{equation}
p(\mathfrak{D}|\rho)=\frac{\mathcal{L}(\mathfrak{D};\rho)}{\int(\D\tau_\textsc{d})\pi(\rho)\mathcal{L}(\mathfrak{D};\rho)}\,.
\end{equation}
The functional $\mathcal{L}(\mathfrak{D};\rho)$ gives the likelihood of a state $\rho$ yielding the measurement data $\mathfrak{D}$. The prior probability distribution $\pi(\rho)$, on the other hand, reflects the prior knowledge\index{prior information (knowledge)} one has about the source. One can define the \emph{prior}\index{prior $(\D\rho)$ (integral measure)} $(\D\rho)\equiv(\D\tau_\textsc{d})\pi(\rho)$\label{symbol:drho}. After inserting all the necessary elements, the objective functional is given by
\begin{equation}
\overline{\mathfrak{C}}(\hat\rho)=\sum_\mathfrak{D}\frac{\int(\D\rho)\mathcal{L}(\mathfrak{D};\rho)\mathfrak{C}(\rho,\hat\rho)}{\int(\D\rho')\mathcal{L}(\mathfrak{D};\rho')}\,.
\end{equation}

To proceed, we need to decide on the form of $\mathfrak{C}(\rho,\hat\rho)$, for the estimator $\hat\rho$ strongly depends on the cost functional. A very typical functional\index{cost functional!-- quadratic form}
\begin{equation}
\mathfrak{C}_1(\rho,\hat\rho)=\frac{\tr{\left(\frac{\rho+\mathfrak{H}}{1+\tr{\mathfrak{H}}}-\hat\rho\right)\mathfrak{G}\left(\frac{\rho+\mathfrak{H}}{1+\tr{\mathfrak{H}}}-\hat\rho\right)}}{2\lVert\mathfrak{G}\rVert_2}\leq 1\,,
\end{equation}
defined by the positive operators $\mathfrak{G}$ and $\mathfrak{H}$\label{symbol:ghop}, can be used as the cost functional and this quantifies a ``distance'' between $\rho$ and $\hat\rho$. Here $\lVert \mathfrak{G}\rVert_2$\label{symbol:2norm} refers to the operator 2-norm\index{2-norm (operator)} of $\mathfrak{G}$ defined as
\begin{equation}
\lVert \mathfrak{G}\rVert_2=\max_{\ket{y}\neq 0}\frac{\sqrt{\bra{y}\mathfrak{G}^\dagger \mathfrak{G}\ket{y}}}{\sqrt{\langle y|y\rangle}}\,.
\end{equation}
This is equal to the largest eigenvalue of $\mathfrak{G}\geq 0$, since for any ket $\ket{y}$\label{symbol:ket},
\begin{align}
&\frac{\sqrt{\bra{y}\mathfrak{G}^2\ket{y}}}{\sqrt{\langle y|y\rangle}}\nonumber\\
= &\,\sqrt{\tr{\mathfrak{G}^2\frac{\ket{y}\bra{y}}{\langle y|y\rangle}}}\nonumber\\
= &\,\sqrt{\sum_jg^2_j|\langle g_j|\,\,\,\rangle|^2}\quad\left(\ket{\,\,\,}\equiv\frac{\ket{y}}{\sqrt{\langle y|y\rangle}}\right)\nonumber\\
\leq &\,g_\text{max}\underbrace{\sqrt{\sum_j|\langle g_j|\,\,\,\rangle|^2}}_{=1}\nonumber\\
= &\,g_\text{max}\,.
\end{align}
In the derivation, the fact that $0\leq\mathfrak{G}=\sum_j\ket{g_j}g_j\bra{g_j}$ is exploited.

To show that $\mathfrak{C}_1(\rho,\hat\rho)$ is indeed bounded from above by 1, we note that
\begin{align}
&\tr{\left(\frac{\rho+\mathfrak{H}}{1+\tr{\mathfrak{H}}}-\hat\rho\right)\mathfrak{G}\left(\frac{\rho+\mathfrak{H}}{1+\tr{\mathfrak{H}}}-\hat\rho\right)}\nonumber\\
=&\,\tr{\left(\frac{\rho+\mathfrak{H}}{1+\tr{\mathfrak{H}}}-\hat\rho\right)^2\mathfrak{G}}\nonumber\\
=&\,\tr{\left(\frac{\rho+\mathfrak{H}}{1+\tr{\mathfrak{H}}}-\hat\rho\right)^2}\tr{\frac{\left(\frac{\rho+\mathfrak{H}}{1+\tr{\mathfrak{H}}}-\hat\rho\right)^2}{\tr{\left(\frac{\rho+\mathfrak{H}}{1+\tr{\mathfrak{H}}}-\hat\rho\right)^2}}\mathfrak{G}}\nonumber\\
\leq &\,\tr{\left(\frac{\rho+\mathfrak{H}}{1+\tr{\mathfrak{H}}}-\hat\rho\right)^2}\lVert\mathfrak{G}\rVert_2\nonumber\\
\leq &\,\left[\tr{\left(\frac{\rho+\mathfrak{H}}{1+\tr{\mathfrak{H}}}\right)^2}+\tr{\hat\rho^2}\right]\lVert\mathfrak{G}\rVert_2\nonumber\\
\leq &\,2\lVert\mathfrak{G}\rVert_2\,.
\end{align}
In establishing the first inequality, the simple identity $\tr{\rho \mathfrak{G}}\leq\text{largest eigenvalue of }\mathfrak{G}=\lVert \mathfrak{G}\rVert_2$ is used. This general quadratic form\index{quadratic form} $\mathfrak{C}_1(\rho,\hat\rho)$ has a unique minimum as long as $\mathfrak{G}\geq 0$. Such a functional gives non-zero cost for $\hat\rho\neq\rho$ and the special case $\mathfrak{G}=1$, $\mathfrak{H}=0$ yields the familiar square of the normalized Hilbert-Schmidt distance (David Hilbert\index{David Hilbert} and Erhard Schmidt\index{Erhard Schmidt}). An extreme case of such a cost functional\index{cost functional!-- delta function} is given by
\begin{equation}
\mathfrak{C}_2(\rho,\hat\rho)=-\delta(\rho-\hat\rho)\,,
\end{equation}
with which a singularly large reduction in cost is offered when $\rho=\hat\rho$ and no reduction is given otherwise.

With $\mathfrak{C}(\rho,\hat\rho)=\mathfrak{C}_1(\rho,\hat\rho)$, the variation $\updelta\mathfrak{C}_1(\rho,\hat\rho)$ is
\begin{equation}
\updelta\mathfrak{C}_1(\rho,\hat\rho)=-\frac{\tr{\updelta\hat\rho\left[\mathfrak{G}\left(\frac{\rho+\mathfrak{H}}{1+\tr{\mathfrak{H}}}\right)+\left(\frac{\rho+\mathfrak{H}}{1+\tr{\mathfrak{H}}}\right)\mathfrak{G}\right]}}{2\lVert\mathfrak{G}\rVert_2}\,.
\end{equation}
The total variation $\updelta\overline{\mathfrak{C}}_1(\hat\rho)$ works out to be
\begin{align*}
\updelta\overline{\mathfrak{C}}_1(\hat\rho)&=-\frac{1}{2\lVert\mathfrak{G}\rVert_2}\sum_\mathfrak{D}\tr{\updelta\hat\rho\frac{\int(\D\rho)\mathcal{L}(\mathfrak{D};\rho)\left[\mathfrak{G}\left(\frac{\rho+\mathfrak{H}}{1+\tr{\mathfrak{H}}}\right)+\left(\frac{\rho+\mathfrak{H}}{1+\tr{\mathfrak{H}}}\right)\mathfrak{G}\right]}{\int(\D\rho')\mathcal{L}(\mathfrak{D};\rho')}}\\
&=-\frac{1}{2\lVert\mathfrak{G}\rVert_2}\sum_\mathfrak{D}\tr{\updelta\hat\rho\left[\mathfrak{G}(\hat\rho_\text{B}-\hat\rho)+(\hat\rho_\text{B}-\hat\rho)\mathfrak{G}\right]}\,,
\end{align*}
where
\begin{equation}
\hat\rho_\text{B}(\mathfrak{H})=\frac{1}{\int(\D\rho')\mathcal{L}(\mathfrak{D};\rho')}\int(\D\rho)\mathcal{L}(\mathfrak{D};\rho)\frac{\rho+\mathfrak{H}}{1+\tr{\mathfrak{H}}}\,.
\label{bayes_est}
\end{equation}
Since minimizing $\overline{\mathfrak{C}}_1$ requires that $\updelta\overline{\mathfrak{C}}_1(\hat\rho)=0$, we thus have $\hat\rho=\hat\rho_\text{B}(\mathfrak{H})$\label{symbol:rhobayes}. The statistical operator $\hat\rho_\text{B}(\mathfrak{H})$ is known as the Bayesian estimator\index{Bayesian} (Thomas Bayes\index{Thomas Bayes}) of $\rho_\text{true}$ for a given operator $\mathfrak{H}$. A common variant of the Bayesian estimator\index{Bayesian} \cite{bayesian1,bayesian2,bayesian3} is defined as $\hat\rho_\text{B}=\hat\rho_\text{B}(0)$. In general, the integral average strongly depends on the definition of $(\D\rho)$, which has no definite form whatsoever even when some constraints are imposed on $(\D\tau_\textsc{d})$. For example, when $D=2$ and spherical coordinates\index{spherical coordinates} $(r,\vartheta,\varphi)$ are used to parameterize the Bloch vector of $\rho$, the constraint of unitary invariance on $(\D\rho)$ fixes $(\D\tau_2)=\D\Omega\,\D r=\D\varphi\,\D\vartheta\sin\vartheta\,\D r$, but $\pi(\rho)=\pi(r)$ can still take any function of the variable $r$. In this sense, there is an element of arbitrariness in the choice of $(\D\rho)$. Moreover, for a fixed form of $(\D\rho)$, the operator integral can be computationally difficult.

A more straightforward estimation scheme would be to consider $\mathfrak{C}(\rho,\hat\rho)=\mathfrak{C}_2(\rho,\hat\rho)$. The corresponding expression for $\overline{\mathfrak{C}}_2(\hat\rho)$ then simplifies to
\begin{equation}
\overline{\mathfrak{C}}_2(\hat\rho)=-\sum_\mathfrak{D}\frac{\mathcal{L}(\mathfrak{D};\hat\rho)}{\int(\D\rho')\mathcal{L}(\mathfrak{D};\rho')}\,.
\end{equation}
Thus, minimizing $\overline{\mathfrak{C}}_2(\hat\rho)$ amounts to looking for the estimator $\hat\rho=\hat\rho_\text{ML}$\label{symbol:rhoml} that maximizes the likelihood functional $\mathcal{L}(\mathfrak{D};\hat\rho)$. This estimator is the \emph{maximum-likelihood} (ML)\index{maximum-likelihood (ML)} estimator. In other words, to estimate $\rho_\text{true}$ whilst minimizing the objective functional $\overline{\mathfrak{C}}_2(\hat\rho)$ after an experiment, we need a scheme to search for a positive operator $\hat\rho_\text{ML}$ of unit trace such that the likelihood functional $\mathcal{L}(\mathfrak{D};\rho)$ takes the largest value within the admissible space of quantum states $\rho$. There is an asymptotic connection between $\hat\rho_\text{B}$ and $\hat\rho_\text{ML}$. That is, when $N$ is sufficiently large, the likelihood functional peaks very sharply around the maximum $\rho=\hat\rho_\text{ML}$ ($\mathcal{L}(\mathfrak{D};\rho)\rightarrow\delta(\rho-\hat\rho_\text{ML})$) and, from Eq.~(\ref{bayes_est}), it follows that $\hat\rho_\text{B}\rightarrow\hat\rho_\text{ML}$.

In a quantum-state tomography experiment, one can, in principle, measure $N$ copies of quantum systems using detectors with perfect detection efficiencies described by a POM $\sum_j\Pi_j=1$, with $j$ running over all detectors. The measurement data $\mathfrak{D}=\{n_1,n_2,n_3,\ldots\}$ is a list of detection outcome occurrences $n_j$ such that $\sum_jn_j=N$\label{symbol:nj}. One may also define the corresponding set of measurement frequencies $f_j=n_j/N$\label{symbol:freq}. For simplicity, we shall consider the POM to be informationally complete\index{informationally complete}. This means that there are $D^2$ linearly independent\index{linearly independent} outcomes in the POM that span the space of $D$-dimensional statistical operators. Therefore, this type of POM fully characterizes the source and maximal information can be extracted from the measurement data to reconstruct $\rho_\text{true}$ \emph{uniquely}\index{unique estimator}. Since the detection of one copy is independent of another, the detection occurrences $n_j$ follow a multinomial\index{multinomial} distribution and so the corresponding likelihood functional\index{likelihood functional!-- perfect measurements} for this scenario is
\begin{equation}
\mathcal{L}(\{n_j\};\rho)=\prod_j p_j^{n_j}=\left(\prod_j p_j^{f_j}\right)^N\,,
\label{simple_like}
\end{equation}
with $p_j=\tr{\rho\Pi_j}$\label{symbol:prob}.

One can construct an operator that maximizes $\mathcal{L}(\{n_j\};\rho)$ whilst paying no heed to the positivity constraint. To do this, we introduce a transposition mapping on a given operator $\Psi=\ket{a}\gamma\bra{b}$ of complex $a$, $b$ and $\gamma$ into an extended Hilbert space \cite{tight_ic}:
\begin{equation}
\Psi=\ket{a}\gamma\bra{b}\longmapsto\ket{a}\ket{b}\gamma\equiv\sket{\Psi}\,.
\end{equation}
The notation $\sket{\Psi}$ denotes a \emph{superket}\label{symbol:sket}\index{superket}. It is a ket that lives in an extended $D^2$-dimensional Hilbert space and is derived from an operator in a $D$-dimensional Hilbert space. Analogously to operators, one can define a $D^2$-dimensional \emph{superoperator}\index{superoperator} $\sket{\Psi}\sbra{\Phi}$ living in this extended Hilbert space. The simple identity
\begin{equation}
\superinner{\Psi}{\Phi}=\tr{\Psi^\dagger\Phi}
\end{equation}
follows from these notations.

Under this formalism, we can systematically study the linear independence of the POM outcomes. The first step is to note that for a set of $N_0$ linearly-independent POM outcomes, if the equation
\begin{equation}
\sum_j\sket{\Pi_j}\,c_j=0
\label{linindep}
\end{equation}
is to be satisfied for a given vector $\vec{c}=\transpose{(c_1,c_2,\ldots,c_{N_0})}$, then $\vec{c}$ must be zero since none of the outcomes can be expressed as a linear combination of the rest. In vector notations, Eq.~(\ref{linindep}) amounts to the scalar product relation
\begin{equation}
\begin{pmatrix}
\sket{\Pi_1},\sket{\Pi_2},\ldots,\sket{\Pi_{N_0}}
\end{pmatrix}
\begin{pmatrix}
c_1\\
c_2\\
\vdots\\
c_{N_0}
\end{pmatrix}=0\,.
\label{linindep2}
\end{equation}
Defining the positive matrix $\mathfrak{M}$ with matrix elements
\begin{equation}
\mathfrak{M}_{jk}\equiv\big<\!\!\big<\Pi_j\big|\Pi_k\big>\!\!\big>=\tr{\Pi_j\Pi_k}\,,
\end{equation}\label{symbol:mgram}The statement in (\ref{linindep2}) implies that the only solution to the matrix equation $\mathfrak{M}\cdot\vec{c}=0$ is $\vec{c}=0$. In the language of linear algebra, we say that the null space of $\mathfrak{M}$ has dimension zero. It follows that the rank of $\mathfrak{M}$ is $N_0$. We have thus constructed a positive matrix $\mathfrak{M}$ that has $N_0$ positive eigenvalues out of a set of $N_0$ linearly independent superkets $\sket{\Pi_j}$. This matrix is known as the Gram matrix (J{\o}rgen Pedersen Gram\index{J{\o}rgen Pedersen Gram}). The largest value of $N_0$ is $D^2$ since this is the maximum number of linearly independent\index{linearly independent} operators spanning the space of Hermitian operators as a basis\index{operator basis}. Therefore, a POM contains the maximal set of $D^2$ linearly independent outcomes if the corresponding Gram matrix $\mathfrak{M}$ has a rank of $D^2$.
\setlength{\parskip}{0pt}

One can also define the \emph{frame superoperator}\index{frame superoperator}
\begin{equation}
\mathcal{F}=\sum_j\sket{\Pi_j}\sbra{\Pi_j}\,.
\label{framesuperop}
\end{equation}\label{symbol:framesuperop}With this, an equivalent criterion for a set of informationally complete\index{informationally complete} POM outcomes $\Pi_j$ is that the superoperator\index{superoperator} $\mathcal{F}$ is invertible. There exist \emph{dual superkets}\index{dual}\index{superket} $\sket{\Theta_j}$\label{symbol:thetaj} of $\sket{\Pi_j}$ with the property
\begin{equation}
\sum_j\sket{\Pi_j}\sbra{\Theta_j}=\mathcal{I}=\sum_j\sket{\Theta_j}\sbra{\Pi_j}\,,
\label{dual_pom}
\end{equation}
where $\mathcal{I}$\label{symbol:identitysuperop} is the identity superoperator\index{superoperator}. The dual\index{dual} property is elucidated by the following equalities:
\begin{align*}
p_j&=\tr{\rho\Pi_j}\\
&=\sbra{\rho}\mathcal{I}\sket{\Pi_j}\\
&=\sum_k\superinner{\rho}{\Pi_k}\superinner{\Theta_k}{\Pi_j}\,.
\end{align*}
Since the final equality is always true for any $\rho$, it follows that $\superinner{\Theta_k}{\Pi_j}=\tr{\Theta_k\Pi_j}=\delta_{jk}$. Using Eq.~(\ref{framesuperop}), the dual\index{dual} superkets\index{superket} can be defined as
\begin{equation}
\sket{\Theta_j}=\mathcal{F}^{-1}\sket{\Pi_j}
\label{dual_pom2}
\end{equation}
and it is straightforward to verify that Eq.~(\ref{dual_pom}) is immediately satisfied. If, in addition, the number of $\Pi_j$s is exactly $D^2$ (\emph{minimal} POM), then the dual\index{dual} superkets\index{superket} $\sket{\Theta_j}$ are uniquely defined as in Eq.~(\ref{dual_pom2}). For \emph{overcomplete} measurements, there is more than one way of defining these dual\index{dual} superkets\index{superket} and the $\sket{\Theta_j}$s in Eq.~(\ref{dual_pom2}) serve as the canonical dual superkets\index{superket}. As an example, we consider a $D$-dimensional \emph{symmetric informationally complete POM}~(SIC POM) \cite{alex,appleby1,appleby2,englert,renes,sicpom,tight_ic}\index{symmetric informationally complete POM (SIC POM)} whose subnormalized rank-1 outcomes $\Pi_j$, that is $\tr{\Pi_j}=1/D$, are such that
\begin{equation}
\superinner{\Pi_j}{\Pi_k}=\tr{\Pi_j\Pi_k}=\frac{D\delta_{jk}+1}{D^2(D+1)}\,.
\label{sicpom}
\end{equation}
The corresponding dual\index{dual} superkets\index{superket} for this POM can be shown (see Appendix \ref{chap:dual_sic}) to be
\begin{equation}
\sket{\Theta_j}=\sket{\Pi_j}D(D+1)-\sket{1}\,.
\end{equation}

With all the necessary tools in place, we can now define the operator that maximizes the likelihood functional over \emph{all} Hermitian operators:
\begin{equation}
\hat\sigma=\sum_jf_j\Theta_j\,.
\label{li_est}
\end{equation}
To verify that this is indeed the solution, we note that $p_j=\tr{\hat\sigma\Pi_j}=f_j$ are the solutions that maximize the likelihood functional in Eq.~(\ref{simple_like}). A simple calculation shows that
\begin{equation*}
\tr{\hat\sigma\Pi_j}=\sum_kf_k\tr{\Pi_j\Theta_k}=\sum_kf_k\delta_{jk}=f_j\,.
\end{equation*}
Alternatively, the estimator $\hat\sigma$\label{symbol:sigmahat} in Eq.~(\ref{li_est}), also known as the \emph{linear-inversion estimator}, can be obtained by directly inverting the set of $D^2$ constraints $\tr{\hat\sigma\Pi_j}=f_j$ for minimal informationally complete data. An essential tool for linear-inversion is a complete set of Hermitian, trace-orthonormal\index{trace-orthonormal} basis operators\index{operator basis} $\Gamma_j=\Gamma^\dagger_j$\label{symbol:gammaj} such that $\tr{\Gamma_j\Gamma_k}=\delta_{jk}$. By ``complete'', we mean that the superkets $\sket{\Gamma_j}$ satisfy the completeness relation
\begin{equation}
\sum_j\sket{\Gamma_j}\sbra{\Gamma_j}=\mathcal{I}\,.
\end{equation}
With this basis, one can express the operators $\hat\sigma=\sum_kt_k\Gamma_k$ and $\Pi_j=\sum_kc_{jk}\Gamma_k$ in terms of $\Gamma_j$. The coefficients $t_k$ can thereafter be obtained by inverting the system of linear equations
\begin{equation}
f_j=\sum_kc_{jk}t_k\,.
\end{equation}

The estimator $\hat\sigma$ is the ML statistical operator we seek if $\hat\sigma\geq 0$ for the measurement data. We say that $\hat\sigma$ is an \emph{unbiased} estimator\index{unbiased estimator} for $\rho_\text{true}$ since the operator $\overline{\hat\sigma}=\sum_j\overline{f_j}\Theta_j=\sum_jp^\text{true}_j\Theta_j=\rho_\text{true}$\label{symbol:probtrue}. This means that the set of all possible estimators $\hat\sigma$, for a given $N$, forming an \emph{uncertainty hyper-ellipsoid}\index{uncertainty hyper-ellipsoid} is such that the operator \emph{centroid}\index{centroid (operator)} of the set is $\rho_\text{true}$. Because of this fact, the estimator $\hat\sigma$ is generally not a positive operator. Geometrically, part of the boundary of the uncertainty hyper-ellipsoid around $\rho_\text{true}$ that contains all estimators $\hat\sigma$ can lie outside the state space for finite $N$. As $N$ increases, the hyper-ellipsoid shrinks to a point in the state space when $N$ becomes infinite. In other words, as long as $N$ is finite, if the true state lies on the boundary, then no matter how small this hyper-ellipsoid is, there will always be estimators that are not positive. For them, it follows that the true peak of $\mathcal{L}(\mathfrak{D};\rho)$ lies outside the state space and the resulting positive ML estimator $\rho_\text{ML}$ that maximizes $\mathcal{L}(\mathfrak{D};\rho)$ inside the state space must necessarily be rank-deficient. In this case, there is no analytical expression for the positive estimator and numerical methods are needed to look for this estimator. The positive ML estimator, like $\hat\sigma$, is also a \emph{consistent} estimator\index{consistent estimator}, which is defined by the property that $\hat\rho_\text{ML}$ approaches $\rho_\text{true}$ as $N$ increases\cite{prod_sic,qst_huangjun}.

\subsection{Uncertainties in quantum estimation}

The usual distance functional
\begin{equation}
\mathcal{C}_\text{H-S}(\rho_\text{true}-\hat\rho)=\overline{\tr{(\rho_\text{true}-\hat\rho)^2}}\,,
\end{equation}\label{symbol:covariance}reminiscent of the Hilbert-Schmidt distance\index{Hilbert-Schmidt distance}, is a common measure of the average deviation of an estimator $\hat\rho$ away from the true statistical operator $\rho_\text{true}$ and is known as the \emph{covariance} of $\rho_\text{true}$ and $\hat\rho$. To evaluate this functional, we express the operators $\rho_\text{true}=\sum_jr^\text{true}_j\Gamma_j$\label{symbol:rtrue} and $\hat\rho=\sum_jt_j\Gamma_j$ in terms of a set of Hermitian, trace-orthonormal\index{trace-orthonormal} basis operators\index{basis operators} $\Gamma_j$. The resulting functional becomes
\begin{equation}
\mathcal{C}_\text{H-S}\left(\vec{t},\vec{r}_\text{true}\right)=\overline{\left(\vec{t}-\vec{r}_\text{true}\right)\cdot\left(\vec{t}-\vec{r}_\text{true}\right)}\,,
\end{equation}\label{symbol:tvec}where $\overline{\,\vec{t}\,\,}=\vec{r}_\text{true}$\label{symbol:rtruevec}. The corresponding \emph{dyadic}
\begin{equation}
\overleftrightarrow{\mathcal{C}}\left(\vec{t},\vec{r}_\text{true}\right)=\overline{\left(\vec{t}-\vec{r}_\text{true}\right)\left(\vec{t}-\vec{r}_\text{true}\right)}
\end{equation}\label{symbol:dyadicsym}is known as the \emph{covariance dyadic}\label{symbol:covariancedyadic}\index{covariance dyadic} and is positive.

More generally, the covariance dyadic\index{covariance dyadic} describes the \emph{mean squared-error}\index{mean squared-error} between $\hat\rho$ and $\rho_\text{true}$ in terms of their respective coefficients. The average of a function $f(\mathfrak{D})$ of the data $\mathfrak{D}$ is given by
\begin{equation}
\overline{f(\mathfrak{D})}=\sum_\mathfrak{D}f(\mathfrak{D})=\int(\D\mathfrak{D})p(\mathfrak{D}|\rho)\,f(\mathfrak{D})\,.
\end{equation}

There exists a lower bound for the covariance dyadic\index{covariance dyadic} and to calculate it, we assume that $\hat\rho$ is unbiased, which as a consequence need not be positive, and note that
\begin{align}
\frac{\partial}{\partial\vec{r}}\overline{\,\vec{t}\,\,}\,\Bigg|_{\vec{r}\,=\,\vec{r}_\text{true}}&=\frac{\partial}{\partial\vec{r}}\vec{r}\,\Bigg|_{\vec{r}\,=\,\vec{r}_\text{true}}\nonumber\\
\Rightarrow \int(\D\mathfrak{D})\frac{\partial}{\partial\vec{r}}\,p(\mathfrak{D}|\rho)\,\vec{t}\,\Bigg|_{\vec{r}\,=\,\vec{r}_\text{true}}&=\overleftrightarrow{1}\nonumber\\
\Rightarrow \int(\D\mathfrak{D})p(\mathfrak{D}|\rho)\left[\frac{\partial}{\partial\vec{r}}\log\left(p(\mathfrak{D}|\rho)\right)\right]\vec{t}\,\Bigg|_{\vec{r}\,=\,\vec{r}_\text{true}}&=\overleftrightarrow{1}\,,
\label{constraint1}
\end{align}\label{symbol:deedeer}where $\overleftrightarrow{1}$ is the unit dyadic\label{symbol:1dyadic}, and
\begin{align}
\int(\D\mathfrak{D})\frac{\partial}{\partial\vec{r}}\,p(\mathfrak{D}|\rho)\,\Bigg|_{\vec{r}\,=\,\vec{r}_\text{true}}&=\frac{\partial}{\partial\vec{r}}\,1\,\Bigg|_{\vec{r}\,=\,\vec{r}_\text{true}}\nonumber\\
\Rightarrow \int(\D\mathfrak{D})p(\mathfrak{D}|\rho)\left[\frac{\partial}{\partial\vec{r}}\log\left(p(\mathfrak{D}|\rho)\right)\right]\vec{r}\,\Bigg|_{\vec{r}\,=\,\vec{r}_\text{true}}&=0\,.
\label{constraint2}
\end{align}
Combining Eqs.~(\ref{constraint1}) and (\ref{constraint2}), we have
\begin{equation}
\int(\D\mathfrak{D})p(\mathfrak{D}|\rho)\left[\frac{\partial}{\partial\vec{r}}\log\left(p(\mathfrak{D}|\rho)\right)\right]\left(\vec{t}-\vec{r}\right)\,\Bigg|_{\vec{r}\,=\,\vec{r}_\text{true}}=\overleftrightarrow{1}\,.
\label{comb_constraints}
\end{equation}
Multiplying the vectors $\vec{x}$ and $\vec{y}$ respectively on the left and right of Eq.~(\ref{comb_constraints}) gives
\begin{equation}
\int(\D\mathfrak{D})p(\mathfrak{D}|\rho)\left[\vec{x}\cdot\frac{\partial}{\partial\vec{r}}\log\left(p(\mathfrak{D}|\rho)\right)\right]\left(\vec{t}-\vec{r}\right)\cdot\vec{y}\,\Bigg|_{\vec{r}\,=\,\vec{r}_\text{true}}=\vec{x}\cdot\vec{y}\,.
\end{equation}
By the Cauchy-Schwarz\index{Cauchy-Schwarz inequality} inequality (Baron Augustin-Louis Cauchy\index{Augustin-Louis Cauchy} and Karl Hermann Amandus Schwarz\index{Karl Hermann Amandus Schwarz}),
\begin{align}
\lvert\vec{x}\cdot\vec{y}\rvert^2&=\int(\D\mathfrak{D})p(\mathfrak{D}|\rho)\left[\vec{x}\cdot\frac{\partial}{\partial\vec{r}}\log\left(p(\mathfrak{D}|\rho)\right)\right]\left(\vec{t}-\vec{r}\right)\cdot\vec{y}\,\Bigg|_{\vec{r}\,=\,\vec{r}_\text{true}}\nonumber\\
&\leq\int(\D\mathfrak{D})p(\mathfrak{D}|\rho)\,\vec{x}\cdot\left[\frac{\partial}{\partial\vec{r}}\log\left(p(\mathfrak{D}|\rho)\right)\right]\left[\frac{\partial}{\partial\vec{r}}\log\left(p(\mathfrak{D}|\rho)\right)\right]\cdot\vec{x}\,\Bigg|_{\vec{r}\,=\,\vec{r}_\text{true}}\nonumber\\
&\quad\times\int(\D\mathfrak{D})p(\mathfrak{D}|\rho)\,\vec{y}\cdot\left(\vec{t}-\vec{r}_\text{true}\right)\left(\vec{t}-\vec{r}_\text{true}\right)\cdot\vec{y}\nonumber\\
&=\vec{x}\cdot\underbrace{\overline{\left[\frac{\partial}{\partial\vec{r}}\log\left(p(\mathfrak{D}|\rho)\right)\right]\left[\frac{\partial}{\partial\vec{r}}\log\left(p(\mathfrak{D}|\rho)\right)\right]}\,\Bigg|_{\vec{r}\,=\,\vec{r}_\text{true}}}_{\equiv\, \overleftrightarrow{F} \text{(Fisher's information dyadic)}}\cdot\,\vec{x}\nonumber\\
&\quad\times\vec{y}\cdot\underbrace{\overline{\left(\vec{t}-\vec{r}_\text{true}\right)\left(\vec{t}-\vec{r}_\text{true}\right)}}_{=\,\overleftrightarrow{\mathcal{C}}\left(\vec{t},\vec{r}_\text{true}\right)}\cdot\,\vec{y}\,,
\end{align}
where $F$\label{symbol:fisherdyadic} is the Fisher's information dyadic\index{Fisher's information dyadic} (Sir Ronald Aylmer Fisher\index{Ronald Aylmer Fisher}). A substitution of $\vec{x}=\overleftrightarrow{F}^{-1}\cdot\vec{y}$ gives the inequality
\begin{equation}
\vec{y}\cdot \overleftrightarrow{F}^{-1}\cdot\vec{y}\leq\vec{y}\cdot\overleftrightarrow{\mathcal{C}}\left(\vec{t},\vec{r}_\text{true}\right)\cdot\vec{y}\,,
\end{equation}
which is satisfied for \emph{any} $\vec{y}$. This implies that
\begin{equation}
\overleftrightarrow{F}^{-1}\leq\overleftrightarrow{\mathcal{C}}\left(\vec{t},\vec{r}_\text{true}\right)\,.
\end{equation}
The inequality presented above is the famous Cram{\'e}r-Rao inequality\index{Cram{\'e}r-Rao!-- inequality} (Harald Cram{\'e}r\index{Harald Cram{\'e}r} and Calyampudi Radhakrishna Rao\index{Calyampudi Radhakrishna Rao}) for unbiased estimation. It tells us that the lowest mean squared-error\index{mean squared-error} $\tr{\overleftrightarrow{\mathcal{C}}\left(\vec{t},\vec{r}_\text{true}\right)}$ is given by $\tr{\overleftrightarrow{F}^{-1}}$.

It is interesting to study the asymptotic expression for the Fisher's dyadic $\overleftrightarrow{F}$ when $N$ is large. To begin, we note that for sufficiently large $N$, the Central Limit Theorem\index{Central Limit Theorem} tells us that \cite{quantum_diag} the conditional probability distribution
\begin{equation}
p(\mathfrak{D}|\rho)=\frac{1}{\sqrt{(2\pi)^{D^2}\det\left\{\overleftrightarrow{\mathcal{C}}_\text{ML}\right\}}}\,\text{exp}\left\{-\frac{1}{2}\left(\vec{r}-\vec{t}_\text{ML}\right)\cdot\overleftrightarrow{\mathcal{C}}_\text{ML}^{-1}\cdot\left(\vec{r}-\vec{t}_\text{ML}\right)\right\}\,
\end{equation}\label{symbol:tmlvec}\label{symbol:covariancedyadicml}takes a Gaussian form (Johann Carl Friedrich Gauss\index{Johann Carl Friedrich Gauss}), where $\vec{t}_\text{ML}$ is the vector of coefficients for $\hat\rho_\text{ML}$. With this,
\begin{align*}
\frac{\partial}{\partial\vec{r}}\log\left(p(\mathfrak{D}|\rho)\right)&=-\frac{1}{2}\frac{\partial}{\partial\vec{r}}\left[\left(\vec{r}-\vec{t}_\text{ML}\right)\cdot\overleftrightarrow{\mathcal{C}}_\text{ML}^{-1}\cdot\left(\vec{r}-\vec{t}_\text{ML}\right)\right]\\
&=-\frac{1}{2}\left\{\left[\overleftrightarrow{\mathcal{C}}_\text{ML}^{-1}+\transpose{\left(\overleftrightarrow{\mathcal{C}}_\text{ML}^{-1}\right)}\right]\cdot\left(\vec{r}-\vec{t}_\text{ML}\right)\right\}\\
&=\overleftrightarrow{\mathcal{C}}_\text{ML}^{-1}\cdot\left(\vec{r}-\vec{t}_\text{ML}\right)\,,
\end{align*}
and so
\begin{align}
\overleftrightarrow{F}&=\overleftrightarrow{\mathcal{C}}_\text{ML}^{-1}\cdot\underbrace{\overline{\left(\vec{r}-\vec{t}_\text{ML}\right)\left(\vec{r}-\vec{t}_\text{ML}\right)}\,\Bigg|_{\vec{r}\,=\,\vec{r}_\text{true}}}_{=\,\overleftrightarrow{\mathcal{C}}_\text{ML}}\cdot\overleftrightarrow{\mathcal{C}}_\text{ML}^{-1}\nonumber\\
&=\overleftrightarrow{\mathcal{C}}^{-1}_\text{ML}\,.
\label{fisher_crb}
\end{align}
An important lesson learned here is that for large $N$, the unbiased ML estimator $\hat\rho$, on average, approaches the lower bound (Cram{\'e}r-Rao bound)\index{Cram{\'e}r-Rao!-- bound} set by the Cram{\'e}r-Rao inequality\index{Cram{\'e}r-Rao!-- inequality} asymptotically. The unbiased ML estimator is thus said to be an \emph{efficient} estimator\index{efficient estimator}, that is, no other unbiased estimator can achieve a lower asymptotic mean squared-error\index{mean squared-error}. When the positivity constraint is taken into account, the Cram{\'e}r-Rao inequality will be modified to accomodate the constraint \cite{constrained_crb,constrained_ml} and it was shown that the corresponding constrained ML estimator is efficient in terms of the constrained Cram{\'e}r-Rao bound.

Equation~(\ref{fisher_crb}) provides an operational way to compute the uncertainties of a real quantity $\mathfrak{q}\equiv\tr{\hat\rho_\text{ML}Q}$, where
\begin{equation}
\overline{\mathfrak{q}}=\overline{\tr{\hat\rho_\text{ML}Q}}=\tr{\rho_\text{true}Q}=\mathfrak{q}_\text{true}\,.
\end{equation}
The corresponding Hermitian operator $Q$ can be similarly expressed in terms of the set of operator basis\index{operator basis} $\{\Gamma_j\}$ such that $Q=\sum_jq_j\Gamma_j$. Note that its variance
\begin{align*}
\left(\Delta \mathfrak{q}\right)^2=\overline{\left(\mathfrak{q}-\mathfrak{q}_\text{true}\right)^2}&=\overline{\tr{\left(\hat\rho_\text{ML}-\rho_\text{true}\right)Q}^2}\\
&=\overline{\left[\vec{q}\cdot\left(\vec{t}_\text{ML}-\vec{r}_\text{true}\right)\right]^2}\\
&=\vec{q}\cdot\overline{\left(\vec{t}_\text{ML}-\vec{r}_\text{true}\right)\left(\vec{t}_\text{ML}-\vec{r}_\text{true}\right)}\cdot\vec{q}\\
&=\vec{q}\cdot\overleftrightarrow{\mathcal{C}}_\text{ML}\cdot\vec{q}\,.
\end{align*}
Equation~(\ref{fisher_crb}) tells us that $\overleftrightarrow{\mathcal{C}}_\text{ML}$ is the inverse of the Fisher's information dyadic\index{Fisher's information dyadic} $\overleftrightarrow{F}$ for sufficiently large $N$. This leads to \cite{quantum_diag}
\begin{equation}
\left(\Delta \mathfrak{q}\right)^2=\vec{q}\cdot\overleftrightarrow{F}^{-1}\cdot\vec{q}\,.
\end{equation}
Hence the meaning of the Fisher's information dyadic\index{Fisher's information dyadic} is quite clear for large $N$: it directly quantifies the uncertainty of the real quantity $q$ and carries the same amount of information as $\overleftrightarrow{\mathcal{C}}_\text{ML}$. If $\overleftrightarrow{\mathcal{C}}_\text{ML}$ is non-invertible, then $\overleftrightarrow{F}$ will carry information in the support of $\overleftrightarrow{\mathcal{C}}_\text{ML}$. 

\section{Informationally complete quantum state estimation}\index{informationally complete}\index{maximum-likelihood (ML)}

\subsection{Steepest-ascent (direct-gradient) algorithm}\index{steepest-ascent method}

Suppose an informationally complete POM, consisting of $D^2$ linearly independent\index{linearly independent} outcomes, is used to reconstruct an unknown true state $\rho_\text{true}$ of dimension $D$. The detection of $N$ copies of quantum systems yields a multinomial statistic\index{multinomial} for the measured number of occurrences\index{number of occurrences} $n_j$ for every outcome $\Pi_j$, and the corresponding likelihood functional $\mathcal{L}(\{n_j\};\rho)$ is given in Eq.~(\ref{simple_like}). To look for $\hat\rho_\text{ML}$ numerically, we first vary the log-likelihood\index{log-likelihood functional!-- perfect measurements} $\log\mathcal{L}(\{n_j\};\rho)$ and obtain
\begin{equation}
\updelta\log\mathcal{L}(\{n_j\};\rho)=\sum_{j}{f_j\frac{\updelta p_j}{p_j}}=\tr{R\updelta\rho}\,,
\end{equation}
where
\begin{equation}
R=\sum_{j}\frac{f_j}{p_j}\Pi_j\,.
\label{rop}
\end{equation}\label{symbol:rmatrix}

Note that maximizing the likelihood functional requires $\updelta\log\mathcal{L}(\{n_j\};\rho)=0$. To increase the value of $\updelta\log\mathcal{L}(\{n_j\};\rho)$ when the maximal value of $\mathcal{L}(\{n_j\};\rho)$ is not reached, we need to look for a suitable variation $\updelta\rho$ such that $\updelta\log\mathcal{L}(\{n_j\};\rho)>0$ while maintaining the positivity of $\rho$.

We begin by parameterizing the statistical operator
\begin{equation}
\rho=\frac{\mathcal{A}^\dagger \mathcal{A}}{\tr{\mathcal{A}^\dagger \mathcal{A}}}
\label{aux_param}
\end{equation}\label{symbol:auxa}
with an auxiliary complex operator $\mathcal{A}$. Under this parametrization,
\begin{align*}
\updelta\rho &=\frac{\updelta \mathcal{A}^\dagger \mathcal{A}+\mathcal{A}^\dagger\updelta \mathcal{A}}{\tr{\mathcal{A}^\dagger \mathcal{A}}}-\frac{\mathcal{A}^\dagger \mathcal{A}}{\tr{\mathcal{A}^\dagger \mathcal{A}}^2}\tr{\updelta \mathcal{A}^\dagger \mathcal{A}+\mathcal{A}^\dagger\updelta \mathcal{A}}\\
&=\frac{\updelta \mathcal{A}^\dagger \mathcal{A}+\mathcal{A}^\dagger\updelta \mathcal{A}}{\tr{\mathcal{A}^\dagger \mathcal{A}}}-\rho\,\frac{\tr{\updelta \mathcal{A}^\dagger \mathcal{A}+\mathcal{A}^\dagger\updelta \mathcal{A}}}{\tr{\mathcal{A}^\dagger \mathcal{A}}}\,.
\end{align*}
It follows that,
\begin{align}
\updelta\log\mathcal{L}(\{n_j\};\rho)&=\tr{R\left(\frac{\updelta \mathcal{A}^\dagger \mathcal{A}+\mathcal{A}^\dagger\updelta \mathcal{A}}{\tr{\mathcal{A}^\dagger \mathcal{A}}}-\rho\,\frac{\tr{\updelta \mathcal{A}^\dagger \mathcal{A}+\mathcal{A}^\dagger\updelta \mathcal{A}}}{\tr{\mathcal{A}^\dagger \mathcal{A}}}\right)}\nonumber\\
&=\tr{\updelta \mathcal{A}\frac{R\mathcal{A}^\dagger-\tr{R\rho}\mathcal{A}^\dagger}{\tr{\mathcal{A}^\dagger \mathcal{A}}}+\updelta \mathcal{A}^\dagger\frac{\mathcal{A}R-\tr{R\rho}\mathcal{A}}{\tr{\mathcal{A}^\dagger \mathcal{A}}}}\,\nonumber\\
&=\frac{1}{\tr{\mathcal{A}^\dagger \mathcal{A}}}\tr{\updelta \mathcal{A}\left[\left(R-1\right)\mathcal{A}^\dagger\right]+\updelta \mathcal{A}^\dagger\left[\mathcal{A}\left(R-1\right)\right]}\,,
\label{variation1}
\end{align}
In deriving Eq.~(\ref{variation1}), the identity $\tr{R\rho}=1$ is invoked. By setting $\updelta\log\mathcal{L}(\{n_j\};\rho)=0$, we arrive at the extremal equation for the positive ML estimator $\hat\rho_\text{ML}$ \cite{rehacek1,qstateest}:
\begin{equation}
R_\text{ML}\hat\rho_\text{ML}=\hat\rho_\text{ML}R_\text{ML}=\hat\rho_\text{ML}\,,
\label{extcon}
\end{equation}
where the operator $R_\text{ML}$ is the operator $R$, defined in Eq.~(\ref{rop}), evaluated with the ML estimator $\hat\rho_\text{ML}$.

One way of ensuring a positive $\updelta\log\mathcal{L}(\{n_j\};\rho)$ every step is to note that the definition of the variation of $\log\mathcal{L}(\{n_j\};\rho)$, in the form of a trace equation, is given by
\begin{equation}
\updelta\log\mathcal{L}(\{n_j\};\rho)=\tr{\updelta \mathcal{A}\Bigg(\frac{\partial \log\mathcal{L}(\{n_j\};\rho)}{\partial \mathcal{A}}\Bigg)^\dagger+\updelta \mathcal{A}^\dagger\Bigg(\frac{\partial \log\mathcal{L}(\{n_j\};\rho)}{\partial \mathcal{A}^\dagger}\Bigg)^\dagger}\,,
\label{total_var_op}
\end{equation}
where the partial derivative $\partial \log\mathcal{L}(\{n_j\};\rho)/\partial \mathcal{A}=\mathcal{A}(R-1)/\tr{\mathcal{A}^\dagger \mathcal{A}}$. Noting that the gradient of $\log\mathcal{L}(\{n_j\};\rho)$, which we now define to be a two-component vector $\vec{\partial}\log\mathcal{L}(\{n_j\};\rho)$\label{symbol:partialvec}, is given by
\begin{equation}
\vec{\partial}\log\mathcal{L}(\{n_j\};\rho)=\begin{pmatrix}
\partial \log\mathcal{L}(\{n_j\};\rho)/\partial \mathcal{A}\\
\partial \log\mathcal{L}(\{n_j\};\rho)/\partial \mathcal{A}^\dagger
\end{pmatrix},
\end{equation}
we can enforce the variation of $\vec{Z}\equiv\transpose{\left(\mathcal{A},\mathcal{A}^\dagger\right)}$\label{symbol:zvec} to follow the direction of the steepest ascent up to the global maximum. In other words,
\begin{equation}
\updelta\vec{Z}=
\begin{pmatrix}
\updelta \mathcal{A}\\
\updelta \mathcal{A}^\dagger
\end{pmatrix}
\equiv\frac{\epsilon}{2}\begin{pmatrix}
\mathcal{A}(R-1)\\
(R-1)\mathcal{A}^\dagger
\end{pmatrix}
\propto\vec{\partial}\log\mathcal{L}(\{n_j\};\rho),
\end{equation}
where $\epsilon$\label{symbol:epsilon} is a small positive parameter. Correspondingly,
\begin{equation}
\updelta \log\mathcal{L}(\{n_j\};\rho)=\epsilon\,\tr{(R-1)\rho(R-1)}\,.
\label{incf}
\end{equation}
Thus, one derives the iterative equation, in discrete form, as
\begin{equation}
\rho_{k+1}=\frac{\Big[1+\frac{\epsilon}{2}\Big(R_k-1\Big)\Big]\rho_k\Big[1+\frac{\epsilon}{2}\Big(R_k-1\Big)\Big]}{\tr{\Big[1+\frac{\epsilon}{2}\Big(R_k-1\Big)\Big]\rho_k\Big[1+\frac{\epsilon}{2}\Big(R_k-1\Big)\Big]}}\,,
\label{algoold}
\end{equation}
which is precisely the iterative equation for the ML scheme established in \cite{rehacek1,qstateest}. It is now clear that ML is actually the method of steepest-ascent to search for $\hat\rho_\text{ML}$. Since this enhanced algorithm attempts to reach the global maximum by directly following the gradient, this method can also be called the \emph{direct-gradient method}\index{direct-gradient method|see{steepest-ascent method}} (ML-DG). Hence, given the above iterative equation, one can attempt to obtain the ML estimator that gives the global maximum of $\log\mathcal{L}(\{n_j\};\rho)$. Numerically, the estimator $\hat\rho_\text{ML}\equiv\rho_\kappa$ is taken such that $\tr{\lvert\left(R_\kappa-1\right)\rho_\kappa\rvert}\leq\varepsilon$\label{symbol:epsilonprec}, where $\tr{\lvert M\rvert}=\tr{\sqrt{M^\dagger M}}$ is the \emph{trace norm}\index{trace norm} for an operator $M$ and $\varepsilon$ is a pre-chosen precision and \emph{must not be confused with the small parameter $\epsilon$}. One can also introduce an enhancement in the rate of convergence to this scheme by attempting to optimize the value of $\epsilon$ in each step of the iteration so that the log-likelihood functional is maximized efficiently. This procedure is usually known as the \emph{line search} and can be done in various ways.

Outlined below is the iterative algorithm for the ML estimation scheme\index{algorithm} \cite{rehacek1,prod_sic}.
\begin{center}
\colorbox{light-gray}{\begin{minipage}[c]{12cm}
  \uline{\textbf{ML algorithm using the steepest-ascent method (ML-DG)}}\\
  Starting from the maximally-mixed state $\rho_1=1/D$, with $k=1$ and a small fixed value of $\epsilon$,
  \begin{enumerate}
	\item Compute $R_k$;
	\begin{itemize}
	\item Escape from loop if $\tr{|R_k\rho_k-\rho_k|}\leq\varepsilon$;
 	\item Otherwise, proceed to the following steps.
 	\end{itemize}
    \item Carry out the \textbf{line search} procedure:
    \begin{itemize}
	\item Use two trial values of $\epsilon_k$ to compute two $\rho_{k+1}$s and determine the value of the likelihood $\mathcal{L}(\{n_j\};\rho_{k+1})$ for both.
    \item Combine these two with $\mathcal{L}(\{n_j\};\rho_k)$, which was in fact obtained from $\epsilon_k=0$, and compute a quadratic function of $\epsilon_k$ that interpolates between the three support values.
    \item Find the $\epsilon_k$ value for which the quadratic function assumes its maximum.
    \end{itemize}
    \item Use this maximizing $\epsilon_k$ to evaluate the new $\rho_{k+1}$ using Eq.~(\ref{algoold}), with $\epsilon$ replaced by $\epsilon_k$.
    \item Set $k=k+1$ and repeat the iteration from the beginning.
\end{enumerate}
  \end{minipage}}
  \end{center}
The optimization of $\epsilon_k$ introduced here is practical since the exact maximum of $\epsilon_k$ is in general hard to compute. Such a line search optimization can in principle expedite the search of $\hat\rho_\text{ML}$. However, when $D$ and the number of POM outcomes are huge, such a procedure becomes impractical since it involves the evaluation of very many large matrices, which can be very computationally expensive. In this case, a fixed value of $\epsilon_k=\epsilon$ is used instead. 

\subsection{Conjugate-gradient algorithm}
\label{subsec:conjgrad}

The steepest-ascent, or direct-gradient, method seeks the extremal solution of a given function by following the path of its steepest gradient in the space of parameters. It may happen that in an iterative step, the path is quite parallel to another one taken in one of the previous iterative steps. In other words, the iterated answer traces out a ``zig-zag'' path in the space of parameters as it approaches the true extremal point. This causes a considerable retardation in the iteration if the precision $\varepsilon$ is chosen too small. Another alternative to this method is the approach of \emph{conjugate gradient}. Initially developed for real quadratic\index{quadratic form} objective functions of the form
\begin{equation}
f(\vec{z})=-\frac{1}{2}\vec{z}\cdot\!\overleftrightarrow{A}\!\cdot\vec{z}+\vec{b}\cdot\vec{z}\,,
\label{quad_form}
\end{equation}
where the real dyadic $\overleftrightarrow{A}\geq 0$ and the dimensionality of the real vectors is $n$, the conjugate-gradient\index{conjugate-gradient method} (CG) iteration takes a path which ``circulates'' directly to the extremal point $z=z_\text{max}$ in exactly $n$ iterative steps. Technically speaking, the search directions $\vec{h}_k$\label{symbol:hk} taken in the $k$-th step is such that $\vec{h}_k\cdot\!\overleftrightarrow{A}\!\cdot\vec{h}_l=0$ for $j\neq k$. This \emph{conjugacy} property is where the name of this approach is derived. One can obtain a complete set of conjugate direction vectors using the \emph{Gram-Schmidt conjugation}\index{Gram-Schmidt conjugation} strategy (J{\o}rgen Pedersen Gram\index{J{\o}rgen Pedersen Gram} and Erhard Schmidt\index{Erhard Schmidt}), a modified orthonormalization technique which accounts for the conjugacy property. However, this strategy ultimately requires all direction vectors to be stored into memory, since a linear combination of all the previously computed direction vectors is required to compute the next one. Such a procedure can be computationally expensive for large $n$.

In the CG method, the gradient vectors $\vec{\nabla}_nf(\vec{z}_k)$ for every $\vec{z}_k$ are used to compute the set of conjugate direction vectors $\vec{h}_k$. Here, $\vec{\nabla}_n$ is the $n$-dimensional gradient vector and $\vec{\nabla}_nf(\vec{z}_k)=\vec{b}-\overleftrightarrow{A}\!\cdot\vec{z}_k$. With this substitution, the linear combination of $\vec{h}_k$s in the Gram-Schmidt conjugation procedure becomes just a single term\footnote{Please consult Ref.~\cite{cgref} for the technical details and graphs.} and so there is no longer a need to store all the previously computed direction vectors. In each step, the conjugate direction vector $\vec{h}_k$ and $\vec{z}_k$ are thus generated pairwise.

The CG algorithm\index{algorithm} for the quadratic form\index{quadratic form} in Eq.~(\ref{quad_form}) is outlined below:
\begin{center}
\colorbox{light-gray}{\begin{minipage}[c]{12cm}
  \textbf{\uline{CG method for quadratic forms}}\\
  Beginning with $\vec{h}_1=\vec{g}_1=\vec{b}-\overleftrightarrow{A}\!\cdot\vec{z}_1$ and $k=1$,
  \begin{enumerate}
  \item Compute $\epsilon_k=\frac{\vec{g}_k\cdot\vec{g}_k}{\vec{h}_k\cdot\!\overleftrightarrow{A}\!\cdot\vec{h}_k}$ and set $\vec{z}_{k+1}=\vec{z}_k+\epsilon_k \vec{h}_k$. This value of $\epsilon_k$ corresponds to the maximum value of $f(\vec{z}_{k+1})$ after a line search procedure.
  \item Set $\vec{g}_{k+1}=\vec{g}_{k}-\epsilon_k\overleftrightarrow{A}\!\cdot\vec{h}_k$.
  \item Set the parameter $t_{k+1}=\frac{\vec{g}_{k+1}\cdot\vec{g}_{k+1}}{\vec{g}_{k}\cdot\vec{g}_{k}}$.
  \item Set $\vec{h}_{k+1}=\vec{g}_{k+1}+t_{k+1}\vec{h}_k$.
  \item Set $k=k+1$ and repeat the iteration from the beginning.
\end{enumerate}
  \end{minipage}}
  \end{center}
Very often, the objective function $f(\vec{z})$ to be maximized is not a simple quadratic form\index{quadratic form} as described in Eq.~(\ref{quad_form}). This introduces a few complications to the simple CG algorithm outlined above. Firstly, the optimal value of $\epsilon_k$ is often not available readily as an analytical expression. Therefore, numerical methods have to be invoked in order to look for the value of $\epsilon_k$ such that $f(\vec{z}_{k+1})$ is maximal. In cases where such a numerical search for the optimal $\epsilon_k$ is computationally expensive, a fixed value of $\epsilon_k$ may be assigned. Secondly, we note that
\begin{equation}
\vec{g}_{k+1}\cdot\vec{g}_{k}=0
\end{equation}
when $f(\vec{z})$ is a quadratic form\index{quadratic form}. This follows from the fact that
\begin{align*}
\vec{g}_{k+1}\cdot\vec{g}_k &= \vec{g}_{k}\cdot\vec{g}_k-\epsilon_k\vec{h}_k\cdot\!\overleftrightarrow{A}\!\cdot\vec{g}_k\\
&=\vec{g}_{k}\cdot\vec{g}_k-\frac{\vec{g}_k\cdot\vec{g}_k}{\vec{h}_k\cdot\!\overleftrightarrow{A}\!\cdot\vec{h}_k}\vec{h}_k\cdot\!\overleftrightarrow{A}\!\cdot\vec{g}_k\\
&=\vec{g}_{k}\cdot\vec{g}_k-\frac{\vec{g}_k\cdot\vec{g}_k}{\vec{h}_k\cdot\!\overleftrightarrow{A}\!\cdot\vec{h}_k}\vec{h}_k\cdot\!\overleftrightarrow{A}\!\cdot\left(\vec{h}_k-t_k\vec{h}_{k-1}\right)\\
&=\vec{g}_{k}\cdot\vec{g}_k-\vec{g}_{k}\cdot\vec{g}_k=0\,.
\end{align*}
Therefore, we have that
\begin{equation}
t_{k+1}=\underbrace{\frac{\vec{g}_{k+1}\cdot\vec{g}_{k+1}}{\vec{g}_{k}\cdot\vec{g}_{k}}}_{\substack{\text{Fletcher-Reeves}\\\text{factor}}}=\underbrace{\frac{\vec{g}_{k+1}\cdot\left(\vec{g}_{k+1}-\vec{g}_k\right)}{\vec{g}_{k}\cdot\vec{g}_{k}}}_{\substack{\text{Polak-Ribi{\` e}re}\\\text{factor}}}\,.
\end{equation}
For a general function $f(\vec{z})$, the two factors are clearly different. It is known that the CG algorithm which uses the Fletcher-Reeves\index{Fletcher-Reeves} factor (Roger Fletcher\index{Roger Fletcher} and Colin Morrison Reeves\index{Colin Morrison Reeves}) converges only when the starting vector $\vec{z}_1$ is close to $\vec{z}_\text{max}$, and that which uses the Polak-Ribi{\` e}re\index{Polak-Ribi{\` e}re} factor (Elijah Polak\index{Elijah Polak} and Gerard Ribi{\` e}re\index{Gerard Ribi{\` e}re}) rarely diverges. This divergence can be prevented by defining the \emph{Polak-Ribi{\` e}re criterion}
\begin{equation}
\gamma_{k+1}=\max\left\{\frac{\vec{g}_{k+1}\cdot\left(\vec{g}_{k+1}-\vec{g}_k\right)}{\vec{g}_{k}\cdot\vec{g}_{k}},0\right\}\,.
\end{equation}\label{symbol:gammakp1}
This implies that when the Polak-Ribi{\` e}re\index{Polak-Ribi{\` e}re} factor becomes negative, the CG algorithm switches back to the DG algorithm. Putting the pieces together\index{algorithm}, we have:
\begin{center}
\colorbox{light-gray}{\begin{minipage}[c]{12cm}
  \textbf{\uline{Polak-Ribi{\` e}re CG method for general objective functions}}\\
  Beginning with $\vec{h}_1=\vec{g}_1=\vec{\nabla}_nf(\vec{z}_1)$ and $k=1$,
  \begin{enumerate}
  \item Compute $\epsilon_k$ using a line search procedure such that $f(\vec{z}_{k}+\epsilon_k\vec{h}_k)$ is maximal and set $\vec{z}_{k+1}=\vec{z}_k+\epsilon_k \vec{h}_k$.
  \item Set $\vec{g}_{k+1}=\vec{\nabla}_nf(\vec{z}_{k+1})$.
  \item Set the parameter $\gamma_{k+1}=\max\left\{\frac{\vec{g}_{k+1}\cdot\left(\vec{g}_{k+1}-\vec{g}_k\right)}{\vec{g}_{k}\cdot\vec{g}_{k}},0\right\}$.
  \item Set $\vec{h}_{k+1}=\vec{g}_{k+1}+\gamma_{k+1}\vec{h}_k$.
  \item Set $k=k+1$ and repeat the iteration from the beginning.
\end{enumerate}
  \end{minipage}}
  \end{center}
The main point of this short discourse is that the above algorithm can be generalized to the space of operators by simply replacing all numerical vectors by vector operators. The inner product of two vector operators $\vec{X}$ and $\vec{Y}$ is defined as $\left<\vec{X},\vec{Y}\right>=\tr{\left(\vec{X}\right)^\dagger\vec{Y}}$, where the trace operation is understood to act on the operators in $\vec{X}$ and $\vec{Y}$. To apply the conjugate-gradient strategy to ML, we first allow the operator vector $\vec{Z}=\transpose{\left(\mathcal{A},\mathcal{A}^\dagger\right)}$ to follow the search direction of the steepest ascent\index{steepest-ascent method}, namely $\updelta\vec{Z}\propto\vec{\partial}\log\mathcal{L}(\{n_j\};\rho)$. Subsequently, $\updelta\vec{Z}$ will follow a series of approximately conjugate search directions defined by the dyadic $\vec{\partial}\vec{\partial}\log\mathcal{L}(\{n_j\};\rho)$ \footnote{To visualize this more vividly, consider a quadratic form\index{quadratic form} of three parameters, contained in the vector $\vec{x}$, given by $f(\vec{x})=-\frac{1}{2}\vec{x}\cdot\!\overleftrightarrow{A}\!\cdot\vec{x}+\vec{b}\cdot\vec{x}$. Then, the three-dimensional gradient $\vec{\nabla}f(\vec{x})=\vec{b}-\overleftrightarrow{A}\!\cdot\vec{x}$ and the search directions $\vec{h}_j$ generated by the conjugate gradient method are related by $\vec{\nabla}\vec{\nabla}f(\vec{x})=-\overleftrightarrow{A}$.}. The standard Polak-Ribi{\` e}re CG method\index{algorithm}, when applied to ML, proceeds as follows:
\begin{center}
\colorbox{light-gray}{\begin{minipage}[c]{12cm}
  \textbf{\uline{ML algorithm using the standard Polak-Ribi{\` e}re CG method (ML-CG)}}\\
  Starting from the parameters $\mathcal{A}=1$, $\vec{Z}_1=\transpose{\left(\mathcal{A},\mathcal{A}^\dagger\right)}$, $\vec{G}_1=\vec{H}_1=\vec{\partial}\log\mathcal{L}(\rho_{1})$ and $k=1$,
\begin{enumerate}
  \item Compute $R_k$;
	    \begin{itemize}
	       \item Escape from loop if $\tr{|R_k\rho_k-\rho_k|}\leq\varepsilon$;
 	       \item Otherwise, proceed to the following steps.
 	    \end{itemize}
  \item Optimize $\epsilon_k$ such that $\mathcal{L}(\{n_j\};\rho_k)$ is maximum using a line search procedure and set $\vec{Z}_{k+1}=\vec{Z}_k+\epsilon_k \vec{H}_k$.
  \item Set $\vec{G}_{k+1}=\vec{\partial}\log\mathcal{L}(\{n_j\};\rho_{k+1})$.
  \item Set the parameter $\gamma_{k+1}=\max\Bigg\{\frac{\big\langle \vec{G}_{k+1},\vec{G}_{k+1}-\vec{G}_k\big\rangle}{\big\langle \vec{G}_k,\vec{G}_k\big\rangle}, 0\Bigg\}$ (Polak-Ribi{\` e}re).
  \item Set $\vec{H}_{k+1}=\vec{G}_{k+1}+\gamma_{k+1}\vec{H}_k$.
  \item Set $k=k+1$ and repeat the iteration from the beginning.
\end{enumerate}
  \end{minipage}}
  \end{center}
%

\begin{figure}[h!]
\centering
\includegraphics[width=0.6\textwidth]{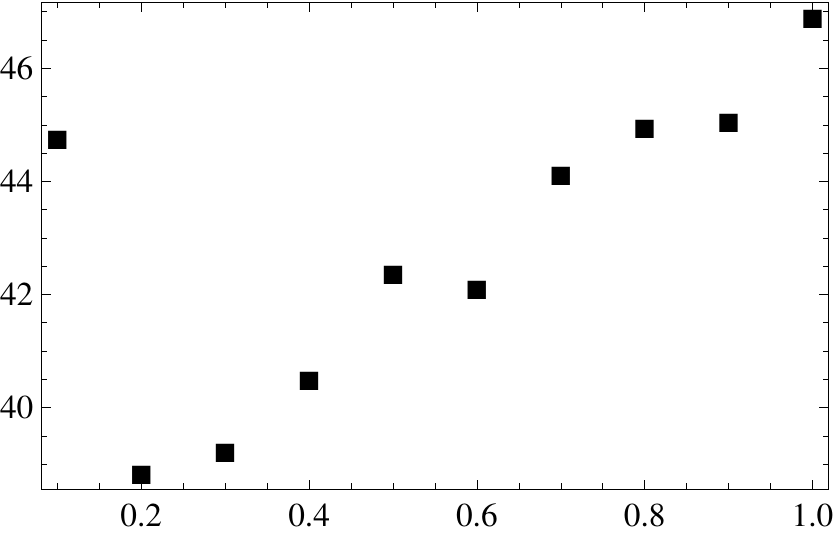}
\put(-235,10){\small\begin{rotate}{90}Average number of steps\end{rotate}}
\put(-100,-8){\small$\xi$}
\caption{\label{fig:prsuperposition}%
Single-qubit state simulated with $10^3$ detection copies over 100 experimental runs. We analyze the performance of ML-CG ($\blacksquare$) in terms of the average number of iterations over the experimental runs. Here the precision $\varepsilon$ is set to $10^{-7}$. In general, lower $\xi$ values can further boost the performance of both schemes.}
\end{figure}

We remind ourselves that the efficiency of the ML-CG method will be higher if the functional to be optimized is very close to a quadratic form\index{quadratic form} in the space of parameters, in which case $\big\langle \vec{G}_{k+1},\vec{G}_k\big\rangle\approx 0$ and $\gamma_{k+1}>0$. Since the likelihood functional $\mathcal{L}(\{n_j\};\rho)$ deviates far away from a quadratic form\index{quadratic form} in $\vec{Z}$, this term can be significant in value, causing $\gamma_{k+1}$ to be constantly reset to 0 and thereby turning the ML-CG method back to steepest-ascent\index{steepest-ascent method}. Hence it is fruitful to consider a new Polak-Ribi{\` e}re\index{Polak-Ribi{\` e}re} criterion, namely
\begin{equation}
\gamma'_{j+1}\rightarrow \max\Bigg\{\frac{\big\langle \vec{G}_{j+1},\vec{G}_{j+1}-\xi\vec{G}_j\big\rangle}{\big\langle \vec{G}_j,\vec{G}_j\big\rangle}, 0\Bigg\}\,,
\end{equation}
where $\xi$\label{symbol:xi} is a suitably chosen parameter, which is less than 1, such that the factor $\xi\big\langle \vec{G}_{k+1},\vec{G}_k\big\rangle$ is small. If $\xi$ is set to 0, corresponding to the Fletcher-Reeves scheme, the ML-CG algorithm may not converge. In general, the optimal value of $\xi$ that gives the minimal average number of iterative steps to achieve a certain numerical precision $\varepsilon$ depends very much on $\rho_\text{true}$. In view of this, we set $\xi=0.5$ for any $\rho_\text{true}$. From hereon, the ML-CG algorithm is defined with the new Polak-Ribi{\` e}re criterion\index{Polak-Ribi{\` e}re}. Figure \ref{fig:prsuperposition} gives a numerical simulation on a single-qubit state $\ket{\,\,\,}=(\ket{0}+\ket{1}\I)/\sqrt{2}$, where $\ket{0}$ and $\ket{1}$ are two orthogonal kets.

To investigate the performance of ML-CG numerically, Monte Carlo simulations are carried out on a unitarily-invariant random ensemble of full-rank two-qubit mixed states. To generate each random mixed state $\rho_\text{true}$, we choose four random normalized kets $\{\ket{\psi_k}\}^3_{k=0}$ and four random complex numbers $\{\alpha_k\}^3_{k=0}$. Then each mixed state is defined as
\begin{equation}
\rho_\text{true}=\sum^3_{k=0}\ket{\psi_k}\frac{|\alpha_k|^\nu}{\sum^3_{k'=0}{|\alpha_k'|^\nu}}\,,
\end{equation}
where $\nu$ is an integer parameter which we vary to obtain random mixed states of varying ranges of purity. To compute the optimal value of $\epsilon_k$ in the $k$th step, we evaluate the likelihood functional at ten different values of $\epsilon_k$ and perform a quadratic curve fitting to obtain the approximate maximum of the likelihood functional. For the POM outcomes, we use the tensor products of the single-qubit SIC POM (also known as the \emph{tetrahedron} measurement\index{tetrahedron measurement}\index{tetrahedron measurement|seealso{symmetric informationally complete POM (SIC POM)}}) subnormalized projectors \cite{prod_sic,qst_huangjun}. These four rank-1 outcomes of the tetrahedron measurement have Bloch vectors defined by
\begin{equation}
\left\{
\vec{a}_1=
\frac{1}{\sqrt{3}}
\begin{pmatrix}
1\\
1\\
1
\end{pmatrix},
\vec{a}_2=
\frac{1}{\sqrt{3}}
\begin{pmatrix}
1\\
-1\\
-1
\end{pmatrix},
\vec{a}_3=
\frac{1}{\sqrt{3}}
\begin{pmatrix}
-1\\
-1\\
1
\end{pmatrix},
\vec{a}_4=
\frac{1}{\sqrt{3}}
\begin{pmatrix}
-1\\
1\\
-1
\end{pmatrix}
\right\}\,.
\label{bloch0}
\end{equation}
This product measurement forms a minimal set of 16 informationally complete POM outcomes.

\begin{figure}[h!]
\centering
\includegraphics[width=0.6\textwidth]{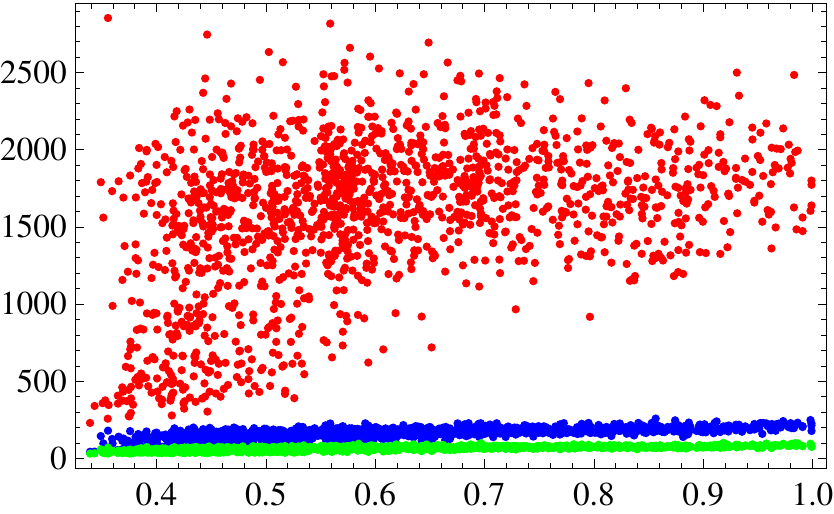}
\put(-225,10){\small\begin{rotate}{90}Average number of steps\end{rotate}}
\put(-113,-8){\small$\tr{\rho^2_\text{true}}$}
\caption{\label{fig:2qubit-iter}%
A total of 1500 random two-qubit full-rank mixed state were simulated with eight thousand detection copies over fifty experimental runs. By fixing the precision $\varepsilon=10^{-7}$, the scatter plots of the average number of iterative steps over the experimental runs for ML-DG with fixed $\epsilon_k=\epsilon$ (ML-DG~I) (Red), ML-DG with optimized $\epsilon_k$ (ML-DG~II) (Blue) and ML-CG (Green) indicate an expected trend. For all the randomly chosen states, ML-CG outperforms ML-DG~II with an average improvement of about 55\%. On average, ML-CG requires about 95\% less number of iterative steps than ML-DG~I for the same precision.}
\end{figure}

\begin{figure}[h!]
\centering
\includegraphics[width=0.6\textwidth]{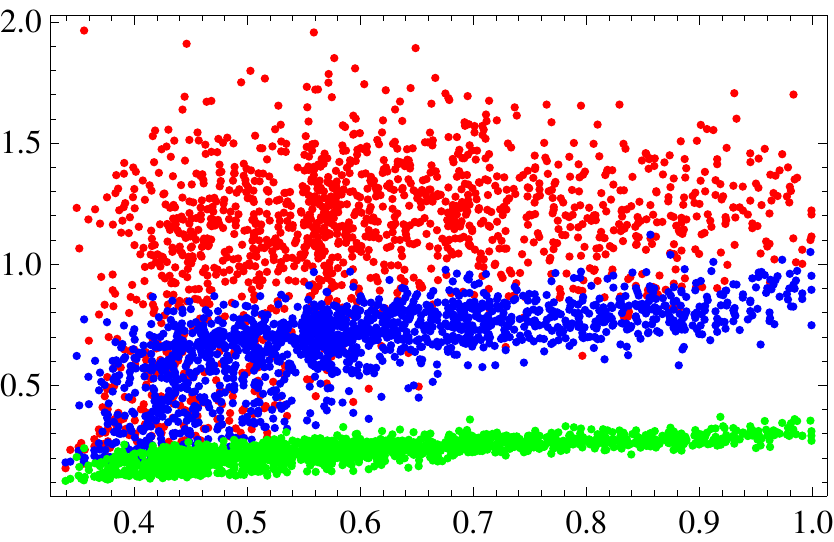}
\put(-225,30){\small\begin{rotate}{90}Average duration /s\end{rotate}}
\put(-113,-8){\small$\tr{\rho^2_\text{true}}$}
\caption{\label{fig:2qubit-time}%
Here is the corresponding plot of the average duration of one complete run of each of the three schemes: ML-DG~I(Red), ML-DG~II (Blue) and ML-CG (Green). The average improvement on which ML-CG outperforms ML-DG~II, in terms of the average duration of one complete run, is about 65\%. The corresponding improvement of ML-CG over ML-DG~I is about 75\%.}
\end{figure}

All simulations are conducted with Mathematica on an Intel i7 Quad Core 2.67~GHz machine. Figures~\ref{fig:2qubit-iter} and \ref{fig:2qubit-time} give the simulated results.
Notice, however, that the improvement, in terms of average duration of each full run, of ML-CG over ML-DG~I is generally smaller than that in terms of the average number of iterations required to complete a full run. The reason lies in the computation of matrix multiplications which can be significant in the conjugate gradient methods as $D$ increases. Nevertheless, ML-CG shows the best average convergence rate for all the randomly generated two-qubit mixed states in terms of both the average number of iterations and average duration compared to all other schemes.

Next, we present two sets of simulation data for four-qubit tomography on the GHZ and W states. Let us emphasize that as the dimension of the Hilbert space increases, the computational cost for evaluating large matrices becomes more significant, especially in the likelihood functional computations required for the quadratic interpolation procedure. This is eminent in four-qubit state estimation. In this case, we also consider performing ML-CG with fixed $\epsilon_k$ to reduce the overall time required to compute a full run of the algorithm.
\begin{table}
\begin{center}
\begin{tabular}{ccccc}
\multicolumn{5}{c}{\underline{\makebox[20em][c]{GHZ state}}%
\rule{0pt}{4ex}}
\\
                       & \,ML-DG~I\, & \, ML-DG~II \, & \, ML-CG~I \, & \,ML-CG~II\,\\
Iterations & 112    & 42                 & 39 & 33                 \\
Duration / s      & 1.069  & 4.416 &     0.688 & 4.113       \\
\hline
\hline
\end{tabular}\\
\begin{tabular}{ccccc}
\multicolumn{5}{c}{\underline{\makebox[20em][c]{W state}}%
\rule{0pt}{4ex}}
\\
                        & \,ML-DG~I\, & \, ML-DG~II \, & \, ML-CG~I \, & \,ML-CG~II\,\\
Iterations & 936    & 133 & 258 & 154                     \\
Duration / s      & 9.288  & 14.330 & 4.759 & 18.123           \\
\hline
\hline
\end{tabular}
\end{center}
\label{GHZ_W}
\caption{Table of the average number of iterations and average duration to complete one full run of the respective iterative schemes for the four-qubit GHZ and W states. The POM for the simulations consists of the tensor products of four single-qubit tetrahedron outcomes. The above illustrates that on average, ML-CG~I, which is ML-CG with fixed $\epsilon_k$, performs better, in terms of the average duration of a full run, than the regular direct and conjugate-gradient schemes with $\epsilon_k$ optimization, even though the average number of iterations can sometimes be significantly reduced using the optimized schemes. The additional time taken for the type II algorithms is mainly due to the heavy matrix evaluations in the line search procedure.}
\end{table}
Finally, we compare the performances of ML-DG and ML-CG by performing quantum state estimation on \emph{one} simulated set of measurement data for an eight-qubit pure state \cite{eight_qubit} with MATLAB. The POM used is the set of $2^{16}=65536$ different tensor products of eight single-qubit tetrahedron outcomes. In practice, it is difficult to store all the 65536 outcomes into memory on a personal computer and so we generate all these outcomes on the fly in each iterative step of the algorithms. In addition, the evaluation of these $256\times 256$ operators is extremely costly. These factors, together, cause a tremendous slowdown in the durations of the algorithms. Hence, the type I algorithms are naturally more practical in this situation than type II algorithms. The simulations show that ML-DG~I takes about 143 hours to complete the run up to a fixed numerical precision $\varepsilon$, whereas ML-CG~I takes about 95 hours to achieve the same precision. Thus, ML-CG~I does in fact offer a more optimistic alternative for quantum state estimation involving quantum systems living in large Hilbert spaces. It is important to note that the conjugate-gradient methodology we have presented in this section is applicable to any algorithm that is based on the steepest-ascent method, as the machineries established are a natural extension to those of steepest-ascent. 

\section{Informationally incomplete quantum state estimation}\index{informationally incomplete}

If the POM used for measurement is informationally complete, then there exists a unique estimator $\hat{\rho}_\text{ML}\geq 0$ that maximizes $\mathcal{L}(\{n_j\};\rho)$. However, if the POM is not informationally complete, then there are infinitely many estimators that maximize $\mathcal{L}(\{n_j\};\rho)$ for a given set of $f_j$s. In fact, because of the concavity\index{concavity} of $\mathcal{L}(\{n_j\};\rho)$, the existence of two such estimators $\hat{\rho}_1$ and $\hat{\rho}_2$ implies the existence of a continuous family of estimators $\hat{\rho}'=\lambda\hat{\rho}_1+(1-\lambda)\hat{\rho}_2$, where $0\leq\lambda\leq 1$. Therefore in order to systematically choose one estimator for statistical prediction, we shall consider the \emph{principle of entropy maximization} (ME)\index{maximum-entropy (ME)} that goes way back to two papers by Edwin Thompson Jaynes~\cite{jaynes1,jaynes2}\index{Edwin Thompson Jaynes} in 1957. In doing so, one can always obtain a unique estimator that maximizes both $\mathcal{L}(\{n_j\};\rho)$ and the von Neumann entropy\index{entropy!-- von Neumann} functional $S(\rho)=-\tr{\rho\log\rho}$\label{symbol:srho} (John von Neumann\index{John von Neumann}). J.~{\v R}eh{\'a}{\v c}ek \emph{et al.} had looked into this ML-assisted ME technique in particular for commuting POM outcomes~\cite{rehacek3} and the photon-number statistics of light~\cite{rehacek4}. This section develops iterative schemes that are applicable for general situations. 

\subsection{General iterative scheme}

The original idea of ML-assisted ME considered by J. {\v R}eh{\'a}{\v c}ek \emph{et al.} involves two steps. The first step is to perform the ML procedure in order to look for the estimators $\hat{\rho}_\text{ML}$ that maximize $\mathcal{L}(\{n_j\};\rho)$ given a fixed set of measured frequencies $f_j$s from a informationally incomplete POM with $K$ outcomes. In this case, there are infinitely many such ML estimators and as a result, the likelihood functional forms a plateau\index{plateau} on the space of statistical operators. The second step is to select the estimator with the maximum value of $S(\rho)$. Such a procedure is equivalent to raising the plateau\index{plateau} into a concave hill so that the resulting estimator chosen gives the globally maximum value. In this way, a unique maximum-likelihood-maximum-entropy (MLME)\index{maximum-likelihood-maximum-entropy (MLME)} estimator\index{unique estimator} can always be obtained for statistical predictions. We do this by considering the Lagrange functional\index{Lagrange functional} (Joseph-Louis Lagrange\index{Joseph-Louis Lagrange}) $\mathcal{D}$\label{symbol:dfunctional} involving the von Neumann entropy functional $S(\rho)$ and the constraints $\tr{\hat{\rho}_\text{ML}\Pi_j}=\tr{\rho\Pi_j}=p_j$ as well as $\tr{\rho}=1$ defined as
\begin{equation}
\mathcal{D}(\rho)=-\tr{\rho\log\rho}+\sum_j\lambda_j\left(p_j-\tr{\hat{\rho}_\text{ML}\Pi_j}\right)+\log\mu\left(\tr{\rho}-1\right)\,,
\end{equation}
where the $\lambda_j$s\label{symbol:lagmult} and $\log\mu$ are the Lagrange multipliers\index{Lagrange multipliers} for the constraints. Varying $\mathcal{D}$ yields
\begin{equation}
\updelta\mathcal{D}(\rho)=\tr{\updelta\rho\left(-\log\rho+\sum_j\lambda_j\Pi_j+\log\mu\right)}\,.
\end{equation}
Thus, setting $\updelta\mathcal{D}(\rho)$ to zero gives the maximum-entropy (ME)\index{maximum-entropy (ME)} estimator of the form
\begin{equation}
\hat{\rho}_\text{ME}=\frac{\E^{\sum_j \lambda_j\Pi_j}}{\tr{\E^{\sum_j \lambda_j\Pi_j}}}\,
\label{maxentstate}
\end{equation}\label{symbol:rhome}which maximizes $S(\rho)$ under the set of constraints, after setting
\begin{equation}
\mu=\frac{1}{\tr{\E^{\sum_j \lambda_j\Pi_j}}}\,.
\end{equation}
The task now is to look for the Lagrange multipliers\index{Lagrange multipliers} using the above constraints. This requires the solutions to a set of nonlinear equations which in general may not be conveniently obtained, especially when the operators $\Pi_j$ do not commute.

An alternative idea is to maximize the likelihood functional $\mathcal{L}(\{n_j\};\rho)$ by optimizing $\lambda_j$ of the estimator in Eq.~(\ref{maxentstate}) so that the resulting MLME estimator $\hat{\rho}_\text{MLME}$\label{symbol:rhomlme} is the one that maximizes $\mathcal{L}(\{n_j\};\rho)$ and is automatically the maximum-entropy\index{maximum-entropy (ME)} estimator. An interesting observation\footnote{Thanks to Dr. Ng Hui Khoon, a research fellow in CQT, for pointing this out.} is that the Lagrange multipliers\index{Lagrange multipliers} are not all independent. This stems from the completeness of the POM $\sum_j\Pi_j=1$ which implies that there are altogether $K-1$ independent constraints for the Lagrange multipliers\index{Lagrange multipliers}. As such one may choose to optimize only $K-1$ Lagrange multipliers\index{Lagrange multipliers}.

Varying $\log\mathcal{L}(\{n_j\};\rho)$ yields
\begin{equation}
\updelta \log\mathcal{L}(\{n_j\};\rho)=N\tr{R\updelta\rho}\,,
\label{varlike}
\end{equation}
where $R\equiv\sum_jf_j\Pi_j/p_j$. Using $\rho$ of the form in Eq.~(\ref{maxentstate}), the variation
\begin{equation}
\updelta\rho=\frac{\updelta O}{\tr{O}}-\rho\frac{\tr{\updelta O}}{\tr{O}}\,,
\label{varrho}
\end{equation}
with $O\equiv\E^{\sum_j \lambda_j\Pi_j}$, involves the variation of $O$ and this is carried out by noting that given an operator $B$,
\begin{equation}
\updelta\E^{B}=\int_0^1\D x\,\E^{(1-x)B}\updelta B\E^{xB}\,.
\end{equation}

Substituting Eq.~(\ref{varrho}) into Eq.~(\ref{varlike}), the resulting variation of the log-likelihood functional $\log\mathcal{L}(\{n_j\};\rho)$ is derived to be
\begin{equation}
\updelta \log\mathcal{L}(\{n_j\};\rho)=N\sum_j\updelta{\lambda_j}\tr{\rho\Pi_j\left(\int^1_0\D x\,\E^{x\sum_j\lambda_j\Pi_j}R\,\E^{-x\sum_j\lambda_j\Pi_j}-1\right)}\,,
\end{equation}
where $N$ is the number of detection copies of the quantum system.
One can immediately find that the derivative of $\log\mathcal{L}(\{n_j\};\rho)$ with respect to $\lambda_j$ is
\begin{equation}
\frac{\partial}{\partial\lambda_j}\log\mathcal{L}(\{n_j\};\rho)=N\tr{\rho\Pi_j\left(\int^1_0\D x\,\E^{x\sum_j\lambda_j\Pi_j}R\,\E^{-x\sum_j\lambda_j\Pi_j}-1\right)}\,.
\end{equation}
Hence the maximal value of $\log\mathcal{L}(\{n_j\};\rho)$ is attained when the extremal equation
\begin{equation}
\int^1_0\D x\,\E^{x\sum_j\hat{\lambda}^\text{MLME}_j\Pi_j}\hat{R}_\text{MLME}\,\E^{-x\sum_j\hat{\lambda}^\text{MLME}_j\Pi_j}=1_{\hat{\rho}_\text{MLME}}
\label{extremal1}
\end{equation}
is satisfied, where $\hat{R}_\text{MLME}\equiv R\left(\hat{\rho}_\text{MLME}\right)$ and $1_{\hat{\rho}_\text{MLME}}$ is the identity operator on the support of $\hat{\rho}_\text{MLME}$. This is of course obvious in hindsight since Eq.~(\ref{extremal1}) is equivalent to the statement
\begin{equation}
\hat{R}_\text{MLME}\,\hat{\rho}_\text{MLME}=\hat{\rho}_\text{MLME}\,\hat{R}_\text{MLME}=\hat{\rho}_\text{MLME}\,
\end{equation}
as in the case of ML.

With the above setting, we can construct an iterative scheme MLME based on the principle of steepest-ascent\index{steepest-ascent method}. Since the $\updelta\lambda_j$s are arbitrary, we can set each variation as follows:
\begin{align}
\updelta\lambda_j&=\epsilon N\partial_j\log\mathcal{L}(\{n_j\};\rho)\equiv\epsilon\frac{\partial}{\partial\lambda_j}\log\mathcal{L}(\{n_j\};\rho)\nonumber\\
&=\epsilon\tr{\rho\Pi_j\left(\int^1_0\D x\,\E^{x\sum_j\lambda_j\Pi_j}R\,\E^{-x\sum_j\lambda_j\Pi_j}-1\right)}\,,
\end{align}
where $\epsilon$ is a positive parameter which defines the step size taken in every iterative step. So now the iteration proceeds by a step of size $\epsilon$ along the direction of the gradient $\partial_j\log\mathcal{L}(\{n_j\};\rho)$ in each step. We thus have the variation $\updelta \log\mathcal{L}(\{n_j\};\rho)$ to be always positive. The MLME scheme is then given by
\begin{center}
\colorbox{light-gray}{\begin{minipage}[c]{12cm}
  \uline{\textbf{Scheme A}}
\begin{align}
\rho_{k+1}&=\frac{\E^{\sum_j \left(\lambda^{(k)}_j+\epsilon\partial_j \log\mathcal{L}(\{n_j\};\rho_k)\right)\Pi_j}}{\tr{\E^{\sum_j \left(\lambda^{(k)}_j+\epsilon\partial_j \log\mathcal{L}(\{n_j\};\rho_k)\right)\Pi_j}}}\,,\\
\partial_j \log\mathcal{L}(\{n_j\};\rho_k)&=N\tr{\rho_k\Pi_j\left(\int^1_0\D x\,\E^{x\sum_j\lambda^{(k)}_j\Pi_j}R_k\,\E^{-x\sum_j\lambda^{(k)}_j\Pi_j}-1\right)}\,.
\label{scheme1}
\end{align}
  \end{minipage}}
  \end{center}
As in the ML iterative scheme, one can always start from the maximally-mixed state. 

The fruit of the above discussion is an iterative scheme that looks for the MLME estimator directly rather than taking the ML-assisted ME approach which involves two steps and a set of nonlinear equations. This iterative scheme is conveniently applicable for general POMs and tomography in any Hilbert space dimension. In general, the CPU time for exponentiating a square matrix is acceptable even for matrices as large as $10\times 10$ using commercial optimized algorithms. The only practical shortcoming in this scheme is the long CPU time required to perform the numerical integration in each iterative step and this can be quite serious as the dimension of the Hilbert space increases. One possible way of circumventing the problem is to approximate the variation of the matrix exponential of an operator $A$ as\footnote{Thanks to Zhu Huangjun who suggested this approximation.}
\begin{equation}
\updelta \mathrm{e}^A\approx\frac{1}{2}(\mathrm{e}^A\delta A+\delta A\,\mathrm{e}^A)\,.
\label{approx}
\end{equation}
Then the direction of ascent in every step of \textbf{Scheme A} is given by
\begin{equation}
\partial_j \log\mathcal{L}(\{n_j\};\rho_k)=N\tr{\rho_k\left(\frac{\Pi_jR_k+R_k\Pi_j}{2}-\Pi_j\right)}\,.
\end{equation}
In this way, the integration procedure can be avoided.

At this point, we would like to make a distinction between this MLME technique and the conventional ME technique\index{maximum-entropy (ME)} \cite{me1,me2}. The ME technique takes the outcome frequencies $f_j$ as the probabilities $p_j$ and tries to search for the positive operator in Eq.~(\ref{maxentstate}) by maximizing $S(\rho)$, subjected to the probability constraints which are mediated by the Lagrange multipliers $\lambda_j$. The fundamental problem with this scheme is that, in general, the $f_j$s cannot be treated as probabilities since they correspond to an operator which is not necessarily positive. This is due to the statistical noise which is inherent in the outcome frequencies arising from measuring finite copies of quantum systems. Therefore, in such cases, the ME technique fails as there simply is no positive operator which is consistent with the measurement data to begin with. The MLME algorithm, on the other hand, looks for the unique MLME estimator\index{unique estimator} by confining the search within the plateau\index{plateau} region inside the Hilbert space. Thus, positivity is ensured. In cases where the $f_j$s are probabilities, both the ME and MLME schemes yield the same estimator by construction since the estimated probabilities $\hat p_j=f_j$ correspond to a statistical operator.

We compare the MLME scheme with the standard ME scheme using the simple example of a trine POM\index{trine POM} defined by the equations
\begin{align}
\Pi_0&=\frac{1}{3}\left(1+\sigma_z\right)\,,\nonumber\\
\Pi_\pm&=\frac{1}{3}\left(1\pm\frac{\sqrt{3}}{2}\sigma_x-\frac{1}{2}\sigma_z\right)\,,
\label{trine}
\end{align}
where the Pauli operators\index{Pauli operators} (Wolfgang Ernst Pauli\index{Wolfgang Ernst Pauli}) $\sigma_x$, $\sigma_y$ and $\sigma_z$ are given by
\begin{equation}
\sigma_x\widehat{=}
\begin{pmatrix}
\;0\;&\;1\;\\
\;1\;&\;0\;
\end{pmatrix},\,
\sigma_y\widehat{=}
\begin{pmatrix}
\;0\;&\;-\I\;\\
\;\I\;&\;0\;
\end{pmatrix},\,\sigma_x\widehat{=}
\begin{pmatrix}
\;1\;&\;0\;\\
\;0\;&\;-1\;
\end{pmatrix}\,.
\end{equation}
A straightforward calculation shows that when $n_0=6$, $n_+=2$ and $n_-=1$ after measuring $N=9$ copies for instance, the standard ME scheme fails as no quantum state has the frequencies $f_0=2/3$, $f_+=2/9$ and $f_-=1/9$ as probabilities. On the other hand, the MLME scheme still gives a positive estimator described by the Bloch vector $(0.194,0,0.981)$ for those frequencies, thus showing its versatility. Only when the frequencies are probabilities giving positive estimators may we use the ME scheme and in this case, the MLME scheme naturally incorporates these constraints.

\subsection{Qubit tomography}

In this example, to benchmark the MLME iterative scheme, qubit tomography simulations are performed using the trine POM defined in Eq.~(\ref{trine}). In this case, no expectation value is measured along the $y$ direction in the three-dimensional Bloch representation. One can easily show that the maximum-entropy\index{maximum-entropy (ME)} estimator $\hat{\rho}_\text{ME}$ for the true state will always be represented by a real and positive matrix by simply minimizing the purity of the estimator since for any qubit statistical operator, a decrease in its purity corresponds to an increase in its entropy. In the simulation, we fix $N=10^6$, $\epsilon=10$ and $\ket{\,\,\,\,}=\left(\ket{0}+\ket{1}\I\right)/\sqrt{2}$, where $\ket{0}\,\widehat{=}\,\transpose{(1,0)}$ and $\ket{1}\,\widehat{=}\,\transpose{(0,1)}$. Therefore
\begin{equation}
\rho_\text{true}\,\widehat{=}\begin{pmatrix}
\texttt{0.5} & \texttt{-0.5 i}\\
\texttt{0.5 i}   & \texttt{0.5}
\end{pmatrix}.
\end{equation}
Since this is an eigenstate of $\sigma_y$, we will ideally have $\langle\sigma_x\rangle =\langle\sigma_z\rangle =0$. This implies that $p_j=\langle\Pi_j\rangle=1/3$ and we expect the final MLME estimator to be the maximally-mixed state.
The MLME \textbf{Scheme A} gives the unique estimator\index{unique estimator}
\begin{equation}
\hat{\rho}_\text{MLME}\,\widehat{=}\begin{pmatrix}
\texttt{ 0.499953} & \texttt{-0.000169978}\\
\texttt{-0.000169978}   & \texttt{ 0.500047}
\end{pmatrix}
\end{equation}
which is consistent with the expected result, under an iteration time of 0.015~s using the approximated gradient expression for a precision $\varepsilon=10^{-7}$ in a particular simulated experimental run on an Intel(R) Core(TM) i7 2.66 GHz computer using Mathematica.

\subsection{Two-qubit tomography}

For simplicity, we consider two different informationally incomplete POMs. The first POM consists of nine outcomes that are tensor products of a pair of qubit trine POM outcomes as in Eq. (\ref{trine}). In this case, there will be expectation values for observables\index{observables} which depend on $\sigma_y$ like $\langle\sigma_x\otimes\sigma_y\rangle$, etc. However, as it turns out, the MLME estimator is still a real statistical operator in the computational basis, with the six expectation values $\tr{\hat{\rho}_\text{MLME}\left(1\otimes\sigma_y\right)}$, $\tr{\hat{\rho}_\text{MLME}\left(\sigma_y\otimes1\right)}$, $\tr{\hat{\rho}_\text{MLME}\left(\sigma_x\otimes\sigma_y\right)}$, $\tr{\hat{\rho}_\text{MLME}\left(\sigma_y\otimes\sigma_x\right)}$, $\tr{\hat{\rho}_\text{MLME}\left(\sigma_y\otimes\sigma_z\right)}$ and $\tr{\hat{\rho}_\text{MLME}\left(\sigma_z\otimes\sigma_y\right)}$ all equal to zero.

\begin{figure}
  \centering
  \subfloat[True state]{\label{fig:2-qubit_true_trine}\includegraphics[width=0.4\textwidth]{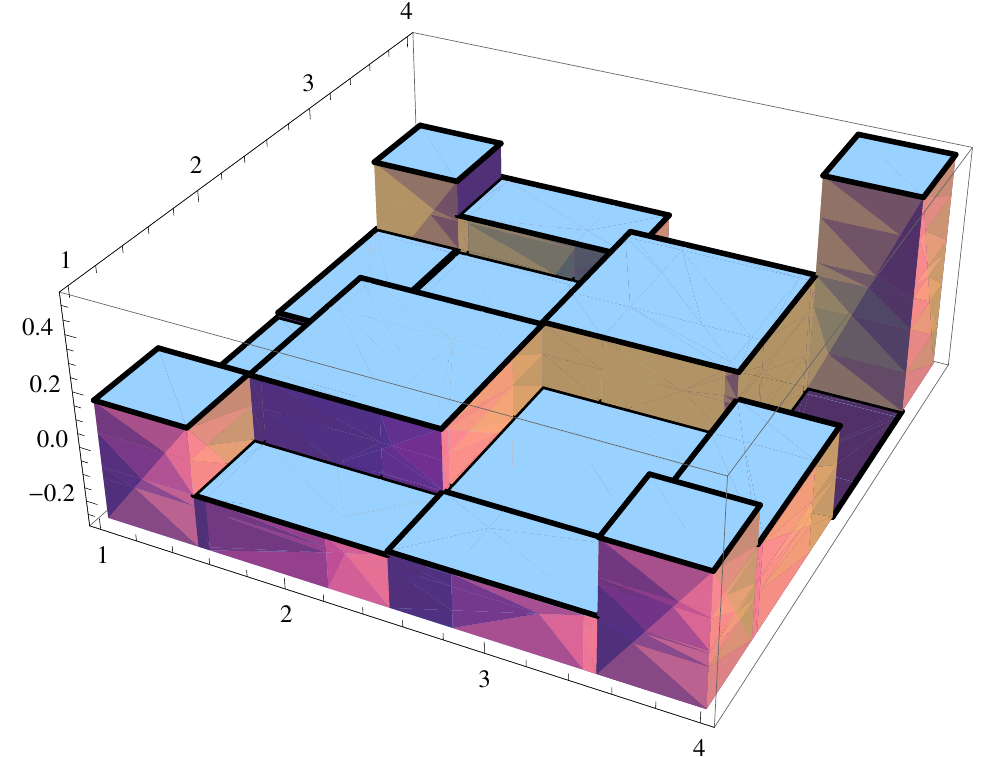}}
  \,\,\,\,\subfloat[Reconstructed state]{\label{fig:2-qubit_maxent_trine}\includegraphics[width=0.4\textwidth]{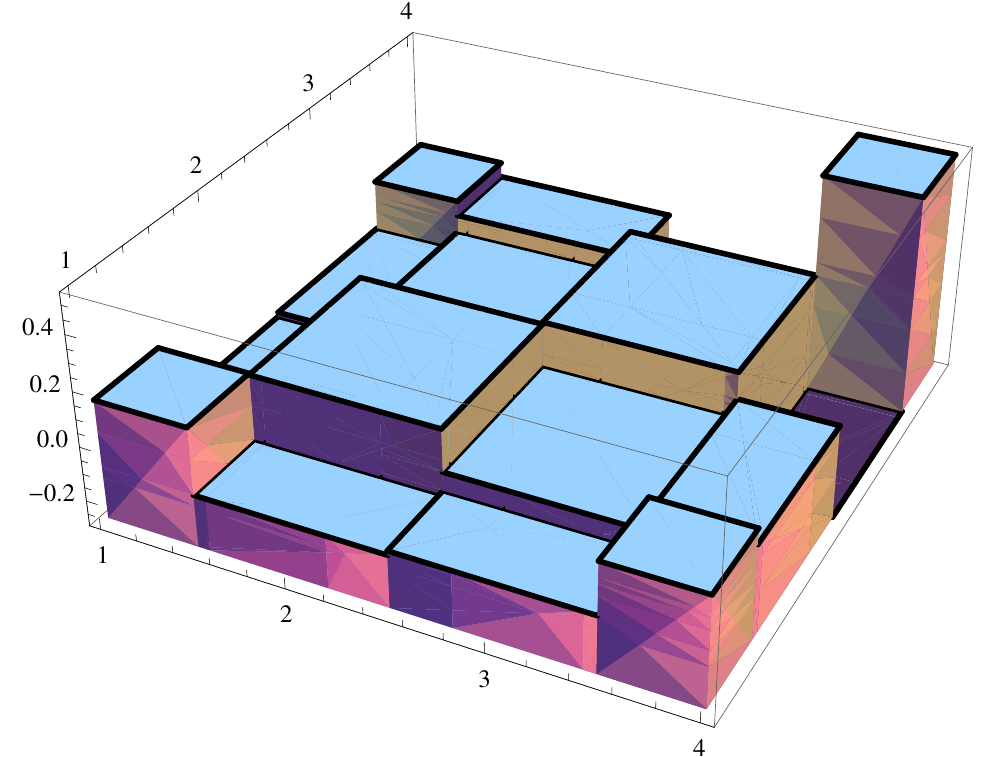}}
  \caption{Two-qubit tomography using joint trine POMs consisting of nine outcomes. A Monte Carlo simulation is performed with the number of detection copies $N=10^6$ on a random true state described by a \emph{real} statistical operator. The vertical axis represents the real matrix elements for both the true and reconstructed statistical operators in the computational basis\index{computational basis}. The horizontal axes respectively represents the row and column labels of the matrices. The trace-class distance $\mathcal{D}_\text{tr}=\tr{|\hat{\rho}_\text{MLME}-\rho_\text{true}|}/2$\label{symbol:distancetraceclass}\index{trace-class distance} is 0.158.}
  \label{fig:trine}
\end{figure}

\begin{figure}
  \centering
  \subfloat[True state]{\label{fig:2-qubit_retrue_randpom}\includegraphics[width=0.4\textwidth]{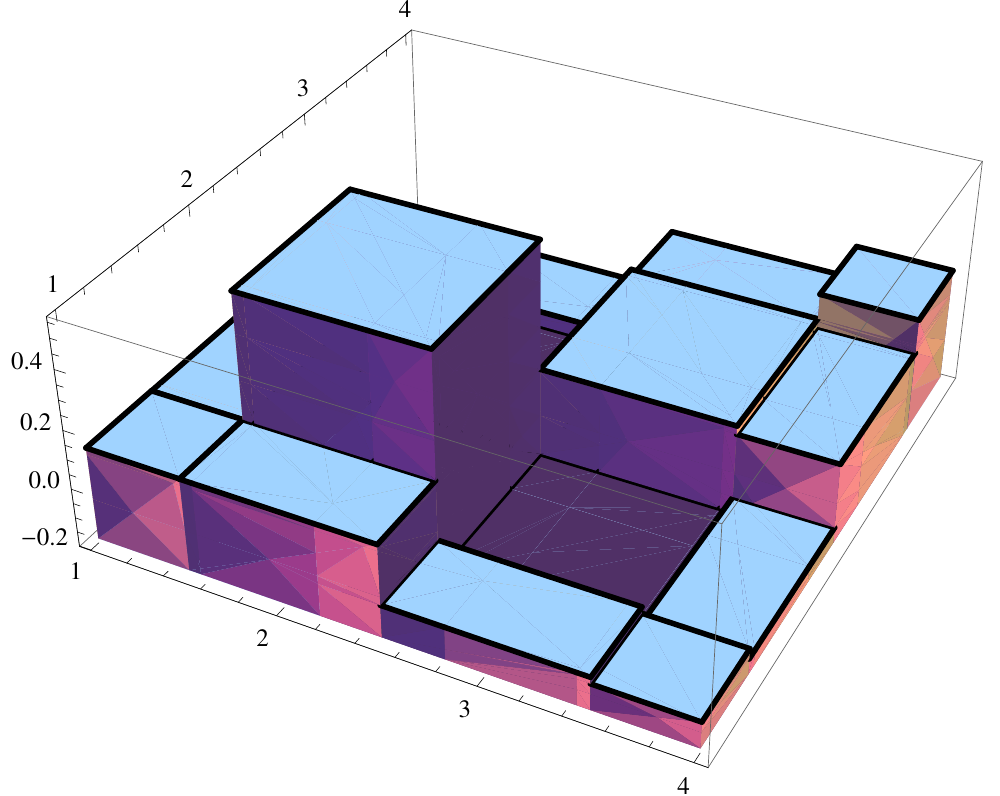}}
  \,\,\,\,\subfloat[Reconstructed state]{\label{fig:2-qubit_remaxent_randpom}\includegraphics[width=0.4\textwidth]{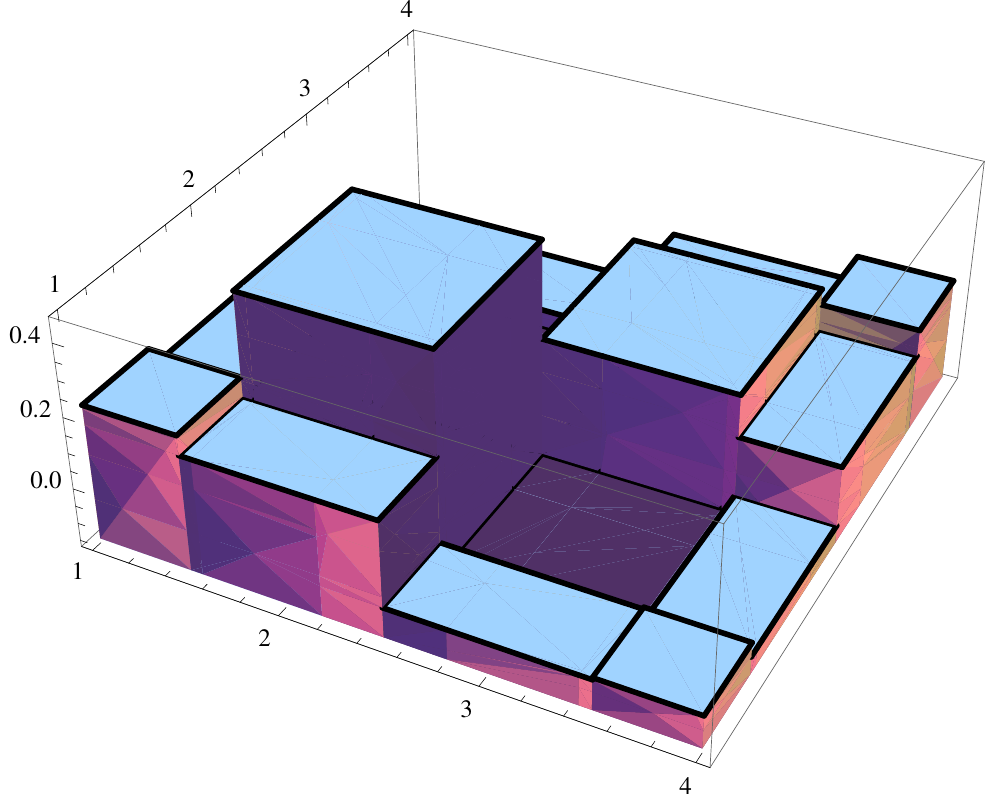}}\\
  \subfloat[True state]{\label{fig:2-qubit_imtrue_randpom}\includegraphics[width=0.4\textwidth]{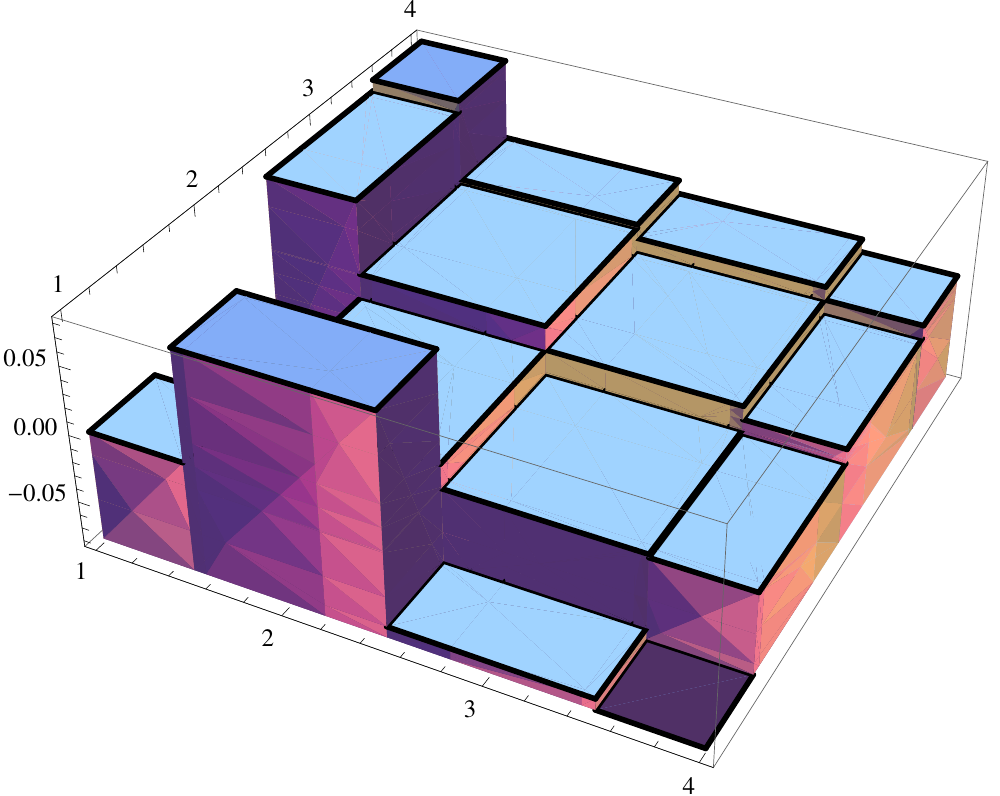}}
  \,\,\,\,\subfloat[Reconstructed state]{\label{fig:2-qubit_immaxent_randpom}\includegraphics[width=0.4\textwidth]{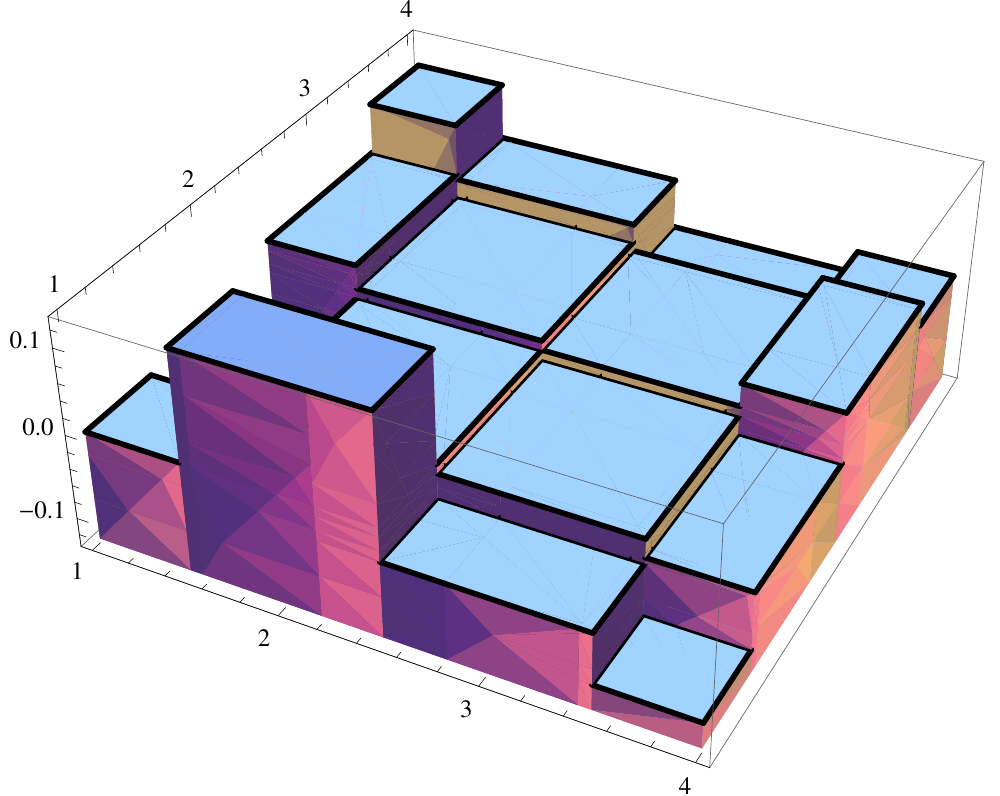}}\\
  \caption{Two-qubit tomography using a random two-qubit POM consisting of nine full-rank outcomes. A Monte Carlo simulation is performed with $N=10^6$ on a random true state represented by a complex positive matrix of unit trace. The vertical axis in each of (a) and (b) represents the real matrix elements of the respective true and reconstructed statistical operators in some computational basis\index{computational basis} and that in each of (c) and (d) represents the respective imaginary matrix elements. In this case $\mathcal{D}_\text{tr}=0.206$.}
  \label{fig:randpom}
\end{figure}

For the second POM, we emphasize the versatility of the MLME scheme by choosing a random POM consisting of nine outcomes by first generating nine random complex operators $B_j$ and then defining
\begin{equation}
\Pi_j=\chi^{-1/2}B^\dagger_jB_j\chi^{-1/2}\,,
\end{equation}
where $\chi=\sum_kB^\dagger_kB_k$. Care has to be taken to ensure that $\chi$ has full rank, which is the typical situation if the operators $B_j$ are randomly chosen. Using this POM, the maximum-entropy estimator is in general a complex statistical operator. The results are shown in Figs. \ref{fig:trine} and \ref{fig:randpom}. The two figures show that the reconstructed statistical operators are in general close to the true statistical operators.

\subsection{Imperfect measurements}\index{imperfect measurements}
\label{subsec:incomp}

In a practical tomography experiment, the detectors used are less than perfect. Typically, detection imperfections can be summarized using a set of positive numbers $\{\eta_j\}$, where $\eta_j<1$ is the detection efficiency\index{detection efficiency} for a particular POM outcome $\Pi_j$. More generally, one can describe a POM with more sophisticated imperfections by introducing the efficiency matrix $M_\eta$\label{symbol:meta}, with positive matrix elements satisfying the inequality
\begin{equation}
\sum_j\eta_{jk}\leq1\,.
\end{equation}
After obtaining these matrix elements through calibration, one can define a new set of outcomes
\begin{equation}
\Pi'_j=\sum_k\eta_{jk}\Pi_k
\end{equation}
such that $G\equiv\sum_k\Pi'_k\leq 1$\label{symbol:gop}\footnote{There are, of course, other types of experimental imperfections, such as the non-uniformity in the thickness of wave plates, the instability of the phase modulator, etc., that can affect the result of state estimation. These imperfections, in principle, can all be accounted for with the POM outcomes $\Pi'_j$.}. Therefore, $\sum_jp_j=\sum_j\tr{\rho\Pi'_j}\leq1$.

As a consequence to these imperfections, we would not know the true total number of copies $N_\text{true}$\label{symbol:numcopiestrue} that have reached all detectors. Denoting the total number of copies registered by the imperfect detectors by $N$, we can write down the likelihood functional for this scenario, with no emphasis on any particular sequence of detector clicks, as
\begin{equation}
\mathcal{L}'(\{n_j\};\rho)=\frac{N_\text{true}!}{N!\left(N_\text{true}-N\right)!}\left(\prod_jp_j^{n_j}\right)\left(1-\sum_{j'}p_{j'}\right)^{N_\text{true}-N}\,.
\label{likeincomp1}
\end{equation}\label{symbol:likelihoodimp}
where the indices here run over all outcomes and we define $\eta\equiv\sum_kp_k\leq 1$\label{symbol:eta} to be the \emph{overall detection efficiency}\index{detection efficiency!-- overall}. The multinomial\index{multinomial} factor takes into account all possible sequences of having $N$ detected copies out of the total of $N_\text{true}$ copies sent to all detectors. Using Stirling's formula\index{Stirling's formula} (James Stirling\index{James Stirling}) $\log N!\approx N\log N-N$, the log-likelihood can be simplified to
\begin{align}
\log\mathcal{L}'(\{n_j\};\rho)&=N\sum_jf_j\log p_j+(N_\text{true}-N)\log(1-\eta)\nonumber\\
&+\log \left(\frac{N_\text{true}^{N_\text{true}}}{N^N(N_\text{true}-N)^{N_\text{true}-N}}\right)\,,
\end{align}
where $\sum_jf_j=\sum_jn_j/N=1$.

Performing the variation, we have
\begin{align}
\updelta \log\mathcal{L}'(\{n_j\};\rho) =& N\sum_j\frac{f_j}{p_j}\updelta p_j-\frac{N_\text{true}-N}{1-\eta}\sum_j\updelta p_j\nonumber\\
&+\updelta N_\text{true}\log(1-\eta) +\updelta N_\text{true}\log N_\text{true}-\updelta N_\text{true}\log(N_\text{true}-N)\nonumber\\
=&\sum_j\left(N\frac{f_j}{p_j}-\frac{N_\text{true}-N}{1-\eta}\right)\updelta p_j+\updelta N_\text{true}\log\left(\frac{(1-\eta)N_\text{true}}{N_\text{true}-N}\right)\nonumber\\
=&\tr{\left(NR-\frac{N_\text{true}-N}{1-\eta}G\right)\updelta\rho}+\updelta N_\text{true}\log\left(\frac{(1-\eta)N_\text{true}}{N_\text{true}-N}\right)\,.
\end{align}
Setting $\updelta \log\mathcal{L}'(\{n_j\};\rho)$ to zero, i.e. maximizing $\log\mathcal{L}'(\{n_j\};\rho)$, requires the derivative
\begin{equation}
\frac{\partial}{\partial N_\text{true}}\log\mathcal{L}'(\{n_j\};\rho)=\log\left(\frac{(1-\eta)N_\text{true}}{N_\text{true}-N}\right)
\end{equation}
to be independently zero. This implies that the extremal equation
\begin{equation}
N_\text{true}=\frac{N}{\eta}
\label{extm}
\end{equation}
has to be satisfied, which is a rather natural statement since the likely number of copies that are actually received by the imperfect detectors is, of course, the true total number multiplied by the overall detection efficiency\index{detection efficiency!-- overall} that is less than one. Then the resulting expression for $\updelta \log\mathcal{L}'(\{n_j\};\rho)$ further simplifies to
\begin{equation}
\updelta \log\mathcal{L}'(\{n_j\};\rho)=N\tr{\left(R'-\frac{1}{\eta}G\right)\updelta\rho}\,,
\label{extr1}
\end{equation}
where $R'=\sum_jf_j\Pi'_j/p_j$ and the operator $-G/\eta$ accounts for inefficient detections.

We may naively make use of Eq.~(\ref{maxentstate}) to derive the following scheme:
\begin{align}
\rho_{k+1}&=\frac{\E^{\sum_j \left(\lambda^{(k)}_j+\epsilon\partial_j \log\mathcal{L}'(\{n_j\};\rho_k)\right)\Pi_j}}{\tr{\E^{\sum_j \left(\lambda^{(k)}_j+\epsilon\partial_j \log\mathcal{L}'(\{n_j\};\rho_k)\right)\Pi_j}}}\,,\\
\partial_j \log\mathcal{L}'(\{n_j\};\rho_k)&=N\text{tr}\Bigg\{\rho_k\Pi_j\int^1_0\D x\,\E^{x\sum_j\lambda^{(k)}_j\Pi_j}\left(R_k-\frac{1}{\eta^{(k)}}G\right)\,\E^{-x\sum_j\lambda^{(k)}_j\Pi_j}\Bigg\}\,,
\label{pscheme2}
\end{align}
with the index $j$ running over all POM outcomes. However, it turns out that there are many different sets of probabilities $p_j$ that maximize $\log\mathcal{L}'(\{n_j\};\rho)$ for a fixed set of measured data and hence multiple MLME estimators. We first note that the log-likelihood functional $\log\mathcal{L}(\{n_j\};\rho)$, after an application of the Stirling's formula on the factorials, is a concave function\index{concave function or functional}\index{log-likelihood functional!-- imperfect measurements} in $p_j$ since
\begin{equation}
\log\mathcal{L}'(\{n_j\};\rho)=\sum_jn_j\log\left(\frac{p_j}{\sum_kp_k}\right)
\label{loglikeincomp2}
\end{equation}
and each logarithmic term in the sum is concave in $p_j$. Hence concavity is not the cause of the existence of non-unique extremal $p_j$s. To identify the root of the problem, we look at the derivatives of $\log\mathcal{L}'(\{n_j\};\rho)$ by differentiating Eq.~(\ref{loglikeincomp2}) with respect to $p_j$, i.e.
\begin{equation}
\frac{\partial}{\partial p_j}\log\mathcal{L}'(\{n_j\};\rho)=\frac{n_j}{p_j}-\frac{N}{\sum_kp_k}\,.
\end{equation}
Then an extremal solution of $p_j$ satisfy the above equations with $\partial\log\mathcal{L}'(\{n_j\};\rho)/\partial p_j=0$ inasmuch as
\begin{equation}
\frac{p_j}{f_j}=\sum_kp_k\,.
\label{soleqn}
\end{equation}

For $K-1$ detected POM outcomes, there are altogether $K-2$ independent equations and one normalization constraint for the \emph{full} set of $p_j$s. From Eq.~(\ref{soleqn}), it is clear that the total number of available equations which are independent is $K-1$ and thus, there exist infinitely many solutions for these reduced set of equations, for the number of independent variables is now more than the number of independent equations. A simple example is a set of three POM outcomes, with $n_3=f_3=0$ for the third outcome. Then the only independent equation involving the probabilities is
\begin{equation}
\frac{p_1}{f_1}=\frac{p_2}{f_2}
\end{equation}
and hence, there are infinitely many solutions of $p_1$, $p_2$ and $p_3=1-p_1-p_2$ which maximize $F$.

In other words, we have infinitely many sets of solutions for $p_j$, with each set giving rise to a unique MLME estimator\index{unique estimator}. The task is then to select the MLME estimator that has the highest entropy out of the continuous family of MLME estimators. To do this, we first realize that Eq.~(\ref{soleqn}) simply implies that the ratio $p_j/f_j$ equals a constant value for $1\leq j\leq K-1$. Hence a scaling transformation on a reference set of solutions $p^\text{ML}_{j,0}$ with a continuous parameter $\alpha$ such that
\begin{equation}
p^\text{ML}_{j,0}\rightarrow\alpha p^\text{ML}_{j,0}\,,\,1\leq j\leq K-1
\end{equation}
also gives another set of solutions which satisfy Eq.~(\ref{soleqn}). The resulting maximum entropy estimator is obtained by varying the Lagrange functional\index{Lagrange functional}
\begin{align}
\mathcal{D}(\rho)=&-\tr{\rho\log\rho}+\sum^{K-1}_{j=1}\lambda_j\left(p_j-\alpha p^\text{ML}_{j,0}\right)+\lambda_K\left[p_K-\left(1-\alpha\sum^{K-1}_{j=1} p^\text{ML}_{j,0}\right)\right]\,\nonumber\\
&+\log\mu\left(\tr{\rho}-1\right)
\end{align}
and later setting the variation zero. In this way, the parameter $\alpha$ is optimized to give an estimator with the highest entropy among the family of MLME estimators. It follows that the extremal equation, after a variation in $\alpha$, is given by
\begin{equation}
\lambda_K=\sum^{K-1}_{j=1}\beta_j\lambda_j\,,
\end{equation}
where
\begin{equation}
\beta_j=\frac{p^\text{ML}_{j,0}}{\sum^{K-1}_{k=1}p^\text{ML}_{k,0}}\,.
\end{equation}
This implies that
\begin{equation}
\hat{\rho}_\text{ME}=\frac{\E^{\sum_j{\lambda_j\left(\Pi_j+\beta_j\Pi_K\right)}}}{\tr{\E^{\sum_j{\lambda_j\left(\Pi_j+\beta_j\Pi_K\right)}}}}\,.
\label{newme}
\end{equation}

Taking the ME estimator of the form in Eq.~(\ref{newme}), one can derive an iterative scheme to maximize $\log\mathcal{L}(\{n_j\};\rho)$ which is given by
\begin{center}
\colorbox{light-gray}{\begin{minipage}[c]{12cm}
\uline{\textbf{Scheme B}}
\begin{align}
\rho_{k+1}=&\frac{\E^{\sum_j \left(\lambda^{(k)}_j+\epsilon\partial_j \log\mathcal{L}'(\{n_j\};\rho_k)\right)\left(\Pi_j+\beta_j\Pi_K\right)}}{\tr{\E^{\sum_j \left(\lambda^{(k)}_j+\epsilon\partial_j \log\mathcal{L}'(\{n_j\};\rho_k)\right)\left(\Pi_j+\beta_j\Pi_K\right)}}}\,,\\
\partial_j \log\mathcal{L}'(\{n_j\};\rho_k)=&\,N\text{tr}\Bigg\{\rho_k\Pi_j\int^1_0\D x\,\Bigg[\E^{x\sum_j\lambda^{(k)}_j\left(\Pi_j+\beta_j\Pi_K\right)}\left(R_k-\frac{1}{\eta^{(k)}}G\right)\nonumber\\
&\quad\quad\quad\quad\quad\quad\quad\quad\times\E^{-x\sum_j\lambda^{(k)}_j\left(\Pi_j+\beta_j\Pi_K\right)}\Bigg]\Bigg\}\,,
\label{scheme2}
\end{align}
  \end{minipage}}
  \end{center}
where the extremal equation to be satisfied by $\hat{\rho}_\text{MLME}$ is
\begin{equation}
\hat{R}_\text{MLME}\hat{\rho}_\text{MLME}=\frac{1}{\eta}G\hat{\rho}_\text{MLME}.
\end{equation}
With the approximation supplied by Eq.~(\ref{approx}), the gradient can be approximated to
\begin{align}
&\partial_j \log\mathcal{L}(\{n_j\};\rho_k)\nonumber\\
=&\,N\text{tr}\Bigg\{\rho_k\frac{\left(\Pi_j+\beta_j\Pi_K\right)\left(R_k-\frac{1}{\eta^{(k)}}G\right)+\left(R_k-\frac{1}{\eta^{(k)}}G\right)\left(\Pi_j+\beta_j\Pi_K\right)}{2}\Bigg\}\,.
\end{align}
Since this scheme is independent on the choice of $p^\text{ML}_{j,0}$ , one may first perform ML starting from the maximally-mixed state and make use of the resulting set of ML probabilities to carry out \textbf{Scheme B}.

\begin{figure}[h!]
  \centering
  \includegraphics[width=0.6\textwidth]{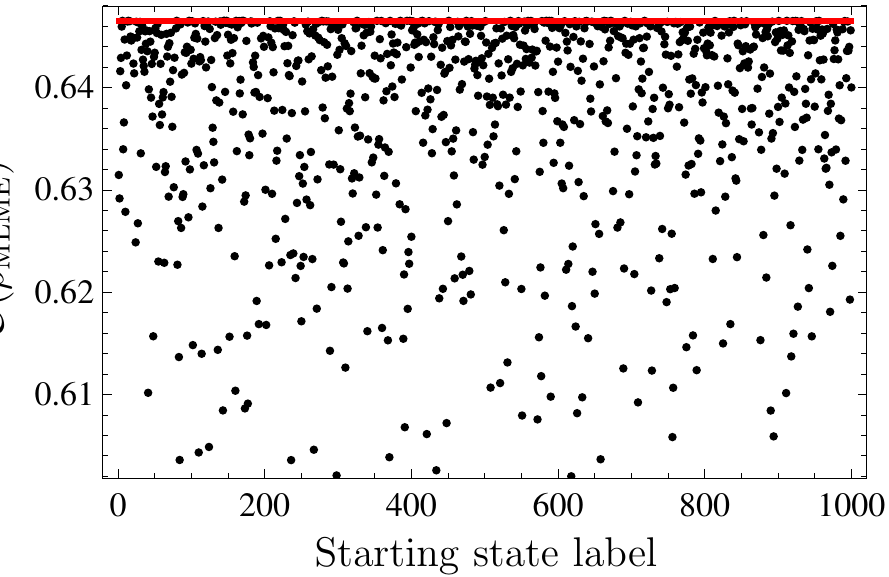}
  \caption{A simulated experiment on a random state, in which 5000 qubits were measured using a random imperfect two-outcome POM. The plot markers, which are indicated by dots, represent the entropies of the MLME estimators generated by the naive scheme starting from random states in the uniform distribution with respect to the Hilbert-Schmidt measure. $10^3$ such estimators were computed. The thick solid line represents the entropy of the MLME estimator generated by \textbf{Scheme B}.}
  \label{fig:unique}
\end{figure}

To demonstrate the results of the scheme, we first ran a single simulated experiment involving the measurement of 5000 copies of qubits prepared in a random state using of a random three-outcome POM, with one of the POM outcomes not registering any qubit. Post-processing the data with \textbf{Scheme B} indeed gives the MLME estimator which has the highest entropy among all other estimators generated using the former naive scheme by varying the starting state for each iteration. The result is shown in Fig.~\ref{fig:unique}.

\begin{figure}[h!]
  \centering
  \includegraphics[width=0.6\textwidth]{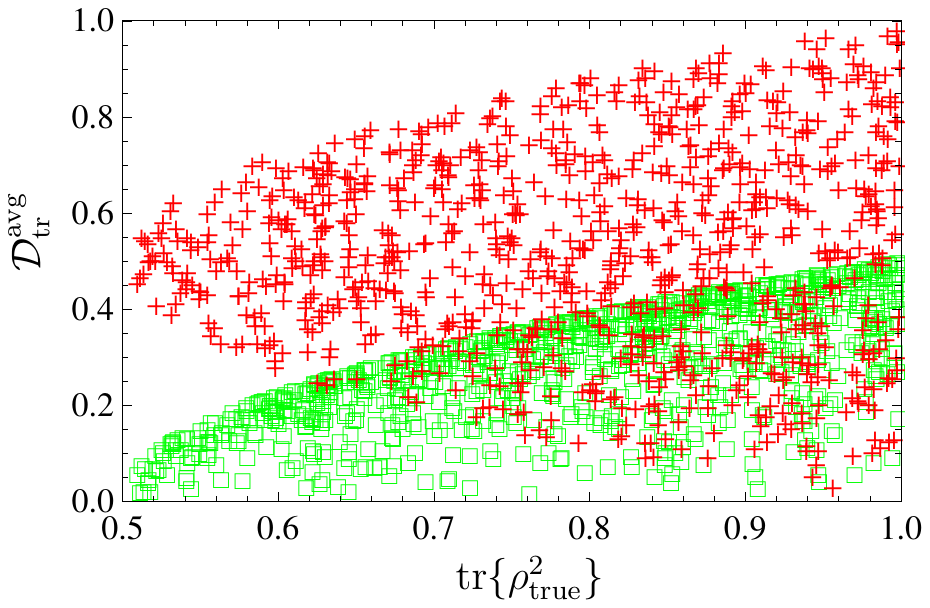}
  \caption{A comparison of two different schemes for a fixed random incomplete POM with $10^3$ random qubit true states distributed uniformly with respect to the Hilbert-Schmidt measure. Fifty experiments were simulated for every true state, with $N=5000$ for each experiment, and the average trace-class distance $\mathcal{D}^\text{avg}_\text{tr}$ was computed. The entire simulation was done with a set of randomly generated, informationally incomplete POM consisting of three outcomes. A POM outcome was discarded to simulate the situation in which two functioning detectors out of the three are registering the qubits. The plot markers denoted by ``$+$'' represent reconstructed states using \textbf{Scheme A}, and those denoted by ``$\square$'' represent the reconstructed states using \textbf{Scheme B}. The missing probabilities estimated by the reconstructed states using the \textbf{Scheme B} are typically closer to the missing frequencies that would be measured if the discarded detector was functioning compared to those estimated by the reconstructed states using \textbf{Scheme A}. About 80\% of the total number of true states respond better under the second scheme.}
  \label{fig:incomp_comp}
\end{figure}

Fig.~\ref{fig:incomp_comp} compares the performances of \textbf{Scheme A}, with which we search for the MLME estimator by assuming that the measured data $n_j$ are all we have while ignoring the possible missing data, in qubit tomography using the trace-class distance
\begin{equation}
\mathcal{D}_\text{tr}=\frac{1}{2}\tr{|\hat{\rho}_\text{MLME}-\rho_\text{true}|}\,
\end{equation}
as the figure of merit to quantify the distance between $\hat{\rho}_\text{MLME}$ and $\rho_\text{true}$. The lesson here is that if one neglects the consequence of imperfect measurements in performing state reconstruction, the quality of the resulting reconstructed state estimator will typically be much lower than that obtained from a scheme which accounts for this imperfection.

In a typical experiment, all detectors are controlled to have the same efficiency $\eta_{jk}=\eta_0\,\delta_{jk}$. In this special setting, the operator $R'$ in Eq.~(\ref{extr1}) further simplifies to
\begin{equation}
R'=R+\frac{1-\eta_0}{\eta_0}\,.
\label{sextr2}
\end{equation}
Incidently, the term that is a multiple of the identity operator does not affect the likelihood maximization procedure at all, and we will obtain exactly \textbf{Scheme A} for the incomplete set of data. In other words, since all the detectors have indistinguishable efficiencies, we can consider this special setting as the situation in which the observer has a \emph{complete} set of measurement data that is less than that for the case when all detectors have 100\% efficiency.


\subsection{A new perspective}
\label{subsec:new_perspective}

Previously, we described the original idea of the ML-assisted ME procedure for a set of informationally incomplete measurements in a given quantum tomography experiment, that is the selection of the most-likely state estimator with the highest von Neumann entropy as the least-biased state estimator. Such a procedure usually involves complicated systems of non-linear equations which are especially hard to solve for non-commuting measurement operators.

We then established novel and more feasible schemes, via the steepest-ascent approach, which are suitable for any set of measurement operators, to obtain the same result by maximizing the likelihood functional over the space of statistical operators, with each operator assuming the form that maximizes the von Neumann entropy functional for a fixed set of probabilities. This MLME approach, which effectively condenses the ML and ME optimization procedures into one, can in fact be slow. This is due to the fact that the proposed MLME algorithm proceeds along a search path that deviates away from steepest-ascent because of the approximation in Eq.~(\ref{approx}).

In the subsequent sections, we establish more efficient MLME algorithms by viewing the problem of MLME in a different perspective \cite{mlme,mlme_comprehensive}. We then apply these new algorithms to several different situations.

\subsubsection{A new algorithm for perfect measurements}
\label{subsec:algo_perfect}
Assuming that the measurement detections are perfect, one can consider the optimization of the \emph{normalized log-likelihood functional} $\log(\mathcal{L}(\{n_j\};\rho))/N$, with $\mathcal{L}(\{n_j\};\rho)$ defined in Eq.~(\ref{simple_like}). The motivation for introducing the normalization will become clear soon. The MLME scheme\index{maximum-likelihood-maximum-entropy (MLME)} can then be perceived as a standard constrained optimization problem: maximize $\log(\mathcal{L}(\{n_j\};\rho))/N$ subjected to the constraint that $S(\rho)$ takes the maximal value $S_\text{max}$. The Lagrange functional for this optimization problem is defined as
\begin{equation}
\mathcal{I}(\lambda;\rho)=\lambda\left(S(\rho)-S_\text{max}\right)+\frac{1}{N}\log\mathcal{L}(\{n_j\};\rho)\,,
\label{newinfo}
\end{equation}\label{symbol:infofunctional}
where $\lambda$ is the Lagrange multiplier corresponding to the constraint for $S(\rho)$. This is equivalent to maximizing $S(\rho)$ with the constraint that $\log(\mathcal{L}(\{n_j\};\rho))/N$ is maximal, as discussed previously. We denote the estimator that maximizes $\mathcal{I}(\lambda;\rho)$ by $\hat\rho_{\text{I},\lambda}$\label{symbol:rhoestinfo}. Incidently, as a result of the normalization of $\log(\mathcal{L}(\{n_j\};\rho))$, the functional $\mathcal{I}(\lambda;\rho)$ is a sum of two different types of entropy, up to an irrelevant additive constant $\sum_jf_j\log f_j$: the von Neumann entropy $S(\rho)$ that quantifies the ``lack of information'', and the \emph{negative} of the \emph{relative entropy}\index{entropy!-- relative} $S(\{f_j\}|\{p_j\})=\sum_jf_j\log(f_j/p_j)$\label{symbol:srel} that quantifies the ``gain of information'' from the measurement data. The scheme can now be interpreted as a simultaneous optimization of two complementary aspects of information, with an appropriately assigned constant relative weight $\lambda$. In addition, the normalization of $\log\mathcal{L}(\{n_j\};\rho)$ renders the optimal value of $\lambda$ to be independent of $N$.

\begin{figure}
\centering
\includegraphics[width=0.8\textwidth]{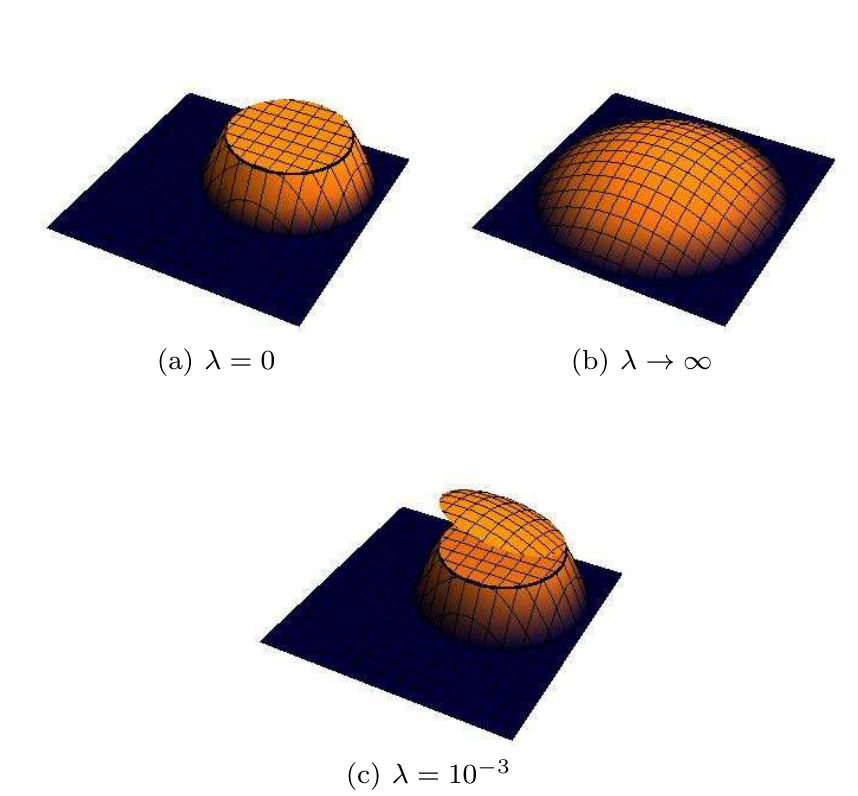}
  \caption{Schematic diagrams of $\mathcal{I}(\lambda,\rho)$ on the space of statistical operators. The maximally-mixed state resides at the center of the square base which represents the Hilbert space. At the extremal points of $\lambda$, $\mathcal{I}(\lambda=0;\rho)=\log(\mathcal{L}(\{n_j\};\rho))/N$, with a convex plateau\index{plateau} at the maximal value, and $\mathcal{I}(\lambda\rightarrow\infty;\rho)=\lambda S(\rho)$. Plot (c) shows the functional with an appropriate choice of value for $\lambda$ for MLME. An additional hill-like structure resulting from $S(\rho)$ is introduced over the plateau\index{plateau}, so that the estimator with the largest entropy can be selected from the convex set of ML estimators within the plateau\index{plateau}.}
  \label{fig:geom}
\end{figure}

When $\lambda=0$, we recover the Lagrange functional for the log-likelihood functional alone. Owing to the informational incompleteness of the measurement data, there exists a convex plateau\index{plateau} structure for the log-likelihood functional. As $\lambda\rightarrow\infty$, the von Neumann entropy becomes increasingly more significant and the resulting estimator $\hat\rho_{\text{I},\lambda\rightarrow\infty}$ approaches the maximally-mixed state $1/D$. Naturally, when $\lambda$ takes on a very small positive value, the contribution from $\lambda S(\rho)$ becomes much smaller than $\log(\mathcal{L}(\{n_j\};\rho))/N$ and the effect of the von Neumann entropy functional is only significant over the plateau\index{plateau} region in which the likelihood is maximal. Figure~\ref{fig:geom} illustrates all the aforementioned points.
This means that, in general, $\lambda$ should be chosen so small that $S(\hat\rho_{\text{I},\lambda})$ takes a value that is very close to the minimum, and below which there are only very slight changes in the two entropy functionals. The methodology to select an appropriate value of $\lambda$ will be discussed in \S\ref{subsec:app}.

Let us derive the iterative algorithm for maximizing $\mathcal{I}(\lambda\rightarrow 0;\rho)$ with respect to $\rho$. After varying $\mathcal{I}(\lambda\rightarrow 0;\rho)$, we have
\begin{equation}
\updelta\mathcal{I}(\lambda\rightarrow 0;\rho)=-\lambda\,\tr{\updelta\rho\log\rho}+\sum_j\frac{f_j}{p_j}\updelta p_j\,.
\label{var1}
\end{equation}
The variations $\updelta p_j$, or $\updelta\rho$, have to be such that $\rho$ stays positive after these variations. With the help of the parametrization in Eq.~(\ref{aux_param}), we find that
\begin{equation}
\updelta\mathcal{I}(\lambda\rightarrow 0;\rho)=\tr{\frac{\updelta \mathcal{A}^\dagger \mathcal{A}}{\tr{\mathcal{A}^\dagger \mathcal{A}}}{\mathfrak{R}}+{\mathfrak{R}}\frac{\mathcal{A}^\dagger \updelta \mathcal{A}}{\tr{\mathcal{A}^\dagger \mathcal{A}}}}\,,
\label{var2}
\end{equation}
where
\begin{equation}
{\mathfrak{R}}=R-1-\lambda\left(\log\rho-\tr{\rho\log\rho}\right)
\end{equation}
with
\begin{equation}
R=\sum_j\frac{f_j}{p_j}\Pi_j\,.
\end{equation}

When $\mathcal{I}(\lambda\rightarrow 0;\rho)$ is maximal, we have $\updelta\mathcal{I}(\lambda\rightarrow 0;\rho)=0$ and the extremal equations
\begin{equation}
\rho\,{\mathfrak{R}}={\mathfrak{R}}\rho=0
\label{exteqn1}
\end{equation}
are satisfied. Therefore, to solve these extremal equations numerically, we iterate the equation
\begin{equation}
\rho_\text{k+1}=\frac{\left(\mathcal{A}^\dagger_k+\updelta \mathcal{A}^\dagger_k\right)\left(\mathcal{A}_k+\updelta \mathcal{A}_k\right)}{\tr{\left(\mathcal{A}^\dagger_k+\updelta \mathcal{A}^\dagger_k\right)\left(\mathcal{A}_k+\updelta \mathcal{A}_k\right)}}\,
\label{itereqn1}
\end{equation}
starting from some statistical operator $\rho_1$, until $k=k'$ such that the norm of $\rho_{k'} {\mathfrak{R}}_{k'}$ is less than some pre-chosen value. We then take $\hat\rho_\text{MLME}\equiv\rho_{k'}$ as the MLME estimator. Maximizing $\mathcal{I}(\lambda\rightarrow 0;\rho)$ will require $\updelta\mathcal{I}(\lambda\rightarrow 0;\rho)$ to be positive whenever $\mathcal{I}(\lambda\rightarrow 0;\rho)$ is less than the maximal value. A straightforward way to enforce positivity is to set
\begin{equation}
\updelta \mathcal{A}_k\equiv\left(\updelta \mathcal{A}^\dagger_k\right)^\dagger\equiv\epsilon \mathcal{A}_k{\mathfrak{R}}_k\propto\epsilon\frac{\partial \mathcal{I}(\lambda;\rho)}{\partial \mathcal{A}_k}\,,
\label{itereqn2}
\end{equation}
with $\epsilon$ being a small positive constant. This is the \emph{steepest-ascent} method. We have thus established a numerical MLME scheme as a set of iterative equations (\ref{itereqn1}) and (\ref{itereqn2}) to search for the MLME estimator using the measurement data obtained from perfect measurement detections. More compactly, the relevant iterative equations are
\begin{center}
\colorbox{light-gray}{\begin{minipage}[c]{12cm}
  \uline{\textbf{New MLME iterative equations for perfect measurements}}
  \begin{align}
  \rho_\text{k+1}&=\frac{\left(1+\epsilon {\mathfrak{R}}_k\right)\rho_k\left(1+\epsilon {\mathfrak{R}}_k\right)}{\tr{\left(1+\epsilon {\mathfrak{R}}_k\right)\rho_k\left(1+\epsilon {\mathfrak{R}}_k\right)}}\,,\nonumber\\
  {\mathfrak{R}}_k&=R_k-1-\lambda\left(\log\rho_k-\tr{\rho_k\log\rho_k}\right)\,.
  \end{align}
  \end{minipage}}
  \end{center}

There exists an interesting structure in these MLME estimators and to explore it, one needs some knowledge on the structure of the POM used and its influence on the $D$-dimensional Hilbert space. Suppose a set of $K$ POM elements $\Pi_j$ are informationally incomplete. A consequence of this is that the number of linearly independent\index{linearly independent} $\Pi_j$s is less than $D^2$. As discussed in \S\ref{sec:qseprelim}, to determine their linear independence, we can look for the eigenvalues of the $K\times K$ Gram matrix $\mathfrak{M}$ whose matrix elements are defined as
\begin{equation}
\mathfrak{M}_{jk}=\tr{\Pi_j\Pi_k}\,.
\end{equation}
Thus, a set of informationally incomplete $\Pi_j$s acting on the $D$-dimensional Hilbert space is such that the number of positive eigenvalues of $\mathfrak{M}$, denoted by $n_{>0}$\label{symbol:numlindep}, is less than $D^2$. Any $D$-dimensional positive operator can be represented by a set of $D^2$ Hermitian basis operators\index{operator basis} $\Gamma_j$ satisfying the trace-orthonormality\index{trace-orthonormal} condition $\tr{\Gamma_j\Gamma_k}=\delta_{jk}$. For dimension two, an example of such a basis\index{operator basis} is the the familiar set of four operators $1/\sqrt{2}$,  $\sigma_x/\sqrt{2}$, $\sigma_y/\sqrt{2}$ and $\sigma_z/\sqrt{2}$. Once the number of independent measurement outcomes $n_{>0}$ is known, one can construct a set $\{\Gamma_j\}_{j=1}^{n_{>0}}$ of $n_{>0}$ trace-orthonormal\index{trace-orthonormal} Hermitian basis operators\index{operator basis} directly from the $K$ POM elements. In other words, each of the $K$ POM elements can be expressed as a linear combination of the $n_{>0}$ basis operators\index{operator basis}
\begin{equation}
\Pi_j=\sum^{n_{>0}}_{k=1}a_{jk}\Gamma_k\,,
\label{pom_recon}
\end{equation}
where all coefficients $a_{jk}$ are real. This implies that the $n_{>0}$-dimensional subspace is spanned by the basis operators\index{operator basis} that \emph{uniquely} specify the POM outcomes. We will coin this subspace the \emph{measurement subspace}\index{measurement subspace}. The rest of the $D^2-n_{>0}$ Hermitian basis operators\index{operator basis}, which are trace-orthonormal\index{trace-orthonormal} to the previous set and span the subspace, that is complement to the measurement subspace can also be constructed.

Suppose a state estimator $\hat{\rho}_\text{ML}$ is generated using the ML procedure on a set of measurement data obtained from the POM outcomes $\Pi_j$. We can represent this estimator by a set of Hermitian trace-orthonormal\index{trace-orthonormal} basis operators\index{operator basis} inasmuch as
\begin{equation}
\hat{\rho}_\text{ML}=\underbrace{\sum^{n_{>0}}_{k=1}c^\text{ML}_k\Gamma_k}_{\equiv\tilde\rho_\text{ML}}+\underbrace{\sum^{D^2}_{k=n_{>0}+1}c^\text{ME}_k\Gamma_k}_{\equiv\tilde\rho_\text{ME}}\,.
\label{rho_decomp}
\end{equation}
The part $\tilde\rho_\text{ML}$ resides in the measurement subspace, which is spanned by the measurement outcomes $\Pi_j$ giving the measurement data, and is uniquely fixed for all ML estimators by the ML procedure for the same set of measurement data. The part $\tilde{\rho}_{\mathrm{ME}}$ resides in the complementary subspace, which is orthogonal to the measurement subspace, and thus does not contribute to the $p_j$s. In other words, $\tr{\tilde\rho_\text{ME}\Pi_j}=0$ and this can imply the existence of a family of $\tilde\rho_\text{ME}$s that gives the same set of ML probabilities as long as the $\hat{\rho}_\text{ML}$s are positive.

Therefore, the MLME scheme can be understood as an optimization over the complementary subspace to maximize $S(\rho)$ under the constraint $\hat{\rho}_\text{MLME}\geq 0$. However, one notes that only certain sets of $c^\text{ME}_j$s are allowed during the optimization in order to obey this positivity constraint. This is especially important when $\hat{\rho}_\text{MLME}$ is rank deficient and lies on the boundary of the state space. Geometrically, the plateau\index{plateau} of most-likely states is generally a much smaller subspace contained in the complementary subspace. In some cases, this plateau\index{plateau} contains a \emph{single} ML estimator because of the positivity constraint even when the measurements are informationally incomplete. In general, the boundary of the plateau\index{plateau} is complicated and deserves further study.

\subsubsection{A new algorithm for imperfect measurements}\index{imperfect measurements}
\label{subsec:algo_imperfect}
In actual experiments, as discussed previously, the measurement detections will usually be imperfect in the sense that the detection efficiency\index{detection efficiency} $\eta_j\leq1$ of a particular measurement outcome $\Pi_j$ is less than unity. In this case, the overall outcome probabilities
\begin{equation}
p_j=\tr{\rho\Pi'_j}
\end{equation}
will not sum to unity. Hence, we have a set of POM with outcomes $\Pi'_j\equiv\eta_j\Pi_j$\label{symbol:pijprime} such that $G'\equiv\sum_j\Pi'_j\leq1$. A consequence of this is that the true total number $N_\text{true}$ of copies received is not known, since only $N<N_\text{true}$ are detected ($N=N_\text{true}$ when all $\eta_j=1$ as in \S\ref{subsec:algo_perfect}).

From \S\ref{subsec:incomp}, the correct form of the likelihood functional\index{likelihood functional!-- imperfect measurements} for this situation is given by
\begin{equation}
\mathcal{L}'(\{n_j\};\rho)=\prod_j\left(\frac{p_j}{\eta}\right)^{n_j}
\label{rel_like}
\end{equation}
up to an irrelevant multiplicative factor, with its corresponding logarithmic variation
\begin{equation}
\updelta\log\mathcal{L}'(\{n_j\};\rho)=N\tr{\left(R'-\frac{G'}{\eta}\right)\updelta\rho}\,
\label{varlike2}
\end{equation}
with $R'=\sum_jf_j\Pi'_j/p_j$. The additional term $-\updelta\rho\,G'/\eta$ in the argument of the trace accounts for copies that have escaped detection.

Defining $\mathcal{I}(\lambda\rightarrow 0;\rho)$ for the new POM and its $\mathcal{L}'(\{n_j\};\rho)$ in Eq.~(\ref{rel_like}), one can derive the iterative equations
\begin{center}
\colorbox{light-gray}{\begin{minipage}[c]{12cm}
  \uline{\textbf{New MLME iterative equations for imperfect measurements}}
\begin{align}
\rho_{k+1}&=\frac{\left(1+\epsilon \mathfrak{R}'_k\right)\rho_k\left(1+\epsilon\mathfrak{R}'_k\right)}{\tr{\left(1+\epsilon\mathfrak{R}'_k\right)\rho_k\left(1+\epsilon \mathfrak{R}'_k\right)}}\,,\nonumber\\
\mathfrak{R}'_k&=R'_k-\frac{G'}{\eta^{(k)}}-\lambda\left(\log\rho_k-\tr{\rho_k\log\rho_k}\right)\,,
\label{iteralgo2}
\end{align}
  \end{minipage}}
  \end{center}
with $\eta^{(k)}=\sum_jp^{(k)}_j$. We note that more efficient algorithms, using the conjugate-gradient method, can be derived from these steepest-ascent algorithms using the machineries introduced in \S\ref{subsec:conjgrad}.

\subsubsection{Applications}
\label{subsec:app}
\begin{flushleft}
\uline{\textbf{Homodyne detection tomography}}\index{homodyne detection tomography}
\end{flushleft}
To discuss the methodology of choosing $\lambda$, we shall apply the MLME scheme to homodyne detection tomography, a technique which is used to reconstruct quantum states of light \cite{homodyne1,homodyne2,homodyne3}. This is typically done by measuring a POM which resembles a set of eigenstate projectors $\ket{x_\vartheta}\bra{x_\vartheta}$\label{symbol:xthetastate} of quadrature\index{quadrature operators} operators $X_\vartheta=X\cos\vartheta+P\sin\vartheta$ for various $\vartheta$ values, where $X$\label{symbol:xpos} and $P$\label{symbol:pmom} are respectively the position\index{position quadrature operator} and momentum\index{momentum quadrature operator} quadrature operators and $x$ and $\vartheta$ are parameters specifying these projectors. Introducing the standard annihilation operator $A=(X+\I P)/\sqrt{2}$\label{symbol:ann}, we have
\begin{equation}
X_\vartheta=\frac{A\E^{-\I\vartheta}+A^\dagger\E^{\I\vartheta}}{\sqrt{2}}\,.
\end{equation}
To facilitate the numerical simulations with the eigenkets $\ket{x_\vartheta}$, the corresponding quadrature wave functions\index{quadrature wave functions} $\left<x_\vartheta|n\right>$\label{symbol:nstate} in the Fock representation\index{Fock states or representation} are needed. To obtain these wave functions, we first note that the product of $A$ and the function $f(A^\dagger A)$ satisfies the relation $A\,f(A^\dagger A)=f(A^\dagger A+1)\,A$ since, for any Fock ket $\ket{n}$,
\begin{align}
A\,f(A^\dagger A)\ket{n}&=f(n)\,A\ket{n}\nonumber\\
&=\sqrt{n-1}\,f(n)\ket{n-1}\nonumber\\
&=\sqrt{n-1}\,f(A^\dagger A+1)\ket{n-1}\nonumber\\
&=f(A^\dagger A+1)\,A\ket{n}\,.
\end{align}
From this relation, we realize that
\begin{equation}
X_\vartheta=\E^{-\I\vartheta A^\dagger A}X\E^{\I\vartheta A^\dagger A}\,,
\end{equation}
and its corresponding quadrature eigenket $\ket{x_\vartheta}$ is thus obtained via a unitary transformation
\begin{equation}
\ket{x_\vartheta}=\E^{-\I\vartheta A^\dagger A}\ket{x}
\end{equation}
of the corresponding eigenket $\ket{x}$ of the position quadrature operator $X$. Hence, in the Fock representation\index{Fock states or representation}, the corresponding quadrature wave functions\index{quadrature wave functions} are given by
\begin{align}
\left<n|x_\vartheta\right>&=\E^{-\I n\vartheta}\left<n|x\right>\nonumber\\
&=\frac{1}{\pi^{-1/4}\sqrt{2^n\,n!}}\,\E^{-\I n\vartheta}\E^{-x^2/2}\,H_n(x)\,,
\end{align}
where $H_n(x)$ are the Hermite polynomials\index{Hermite polynomials} (Charles Hermite\index{Charles Hermite}) of degree $n$.

\begin{figure}[h!]
  \centering
  \includegraphics[width=0.6\textwidth]{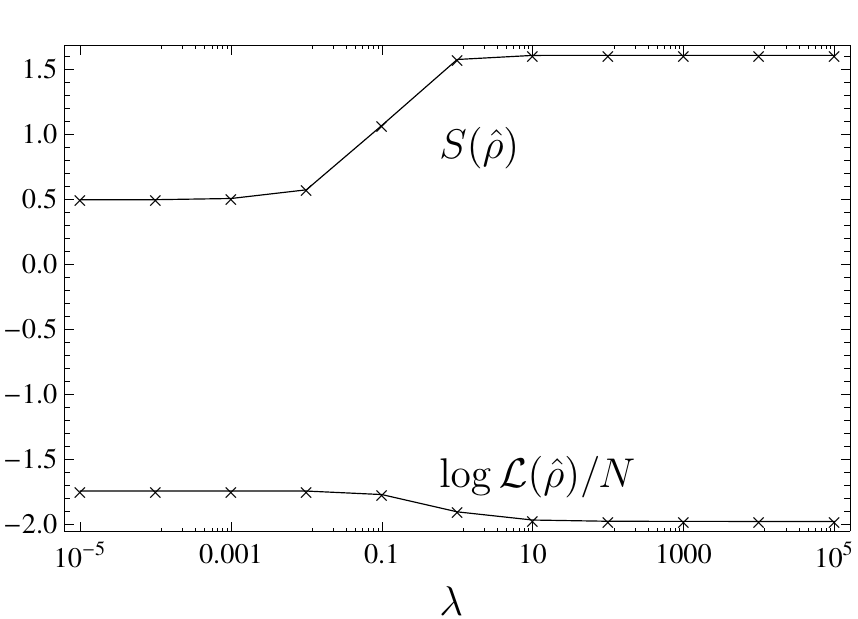}
  \caption{A simulation on quantum tomography on a randomly generated mixed state of light in the five--dimensional Fock space\index{Fock states or representation}. In this plot, the number of copies of quantum systems measured is fixed at $N=10^4$. A choice of 20 quadrature eigenstates made up of four different $\vartheta$ settings, with five $x$ values corresponding to each setting, which are projected onto this space was used and state estimators are constructed for different values of $\lambda$. As $\lambda$ decreases, both the entropy and likelihood functionals approach their respective optimal values obtained from MLME (i.e. when $\lambda\rightarrow 0$). When $\lambda$ is zero, there is a convex set of estimators giving the optimal likelihood value. For very large $\lambda$ values, the estimators approach the maximally-mixed state and hence $S(\rho)$ approaches the maximal value $\log 5$.}
  \label{fig:behavior}
\end{figure}

It is clear that a finite set of such measurements is never informationally complete in the infinite-dimensional Hilbert space and thus the MLME scheme is necessary to obtain a unique estimator\index{unique estimator}. Figure~\ref{fig:behavior} shows the dependence of $\log\left(\mathcal{L}(\hat\rho)\right)/N$ and $S(\hat\rho)$ on $\lambda$ such that $\updelta\mathcal{I}(\lambda\rightarrow 0;\hat\rho)=0$. In practice, $\lambda$ can be chosen from a range near zero, within which $\log\left(\mathcal{L}(\hat\rho)\right)/N$ and $S(\hat\rho)$ remain almost constant.

Homodyne detection tomography is commonly used not only in quantum tomography on the true state, but also in quantum diagnostics where a given true state is to be classified as being classical/non-classical or separable/entangled. With the help of the coherent states\index{coherent states} $\ket{\alpha}\bra{\alpha}$, the following decomposition
\begin{equation}
\rho=\frac{1}{\pi}\int(\D\alpha')\,\ket{\alpha'}P(\alpha')\bra{\alpha'^*}\,
\label{gsprep}
\end{equation}
for a state $\rho$ can be used to distinguish classical states from non-classical ones, where the function $P(\alpha')$\label{symbol:pglauber} is known as the \emph{Glauber-Sudarshan $P$ function}\index{Glauber-Sudarshan $P$ function} (Roy Jay Glauber\index{Roy Jay Glauber} and Ennackal Chandy George Sudarshan\index{Ennackal Chandy George Sudarshan}) of the complex parameter $\alpha'$\label{symbol:alpha}. Using this decomposition, we define the state $\rho$ to be a classical state if $P(\alpha)$ is positive for all $\alpha$, and only then: that is, $\rho$ is a statistical mixture of coherent states\index{coherent states}. Otherwise, $\rho$ is non-classical. The symbol $(\D \alpha)$ denotes the integral measure over the real and imaginary parts of the complex variable $\alpha$.

\begin{figure}[h!]
  \centering
  \includegraphics[width=0.6\textwidth]{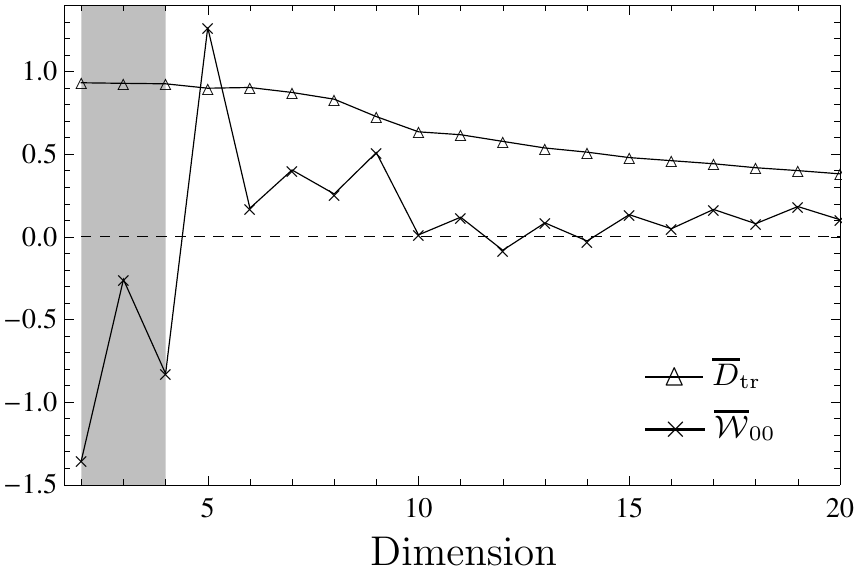}
  \caption{A simulation on quantum tomography on a randomly generated mixed state $\rho_\text{true}$ of light in the 20--dimensional Fock space\index{Fock states or representation} with a slightly positive $\mathcal{W}_{00}=0.141\,$. $\overline D_\text{tr}$ and $\overline{\mathcal{W}}_{00}$ respectively denote the trace-class distance between the reconstructed estimator and the true state and the Wigner functional at the phase space origin, both averaged over 50 experiments with $N=10^4$. The same set of 20 quadrature eigenstates as in Fig.~\ref{fig:behavior}, projected onto this space was used and this set of measurements is informationally complete in the \mbox{two-,} three-, and four-dimensional Fock subspaces (shaded region). The values $\overline{\mathcal{W}}_{00}$ and $\overline D_\text{tr}$ were obtained by ML \cite{homodyne1,homodyne2,homodyne3} in subspaces of dimensions two to four, and by the MLME scheme in dimensions greater than four. The plot shows a strong dependence of $\overline{\mathcal{W}}_{00}$ and $\overline D_\text{tr}$ on the subspace dimension. In this case, it is obvious that the negativity of $\overline{\mathcal{W}}_{00}$ inferred by a reconstruction in a subspace too small is just an artifact of the truncations\index{state-space truncation}. Also, $\overline D_\text{tr}$ decreases as the reconstruction subspace\index{reconstruction subspace} increases in dimension. This demonstrates the advantages of the MLME scheme over the ML method.}
  \label{fig:negativity}
\end{figure}

One very popular way to represent the measurement data obtained in a typical homodyne experiment is by means of the Wigner functional (Eugene Paul Wigner\index{Eugene Paul Wigner}) of $\rho$ defined as
\begin{equation}
\mathcal{W}(x,p)=\int\D y\,\E^{\I py}\bra{x-\frac{y}{2}}\rho\ket{x+\frac{y}{2}}\,.
\end{equation}\label{symbol:wigner}This functional is a quasi-probability density functional that maps the statistical operator onto the phase space (see Ref.~\cite{wigner}) and has many nice properties that are symmetric with respect to the phase space variables $x$ and $p$. In addition, this functional can be used to determine if a state $\rho$ is non-classical. To see this, we note the coherent-state representation of $\rho$ defined in Eq.~(\ref{gsprep}) and the expression for the wave function of the ket $\ket{\alpha}$ given by
\begin{equation}
\left<x|\alpha'\right>=\left<\alpha'^*|x\right>^\dagger=\frac{1}{\pi^{\frac{1}{4}}}\,\E^{-\frac{1}{2}x^2+\sqrt{2}\,x\alpha'-\frac{1}{2}\alpha'^2-\frac{1}{2}|\alpha'|^2}\,.
\end{equation}
Using these equations,
\begin{align}
\mathcal{W}(x,p)&=\int(\D\alpha')\,P(\alpha')\int\D y\,\E^{\I py}\left<x-\frac{y}{2}\bigg|\alpha'\right>\left<\alpha'^*\bigg|x+\frac{y}{2}\right>\nonumber\\
&=\frac{\E^{-x^2}}{\pi^{\frac{3}{2}}}\int(\D\alpha')\,\Bigg\{P(\alpha')\,\E^{\sqrt{2}\,x(\alpha'+\alpha'^*)-\frac{1}{2}(\alpha'^2+\alpha'^{*\,2})-|\alpha'|^2}\nonumber\\
&\quad\times\underbrace{\int\D y\,\E^{-\frac{y^2}{4}+y\left(\frac{\alpha'^*-\alpha'}{\sqrt{2}}+\I p\right)}}_{=2\sqrt{\pi}\,\text{exp}\left(\left(\frac{\alpha'^*-\alpha'}{\sqrt{2}}+\I p\right)^2\right)}\Bigg\}\nonumber\\
&=\frac{2\,\E^{-x^2-p^2}}{\pi}\int(\D\alpha')\,P(\alpha')\,\E^{-2|\alpha'|^2+2\sqrt{2}\,\text{Re}\{(x-\I p)\,\alpha'\}}\,.
\end{align}
Since the exponentials are always positive, any non-positivity of $\mathcal{W}(x,p)$ must originate from a non-positive $P(\alpha')$. The converse is in general not true, however, as there are non-classical quantum states that give positive Wigner functions. A naive quantity that is often investigated as an indication of whether an unknown true state is non-classical is the value of the Wigner functional at the phase space origin\index{Wigner functional!-- at phase space origin} evaluated with a reconstructed estimator $\hat\rho$ for the unknown true state. This is defined as $\mathcal{W}_{00}\equiv\mathcal{W}(0,0)=2\tr{\hat\rho \mathcal{P}}$\label{symbol:w00}\label{symbol:parity}, with the \emph{parity operator}\index{parity operator} $\mathcal{P}=\int\mathrm{d}x\,\ket{x}\bra{-x}$. In the Fock representation\index{Fock states or representation}, the parity operator becomes
\begin{align}
\mathcal{P}&=\int\mathrm{d}x\,\ket{x}\bra{-x}\nonumber\\
&=\sum^\infty_{n=0}\int\mathrm{d}x\,\ket{x}\underbrace{\left<-x|n\right>}_{=\,(-1)^n\left<x|n\right>}\bra{n}\nonumber\\
&=\sum^\infty_{n=0}\ket{n}(-1)^n\bra{n}\\
&=(-1)^{A^\dagger A}\,
\end{align}
due to the property of the Hermite polynomials contained in the complex function $\left<x|n\right>$. To obtain an estimator $\hat\rho$, one would need to choose a subspace\index{state-space truncation} from the infinite-dimensional Hilbert space in which the reconstruction procedure is tractable. This means that the value of $\mathcal{W}_{00}$ will depend on this truncation, which in turn relies on the prior knowledge\index{prior information (knowledge)} one has about the true state. Using the new MLME scheme, we perform a simulation, shown in Fig.~\ref{fig:negativity}, to illustrate this dependence.

\begin{figure}
\centering
\includegraphics[width=0.8\textwidth]{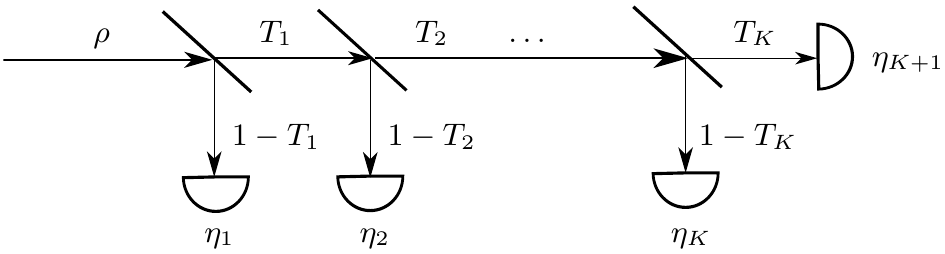}
  \caption{A schematic diagram representing the time-multiplexed setup with $K+1$ output ports. The $T_j$s\label{symbol:transmission} are the respective transmission probabilities for the $j$th beam splitter. The overall efficiency for, say, the $k$th port is given by $\tilde{\eta}_k=\eta_k(1-T_k+T_{K+1}\delta_{k,K+1})\prod^{k-1}_{j=1} T_j$.}
  \label{fig:tmd_diag}
\end{figure}

If the true state lies outside the subspace of interest, then the estimated value of $\mathcal{W}_{00}$ can drastically deviate from the true value. It is clear that a truncation of the Hilbert space into a smaller reconstruction subspace\index{reconstruction subspace}\index{state-space truncation} can lead to diagnostics which are highly incompatible with the true result. So, if one is interested in performing an objective quantum tomography experiment on a given collection of identically-prepared quantum systems with some prior knowledge\index{prior information (knowledge)} regarding its true state, an option would be to reconstruct the MLME estimator in the largest possible subspace based on this prior knowledge\index{prior information (knowledge)}. By enlarging the reconstruction subspace\index{reconstruction subspace}, many more admissible states are taken into consideration and more reliable state estimations and quantum diagnostics can thus be performed. We now have an operational reconstruction scheme that combines our knowledge and ignorance about the unknown true state to give us a unique state estimator\index{unique estimator} in an objective way.
\begin{flushleft}
\uline{\textbf{Time-multiplexed detection tomography}}\index{time-multiplexed detection (TMD) tomography}
\end{flushleft}
Next, we apply the MLME technique to simulation experiments on \emph{time-multiplexed detection} (TMD) tomography \cite{tmd1,tmd2}. For experiments of this type, photon pulses\index{photon pulses} of a particular quantum state, where each pulse is a wave packet containing a few photons, are sent through a series of beam splitters\index{beam splitter}\footnote{The word ``beam splitter'', used in this context, represents a class of possible apparatuses used to split photon pulses, which includes conventional beam splitters, optical fibers\index{optical fibers}, etc.}, each associated with a certain transmission probability\index{transmission probability}. Behind each of the output ports of such a series is a single-photon detector that either registers a click from an incoming split photon pulse, with some detection efficiency\index{detection efficiency}, or does nothing. Thus, each output port has a certain overall efficiency $\tilde{\eta}_j$ which is related to the relevant transmission probabilities and detection efficiency\index{detection efficiency} (See Fig.~\ref{fig:tmd_diag}).

As a consequence of this, the POM outcomes
\begin{equation}
\Pi_{j}=\sum_n\ket{n}c_{jn}\bra{n}
\end{equation}
will be a mixture of Fock\index{Fock states or representation} states, with the coefficients $c_{jn}$ related to $\eta_j$ \cite{fiberloop}. If there are $N_\text{ports}$\label{symbol:nports} output ports, where \emph{all} $\eta_j$s are different, there will be $2^{N_\text{ports}}$ distinct POM outcomes that arise from the binary nature of the single-photon detectors. In addition, $\sum^{2^{N_\text{ports}}}_{j=1}\Pi_j=1$ since the $2^{N_\text{ports}}$ binary sequences of detection configurations constitute all possible events. These POM outcomes commute and a measurement of these outcomes only gives information about the diagonal entries of the statistical operator of the true state in the Fock\index{Fock states or representation} basis. In order to obtain information about the off-diagonal entries, one can, for instance, displace the current set of $2^{N_\text{ports}}$ POM outcomes in phase space with some complex value $\alpha_k$ away from the origin using the displacement operator
\begin{equation}
\mathcal{D}(\alpha_k)=\E^{\alpha_k A^\dagger-\alpha^*_k A}\,.
\end{equation}\label{symbol:dispop}
Then, the new set of outcomes
\begin{equation}
\Pi_j(\alpha_k)=\frac{1}{\mathcal{N}}\mathcal{D}(\alpha_k)\Pi_j\mathcal{D}^\dagger(\alpha_k)\,,
\end{equation}
with $\mathcal{N}$ being the total number of such displaced set of $2^{N_\text{ports}}$ outcomes, do not commute with the undisplaced set. These displaced outcomes are suitable for a measurement that is designed to obtain information about the unknown true state by sampling over multiple $\alpha_k$s. Experimentally, these displaced POM outcomes can be realized with unbalanced homodyne detection \cite{unbalanced}.

In the simulations, four output ports, corresponding to a total of $2^4=16$ POM outcomes, are considered. Two different true states are selected to illustrate the results of MLME. The first true state is chosen to be a stationary state\index{stationary state of a laser} of a laser given by
\begin{equation}
\rho_\text{ss}=\E^{-\mu}\sum^\infty_{n=0}\ket{n}\frac{\mu^n}{n!}\bra{n}\,,
\label{laser_ss}
\end{equation}\label{symbol:rhoss}where $\mu$ defines the mean number of photons \cite{laserss}. For the second true state, the state \mbox{$\rho_{\alpha'}=\ket{\,\,\,}_{\alpha'}{\vphantom{\bra{\,\,\,}}}_{\alpha'}\!\bra{\,\,\,}$}, where
\begin{equation}
\ket{\,\,\,}_{\alpha'}=\frac{\ket{\alpha'}+\ket{-\alpha'}}{\sqrt{2\left(1+\E^{-2|\alpha'|^2}\right)}}\,
\label{schro_cat}
\end{equation}
is the superposition of the coherent states\index{coherent states} $\ket{\alpha'}$ and $\ket{-\alpha'}$, is chosen. Statistical operators are first reconstructed from the simulated data. For this reconstruction, one has to decide on the dimension $D_\text{sub}$\label{symbol:dsub} of the truncated Hilbert space for the reconstructions. This procedure, also commonly known as \emph{state-space truncation}, depends on the prior information\index{prior information (knowledge)} about the unknown state. In our case, suppose one knows that the mean number of photons of the source is $\mu\approx 4$, which is the value assigned in the simulation. Then, one may anticipate that all the relevant information about the true state should be contained in a Hilbert space of a dimension which is close to $\mu$. In fact, it is a common practice to choose $D_\text{sub}$, compatible with this information, such that the displaced operators form an informationally complete POM. Then, the standard ML method can be applied to state estimation. We shall compare the result of this approach with another, perhaps more objective, methodology in which we select a larger subspace compatible with this prior information\index{prior information (knowledge)} and estimate the state with MLME.

To represent the reconstructed statistical operators $\hat\rho_\text{sub}$, the Wigner functions\index{Wigner functional} $\mathcal{W}(x,p)$ of the dimensionless position and momentum quadrature values, $x$ and $p$ respectively, are calculated in accordance with\footnote{Refer to Appendix \ref{chap:wigner} for its derivation.}
\begin{align}
&\mathcal{W}(x,p)\nonumber=2\E^{-|\alpha|^2}\sum^{D_\text{sub}-1}_{m=0}\sum^{D_\text{sub}-1}_{n=0}\bra{m}\hat\rho_\text{sub}\ket{n}\nonumber\\
\times &\left[(-1)^{n_<}\sqrt{\frac{2^{n_>}n_<!}{2^{n_<}n_>!}}(x+\I^{\,\text{sgn}(n-m)} p)^{n_>-n_<}L_{n_<}^{(n_>-n_<)}\left(2\,|\alpha|^2\right)\right]\,,
\label{wigner}
\end{align}
where $L_{n}^{(\nu)}(y)$ is the degree-$n$ \emph{associated Laguerre polynomial}\label{symbol:laguerre}\index{Laguerre polynomials (associated)} (Edmond Nicolas Laguerre\index{Edmond Nicolas Laguerre}) in $y$ of order $\nu$ and $\alpha=x+\I p$, for all the statistical operators. Here, we define $n_<\equiv \min\{m,n\}$ and $n_>\equiv \max\{m,n\}$.

\begin{figure}
\centering
\includegraphics[width=0.8\textwidth]{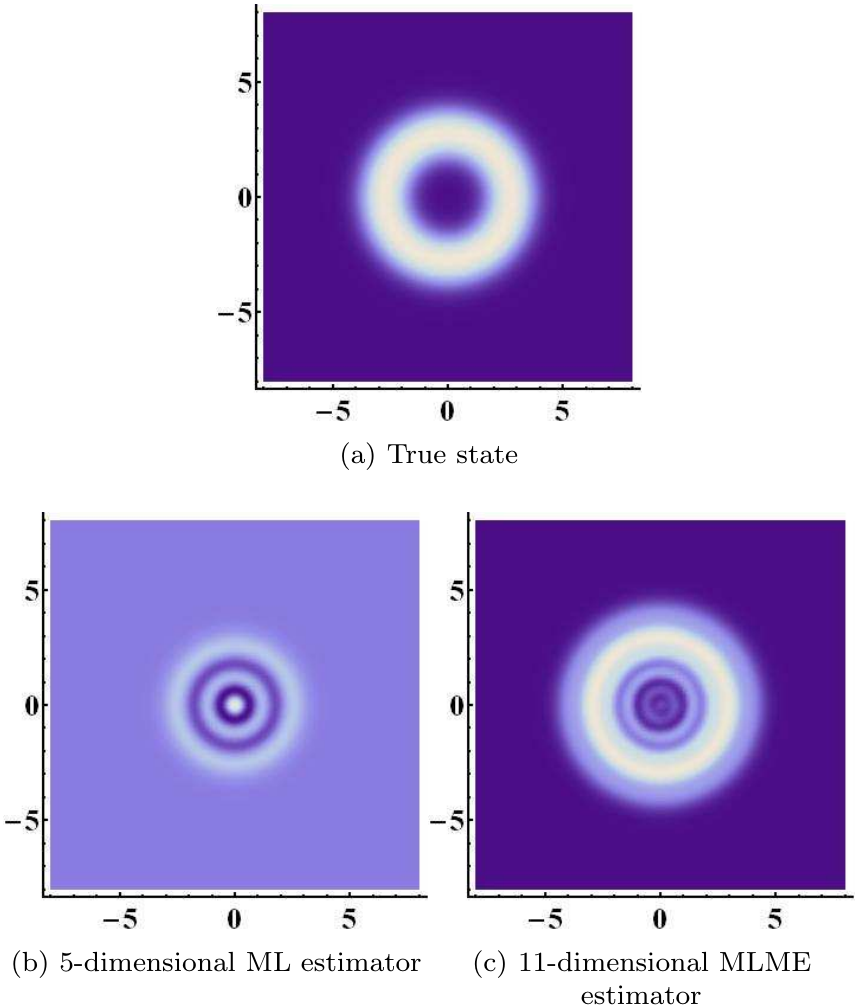}
  \caption{Density plots of the Wigner functions, in phase space, of various statistical operators for (a) the true state (20-dimensional stationary state of a laser, $\mu=4$) with $\tilde\tau\approx 0.394$, (b) the 5-dimensional ML estimator with $\tilde\tau\approx 0.921$ and (c) the 11-dimensional MLME estimator with $\tilde\tau\approx 0.489$. Here, brighter regions indicate the locations of larger Wigner function values, and vice versa. The statistical operator for (b) is obtained using ML by assuming a 5-dimensional subspace in which the displaced POM outcomes are informationally complete. The statistical operator for (c) is obtained by assuming a larger subspace of dimension 11 using MLME. Numerous artificial non-classical features of the ML estimator, a signature of its highly oscillatory Wigner function, are manifested as an abnormally large value of $\tilde\tau$, an inevitable byproduct of state-space truncation. One can see that with MLME, extraneous artifacts of the Wigner function resulted from such a truncation can be largely removed.}
  \label{fig:tmd_density}
\end{figure}
\begin{figure}
\centering
\includegraphics[width=0.8\textwidth]{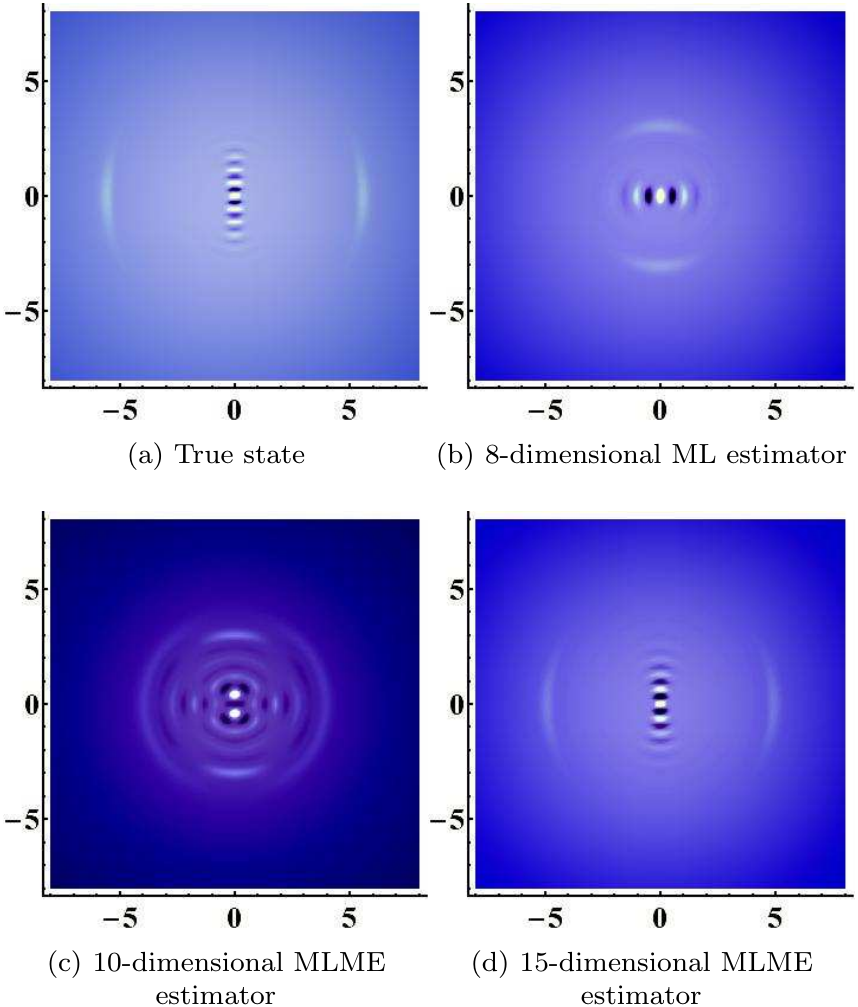}
  \caption{Density plots of the Wigner functions, in phase space, of various statistical operators for (a) the true state ($\rho_{\alpha'}$, $\alpha'=5$), (b) the 8-dimensional ML estimator, (c) the 10-dimensional and (d) 15-dimensional MLME estimators. In this case, the Wigner function of the ML estimator differs greatly from that of the true state, an example of misleading information obtained via state-space truncation. A transition in the structure of the Wigner function occurs at $D_\text{sub}=10$, with the MLME estimator for $D_\text{sub}=15$ giving a more accurate estimated picture of the Wigner function of the true state.}
  \label{fig:tmd_density_cat}
\end{figure}

To quantify the non-classicality\index{non-classicality} of the statistical operators, we make use of the concept of \emph{non-classicality depth}\index{non-classicality!-- depth} introduced in Ref.~\cite{nonclassical_depth}. Let us define the function
\begin{equation}
\mathcal{R}(\alpha;\tau)=\frac{1}{\pi\tau}\int (\D w)^2\,\text{exp}\left(-\frac{|\alpha/\sqrt{2}-w|^2}{\tau}\right)P(w)\,,
\end{equation}\label{symbol:rfunc}where $w$ is a complex variable, $P(w)$ is the Glauber-Sudarshan $P$ function, and the parameter $\tau$ is in the range $0\leq\tau\leq1$. From the above definition, it follows that $\mathcal{R}(\alpha;\tau)$ is a continuous interpolating function of $\tau$ from the typically singular, as well as non-positive, $P(\alpha/\sqrt{2})$ ($\tau\rightarrow 0$), to the Wigner function $\mathcal{W}(\alpha)$ ($\tau=1/2$), and finally to the positive \emph{Husimi $\mathcal{Q}$ function}\label{symbol:qfunc}\index{Husimi $\mathcal{Q}$ function} (K{\^o}di Husimi\index{K{\^o}di Husimi}) $\mathcal{Q}(\alpha/\sqrt{2})=\bra{\alpha/\sqrt{2}}\rho\ket{\alpha/\sqrt{2}}$ ($\tau\rightarrow 1$). The non-classicality depth\index{non-classicality!-- depth} is then defined as the smallest value $\tau=\tilde\tau$\label{symbol:taudepth}, above which $\mathcal{R}(\alpha;\tau)\geq0$. Any mixture of coherent states\index{coherent states} is therefore a classical state since, in this case, $\tilde\tau=0$. A quantum state with $\tilde\tau>0$ is a non-classical state. This measure of non-classicality\index{non-classicality} captures the non-classical nature of quantum states through a one-parameter family of functions, which can otherwise be invisible to measures involving a fixed value of $\tau$, such as the conventional negativity of the Wigner function. This non-classicality depth is but one of a few approaches for quantifying the non-classicality of quantum states and we will, without fixating on this quantity, adopt it as an appropriate measure that is not worse than other proposals. The generalization of Eq.~\eqref{wigner}\footnote{Refer to Appendix \ref{chap:nonclass_depth} for its derivation.} to arbitrary values of $\tau$,
\begin{align}
&\mathcal{R}(x,p;\tau)=\frac{\E^{-\frac{|\alpha|^2}{2\tau}}}{\tau}\sum^{D_\text{sub}-1}_{m=0}\sum^{D_\text{sub}-1}_{n=0}\bra{m}\hat\rho_\text{sub}\ket{n}\nonumber\\
\times \Bigg[&\,(-1)^{n_<}\sqrt{\frac{n_<!}{n_>!}}\left(\frac{1-\tau}{\tau}\right)^{n_>}\left(\frac{x+\I^{\,\text{sgn}(n-m)} p}{\sqrt{2}(1-\tau)}\right)^{n_>-n_<}L_{n_<}^{(n_>-n_<)}\left(\frac{|\alpha|^2}{2\tau(1-\tau)}\right)\Bigg]\,,
\label{nonclass_eq}
\end{align}
is useful for the numerical computation of $\tilde\tau$. For the truncated version
\begin{equation}
\rho^\text{sub}_\text{ss}=\frac{1}{\sum^{D_\text{sub}-1}_{n=0}\frac{\mu^n}{n!}}\,\sum^{D_\text{sub}-1}_{n=0}\ket{n}\frac{\mu^n}{n!}\bra{n}
\label{laser_ss_sub}
\end{equation}
of the stationary state in Eq.~(\ref{laser_ss}), taking , Eq.~(\ref{nonclass_eq}) simplifies to
\begin{align}
&\mathcal{R}_{\text{ss}}(x,p;\tau)\nonumber\\
=&\,\frac{\E^{-\frac{|\alpha|^2}{2\tau}}}{\tau\,\sum^{D_\text{sub}-1}_{n=0}\frac{\mu^n}{n!}}\,\sum^{D_\text{sub}}_{n=0}(-1)^n\,\frac{\mu^n}{n!}\left(\frac{1-\tau}{\tau}\right)^nL_n\left(\frac{|\alpha|^2}{2\tau(1-\tau)}\right)\,.
\end{align}

The performances of both MLME and the standard ML method on the true states defined in Eqs.~(\ref{laser_ss}) and (\ref{schro_cat}) are illustrated by the Wigner function plots of the respective statistical operators obtained from both methods. These are shown in Figs.~\ref{fig:tmd_density} and \ref{fig:tmd_density_cat}. The respective non-classicality depths\index{non-classicality!-- depth} are also computed for Fig.~\ref{fig:tmd_density}. For the state $\rho_{\alpha'}$, all the reconstructed statistical operators are highly non-classical, with $\tilde\tau=1$ \cite{telenonclass} for all them. Rather than comparing the $\tilde\tau$ values, the structure of the Wigner functions for various reconstruction subspaces\index{reconstruction subspace} will be briefly analyzed instead in Fig.~\ref{fig:tmd_density_cat}.

\begin{flushleft}
\uline{\textbf{Light-beam tomography}}\index{light beam tomography}
\end{flushleft}

\begin{figure}[h!]
\centering
\includegraphics[width=0.6\textwidth]{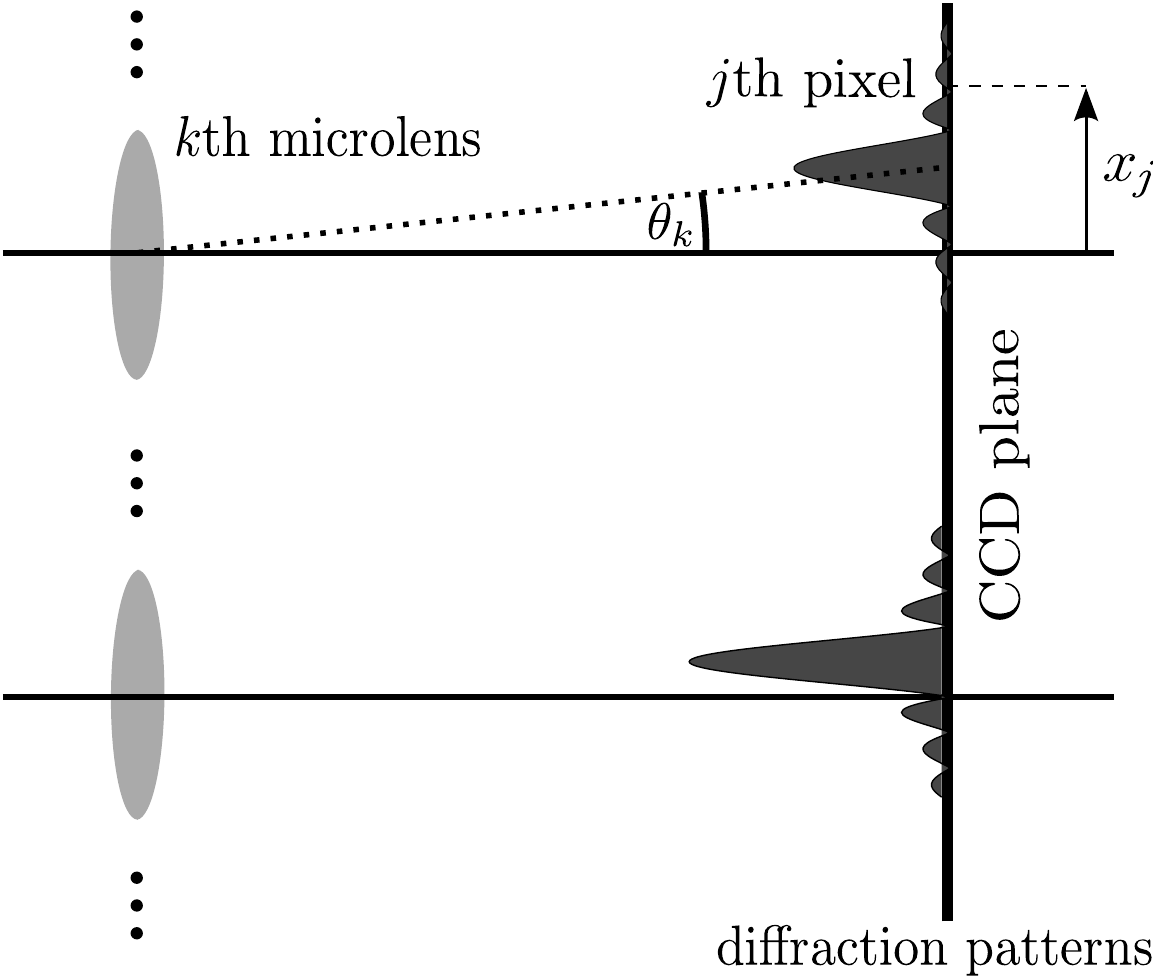}
\caption{Schematic diagram of the diffraction patterns of an incoming light beam that is obtained from a SH wave front sensor. The light beam is transformed by an array of microlenses (apertures). A CCD camera is placed at the rear focal plane\index{focal plane} of the array. The measurement data consist of the measured intensities of the beam. The intensity\index{intensity} at the $j$th pixel\index{pixel}, located at position\index{position} $x_j$, behind the $k$th microlens aperture is denoted by $I_k(x_j)$.\label{fig:sh}}
\end{figure}

Finally, we make use of the MLME algorithm to reconstruct states of classical light beams that are measured using the Shack-Hartmann\index{Shack-Hartmann (SH)} (SH) wave front\index{wave front} sensor (Roland Shack\index{Roland Shack} and Johannes Franz Hartmann\index{Johannes Franz Hartmann}). An incoming light beam is transformed by a regular array of \emph{microlens} apertures\index{microlens apertures or microlenses} and detected in its rear focal plane\index{focal plane} by a charge-coupled device\index{charge-coupled device (CCD)} (CCD) camera (see Fig.~\ref{fig:sh}). A plane wave\index{plane wave} traversing in the transverse plane of the SH sensor gives rise to a detection, where the individual diffraction\index{diffraction} patterns are centered at the corresponding optical centers of the microlenses\index{microlens apertures or microlenses}. For a distorted wave front\index{wave front}, the observed diffraction\index{diffraction} pattern behind the $k$th microlens aperture\index{microlens apertures or microlenses} will be deflected by an angle $\theta_k$. Since the set of angles $\theta_k$ is related to the local wave front\index{wave front} tilts with respect to the transverse plane of the SH sensor, the shape of the wave front\index{wave front} can be inferred. Clearly, this standard technique of wave front\index{wave front} reconstruction fails in the presence of imperfect coherence, where the notions of ``wave front''\index{wave front} and ``optical phase'' are no longer well-defined and a more general description of the state of the light beam is necessary.

Recently, an alternative theory for SH detection, based on the principles of quantum state tomography, has been introduced. It was shown that a complete characterization of a beam of light is possible from the measurement data obtained with the SH sensor under certain assumptions with regards to the aperture profiles \cite{coherence}. Analogously to quantum states, we can describe a coherent beam\index{coherent beam} (mode), with a complex amplitude\index{complex amplitude} $\psi(x)$\label{symbol:psix}, by a ket $\ket{\psi}$, such that $\psi(x)=\langle x|\psi\rangle$.

The transformation of the complex amplitude $\psi(x)$ of an incoming light beam, which is propagating from the $k$th microlens aperture to the SH sensor, can be described by the linear transformation \cite{fourieroptics}
\begin{equation}
\psi^{(k)}_\text{prop}(x)=\mathcal{T}^{(k)}_\text{prop}\bigl(\psi(x)\bigr)\,.
\end{equation}\label{symbol:psipropx}With the identity
\begin{equation}
\psi(x)=\int\D x'\,\delta(x-x')\psi(x')\,,
\end{equation}
the complex amplitude $\psi^{(k)}_\text{prop}(x)$, after propagation, is given by
\begin{equation}
\psi^{(k)}_\text{prop}(x)=\int\D x'\,\mathfrak{h}_k(x-x')\psi(x')\,,
\label{propeqn}
\end{equation}
where $\mathfrak{h}_k(x-x')=\mathcal{T}^{(k)}_\text{prop}\bigl(\delta(x-x')\bigr)$\label{symbol:hkx} is the \emph{impulse response function}\index{impulse response function} of the $k$th microlens\index{microlens apertures or microlenses} aperture, which describes the free propagation\index{propagation} of the beam from the aperture to the SH sensor. Apart from wave propagation that is energy-conserving, there is an additional effect on the wave amplitude as the light beam passes through the microlens aperture that can result in energy attenuation. This is mathematically described by the multiplicative transformation $\psi(x)\rightarrow a_k(x)\psi(x)$, where the aperture function\label{symbol:akx}\index{aperture function} $a_k(x)$ of the $k$th aperture gives the resulting aperture effect on the beam profile. Hence, on the focal plane\index{focal plane} of the $k$th microlens aperture\index{microlens apertures or microlenses} where the SH sensor resides, the final complex amplitude\index{complex amplitude} $\psi'_k(x)$\label{symbol:psixprime} of the beam is given by the convolution integral
\begin{equation}
\psi'_k(x)=\int\D x'\,\mathfrak{h}_k(x-x')a_k(x')\psi(x')\,.
\label{transform}
\end{equation}
Since the detection region of the SH sensor is small, we can compare Eq.~(\ref{propeqn}) with the Fresnel diffraction equation\index{Fresnel diffraction equation} (Augustin-Jean Fresnel\index{Augustin-Jean Fresnel}) for the normalized amplitudes\index{complex amplitude}, that is
\begin{equation}
\psi^{(k)}_\text{prop}(x)=\sqrt{\frac{\zeta}{z}}\,\int\D x'\,\E^{\I\frac{\zeta}{2z}(x-x')^2}\psi(x')\,,
\end{equation}
where $\lambda$ is the wavelength of the beam, $\zeta=2\pi/\lambda$, and irrelevant phase factors are neglected, to conclude that the normalized impulse response function\index{impulse response function} can be defined as
\begin{equation}
\mathfrak{h}_k(x-x')=\sqrt{\frac{\zeta}{z}}\,\E^{\I\frac{\zeta}{2z}(x-x')^2}\,.
\end{equation}
Here, the $z$ direction is taken to be the optical axis\index{optical axis}. It follows that the functions $\mathfrak{h}_k(x-y)$ are orthogonal. That is,
\begin{align}
\int\D x'\,\mathfrak{h}^*_k(x'-x)\mathfrak{h}_k(x'-y)&=\frac{\zeta}{z}\int\D x'\,\E^{-\I\frac{\zeta}{2z}(x'-x)^2}\E^{\I\frac{\zeta}{2z}(x'-y)^2}\nonumber\\
&=\frac{\zeta}{z}\,\E^{-\I\frac{\zeta}{2z}\left(x^2-y^2\right)}\underbrace{\int\D x'\,\E^{\I x'\,\frac{\zeta}{z}(x-y)}}_{=\,\frac{z}{\zeta}\delta(x-y)}\nonumber\\
&=\delta(x-y)\,.
\label{convolution}
\end{align}
More generally, this orthogonality property follows directly from energy conservation\index{energy conservation} of the light field during propagation\index{propagation}. By defining $I^\text{prop}_k(x)$\label{symbol:ipropk} to be the intensity\index{intensity} of the propagated beam from the $k$th aperture, at position\index{position} $x$, to be $I^\text{prop}_k(x)=\big|\psi^{(k)}_k(x)\big|^2$ and $I(x)=\big|\psi(x)\big|^2$ to be the initial intensity at the same position before propagation\index{propagation},
\begin{align}
&\int\D x'\,I^\text{prop}_k(x')\nonumber\\
=&\int\D x'\,\big|\psi^{(k)}_k(x')\big|^2\nonumber\\
=&\int\D x'\int\D x''\,\mathfrak{h}^*_k(x'-x'')\psi^*(x'')\int\D x'''\,\mathfrak{h}_k(x'-x''')\psi(x''')\nonumber\\
=&\int\D x''\int\D x'''\,\psi^*(x'')\psi(x''')\int\D x'\,\mathfrak{h}^*_k(x'-x'')\mathfrak{h}_k(x'-x''')\nonumber\\
=&\int\D x''\,\big|\psi(x'')\big|^2\nonumber\\
=&\int\D x''\,I(x'')\nonumber\\
\Rightarrow&\int\D x'\,\mathfrak{h}^*_k(x'-x'')\mathfrak{h}_k(x'-x''')=\delta(x''-x''')\,.
\end{align}

Suppose now, a generic \emph{partially} coherent beam\index{partially coherent beam} is detected by the SH sensor. We can describe the state of such a beam with a \emph{coherence operator}\index{coherence operator} $\rho_\text{coh}$\label{symbol:rhocoh}. Using a computational basis\index{computational basis} of orthonormal\index{orthonormal} modes $|\psi_n\rangle$, the $D$-dimensional coherence operator $\rho_\text{coh}$ is given by
\begin{equation}
\rho_\text{coh}\,=\,\sum_{mn}\ket{\psi_m}\rho^\text{coh}_{mn}\bra{\psi_n}\,\widehat{=}
\begin{pmatrix}
\rho^\text{coh}_{00}&\cdots&\rho^\text{coh}_{0\,D-1}\\
\vdots&\ddots&\vdots\\
\rho^\text{coh}_{D-1\,0}&\cdots&\rho^\text{coh}_{D-1\,D-1}
\end{pmatrix}\,.
\end{equation}
By defining the aperture operator\index{aperture operator}\label{symbol:ak}
\begin{equation}
\mathfrak{A}_k=\int\D x'\,\ket{x'}a_k(x')\bra{x'}
\end{equation}
for the $k$th microlens aperture and the impulse response operator
\begin{equation}
\mathfrak{U}_k=\int\D x''\int\D x'\,\ket{x'}\mathfrak{h}_k(x'-x'')\bra{x''}
\end{equation}\label{symbol:uk}that is unitary from the orthogonality relation in Eq.~(\ref{convolution}), the representation of the corresponding transformed state $\rho'_\text{coh}$,
\begin{align}
\rho'_\text{coh}&=\,\mathfrak{U}_k\mathfrak{A}_k\,\rho_\text{coh}\,\mathfrak{A}_k\mathfrak{U}^\dagger_k\nonumber\\
&=\sum_{mn}\underbrace{\mathfrak{U}_k\mathfrak{A}_k\ket{\psi_m}}_{\equiv\,\ket{\psi'_m}}\rho^\text{coh}_{mn}\underbrace{\bra{\psi_n}\mathfrak{A}_k\mathfrak{U}_k^\dagger}_{\equiv\,\bra{\psi'_n}}\nonumber\\
&=\sum_{mn}\ket{\psi'_m}\rho^\text{coh}_{mn}\bra{\psi'_n}\,,
\end{align}
on the focal plane\index{focal plane} of the apertures follows from the linearity of optics transformations. The intensity\index{intensity} $I_k(x_j)$\label{symbol:ik} at position\index{position} $x_j$\footnote{In order to talk about a physical position\index{position} ket $\ket{x_j}$, it is important to understand that the specification of $x_j$ comes with a certain \emph{finite} precision. As such, these physical kets now normalize to the Kronecker delta\index{Kronecker delta}, that is $\langle x_j|x_{j'}\rangle=\delta_{jj'}$.} on the focal plane\index{focal plane} of the $k$th aperture is
\begin{equation}
\begin{split}
\label{intensity1}
I_k(x_j)&\equiv\langle x_j|\rho'_\text{coh}|x_j\rangle\\
&=\langle x_j| \bigg(\sum_{mn}\ket{\psi'_{m,j}}\rho^\text{coh}_{mn}\bra{\psi'_{n,k}}\bigg)
|x_j\rangle\\
&=\sum_{mn}\rho^\text{coh}_{mn}\,\psi'_{m,k}(x_j)\,\psi'_{n,k}(x_j)^*\,,
\end{split}
\end{equation}
where $\psi'_{n,k}(x_j)=\langle x_j|\psi'_{n,k}\rangle$ are the complex amplitudes\index{complex amplitude} of the transformed light beam obtained from the amplitudes $\psi_n(x_j)=\langle x_j|\psi_n\rangle$ of Eq.~\eqref{transform}. Since $\rho_\text{coh}$ possesses all the properties of a statistical operator, the MLME technique can be used to estimate the true coherence operator\index{coherence operator} $\rho^\text{true}_\text{coh}$\label{symbol:rhocohtrue} that describes a given light beam. To this end, we need to compute the corresponding POM describing the measurement outcomes of the SH sensor. By relating $I_k(x_j)$ to the corresponding probabilities of the outcomes $\Pi_k(x_j)\widehat{=}\sum_{mn}\ket{\psi'_m}\Pi_{k,nm}(x_j)\bra{\psi'_n}$\label{symbol:pikxj}, we have
\begin{equation}
\label{intensity2}
\begin{split}
I_k(x_j)&=\tr{\rho_\text{coh}\,\Pi_k(x_j)}\\
&=\sum_{mn}\rho^\text{coh}_{mn}\,\Pi_{k,nm}(x_j)\,.
\end{split}
\end{equation}
Comparing Eqs.~\eqref{intensity1} and \eqref{intensity2}, the positive operator describing the detection outcome at the
$j$th pixel\index{pixel} of the CCD\index{charge-coupled device (CCD)} camera behind the $k$th aperture is given by
\begin{equation}
\label{POMelements}
\Pi_{k,nm}(x_j)=\psi'_{m,k}(x_j)\,\psi'_{n,k}(x_j)^*.
\end{equation}

\begin{figure}
\centering
\includegraphics[width=0.9\textwidth]{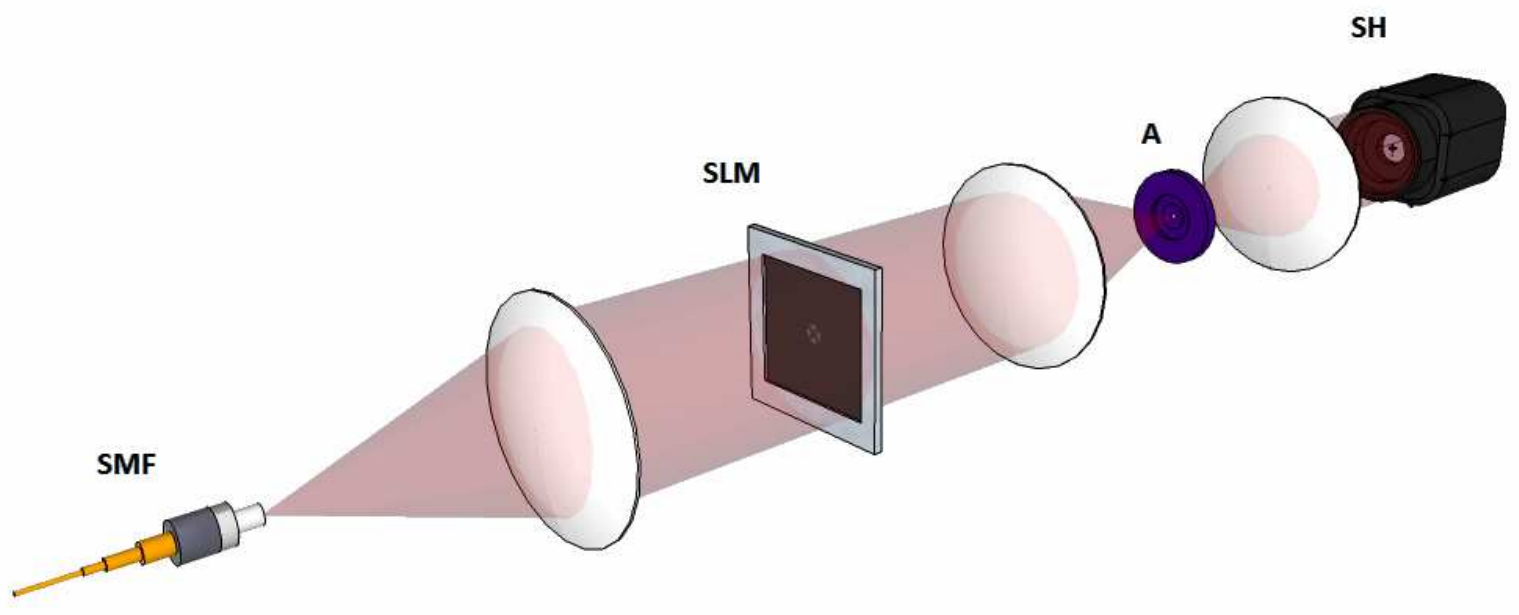}
\caption{Experimental set-up involving a single-mode fiber\index{single-mode fiber} (SMF), a spatial light modulator\index{spatial light modulator} (SLM), an aperture stop (A) and a Shack-Hartmann\index{Shack-Hartmann (SH)} (SH) sensor. \label{fig:setup}}
\end{figure}

In the experiment, a controlled preparation of optical beams is realized using the principles of digital holography\index{digital holography} \cite{setup1,setup2}. Figure~\ref{fig:setup} shows the set-up. The essence of the beam preparation lies in the numerical construction of a digital hologram\index{digital hologram} that is programmed to produce a superposition of a reference plane wave\index{plane wave} and a beam with the true state $\rho^\text{true}_\text{coh}$ of interest. This is achieved with the help of an amplitude spatial light modulator\index{spatial light modulator} (OPTO SLM) with a resolution of 1024$\times$768 pixels\index{pixel}. The hologram\index{digital hologram} is then illuminated by the reference plane wave\index{plane wave} that is considered in the superposition. To approximately produce this plane wave\index{plane wave}, a collimated\index{collimated} Gaussian beam\index{Gaussian beam} is generated by placing the output of a single-mode fiber\index{single-mode fiber} at the focal plane\index{focal plane} of a collimating lens. In this way, the digital hologram\index{digital hologram} can be fully situated at the center of the collimated\index{collimated} Gaussian beam\index{Gaussian beam} of a larger beam waist, where this beam can then be approximated to be a plane wave\index{plane wave} with high accuracy. The resulting diffraction\index{diffraction} spectrum, after illuminating the digital hologram\index{digital hologram} with the collimated\index{collimated} Gaussian beam\index{Gaussian beam}, involves several diffraction\index{diffraction} orders, of which only one contains useful information about $\rho^\text{true}_\text{coh}$. To filter out the unwanted diffraction orders\index{diffraction}, a 4-$f$ optical processor\index{4-$f$ optical processor}, with a small circular aperture stop placed at the rear focal plane\index{focal plane} of the second lens, is used for this purpose (the aperture stop in Fig.~\ref{fig:setup}). The resulting light beam with the state $\rho^\text{true}_\text{coh}$ is then focussed at the rear focal plane\index{focal plane} of the third lens. This completes the preparation stage.

The measurement of the light beam involves a Flexible Optical SH sensor with 128 microlenses\index{microlens apertures or microlenses} that form a hexagonal array. Each microlens\index{microlens apertures or microlenses} has a focal length\index{focal length} of 17.9mm and a hexagonal aperture with a diameter of 0.3mm. The signal at the focal plane\index{focal plane} of the array is detected by a uEye CCD\index{charge-coupled device (CCD)} camera that has a resolution of 640$\times$480 pixels\index{pixel}, with each pixel\index{pixel} being 9.9$\upmu$m$\times$9.9$\upmu$m in dimensions.

The aforementioned set-up is used for generating and analyzing low-order Laguerre-Gaussian (LG)\index{Laguerre-Gaussian modes} modes. The LG\index{Laguerre-Gaussian modes} modes can serve as important resources in quantum information processing \cite{Lgmodes}. In this experiment, only LG\index{Laguerre-Gaussian modes} modes with no radial nodes are considered. Such modes form a one-parameter orthonormal\index{orthonormal} basis, where the modes are specified by the orbital angular momentum quantum number\index{orbital angular momentum!-- quantum number} $l$. In polar coordinates\index{polar coordinates}, the relevant part of the complex amplitude\index{complex amplitude} of a LG\index{Laguerre-Gaussian modes} mode $\text{LG}_l$\label{symbol:lgl}, for a fixed $l$, is given by
\begin{equation}
\langle s,\varphi|\text{LG}_l\rangle\propto\,s^l \E^{\I l \varphi} \E^{-s^2}\,.
\end{equation}
On the other hand, the orbital angular moment operator $L_z$\label{symbol:lz}\index{orbital angular momentum!-- operator $L_z$} in the $z$ direction, in position representation, is given by
\begin{align}
\bra{x,y}L_z&=\bra{x,y}\left(X_xP_y-X_yP_x\right)\nonumber\\
&=\frac{\hbar}{\I}\left(x\,\frac{\partial}{\partial y}-y\,\frac{\partial}{\partial x}\right)\bra{x,y}\,.
\label{lz_pos}
\end{align}
To express the derivatives in Eq.~(\ref{lz_pos}) in terms of polar coordinates, we begin with the parametrization
\begin{align}
x&=s\cos\varphi\,,\nonumber\\
y&=s\sin\varphi\,.
\end{align}
In a compact matrix form, the corresponding variations are then given by
\begin{equation}
\begin{pmatrix}
\updelta x\\
\updelta y
\end{pmatrix}
=\begin{pmatrix}
\cos\varphi & -s\sin\varphi\\
\sin\varphi & s\cos\varphi
\end{pmatrix}
\begin{pmatrix}
\updelta s\\
\updelta \varphi
\end{pmatrix}\,.
\end{equation}
By inverting the matrix equation, we get
\begin{equation}
\begin{pmatrix}
\updelta s\\
\updelta \varphi
\end{pmatrix}
=\frac{1}{s}\begin{pmatrix}
s\cos\varphi & s\sin\varphi\\
-\sin\varphi & \cos\varphi
\end{pmatrix}
\begin{pmatrix}
\updelta x\\
\updelta y
\end{pmatrix}\,.
\label{inv_polar}
\end{equation}
Using the definitions
\begin{align*}
\updelta f&=\left(\updelta x\,\frac{\partial}{\partial x}+\updelta y\,\frac{\partial}{\partial y}\right)f\\
&=\left(\updelta s\,\frac{\partial}{\partial s}+\updelta \varphi\,\frac{\partial}{\partial \varphi}\right)f
\end{align*}
for the total variation of a function $f$ and Eq.~(\ref{inv_polar}), we obtain
\begin{align}
\frac{\partial}{\partial x}&=\cos\varphi\,\frac{\partial}{\partial s}-\sin\varphi\,\frac{1}{s}\,\frac{\partial}{\partial\varphi}\,,\nonumber\\
\frac{\partial}{\partial y}&=\sin\varphi\,\frac{\partial}{\partial s}+\cos\varphi\,\frac{1}{s}\,\frac{\partial}{\partial\varphi}\,.
\end{align}
Hence,
\begin{align*}
\bra{s,\varphi}L_z\ket{\text{LG}_l}&=\frac{\hbar}{\I}\frac{\partial}{\partial\varphi}\langle s,\varphi|\text{LG}_l\rangle\\
&=l\hbar\,\langle s,\varphi|\text{LG}_l\rangle\\
&=\bra{s,\varphi}\,l\hbar\,\ket{\text{LG}_l}
\end{align*}
for all $\bra{s,\varphi}$. This shows that $\ket{\text{LG}_l}$ is an eigenket of $L_z$, implying that each photon, prepared in the state $\ket{\text{LG}_l}\bra{\text{LG}_l}$, carries an orbital angular momentum\index{orbital angular momentum} of $l\hbar$.

For the source of light beams, we would like to prepare the state $\rho^\text{true}_\text{coh}=\rho^\text{sup}_\text{coh}=\ket{\psi_\text{sup}}\bra{\psi_\text{sup}}$, where
\begin{equation}
\label{statetrue}
\ket{\psi_\text{sup}}=\left(\ket{\text{LG}_0}-\ket{\text{LG}_1}\I-\ket{\text{LG}_2}\right)\frac{1}{\sqrt{3}}\,,
\end{equation}
using the OPTO SLM. In the presence of experimental imperfections, however, the true state $\rho^\text{true}_\text{coh}$ prepared this way will not be exactly the same as $\rho^\text{sup}_\text{coh}$. After measuring this beam with the SH sensor, the data are processed using the MLME algorithm in Eq.~(\ref{iteralgo2}) to obtain the estimator $\hat\rho^\text{MLME}_\text{coh}$\label{symbol:rhocohmlme} for $\rho^\text{true}_\text{coh}$, since $G<1$. To quantify the quality of $\hat\rho^\text{MLME}_\text{coh}$, we investigate the \emph{fidelity}\index{fidelity} between $\hat\rho^\text{MLME}_\text{coh}$ and $\rho^\text{sup}_\text{coh}$.

\begin{figure}
\centering
\includegraphics[width=0.8\textwidth]{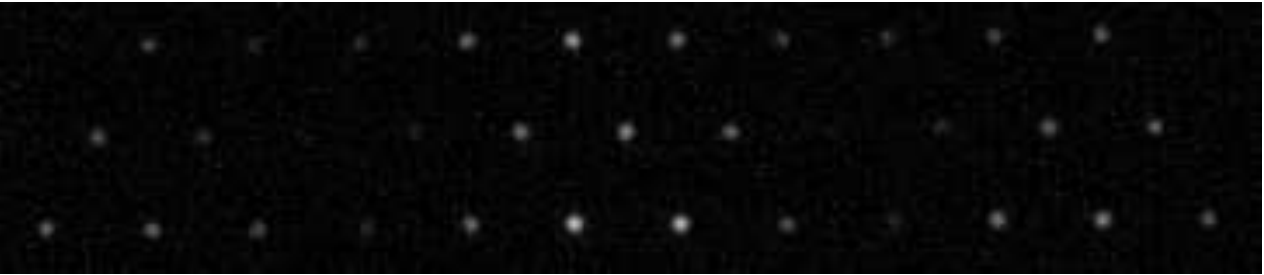}
\caption{CCD\index{charge-coupled device (CCD)} image for the state $\rho^\text{true}_\text{coh}$. The relevant part of the SH readout used for the beam reconstruction is shown. Contributions from the individual SH apertures are indicated by bright spots, with each spot made up of multiple pixels\index{pixel}. Note that the two void regions correspond to the phase singularities of the state $\rho^\text{sup}_\text{coh}$. This hints that $\rho^\text{true}_\text{coh}\approx\rho^\text{sup}_\text{coh}$.
\label{fig:data}}
\end{figure}
Figure~\ref{fig:data} shows the CCD\index{charge-coupled device (CCD)} image for the state $\rho^\text{true}_\text{coh}$. Each aperture gives rise to a bright spot in the CCD\index{charge-coupled device (CCD)} image. To maximize the signal-to-noise ratio\index{signal-to-noise ratio}, only the pixel\index{pixel} with the highest intensity\index{intensity} within each spot is selected as a measurement datum. The set of intensities, corresponding to maximum-intensity\index{intensity} pixels\index{pixel}, constitute the measurement data to be used for state reconstruction. In our case, the corresponding POM consists of $35$ linearly independent outcomes described by Eq.~\eqref{POMelements}. This measurement is, therefore, informationally complete for $D_\text{sub}\leq5$.
\begin{figure}
\centering
\includegraphics[width=0.8\textwidth]{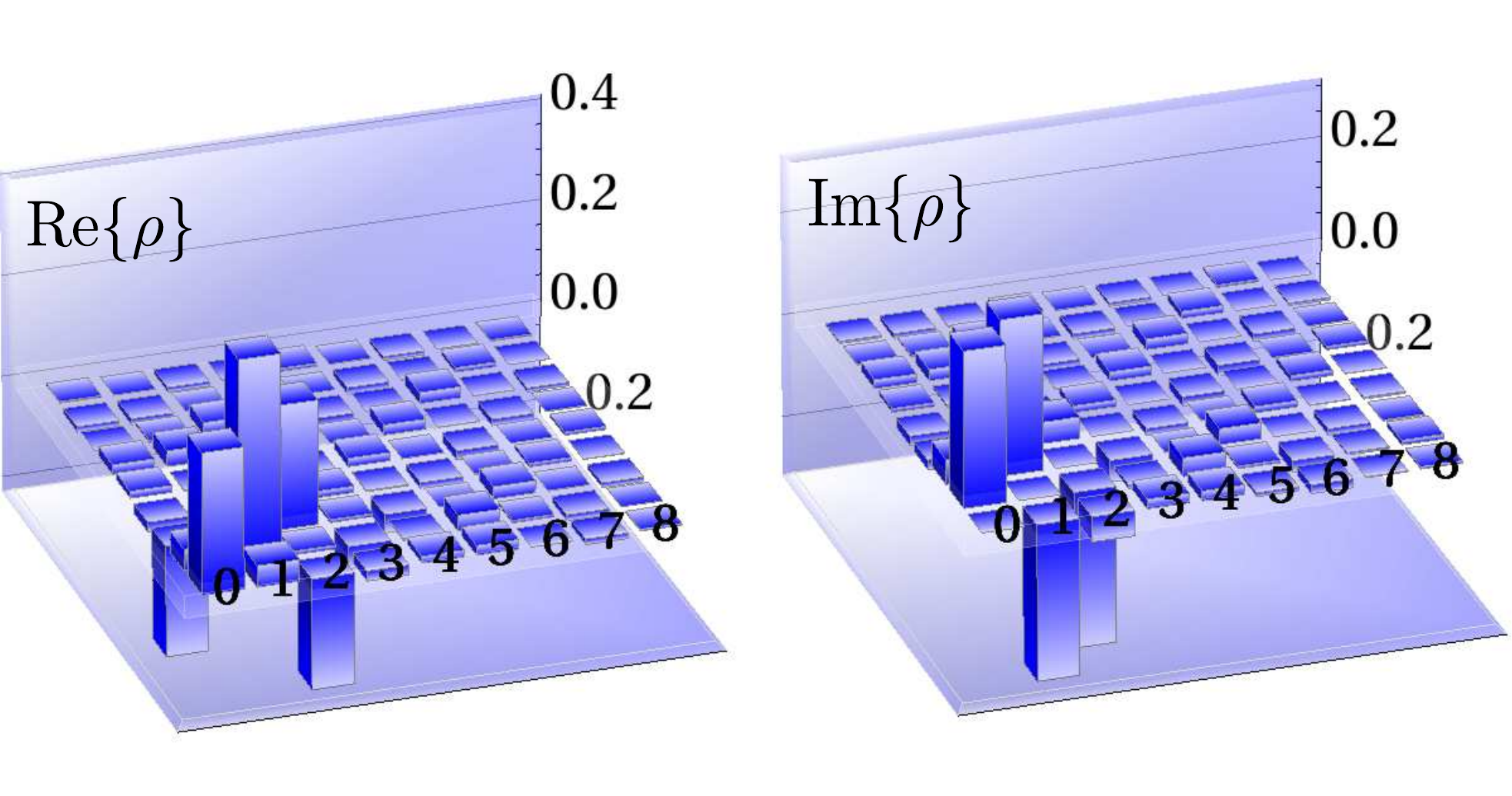}
\caption{MLME state estimation from informationally incomplete data for $D_\text{sub}=9$. The real (left) and imaginary (right) parts of the reconstructed coherence operator $\hat\rho^\text{MLME}_\text{coh}$ are shown. The reconstruction subspace\index{reconstruction subspace} is spanned by the modes $\text{LG}_l$, with $l=0,1,\ldots,8$. In this case, $56$ out of $91$ independent outcomes, required for complete characterization of $\rho^\text{true}_\text{coh}$, are not accessible, yet the MLME estimator $\hat\rho^\text{MLME}_\text{coh}$ is close to $\rho^\text{sup}_\text{coh}$, with a fidelity\index{fidelity} of $92\%$.
\label{fig:recon}}
\end{figure}
In cases where state reconstruction on informationally complete subspaces gives unsatisfactory results, the MLME approach can be used on the informationally incomplete data to give reasonable estimators on a larger subspace, as illustrated in Fig.~\ref{fig:recon}.

So far, the procedure of state-space truncation\index{state-space truncation} is performed in the basis of the $\text{LG}_l$ modes. In this basis, when $\rho^\text{true}_\text{coh}$ is known to be quite close to $\rho^\text{sup}_\text{coh}$, the truncation\index{state-space truncation} of modes of higher orders will not result in a great loss of reconstruction information, as implied by the structure of $\rho^\text{sup}_\text{coh}$ in Eq.~(\ref{statetrue}). The situation will be very different when there is no such prior knowledge about $\rho^\text{true}_\text{coh}$, except for the fact that the possible values of $l$ lie in a certain range. In this situation, there is no appropriate strategy to choose a computational basis\index{computational basis} in which the state-space truncation\index{state-space truncation} can be done effectively and justifiably. More generally, estimating the unknown state $\rho^\text{true}_\text{coh}$ on a truncated subspace can very often result in missing important reconstruction information and this will lead to strongly biased estimators. A remedy for this problem is to perform state reconstruction on a sufficiently large subspace that is compatible with the knowledge about the range of values of $l$.

\begin{figure}[h!]
\centering
\includegraphics[angle=-90,width=0.6\textwidth]{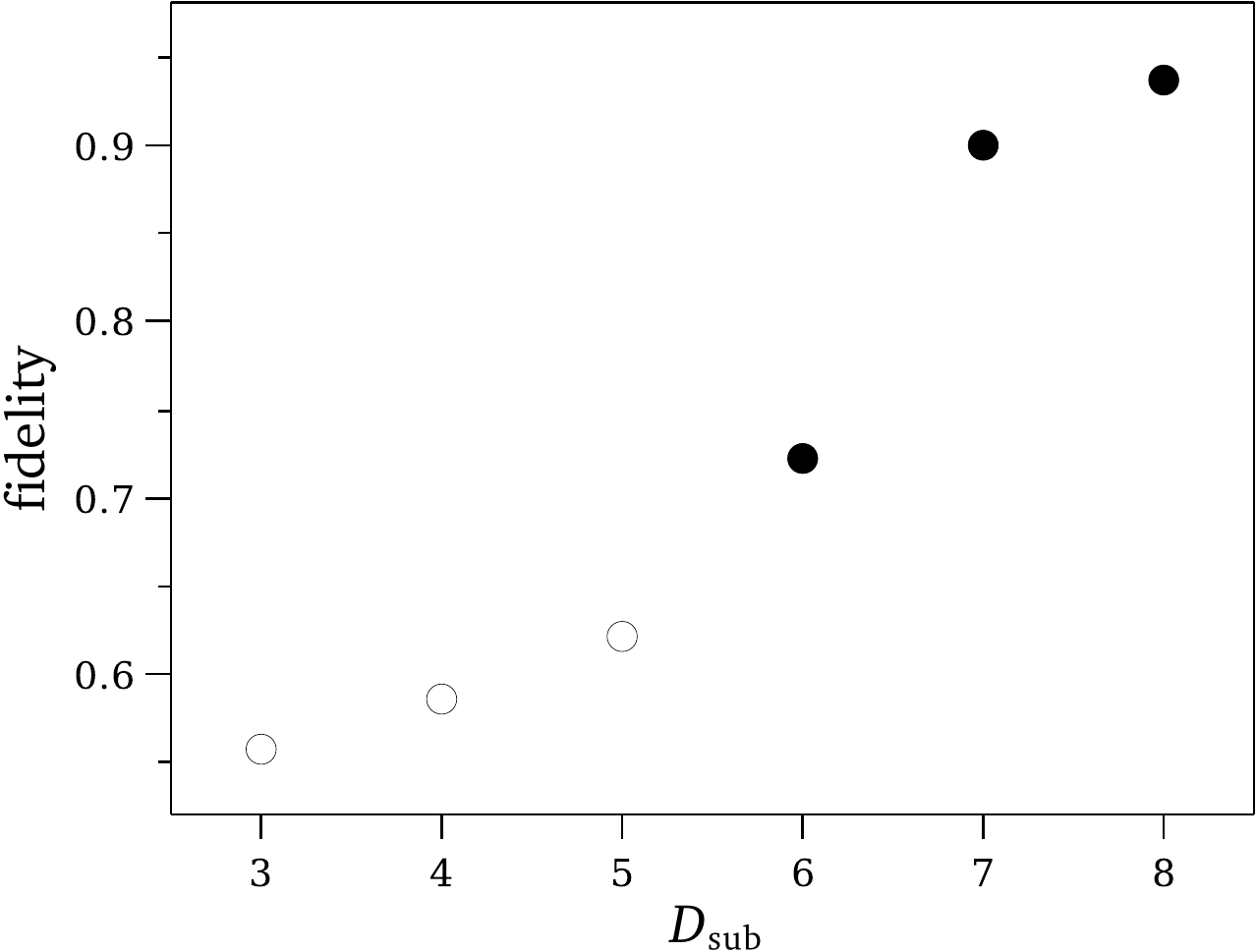}
\caption{Average fidelities, computed over 50 random choices of computational bases, of the estimators for different dimensions $D_\text{sub}$ of the reconstruction subspace\index{reconstruction subspace}. The unfilled (filled) circular plot markers correspond to informationally complete (incomplete) tomography,
respectively.
\label{fig:meanfid}}
\end{figure}

To emphasize this point, we simulate the following scenario:
\begin{itemize}
\item The set of measurement data, obtained from the CCD\index{charge-coupled device (CCD)} image shown in Fig.~\ref{fig:data}, is distributed to $50$ parties. The possible values of $l$ for the true state $\rho^\text{true}_\text{coh}$ are known to lie in the range $l\in[0,7]$.
\item Each party selects a computational basis\index{computational basis} and estimates the state of the beam for $D_\text{sub}=3,4,\ldots,8$ using either the ML (for $D_\text{sub}\leq5$) or the MLME algorithm (for $D_\text{sub}>5$).
\item The reconstructed estimators for the six values of $D_\text{sub}$ are reported by each party and the average fidelity\index{fidelity} of the estimators for every value of $D_\text{sub}$ are calculated.
\end{itemize}
A typical outcome of this scenario is shown in Fig.~\ref{fig:meanfid}. As can be seen, performing state-space truncations\index{state-space truncation} in order to reconstruct $\rho^\text{true}_\text{coh}$ with an informationally complete set of data generally leads to low fidelities in the estimators. Increasing the number of degrees of freedom and using the MLME algorithm to cope with the completeness issue seems to be a much better strategy.\\





\section{Hedged quantum state estimation -- a comparison}\index{hedged maximum-likelihood (HML)}

As strongly advocated by Robin Blume-Kohout\index{Robin Blume-Kohout} \cite{bayesian2,bayesian3}, this method has at least two advantages compared to the maximum likelihood estimation protocol. Firstly the estimator obtained this way is always full-rank, thereby eradicating the problem of zero eigenvalues which are not necessarily justified by a finite number of measurement copies. This is because a zero eigenvalue corresponds to zero probability for a particular outcome in for instance the eigenbasis of the estimator and this requires an extremely high confidence which the measured data cannot give. Secondly, the likelihood functional $\mathcal{L}(\{n_j\};\rho)$ in general has a broad peak over a range of statistical operators. By looking at just the peak of the likelihood functional, one eliminates all other possible states that are close to the maximum. Therefore it is more reliable to take into account all possible states in the vicinity to give an estimator that is much less sensitive to slight changes in the measured data than the maximum likelihood estimator. However it is typically hard to evaluate the integrals and a systematic way of choosing a suitable prior and the volume measure $(\D\rho)$ is unknown \cite{bayesian3}.

Recently, Robin introduced the \emph{hedged likelihood functional}\index{hedged likelihood functional} \cite{HMaxLik} which is given by
\begin{equation}
\mathcal{L}_\text{H}(\{n_j\};\rho)=(\det\rho)^\beta\mathcal{L}(\{n_j\};\rho)\,,
\end{equation}\label{symbol:likelihoodhedged}
where $\beta\approx\frac{1}{2}$. It is analogous to the classical Bayesian method of supplying a Dirichlet-type\index{Dirichlet prior} prior probability distribution (Johann Peter Gustav Lejeune Dirichlet\index{Johann Peter Gustav Lejeune Dirichlet}) and gives the following estimated probabilities
\begin{equation*}
\hat{p}_j=\frac{n_j+\beta}{N+D\beta}
\end{equation*}
when the measurement operators are now projectors of any complete set of orthonormal\index{orthonormal} basis states in the $D$ dimensional Hilbert space. This smooth, unitarily-invariant hedging functional\index{hedging functional} $(\det\rho)^\beta$ was proven to be the unique one for carrying out such a transformation. It is shown that maximizing this functional will result in an estimator which is always full-rank and therefore more compatible with finite number of measurement copies.

In this last section of Chap.~\ref{chap:qse}, we first review some properties of $\mathcal{L}_\text{H}(\{n_j\};\rho)$ which were mentioned in \cite{HMaxLik} using variational methods in \S\ref{subsec:hedged}. Next we will derive an iterative scheme to maximize $\mathcal{L}_\text{H}(\{n_j\};\rho)$ based on the steepest-ascent\index{steepest-ascent method} method in \S\ref{subsec:hmaxlik}. In \S\ref{subsec:tomoinc}, we will discuss informationally incomplete measurements and report some interesting features with regards to the hedged maximum likelihood estimators. In particular, we first prove that given any POM in general, informationally complete or not, the estimator that maximizes $\mathcal{L}_\text{H}(\{n_j\};\rho)$ (HML estimator) is unique\index{unique estimator}. Next we show, by means of qubit tomography simulations, that for some POMs, the HML estimators are actually relatively close to the estimators that maximize both the conventional likelihood and von Neumann entropy functionals simultaneously (MLME estimator) even for relatively small $N$.

\subsection{The hedged likelihood functional}
\label{subsec:hedged}
The main objective is to maximize the concave\index{concave function or functional} hedged likelihood functional $\mathcal{L}_\text{H}(\{n_j\};\rho)$ defined as
\begin{equation}
\mathcal{L}_\text{H}(\{n_j\};\rho)=(\det\rho)^\beta\mathcal{L}(\{n_j\};\rho)\,,\,\mathcal{L}(\{n_j\};\rho)=\prod_j{p_j^{n_j}}\,,
\end{equation}
where the probabilities $p_j=\tr{\rho\Pi_j}$. As always, we can equivalently maximize the log-likelihood functional $\log \mathcal{L}_\text{H}(\{n_j\};\rho)$. Performing a variation on $\log \mathcal{L}_\text{H}(\{n_j\};\rho)$, we have
\begin{align}
\updelta \log \mathcal{L}_\text{H}(\{n_j\};\rho)&=\beta\frac{\updelta\det\rho}{\det\rho}+\updelta\log\mathcal{L}(\{n_j\};\rho)\nonumber\\
&=\tr{\big(\beta\rho^{-1}+NR\big)\updelta\rho}\,,
\label{deltaf}
\end{align}
where $R=\sum_j{f_j\Pi_j/p_j}$. To get the second equality for the first term, we invoke the identity
\begin{equation}
\updelta\det\rho=\left(\det\rho\right)\tr{\rho^{-1}\updelta\rho}\,.
\end{equation}
With the usual parametrization presented in Eq.~(\ref{aux_param}), we obtain the variation
\begin{equation}
\updelta\log \mathcal{L}_\text{H}(\{n_j\};\rho)=\frac{\tr{\big[\beta(\rho^{-1}-D)+N(R-1)\big](\mathcal{A}^\dagger\updelta \mathcal{A} + \updelta\mathcal{A}^\dagger \mathcal{A})}}{\tr{\mathcal{A}^\dagger \mathcal{A}}}\,.
\label{deltafresult}
\end{equation}
By setting $\updelta \log \mathcal{L}_\text{H}(\{n_j\};\rho)= 0$, we arrive at the extremal equation
\begin{equation}
\beta(1-D\hat{\rho}_\text{HML})+N(R_{HML}-1)\hat{\rho}_\text{HML}=0\,,
\label{extremalrho}
\end{equation}\label{symbol:rhohml}
where $R_\text{HML}=\sum_jf_j\Pi_j/\hat{p}_j$ with $\hat{p}_j=\tr{\hat{\rho}_\text{HML}\Pi_j}$.

From the extremal equation in (\ref{extremalrho}), we can recover two properties of $\hat{\rho}_\text{HML}$ which were mentioned in \cite{HMaxLik}. Assuming now that the POM outcomes are projectors of a given set of $D$ orthonormal\index{orthonormal} basis states used to represent $\hat{\rho}_\text{HML}$, i.e.
\begin{equation*}
\Pi_j=\ket{j}\bra{j}\,,\,\hat{\rho}_\text{HML}=\sum_j{\ket{j}\hat{p}_j\bra{j}}.
\end{equation*}
This set of measurements is not informationally complete since the number of measurement outcomes is $D$, which is less than the minimal number $D^2$ required to unambiguously specify a state. Then by direct substitution of the forms of $\Pi_j$ and $\hat{\rho}_\text{HML}$ into Eq.~(\ref{extremalrho}), multiplying $\Pi_k$ on both sides and taking the trace, one can obtain the expression for $\hat{p}_k$ as
\begin{equation}
\hat{p}_k=\frac{n_k+\beta}{N+D\beta}\,,
\label{eigenvalue}
\end{equation}
which is exactly the \textquotedblleft\emph{add $\beta$ rule}\textquotedblright\, that assigns a small non-zero probability for outcomes with zero occurrence in a finite-sample tomography experiment.

To show the next property, that is the eigenvalues of $\hat{\rho}_\text{HML}$ are non-zero for any $\Pi_j$ in general, a transparent approach is to rewrite both Eq.~(\ref{extremalrho}) and its corresponding adjoint statement as
\begin{equation}
\Bigg[D-\frac{N}{\beta}(R_\text{HML}-1)\Bigg]\hat{\rho}_\text{HML}=\hat{\rho}_\text{HML}\,\Bigg[D-\frac{N}{\beta}(R_\text{HML}-1)\Bigg]=1\,.
\label{identity}
\end{equation}
It is now clear that the extremal equation enforces the existence of the inverse of any $\hat{\rho}_\text{HML}$, with
\begin{equation}
\hat{\rho}_\text{HML}^{-1}=D-\frac{N}{\beta}(R_\text{HML}-1)\,.
\end{equation}
This means that for any non-zero $\beta$, $\hat{\rho}_\text{HML}$ is always full-rank. Therefore the peak of $\mathcal{L}_\text{H}(\{n_j\};\rho)$ for any given set of $f_j$s always lies inside the admissible state space. This is consistent with the fact that the hedged likelihood functional goes smoothly to zero on the boundary of the state space.

It was also reported that for most of the mixed states, taking $\beta=1/2$ gives optimal estimation results with respect to some distance measures between the true state $\rho_\text{true}$ and $\hat{\rho}_\text{HML}$. For nearly-pure states, a small value of $\beta$ is needed to achieve good accuracy since now the true states can have eigenvalues that are very close to zero and so large $\beta$ values can result in significant deviations. Keeping in mind that pure states are, strictly speaking, a fiction in practical state preparation, we will set the $\beta=1/2$ in the subsequent analysis.

\subsection{The HML algorithm}
\label{subsec:hmaxlik}\setlength{\parskip}{0pt}
A way of searching for the maximum of the hedged likelihood functional is to start from an arbitrary state, usually the maximally-mixed state $\rho_1=1/D$, and ascend in the direction of the steepest gradient. To determine this direction, we revisit Eq.~(\ref{deltafresult}) and recognize that the two component gradient $\vec{\partial}\log\mathcal{L}_\text{H}$ is given by
\begin{equation}
\vec{\partial}\log\mathcal{L}_\text{H}=\begin{pmatrix}
\partial \log\mathcal{L}_\text{H}/\partial \mathcal{A}\\
\partial \log\mathcal{L}_\text{H}/\partial \mathcal{A}^\dagger
\end{pmatrix},\,\Bigg(\frac{\partial \log\mathcal{L}_\text{H}}{\partial\mathcal{A}}\Bigg)^\dagger=\frac{\partial \log\mathcal{L}_\text{H}}{\partial \mathcal{A}^\dagger}\,,
\end{equation}
where
\begin{equation}
\frac{\partial \log\mathcal{L}_\text{H}}{\partial \mathcal{A}}=\frac{\mathcal{A}\big[\beta(\rho^{-1}-D)+N(R-1)\big]}{\tr{\mathcal{A}^\dagger \mathcal{A}}}\,,
\label{derivative}
\end{equation}
which follows from Eq.~(\ref{total_var_op}).

In order to ensure that $\updelta \log\mathcal{L}_\text{H}$ is always positive in the search process, we can set the variations $\updelta \mathcal{A}$ and $\updelta \mathcal{A}^\dagger$ to be proportional to the respective derivatives $\partial \log\mathcal{L}_\text{H}/\partial \mathcal{A}$ and $\partial \log\mathcal{L}_\text{H}/\partial \mathcal{A}^\dagger$, the steepest-ascent\index{steepest-ascent method} method. Thus, the variation of the two component vector operator $\vec{Z}=\transpose{(\mathcal{A},\mathcal{A}^\dagger)}$ is given by
\begin{equation}
\updelta\vec{Z}=\begin{pmatrix}
\updelta \mathcal{A}\\
\updelta \mathcal{A}^\dagger
\end{pmatrix}
\equiv\frac{\epsilon}{2}\begin{pmatrix}
\mathcal{A}\big[\beta(\rho^{-1}-D)+N(R-1)\big]\\
\big[\beta(\rho^{-1}-D)+N(R-1)\big]\mathcal{A}^\dagger
\end{pmatrix},
\end{equation}
for a small $\epsilon$ parameter. We thus have a simple iterative scheme\index{algorithm} (HML) to look for the extremal state $\hat{\rho}_\text{HML}$ which maximizes the hedged likelihood given an initial statistical operator $\rho_1$, which is given by
\begin{center}
\colorbox{light-gray}{\begin{minipage}[c]{12cm}
\uline{\textbf{HML iterative equations}}
\begin{align}
\rho_\text{k+1}=\frac{\big[1+\Delta_k\big]\rho_k\big[1+\Delta_k\big]}{\tr{\big[1+\Delta_k\big]\rho_k\big[1+\Delta_k\big]}}\,,\\\nonumber
\Delta_k=\frac{\epsilon}{2}\big[\beta(\rho_k^{-1}-D)+N(R_k-1)\big]\,.
\end{align}
  \end{minipage}}
  \end{center}

There exists a slight technical detail in choosing an appropriate $\epsilon$ for the entire iteration. We note that the ratio $\Delta_k/\epsilon$ involves the inverse of $\rho_k$ in every step and is of the order of $N$, which can be significantly large as the number of detected copies increases. Setting $\epsilon$ too large, even to the order of 1, can result in a rank deficient $\rho_k$ that can produce an indeterminate inverse since the iterative equation tends to that of ML for large $N$. By experience, $\epsilon=1/N$ seems to be a wise choice.



\subsection{Informationally incomplete measurements}
\label{subsec:tomoinc}
Typically, we use a set of informationally complete measurement outcomes to infer a positive statistical operator which is compatible with the measured data. One can do this by looking for the unique statistical operator\index{unique estimator} which maximizes the conventional likelihood functional $\mathcal{L}(\{n_j\};\rho)$. We will therefore require at least a minimal set of $D^2$ linearly independent measurement outcomes to obtain a unique estimator\index{unique estimator}. The situation changes when we perform informationally incomplete measurements. As discussed previously, MLME is one method of obtaining a unique and statistically meaningful estimator\index{unique estimator} out of a set of informationally incomplete data.

An interesting property of the estimator $\hat{\rho}_\text{HML}$ is that it is \textbf{always} unique\index{unique estimator} for any given set of measurement outcomes $\Pi_j$ (See Appendix \ref{chap:hedged_unique} for a proof). This implies that regardless of whether a set of measurement outcomes is informationally complete, maximizing the hedged likelihood functional always gives a unique estimator\index{unique estimator}. One can understand this intuitively by drawing analogy from the information functional $\mathcal{I}(\lambda;\rho)$ discussed in \S\ref{subsec:new_perspective}. Then, it is convenient to treat the functional $\beta\log\left(\det\rho\right)$ as an ``entropy-like'' term much like the von Neumann entropy functional $-\tr{\rho\log\rho}=-\log\left(\det\left\{\rho^{-\rho}\right\}\right)$. In this sense, the mechanism of HML is rather similar to that of MLME.

\begin{figure}[h!]
\centering
\includegraphics[width=0.6\textwidth]{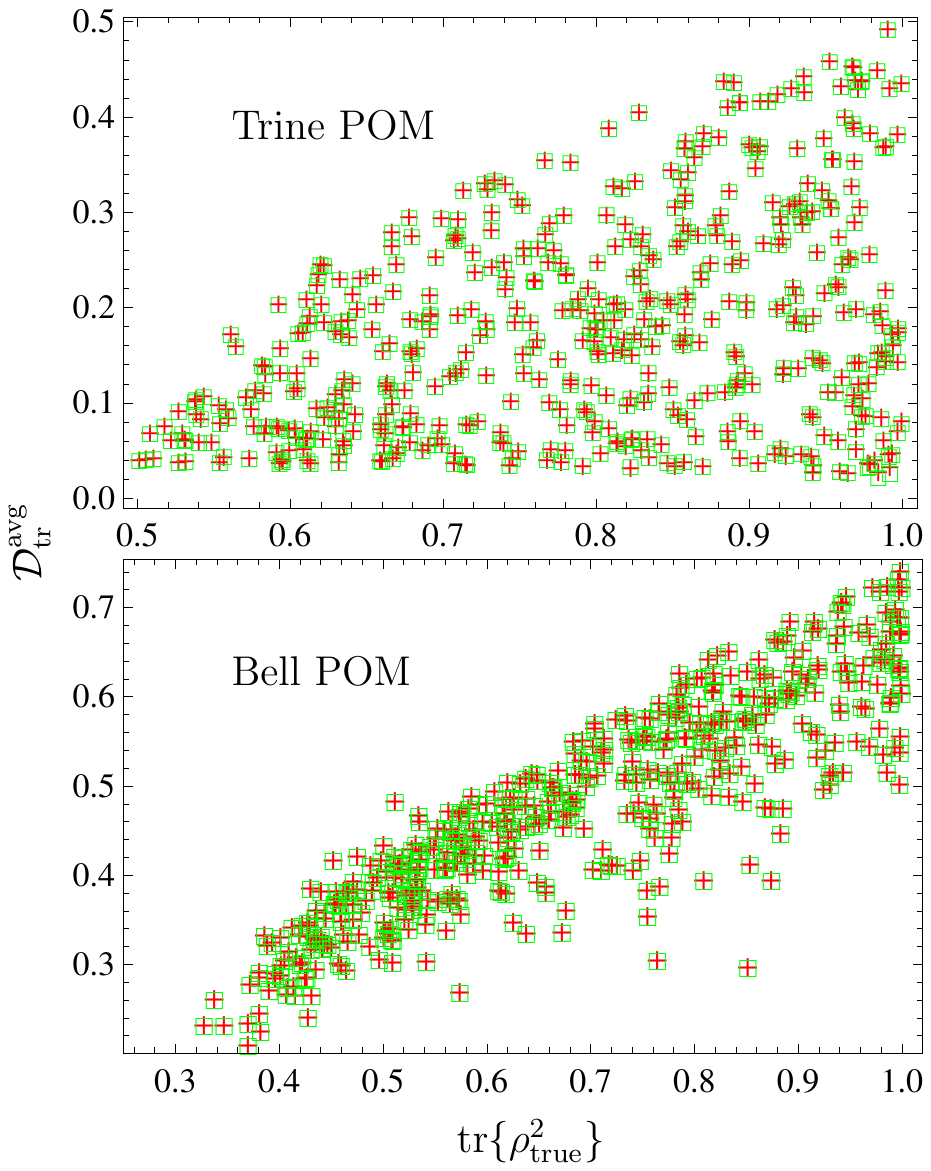}
\caption{\label{fig:perf}%
A numerical comparison between HML and MLME. A total of 500 random true states $\rho_\text{true}$ are generated for each POM.
For every true state, a total of 100 experiments for a fixed $N=500$ were simulated and the average trace-class distance $\mathcal{D}^\text{avg}_\text{tr}$ was plotted. In each plot, for almost all the random states, the estimators $\hat{\rho}_\text{HML}$ ($+$) and $\hat{\rho}_\text{MLME}$ ($\square$) almost coincide on average.}
\end{figure}

The distance between the HML and MLME estimators, defined by the trace-class distance $\mathcal{D}_\text{tr}=\tr{|\hat{\rho}_\text{MLME}-\hat{\rho}_\text{HML}|}/2$, will depend on $\rho_\text{true}$ and the POM outcomes $\Pi_j$. In fact, there are cases in which $\hat{\rho}_\text{HML}$ and $\hat{\rho}_\text{MLME}$ are close to each other for a fixed set of measurement data. We illustrate this point with two examples. In the first example, we consider qubit tomography using the trine POM in Eq.~(\ref{trine}). In the second example, we look at two-qubit tomography using a POM consisting of the four standard Bell state projectors\index{Bell states|see{maximally-entangled states}}\index{maximally-entangled states} defined as
\begin{align}
\ket{\Phi_+}\bra{\Phi_+}&=\frac{1}{2}\bigl(1+\sigma_x\otimes\sigma_x
                           -\sigma_y\otimes\sigma_y
                           +\sigma_z\otimes\sigma_z\bigr)\,,\nonumber\\
\ket{\Phi_-}\bra{\Phi_-}&=\frac{1}{2}\bigl(1-\sigma_x\otimes\sigma_x
                           +\sigma_y\otimes\sigma_y
                           +\sigma_z\otimes\sigma_z\bigr)\,,\nonumber\\
\ket{\Psi_+}\bra{\Psi_+}&=\frac{1}{2}\bigl(1+\sigma_x\otimes\sigma_x
                           +\sigma_y\otimes\sigma_y
                           -\sigma_z\otimes\sigma_z\bigr)\,,\nonumber\\
\ket{\Psi_-}\bra{\Psi_-}&=\frac{1}{2}\bigl(1-\sigma_x\otimes\sigma_x
                           -\sigma_y\otimes\sigma_y
                           -\sigma_z\otimes\sigma_z\bigr)\,.
\end{align}
A relatively small $N=500$ is fixed throughout the simulations. Figure~\ref{fig:perf} shows the results.

In general, the distance between $\hat{\rho}_\text{HML}$ and $\hat{\rho}_\text{MLME}$ for a fixed set of $\Pi_j$s will also depend on the number of detection copies $N$. As $N$ becomes extremely large, the two estimators approach each other for some POMs and in this case, any of the two methods is fine as far as state estimation with these incomplete POMs is concerned. Figure~\ref{fig:perf} shows that in these two examples, even for relatively small $N$, the distance between the two estimators are in general quite small. Hence, the performance of HML and MLME can sometimes be comparable even for a reasonably small number of detection copies. 

\section{Chapter summary}

We have discussed many aspects of quantum state estimation. In introducing the idea of informationally complete state estimation, we established several maximum-likelihood algorithms using steepest-ascent and conjugate-gradient techniques. We showed that the efficiency of the conjugate-gradient algorithms is generally higher than that of the steepest-ascent algorithm. It must be emphasized that the approach to derive the conjugate-gradient maximum-likelihood algorithms is naturally extended to all other algorithms that are based on the steepest-ascent method.

Next, we established maximum-likelihood-maximum-entropy algorithms to deal with informationally incomplete data and finally applied these algorithms to three different types of tomography for state reconstruction of complex quantum states with infinitely many degrees of freedom. An important lesson that can be learnt from this study is that with a limited set of measurement data, reconstructing an unknown quantum state on a heavily truncated Hilbert space, in which the measurement data become informationally complete, using the standard maximum-likelihood technique can give rise to extraneous features in the reconstructed states that arise from the state-space truncation. One straightforward approach to minimize this problem is to apply the maximum-likelihood-maximum-entropy state estimation technique on a larger reconstruction subspace that is compatible with any known prior information about the quantum state. The choice of the dimension of the reconstruction subspace, as well as an appropriate computational basis for the truncation, depends very much on the available prior information and is sometimes more of an art rather than a science for complex quantum systems.

Finally, we derived an iterative algorithm, using the steepest-ascent method, to maximize the hedged likelihood functional that was proposed as a more operational alternative to Bayesian state estimation. We showed that the hedged maximum-likelihood estimator obtained is always unique regardless of the informational completeness of the measurement outcomes, unlike a conventional maximum-likelihood estimator. We also gave numerical plots to show that for some typical single-qubit measurements, the hedged maximum-likelihood estimator is very close to the maximum-likelihood-maximum-entropy estimator for a given set of measurement data on average even for a relatively few number of copies. Hence for practical purposes, one can rely on this new state estimation technique to obtain an estimator that is sufficiently close to the maximum-likelihood-maximum-entropy estimator for some measurements. Otherwise, the hedged maximum-likelihood estimator can still serve as a convenient estimator for the unknown quantum state. 

\cleardoublepage
\chapter{Two-qubit Entanglement Detection with State Estimation}

Entanglement witnesses are Hermitian observables\index{observables} which, when their expectation values are measured, can indicate if a given unknown quantum state is entangled. In this chapter, we discuss another important application, in addition to those discussed in Chapter~\ref{chap:qse}, of the MLME numerical schemes to bipartite entanglement witness measurement.

To this end, we first introduce an unprecedented protocol to measure a family of a particular kind of entanglement witnesses at one go \cite{witbas} in \S\ref{sec:wbm} and \S\ref{sec:witbas_prop}. Such a family of witnesses are known as \emph{optimal witnesses}\index{entanglement witness!-- optimal} \cite{lewenstein}. An entanglement witness is defined as an optimal witness if no other witnesses can detect all entangled states detected by this witness, as well as other entangled states. Next, in \S\ref{sec:witbas_adapt}, we will establish an adaptive strategy to measure these families of witnesses in order to improve the efficiency of entanglement detection. 

\section{Witness bases measurement}
\label{sec:wbm}

A general $K$-partite\index{K@$K$-partite} pure quantum state (describing a composite of $K$ quantum systems) is defined as an entangled state if its ket cannot be written in the form\index{product state}\index{product state|seealso{separable state}}
\begin{equation}
\ket{\,\,\,}_\text{prod}=\ket{\Psi_1}\ket{\Psi_2}\ldots\ket{\Psi_K}\,,
\end{equation}\label{symbol:prodket}a product or factorizable form. More generally, a $K$-partite\index{K@$K$-partite} mixed state is defined to be an entangled mixed state if it cannot be written in the \emph{separable}\index{separable state} form
\begin{equation}
\rho_\text{sep}=\sum_jp_j\rho^{(j)}_1\otimes\rho^{(j)}_2\otimes\ldots\otimes\rho^{(j)}_K\,,
\label{sep_state}
\end{equation}\label{symbol:rhosep}where $\sum_jp_j=1$. By defining $\textsc{t}_l$\label{symbol:tl} to be the partial transpose on the $l$th subsystem, from Eq.~(\ref{sep_state}), it can be readily shown that $\rho_\text{sep}^{\textsc{t}_l}\geq0$. To determine if a given unknown state $\rho_\text{true}$, with a fixed known $K$, is entangled, one can measure the expectation value of a particular kind of Hermitian observable\index{observables}, known as \emph{entanglement witness}\index{entanglement witness}, to obtain some information about the existence of entanglement. Mathematically, an entanglement witness $\mathfrak{W}$\label{symbol:witness} is a Hermitian operator, $\mathfrak{W}^\dagger=\mathfrak{W}$, with the property that $\tr{\rho_\text{sep}\mathfrak{W}}\geq0$ for \emph{all} separable states and $\tr{\rho_\text{ent}\mathfrak{W}}<0$ for at least one entangled state $\rho_\text{ent}$\label{symbol:rhoent}. Thus, for a given unknown state $\rho_\text{true}$, the condition $\langle\mathfrak{W}\rangle=\tr{\rho_\text{true}\mathfrak{W}}<0$ implies that $\rho_\text{true}$ is entangled. However, if $\tr{\rho_\text{true}\mathfrak{W}}\geq0$, no conclusion can be drawn as to whether $\rho_\text{true}$ is entangled or not. Geometrically, measuring the expectation value of an entanglement witness introduces a hyperplane that ``dissects'' the Hilbert space, with the side to which $\tr{\rho_\text{true}\mathfrak{W}}<0$ containing only entangled states.

For $K=2$ (\emph{bipartite}\index{bipartite} systems), a Hermitian operator $O$ is \emph{decomposable} if it can be written as
\begin{equation}
O=O_1^{\textsc{t}_2}+O_2
\end{equation}
in terms of the positive operators $O_1$ and $O_2$. According to Ref.~\cite{lewenstein}, a $D$-dimensional, bipartite\index{bipartite}, optimal decomposable witness\index{entanglement witness!-- optimal}\index{entanglement witness!-- decomposable} is defined as
\begin{equation}
\mathfrak{W}=Q^{\textsc{t}_2}
\label{opt_wit}
\end{equation}
for a given positive operator $Q$ with no product kets in its range. In other words, for any $D$-dimensional ket $\ket{x}$, the resulting non-zero ket $Q\ket{x}$ must be entangled. It is clear that $\tr{\rho_\text{sep}\mathfrak{W}}=\tr{\rho_\text{sep}Q^{\textsc{t}_2}}=\tr{\rho_\text{sep}^{\textsc{t}_2}Q}\geq0$. Throughout the analysis, we fix $D=2^2=4$ for the case of two-qubit quantum systems. One can easily construct such optimal witnesses from pure states, where $Q=\ket{\Psi}\bra{\Psi}$. From the definition given in Eq.~(\ref{opt_wit}), it follows that $\ket{\Psi}\bra{\Psi}$ must be an entangled state. The Schmidt decomposition\index{Schmidt decomposition} of its corresponding ket
\begin{equation}
\ket{\Psi}=\ket{00}\cos\alpha+\ket{11}\sin\alpha
\end{equation}
is useful for subsequent calculations. Evaluating $Q^{\textsc{t}_2}$,
\begin{align}
Q^{\textsc{t}_2}=&\,\left[\left(\ket{00}\cos\alpha+\ket{11}\sin\alpha\right)\left(\cos\alpha\bra{00}+\sin\alpha\bra{11}\right)\right]^{\textsc{t}_2}\nonumber\\
=&\,\ket{00}\left(\cos\alpha\right)^2\bra{00}+\ket{11}\left(\sin\alpha\right)^2\bra{11}\nonumber\\
&\,+\ket{01}\sin\alpha\cos\alpha\bra{10}+\ket{10}\sin\alpha\cos\alpha\bra{01}\nonumber\\
=&\,\ket{00}\left(\cos\alpha\right)^2\bra{00}+\ket{11}\left(\sin\alpha\right)^2\bra{11}\nonumber\\
&\,+\left[\left(\ket{01}+\ket{10}\right)\frac{1}{\sqrt{2}}\right]\sin\alpha\cos\alpha\left[\frac{1}{\sqrt{2}}\left(\bra{01}+\bra{10}\right)\right]\nonumber\\
&\,+\left[\left(\ket{01}-\ket{10}\right)\frac{1}{\sqrt{2}}\right]\left(-\sin\alpha\cos\alpha\right)\left[\frac{1}{\sqrt{2}}\left(\bra{01}-\bra{10}\right)\right]\,.
\label{partial_trans}
\end{align}
The important point of this calculation is to realize that for \emph{any} pure state $\ket{\Psi}\bra{\Psi}$, the eigenkets of $\left(\ket{\Psi}\bra{\Psi}\right)^{\textsc{t}_2}$ are always the same kinds: two product kets $\left\{\ket{00},\ket{11}\right\}$ and two Bell\index{maximally-entangled states} kets $\left\{\left(\ket{01}+\ket{10}\right)/\sqrt{2},\left(\ket{01}-\ket{10}\right)/\sqrt{2}\right\}$.

When we measure the projectors $\ket{00}\bra{00}$, $\ket{11}\bra{11}$, $\ket{\Psi_+}\bra{\Psi_+}$ and $\ket{\Psi_-}\bra{\Psi_-}$, we in fact measure a family of optimal witnesses at one go. Such a progress allows us to search for the ``best'' entanglement witness out of the measured family that has the highest chance of detecting entanglement of $\rho_\text{true}$. We start by defining the witness criterion
\begin{equation}
\min_\alpha\left\{\langle\left(\ket{\Psi}\bra{\Psi}\right)^{\textsc{t}_2}\rangle\right\}\geq0
\end{equation}
which is obeyed by all separable states and is violated for the entangled states that are detected by this family of
witnesses. The minimization means that we are searching for the witness that maximizes the chance of violating the inequality $\langle\left(\ket{\Psi}\bra{\Psi}\right)^{\textsc{t}_2}\rangle\geq0$ in order to detect the presence of entanglement. From Eq.~(\ref{partial_trans}),
\begin{align*}
&\min_\alpha\left\{\langle\left(\ket{\Psi}\bra{\Psi}\right)^{\textsc{t}_2}\rangle\right\}\\
=&\,\min_\alpha\left\{f_1\left(\cos\alpha\right)^2+f_2\left(\sin\alpha\right)^2+(f_3-f_4)\sin\alpha\cos\alpha\right\}\\
=&\,\min_\alpha\Bigg\{\frac{f_1+f_2}{2}+\underbrace{\left(\frac{f_3-f_4}{2}\right)\sin(2\alpha)+\left(\frac{f_1-f_2}{2}\right)\cos(2\alpha)}_{=\frac{1}{2}\sqrt{\left(f_1-f_2\right)^2+\left(f_3-f_4\right)^2}\sin\left(2\alpha+\tan^{-1}\left(\frac{f_1-f_2}{f_3-f_4}\right)\right)}\Bigg\}\\
=&\,\frac{f_1+f_2}{2}-\frac{1}{2}\sqrt{\left(f_1-f_2\right)^2+\left(f_3-f_4\right)^2}\,,
\end{align*}
where $f_1=\langle\ket{00}\bra{00}\rangle$ and $f_2=\langle\ket{11}\bra{11}\rangle$ are the measured frequencies (or expectation values) for the product states, and $f_3=\langle\ket{\Psi_+}\bra{\Psi_+}\rangle$ and $f_4=\langle\ket{\Psi_-}\bra{\Psi_-}\rangle$ are those for the Bell states\index{maximally-entangled states}. So, the witness criterion now reduces to the simple inequality
\begin{equation}
4f_1f_2\geq\left(f_3-f_4\right)^2\,.
\label{wit_criterion}
\end{equation}
Thus, once the frequency data are obtained after measurement, the presence of entanglement can be detected as long as the witness criterion is violated.

\begin{figure}[h!]
  \centerline{\includegraphics[width=0.4\textwidth]{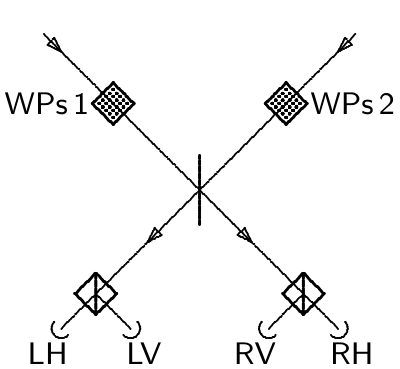}}
\caption{\label{fig:2pol-witness}%
  A linear-optics set-up that offers an experimental implementation of the optimal witness of Eq.~(\ref{opt_wit}) for polarization qubits. Two photons that are indistinguishable by their spatial-spectral properties are simultaneously incident on a half-transparent mirror, photon~1 from the left and photon~2 from the right. They carry one polarization qubit each, with their unknown two-qubit state to be analyzed. After being transmitted through, or reflected off, the half-transparent mirror, the photons are detected behind polarizing beam splitters that reflect vertically polarized photons and transmit horizontally polarized ones. The four detectors LH, LV, RV, and RH must be able to discriminate between one-photon and two-photon events. The four eigenstates of the family of entanglement witnesses are distinguished by different signatures; see Table~\ref{tbl:sign}. By letting the photons pass through polarization changing wave plates\index{wave plates} in the input ports, labeled by WPs\,1 and WPs\,2, one realizes other witnesses that differ from the witness of Eq.~(\ref{opt_wit}) by local unitary transformations.
}
\end{figure}

The projectors $\ket{00}\bra{00}$, $\ket{11}\bra{11}$, $\ket{\Psi_+}\bra{\Psi_+}$ and $\ket{\Psi_-}\bra{\Psi_-}$ form an orthogonal POM. This basis is known as a \emph{witness basis}\index{witness basis} since measuring these projectors amounts to measuring the entire one-parameter family of optimal witnesses. In practice, it is possible to set up an experiment to measure such a family of witnesses using a photon source. Figure~\ref{fig:2pol-witness} illustrates such a set-up and Table~\ref{tbl:sign} explains the measurement outcomes of the set-up in the figure \cite{witbas}.
\begin{table}[h!]
\centering
\begin{tabular}{cc}
Eigenket & Counts (LH,LV,RV,RH)\\  \hline\hline\\
$|\textsc{vv}\rangle$ & (0,2,0,0) or (0,0,2,0)\\[1ex]
$|\textsc{hh}\rangle$ & (2,0,0,0) or (0,0,0,2)\\[2ex]
$\bigl(|\textsc{vh}\rangle+|\textsc{hv}\rangle\bigr)\big/\sqrt{2}$ &
(1,1,0,0) or (0,0,1,1)\\[2ex]
$\bigl(|\textsc{vh}\rangle-|\textsc{hv}\rangle\bigr)\big/\sqrt{2}$ &
(1,0,1,0) or (0,1,0,1)
\end{tabular}
\caption{\label{tbl:sign}%
Signatures of the relevant projectors in witness basis\index{witness basis} measurement set-up for polarization\index{photon polarizations} qubits ($0\widehat{=}\textsc{v}$, $1\widehat{=}\textsc{h}$)\label{symbol:hpolvpol}, detected by the set-up of Fig.~\ref{fig:2pol-witness}, with no wave plates\index{wave plates} in the input ports. As a consequence of the Hong--Ou--Mandel\index{Hong--Ou--Mandel} effect \cite{HOMeffect} (Chung Ki Hong\index{Chung Ki Hong}, Zhe-Yu Ou\index{Zhe-Yu Ou} and Leonard Mandel\index{Leonard Mandel}), the cases (1,0,0,1) and (0,1,1,0) do not occur.}
\end{table}
Another advantage of witness basis\index{witness basis} measurement is that, unlike conventional witness measurement where only the expectation value of $\mathfrak{W}$ is collected for inference, all frequency data are used to perform quantum state estimation to obtain more information about the unknown state. It is therefore desirable to measure an informationally complete set of witness bases, such that if all the witness bases miss the entanglement detection, a full estimation can be performed to identify the unknown quantum state. To construct this informationally complete set of bases, we first think of a single witness basis\index{witness basis} measurement as being equivalent to a measurement of multiple observables\index{observables}. These observables\index{observables} can be decomposed into linear combinations of tensor products of the single-qubit \emph{Weyl} operators\index{Weyl operators} (Hermann Klaus Hugo Weyl\index{Hermann Klaus Hugo Weyl}) since these operators form a complete operator basis\index{operator basis}. As we are dealing with two-qubit systems, the corresponding single-qubit Weyl operators\index{Weyl operators} are defined as
\begin{eqnarray}
  \label{eq:Pauli}
  X\equiv\sigma_x&=&|\textsc{h}\rangle\langle\textsc{v}|
      +|\textsc{v}\rangle\langle\textsc{h}|\,,\nonumber\\
  Y\equiv\sigma_y&=&|\textsc{h}\rangle\I\langle\textsc{v}|
      -|\textsc{v}\rangle\I\langle\textsc{h}|\,,\nonumber\\
  Z\equiv\sigma_z&=&|\textsc{v}\rangle\langle\textsc{v}|
    -|\textsc{h}\rangle\langle\textsc{h}|\,
\end{eqnarray}
in terms of the polarization\index{photon polarizations} basis. Since measuring a two-qubit witness basis\index{witness basis}, which comprises four orthogonal projectors, gives only three independent outcomes, this means that we obtain expectation values of only \emph{three} two-qubit observables\index{observables}. These three observables\index{observables} are
\begin{align}
Z_1Z_2&=\ket{\textsc{h}\textsc{h}}\bra{\textsc{h}\textsc{h}}+\ket{\textsc{v}\textsc{v}}\bra{\textsc{v}\textsc{v}}-\ket{\Psi_+}\bra{\Psi_+}-\ket{\Psi_-}\bra{\Psi_-}\,,\nonumber\\
Z_11+1Z_2&=\ket{\textsc{h}\textsc{h}}2\bra{\textsc{h}\textsc{h}}+\ket{\textsc{v}\textsc{v}}2\bra{\textsc{v}\textsc{v}}\,,\nonumber\\
X_1X_2+Y_1Y_2&=\ket{\Psi_+}2\bra{\Psi_+}-\ket{\Psi_-}2\bra{\Psi_-}\,.
\end{align}
Here, $A_1A_2=A\otimes A$. With this formalism, we are now able to construct an informationally complete set of witness bases. By introducing the \emph{Clifford} unitary operator\index{Clifford!-- unitary operator} (William Kingdon Clifford\index{William Kingdon Clifford}) $C$\label{symbol:cliffordqubit} that permutes the Weyl operators\index{Weyl operators} cyclically,
\begin{equation}
  \label{eq:cyclicC}
  CX=YC\,,\quad CY=ZC\,,\quad CZ=XC\,,
\end{equation}
we can construct an informationally complete set of \emph{six} witness bases. Table~\ref{tbl:SixWitnesses} lists these six witness bases. Note that one inevitably needs an overcomplete set since there may exist a repeated observable from a pair of bases. More details on the structures of informationally complete sets of two-qubit witness bases will be discussed in the next section.
\begin{table}[h!]
\centering
\begin{tabular}{cccc}
& $U^\text{wp}_1$ & $U^\text{wp}_2$ & Observables\index{observables}\\ \hline\hline\\
1 & $1$ & $1$ & $Z_11 + 1Z_2$, $Z_1Z_2$, $X_1X_2 + Y_1Y_2$\\
2 & $1$ & $X$          & $Z_11 - 1Z_2$, $Z_1Z_2$, $X_1X_2 - Y_1Y_2$\\
3 & $C^{\dagger}$ & $C$          & $X_11 + 1Y_2$, $X_1Y_2$, $Y_1Z_2 + Z_1X_2$\\
4 & $C^{\dagger}$ & $XC$         & $X_11 - 1Y_2$, $X_1Y_2$, $Y_1Z_2 - Z_1X_2$\\
5 & $C$          & $C^{\dagger}$ & $Y_11 + 1X_2$, $Y_1X_2$, $Z_1Y_2 + X_1Z_2$\\
6 & $C$         & $XC^{\dagger}$ & $Y_11 - 1X_2$, $Y_1X_2$, $Z_1Y_2 - X_1Z_2$
\end{tabular}
\caption{\label{tbl:SixWitnesses}%
The six witness bases of the kind depicted in Fig.~\ref{fig:2pol-witness} that enable full tomography of the two-qubit state. The second and third columns list the unitary operators $U^\text{wp}_1$\label{symbol:unitarywp} and $U^\text{wp}_2$ that describe the effect of the wave plates\index{wave plates} WPs\,1 and WPs\,2, respectively, on the polarization\index{photon polarizations} of the incoming photons. The fourth column states the three two-qubit operators whose expectation values are determined when the eigenstate basis of the corresponding witness is measured.}
\end{table}
The wave plates\index{wave plates} in the input ports implement the unitary transformations $U^\text{wp}_1$ and $U^\text{wp}_2$ on the polarization\index{photon polarizations} of photons~1 and 2, respectively, and so the incoming two-photon state $\rho_{\mathrm{2ph}}$\label{symbol:rho2ph} is transformed in accordance with
\begin{equation}
  \label{eq:rho-trans}
  \rho_{\mathrm{2ph}}\to
   U^\text{wp}_1\otimes U^\text{wp}_2\,\rho_{\mathrm{2ph}}\bigl(U^\text{wp}_1\otimes U^\text{wp}_2\bigr)^{\dagger}
\end{equation}
before the photons arrive at the half-transparent mirror. In effect, then, the family of optimal witnesses of the transformed witness basis\index{witness basis} is measured rather than the original family. The Clifford operator\index{Clifford!-- unitary operator} $C$ is implemented by wave plates\index{wave plates} that yield the polarization\index{photon polarizations} changes
\begin{eqnarray}
  \label{eq:ConHV}
  |\textsc{v}\rangle&\to&C|\textsc{v}\rangle
  =\bigl(|\textsc{v}\rangle+|\textsc{h}\rangle\bigr)\bigr/\sqrt{2}\,,
\nonumber\\
  |\textsc{h}\rangle&\to&C|\textsc{h}\rangle
  =\mathrm{i}\bigl(|\textsc{h}\rangle-|\textsc{v}\rangle\bigr)\bigr/\sqrt{2}\,,
\end{eqnarray}
possibly accompanied by an irrelevant over-all phase factor. 

\section{Properties of two-qubit informationally complete witness bases}
\label{sec:witbas_prop}

We exhaustively list and investigate the set of informationally complete two-qubit witness bases\index{witness basis} that live in the simplest bipartite\index{bipartite} Hilbert space. Some observations are made regarding the structure and unitary equivalences of these bases.

\subsection{Construction}
\label{subsec:2x2_construct}

We begin by parameterizing an entanglement witness $\mathfrak{W}$ for a two-qubit system with three parameters $(u_1,u_2,a)$, where $a=0,1$ and $u_k=1,2,3$, with $u_1$ and $u_2$ labeling the respective unitary Weyl operators\index{Weyl operators} $U_1$ and $U_2$ for qubits 1 and 2. Since we want to search for informationally complete sets of witness bases\index{witness basis}, a good strategy will be to use a complete set of mutually unbiased bases. For this, we will consider the (ordered) set of order-2 qubit Weyl operators\index{Weyl operators} \{$Z$,$X$,$\I XZ$\}. These operators are order-2 since $Z^2=X^2=(\I XZ)^2=1$. The labels $u_1$ and $u_2$ are each defined to refer to one of the three Weyl operators\index{Weyl operators} in the given order. For instance, $u_1=1\leftrightarrow U_1=Z_1$, $u_1=2\leftrightarrow U_1=X_1$ and $u_1=3\leftrightarrow U_1=\I X_1Z_1$ for this set of order-2 qubit Weyl operators\index{Weyl operators} that refer to qubit 1. The corresponding complementary operators\index{complementary operators} $V_1$ and $V_2$, such that $U_kV_k=-V_kU_k$, will each refer to an operator from a list that is a cyclic permutation of the Weyl operators\index{Weyl operators} given above, that is \{$X$,$\I XZ$,$Z$\}. There is in principle more than one list of complementary\index{complementary operators} Weyl operators\index{Weyl operators} but we shall refer to the aforementioned list unless otherwise stated.

The Schmidt decomposition of a two-qubit pure state is given by
\begin{equation}
\ket{\,\,\,}=\sum^1_{j=0}\ket{j,\,j}\lambda_j\,.
\end{equation}
The decomposable witness defined as
\begin{equation}
\mathfrak{W}(a)=V_2^{a}\left(\ket{\,\,\,}\bra{\,\,\,}\right)^{\textsc{t}_2}V_2^{-a}\,
\end{equation}
can be written as
\begin{equation}
\mathfrak{W}(a)=\sum^1_{j,k=0}\ket{j,\,k+a}\lambda_j\lambda_k\bra{k,\,j+a}\,,
\label{witness}
\end{equation}
where a cyclic shift, effected by the unitary operator $V_2^a$, is applied to the kets of qubit 2 to account for non-unique orbits of witnesses. We recall that any operator can be written as functions of the Weyl operators\index{Weyl operators} since these operators are algebraically complete. This means that any such two-qubit projector $\ket{j,k}\bra{j,k}$ is given by
\begin{equation}
\ket{j,k}\bra{j,k}=\frac{1}{4}\sum^1_{m,n=0}\left[(-1)^{-j} U_1\right]^m\left[(-1)^{-k}U_2\right]^n\,.
\label{projector}
\end{equation}
By expressing $\mathfrak{W}$, given in Eq.~(\ref{witness}), in terms of the Weyl operators\index{Weyl operators} and picking out the four operators that are measurable in a given two-qubit tomography experiment to be
\begin{equation}
\ket{j,k+a}\bra{j,k+a}+\ket{k,j+a}\bra{k,j+a}\, \text{for all}\, j,k
\end{equation}
and
\begin{equation}
\ket{j,k+a}\bra{k,j+a}+\ket{k,j+a}\bra{j,k+a}\,,\,j\neq k\,,
\end{equation}
we arrive at the equations
\begin{align}
&\ket{j,k+a}\bra{j,k+a}+\ket{k,j+a}\bra{k,j+a}\nonumber\\
=&\sum^1_{m,n=0}\left[(-1)^{-jm-kn}+(-1)^{-km-jn}\right](-1)^{-an}U_1^mU_2^n\,,\label{Weyl_eqn1}\\
&\ket{j,k+a}\bra{k,j+a}+\ket{k,j+a}\bra{j,k+a}\nonumber\\
=&\sum^1_{m,n=0}\left[(-1)^{-jm-kn}V_1^{k-j}V_2^{j-k}+(-1)^{-km-jn}V_1^{j-k}V_2^{k-j}\right](-1)^{-an}U_1^mU_2^n\,.
\label{Weyl_eqn2}
\end{align}

To extract the relevant independent observables from Eqs.~(\ref{Weyl_eqn1}) and (\ref{Weyl_eqn2}), we note that
\begin{equation}
V^{-1}_k=V^\dagger_k=V_k
\end{equation}
when the $V_k$s are single-qubit unitary operators and
\begin{equation}
\sum_j\E^{\frac{2\pi\I\,(m-n)j}{D}}=D\delta_{mn}\,,
\end{equation}
where $D=2$ in this case.
From Eq.~(\ref{Weyl_eqn1}),
\begin{align}
O(m',n';a)=&\,\sum^1_{j,k=0}(-1)^{jm'}(-1)^{kn'}\nonumber\\
&\,\times\left\{\sum^1_{m,n=0}\left[(-1)^{-jm-kn}+(-1)^{-km-jn}\right](-1)^{-an}U_1^mU_2^n\right\}\nonumber\\
=&\,4\left[(-1)^{an'}U^{m'}_1U^{n'}_2+(-1)^{am'}U^{n'}_1U^{m'}_2\right]\,.
\end{align}
By looking at different values $m'$ and $n'$, we have
\begin{align}
m'=0,n'=0&\,\rightarrow\,&\,O(0,0;a)&=4\,,\label{o11}\\
m'=0,n'=1\text{ or }m'=1,n'=0&\,\rightarrow\,&\,O(0,1;a)&=O(1,0;a)\nonumber\\
&\,&=4\big[U_1+&(-1)^aU_2\big]\,,\label{o12}\\
m'=1,n'=1&\,\rightarrow\,&\,O(1,1;a)&=8(-1)^aU_1U_2\,,\label{o13}
\end{align}
out of which two observables $U_1+(-1)^aU_2$, $U_1U_2$ can be extracted from Eqs.~(\ref{o12}) and (\ref{o13}) respectively. From Eq.~(\ref{Weyl_eqn2}), we consider all the four possible combinations
\begin{align}
j=0,k=0\,\rightarrow\,&\,\sum^1_{m,n=0}(-1)^{an}U_1^mU_2^n\,,\label{o21}\\
j=0,k=1\text{ or }j=1,k=0\,\rightarrow\,&\,\sum^1_{m,n=0}\left[(-1)^{-m-an}+(-1)^{-(a+1)n}\right]V_1V_2U_1^mU_2^n\nonumber\\
=&\,2V_1V_2\left[1-(-1)^aU_1U_2\right]\,,\label{o22}\\
j=1,k=1\,\rightarrow\,&\,\sum^1_{m,n=0}(-1)^{m+(a+1)n}U_1^mU_2^n\,,\label{o23}
\end{align}
from which the only other independent observable that can be extracted is $V_1V_2\left[1-(-1)^aU_1U_2\right]$. It can be verified that the six sets of three independent observables\index{observables} listed in Table~\ref{tbl:SixWitnesses} are easily obtained from the three simplified observable\index{observables} expressions.

We need a total of 15 linearly independent observables\index{observables} to perform full tomography on a two-qubit state. To search for these sets of informationally complete observables, each of a pre-chosen set of $3\!\times\!3\times\!2=18$ observables\index{observables} is expressed in terms of the 15 Weyl basis operators\index{Weyl operators} $X_1^{p_1}Z_1^{q_1}\otimes X_2^{p_2}Z_2^{q_2}$\footnote{The identity operator is excluded.}, where $p_k$ and $q_k$ each takes the value 0 or 1 and are not simultaneously zero. Next, we form a $18\times 15$ \emph{observable matrix}\index{observable matrix} $M_\text{obs}$\label{symbol:matrixobs}, with each row representing an observable\index{observables} and having phase factor coefficients as matrix entries, to have an informationally complete set of witness bases\index{witness basis}. Thus, for a set of 18 observables\index{observables} $\{O_1,O_2,\ldots,O_{18}\}$, the observable matrix\index{observable matrix} $M_\text{obs}$ satisfies the equation
\begin{equation}
\begin{pmatrix}
O_1\\
O_2\\
\vdots\\
O_{18}
\end{pmatrix}
=\underbrace{\begin{pmatrix}
M_{1,1} & M_{1,2} & \cdots & M_{1,15}\\
M_{2,1} & M_{2,2} & \cdots & M_{2,15}\\
\vdots  & \vdots  & \ddots & \vdots  \\
M_{18,1} & M_{18,2} & \cdots & M_{18,15}\\
\end{pmatrix}}_{=M_\text{obs}}
\begin{pmatrix}
X_1\\
Z_1\\
\vdots\\
X_1Z_1X_2Z_2
\end{pmatrix}\,.
\end{equation}
The task is then to look for the combination of settings $\{u_1,u_2,a\}$ such that $M_\text{obs}$ has 15 non-zero singular values\index{singular values}.

\subsection{Local unitary equivalence}

There are altogether $3\times3\times2=18$ different combinations of triplets $(u_1,u_2,a)$ available to form a set consisting of six distinct triplets. Hence the total number of possible sets is $\binom{18}{6}=18564$, which is tractable enough for us to perform an exhaustive search for all the full-rank sets\footnote{Here, a full-rank set corresponds to an observable matrix $M_\text{obs}$ with 15 non-zero singular values}. Using the list of Weyl operators\index{Weyl operators} given in the previous section, we find that there are altogether 1395 sets that are informationally complete after the numerical search.

These sets are categorized into six classes and within each class, all sets give exactly the same set of singular values\index{singular values} of $M_\text{obs}$. The first class contains only three members which are related by the order-3 qubit Clifford transformation\index{Clifford!-- transformation} $C_1$, defined in Eq.~(\ref{eq:cyclicC}), on the entire reference set. Classes~2 to 6 each comprises a number of families of sets, each of which are generated by the local unitary transformation effected by the operator $V_1$ on a reference set in the family, which amounts to changing the value of $a$. Some of the witness bases\index{witness basis} in a particular set of six are not affected by the transformation. We call a transformation that is effected on $n$ witness bases\index{witness basis} out of the six in a particular set to be an $n$-$V_1$ transformation. Table~\ref{2x2_class} summarizes the results. Another symmetry is that these 1395 sets are invariant under a cyclic permutation of the list of $V_1$ operators.

\begin{table}[h!]
\centering
\begin{tabular}{cccccccc}
Families & $1$-$V_1$ & $2$-$V_1$ & $3$-$V_1$ & $4$-$V_1$ & $5$-$V_1$ & $6$-$V_1$ & Number of sets\\ \hline\hline
\multicolumn{8}{c}{\underline{\textbf{Class 2}}}\\
17 & 4 & 6 & 4 & 1 & 0 & 0 & 272\\
2 & 3 & 3 & 1 & 0 & 0 & 0  & 16\\
\multicolumn{8}{c}{\underline{\textbf{Class 3}}}\\
8 & 4 & 6 & 4 & 1 & 0 & 0  & 128\\
2 & 1 & 3 & 3 & 0 & 0 & 0  & 16\\
\multicolumn{8}{c}{\underline{\textbf{Class 4}}}\\
18 & 2 & 7 & 12 & 7 & 2 & 1 & 576\\
\multicolumn{8}{c}{\underline{\textbf{Class 5}}}\\
18 & 4 & 6 & 4 & 1 & 0 & 0 & 288\\
\multicolumn{8}{c}{\underline{\textbf{Class 6}}}\\
3 & 0 & 15 & 0 & 15 & 0 & 1 & 96
\end{tabular}
\caption{\label{2x2_class}%
The results of local unitary equivalences for sets in Classes~2 to 6. Class~1 contains only three sets which are mutually related by the qubit Clifford transformation\index{Clifford!-- transformation} that permutes the qubit Weyl operators\index{Weyl operators}. The value under the column ``1-$V_1$'', for instance, gives the number of 1-$V_1$ transformations that are performed on a fixed reference set in each of the families that falls in the class. For example, the first row says that for each family out of 17 in Class~2, including the reference set, there exists a total of 16 sets with 15 of them generated by applying various types of $n$-$V_1$ transformations on the reference set in the family. Families with the configuration (4,6,4,1,0,0), for instance, are due to the fact that two witness bases\index{witness basis} in the reference set of every family, having the same $u_1$,$u_2$ settings, are unaffected by the $V_1$ transformations and so there are $\binom{4}{1}=4$ $1$-$V_1$ transformations, $\binom{4}{2}=6$ $2$-$V_1$ transformations, $\binom{4}{3}=4$ $3$-$V_1$ transformations and $\binom{4}{4}=1$ $4$-$V_1$ transformations. Every family in Class 4 has half of the 32 sets equivalent to the other half via an overall $V_1$ transformation on the entire set. For instance, sets that are generated by the $1$-$V_1$ and $5$-$V_1$ transformations on the reference set in a particular family are related via an overall $V_1$ transformation and so on. Half the set generated by the $3$-$V_1$ transformations on the reference set is equivalent to the other half generated by the same type of transformations in the same manner. The entries under the last column includes the reference set in each family. There are 1392 informationally complete sets of witness bases\index{witness basis} out of these five classes. Together with the three sets in Class 3, there is a total of 1395 sets.}
\end{table}

\subsection{A summary}
There exist many full-rank solutions for the two-qubit case and we listed six classes of informationally complete sets of witness bases, with all sets giving the same singular values of $M_\text{obs}$ within each class. These informationally complete sets are invariant under a cyclic permutation of the complementary\index{complementary operators} $V$ operators. Finally we mention that the results presented here are valid for the list of $V$ operators we used, and that the structures may vary if different choices of $V$ operators are taken. For instance, a given $V$ operator remains complementary if the operator $U$ is multiplied to it. So there will be two such complementary operators for every operator $U$. Hence, we have a total of eight different lists of complementary $V$ operators and every list, in general, gives different informationally complete sets and, therefore, different structures. The properties of the witness bases for quantum systems of larger dimensions are still largely unknown at this point. 

\section{Adaptive witness bases measurement with state estimation}\index{adaptive estimation scheme!-- adaptive witness basis measurement}
\label{sec:witbas_adapt}

We now have all the necessary tools to establish an adaptive scheme to measure the witness bases\index{witness basis} in such a way that the number of witness bases\index{witness basis} needed to detect the entanglement of the unknown state $\rho_\text{true}$ is optimized. Each time a witness basis is measured, a set of four frequencies is obtained and this can be used to partially estimate $\rho_\text{true}$ using MLME. Since the MLME estimators are generally mixed states, there is a chance that the purity of a MLME estimator is lower than that of $\rho_\text{true}$, especially when $\rho_\text{true}$ is a nearly-pure state. If the measurement of a witness basis detects the entanglement of this estimator, measuring the same witness basis could very likely detect the entanglement of $\rho_\text{true}$. This is due to the trend that entanglement detection becomes more difficult as the purity of $\rho_\text{true}$ decreases. The extreme cases are the maximally-entangled Bell states\index{maximally-entangled states} and the separable maximally-mixed state.

Defining the operators $\Pi^\text{wit}_1$\label{symbol:pijwit} and $\Pi^\text{wit}_2$ to be the outcomes of the product states, and $\Pi^\text{wit}_3$ and $\Pi^\text{wit}_4$ to be those of the maximally-entangled states for a given witness basis, an adaptive strategy\index{algorithm} based on this idea is as follows:

\begin{center}
\colorbox{light-gray}{\begin{minipage}[c]{12cm}
  \uline{\textbf{Adaptive witness bases measurement}}\\
Starting from $k=1$ and a witness basis,
\begin{enumerate}
  \item Obtain the frequency data by measuring the witness basis;
  \begin{itemize}
    \item If the witness criterion is violated, escape the loop;
    \item Otherwise, proceed to the following steps.
  \end{itemize}
  \item If $k>1$, combine these data with the previous ones and renormalize all the frequencies.
  \item Look for the MLME estimator $\hat\rho_\text{MLME}$ consistent with the total collected data.
  \item For each of the $6-k$ witness bases left, choose the one which gives the minimum value of the function $4p_1p_2-(p_3-p_4)^2$, where $p_j=\tr{\hat\rho_\text{MLME}\Pi^\text{wit}_j}$ are the probabilities of the outcomes $\Pi^\text{wit}_j$ from one of the $6-k$ witness bases, calculated based on the MLME estimator.
  \item Set $k=k+1$ and repeat the iteration from the beginning.
\end{enumerate}
\end{minipage}}
\end{center}

To investigate the performance of this adaptive measurement scheme, we perform simulations for both pure and full-rank mixed two-qubit states respectively.
\begin{figure}[h!]
  \centering
  \subfloat[]{\label{fig:ordered}\includegraphics[width=0.45\textwidth]{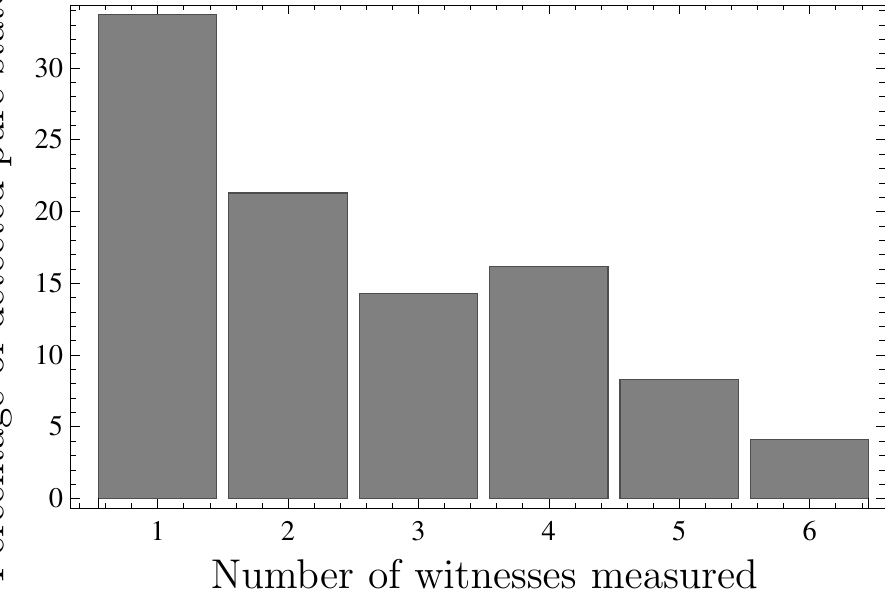}\quad\includegraphics[width=0.45\textwidth]{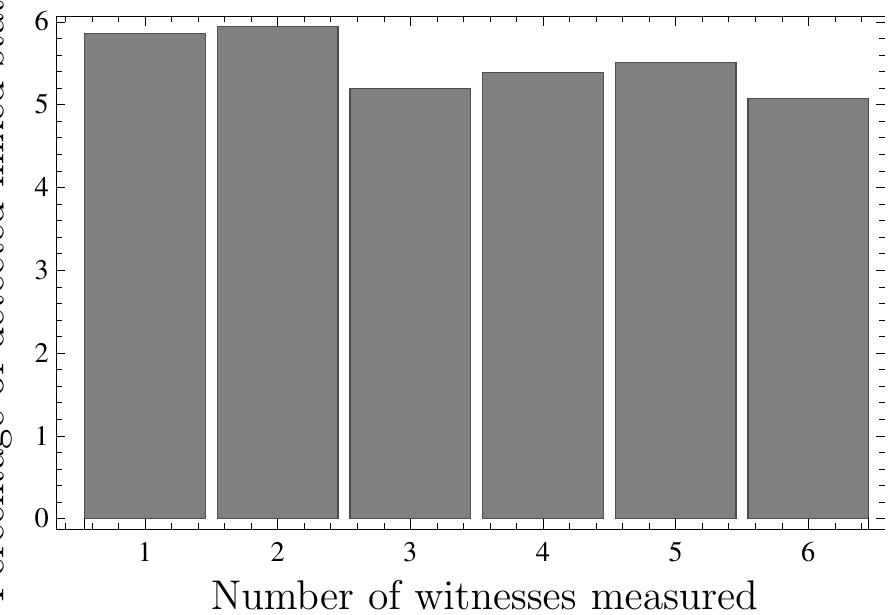}}\\
  \subfloat[]{\label{fig:adapt}\includegraphics[width=0.45\textwidth]{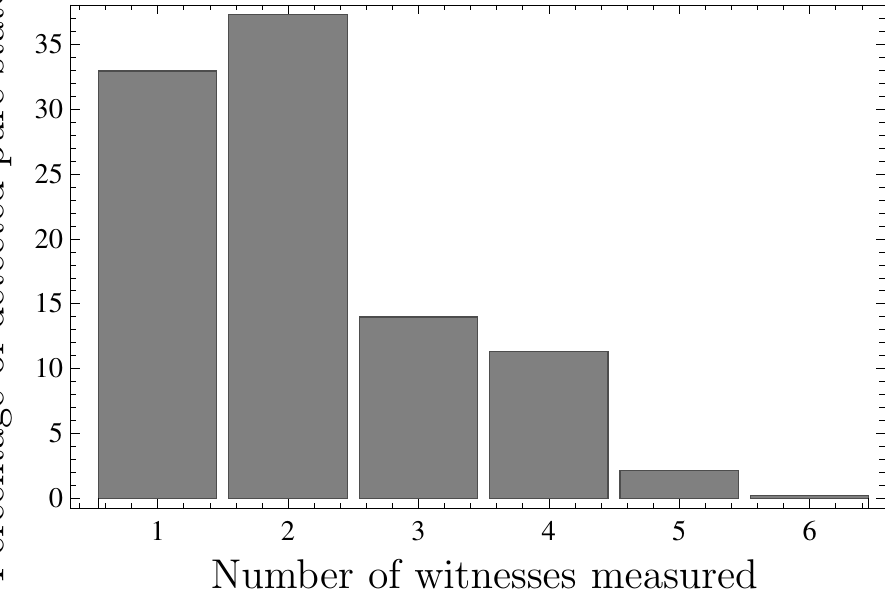}\quad\includegraphics[width=0.45\textwidth]{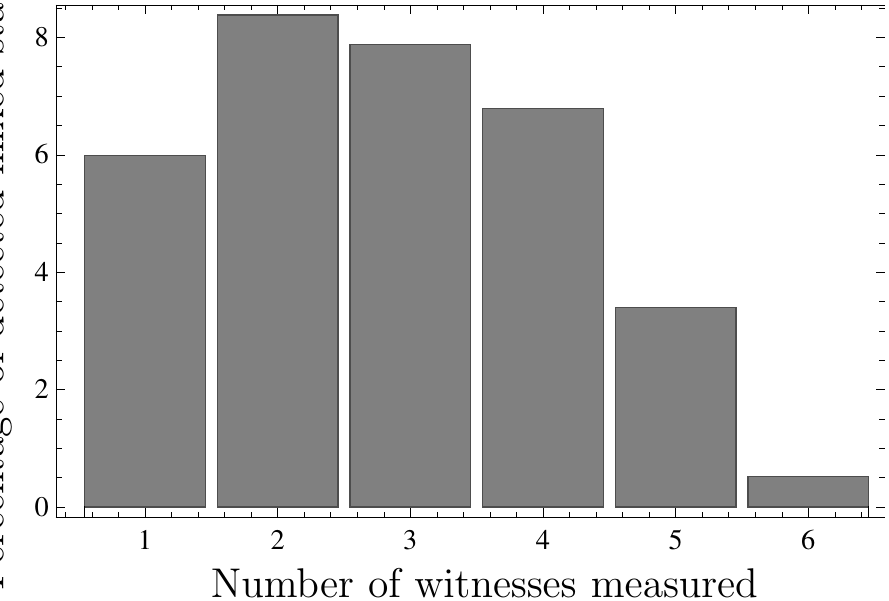}}\\
  \subfloat[]{\label{fig:adaptsep}\includegraphics[width=0.45\textwidth]{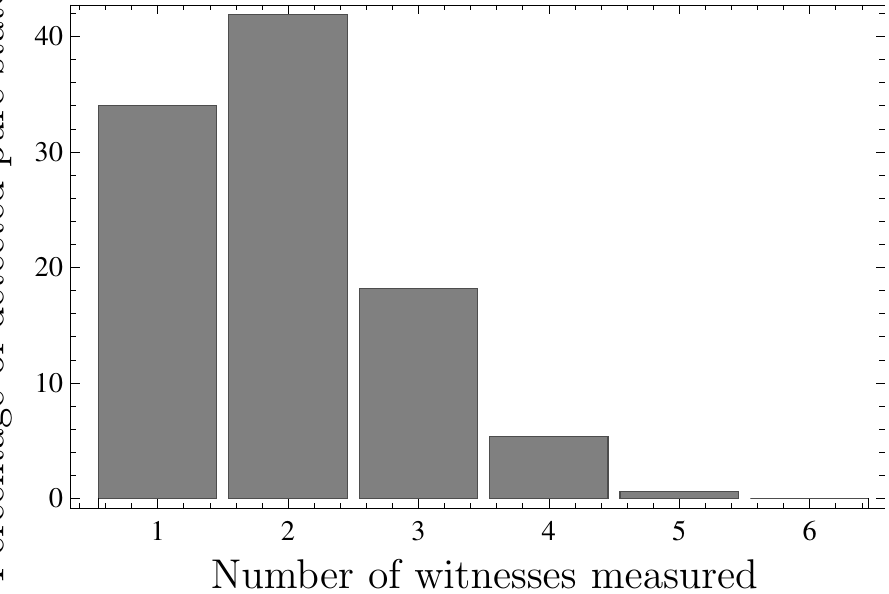}\quad\includegraphics[width=0.45\textwidth]{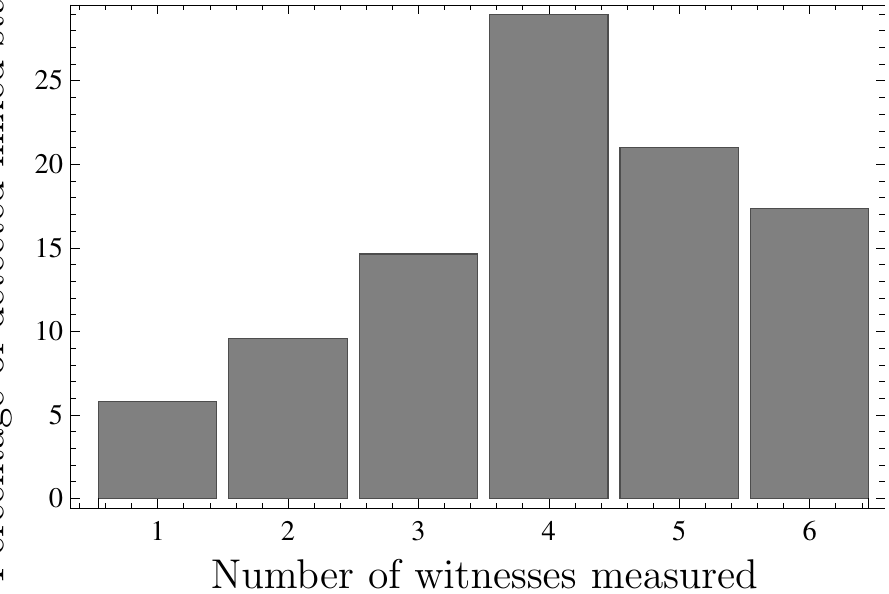}}
  \caption{A simulation on the measurement of the set of six informationally complete two-qubit entanglement witness bases for $10^4$ random two-qubit pure states, as well as full-rank mixed states, with the measurements done for one state at a time.}
  \label{fig:witbas_res}
\end{figure}
Figure~\ref{fig:ordered} shows the percentage of pure and mixed states detected by a specific number of the six witness bases\index{witness basis} in a particular ordering by violating the witness criterion in Eq.~(\ref{wit_criterion}). Figure~\ref{fig:adapt} shows the plot generated using the adaptive strategy for choosing the subsequent witness basis based on the MLME estimator obtained using the accumulated measurement data. Doing so will reduce the mean number of witnesses required to detect entanglement for a given pure state. Note that about 2\% of the random pure states and about 67\% of the random mixed states are undetected by the six witnesses \emph{without performing full tomography} with the aforementioned strategies.

The number of quantum states that are not detected by all the witness bases\index{witness basis} can be further reduced (Fig.~\ref{fig:adaptsep}) by performing one additional step to check if there are separable states in the ML convex set\index{convex set} produced by the accumulated incomplete measurement data after the witness criterion is not violated. The entanglement of a quantum state is considered to be detected when no separable states are present in the convex set\index{convex set}, since subsequent witness basis measurements ultimately reduce the size of the convex set to a single estimator --- the true state $\rho_\text{true}$ for large $N$ --- that was previously inside this larger set.

To perform this search, we maximize the likelihood functional over the space of separable states and compare this maximum value with that obtained by maximizing the same functional over all states. If the former is lower than the latter, this means that the true ML estimators in the convex set\index{convex set} cannot be separable.

The iterative algorithm for the maximization is in fact very similar to that established in Ref.~\cite{ent_sep}. Without going through the derivation, we present the algorithm\index{algorithm} below:

\begin{center}
\colorbox{light-gray}{\begin{minipage}[c]{12cm}
  \uline{\textbf{ML over the space of separable states}}\\
Starting from $k=1$, a fixed small parameter $\epsilon$ and a separable state
\begin{equation*}
\rho_1=\sum^{\geq16}_{l=1}\ket{\varphi^{(1)}_{1,l}}\bra{\varphi^{(1)}_{1,l}}\otimes\ket{\varphi^{(1)}_{2,l}}\bra{\varphi^{(1)}_{2,l}}\,,
\end{equation*}
where $\tr{\rho_1}=1$ and the randomly chosen kets $\ket{\varphi^{(1)}_{1,l}}$ and $\ket{\varphi^{(1)}_{2,l}}$ are subnormalized,
\begin{enumerate}
	\item Compute $R_k$ as in Eq.~(\ref{rop});
	\begin{itemize}
	\item Escape from loop if
    \begin{equation*} \sqrt{\sum^{2}_{m=1}\sum^{\geq16}_l\tr{\left[\left(R'^{\,(k)}_{m,l}-1\right)\ket{\varphi^{(k)}_{m,l}}\bra{\varphi^{(k)}_{m,l}}\right]^2}}\leq\varepsilon\,,
    \end{equation*}
    where
    \begin{equation*}
    R'^{\,(k)}_{m,l}=\frac{\text{tr}_{(m\bmod{2})+1}\left\{R_k\ket{\varphi^{(k)}_{m,l}}\bra{\varphi^{(k)}_{m,l}}\right\}}{\left<\varphi^{(k)}_{m,l}\Big|\varphi^{(k)}_{m,l}\right>}\,;
    \end{equation*}
 	\item Otherwise, proceed to following steps.
 	\end{itemize}
    Compute the new operators
    \begin{equation*}
    \ket{\varphi^{(k+1)}_{m,l}}\bra{\varphi^{(k+1)}_{m,l}}=\left(1+\epsilon R'^{\,(k)}_{m,l}\right)\ket{\varphi^{(k)}_{m,l}}\bra{\varphi^{(k)}_{m,l}}\left(1+\epsilon R'^{\,(k)}_{m,l}\right)
    \end{equation*}
    and
    \begin{equation*}
    \rho_{k+1}=\frac{\sum^{\geq16}_{l=1}\ket{\varphi^{(k+1)}_{1,l}}\bra{\varphi^{(k+1)}_{1,l}}\otimes\ket{\varphi^{(k+1)}_{2,l}}\bra{\varphi^{(k+1)}_{2,l}}}{\sum^{\geq16}_l\left<\varphi^{(k+1)}_{1,l}\Big|\varphi^{(k+1)}_{1,l}\right>\left<\varphi^{(k+1)}_{2,l}\Big|\varphi^{(k+1)}_{2,l}\right>}\,.
    \end{equation*}
    \item Set $k=k+1$ and repeat the iteration from the beginning.
\end{enumerate}
\end{minipage}}
\end{center}

With this additional step, the percentage of undetected pure states is reduced to practically zero (0.01\%) and one needs no more than five witness bases\index{witness basis} to detect entanglement for the rest of the pure states. The improvement is even more dramatic for the mixed states, with a reduction from about 67\% to about 2.7\%. The mean number of witness bases\index{witness basis} needed to detect entanglement for mixed states is higher than that for pure states. This is not surprising, since mixed states generally have lower entanglement and are, therefore, harder to detect. Also, the mixed states are more likely to be separable than the pure states. 

\cleardoublepage
\chapter{Quantum Process Estimation}\index{informationally incomplete}\index{quantum process estimation (tomography)}

\section{Introduction}

Quantum process tomography (QPT) is an important tool to characterize the operation of a given quantum channel\footnote{The words ``quantum process'' and ``quantum channel'' will be used interchangeably.} \cite{qpt1,qpt2,qpt3}. Such a characterization is needed, for example, when one attempts to construct a quantum channel comprising multiple logic gates, each carrying out a specific quantum process. One such quantum channel for entanglement distillation, for instance, would consist of controlled \textsc{not} \textsc{cnot}\index{cnot@\textsc{cnot}} gates. A physical quantum process is described by a completely-positive map $\mathcal{M}$\label{symbol:mapcp}. That is, given a particular input quantum state $\rho_\text{i}$\label{symbol:rhoin} residing in the $\ID$-dimensional Hilbert space $\mathcal{H}$\label{symbol:hhilbert}, the resulting output state $\rho_\text{o}$\label{symbol:rhoout} in the $\OD$-dimensional Hilbert space $\mathcal{K}$\label{symbol:khilbert} is given by
\begin{equation}
\rho_\text{o}=\mathcal{M}\left(\rho_\text{i}\right)=\sum_{m}K_m\rho_\text{i}K^\dagger_m\,,
\label{cptp}
\end{equation}
with the Kraus operators\index{Kraus operators} (Karl Kraus\index{Karl Kraus}) $K_m$\label{symbol:kraus} satisfying the relation $\sum_mK^\dagger_mK_m=1_\mathcal{K}$. The $K_m$s are not unique and any other set of Kraus operators
\begin{equation}
K'_m=\sum_{m'}u_{mm'}K_{m'}\,,
\end{equation}
where the $u_{mm'}$s are the elements of a unitary matrix, also parameterizes the completely-positive map $\mathcal{M}$ \cite{nielsenchuang}.

The idea behind QPT is to estimate such completely-positive maps with measurements. Much like quantum state tomography, the estimation of an unknown quantum process can be perceived as the estimation of a positive \emph{Choi-Jami{\'o}\l kowski operator}\index{Choi-Jami{\'o}{\l}kowski!-- operator or matrix} (Man-Duen Choi\index{Man-Duen Choi} and Andrzej Edmund Jami{\'o}{\l}kowski\index{Andrzej Edmund Jami{\'o}{\l}kowski}) $E_\text{true}$\label{symbol:etrue} that is represented by a $\ID\OD\times\ID\OD$ matrix \cite{choi,jam}. Such an operator contains all accessible information about the quantum process. The standard QPT procedure involves the measurement of multiple copies of $L$\label{symbol:l} different output states, with each output state corresponding to one of the $L$ linearly independent input states $\rho^{(l)}_\text{i}$\label{symbol:rhoinl}, thereby using a POM of, say, $M$\label{symbol:m} outcomes. The unknown operator $E_\text{true}$ is estimated by linear-inversion of the $LM$ measurement frequencies, which consists of $\ID^2\OD^2$ linearly independent constraints. Like the linear-inversion procedure for quantum state estimation, the resulting estimator obtained may not be positive. If that is the case, the estimator cannot be used for statistical predictions. This failure occurs whenever the observed relative frequencies of the measurement outcomes do not have consistent interpretation as probabilities. What is, therefore, called for, is an estimation procedure that ensures a physically meaningful estimator whatever the measurement data may be.

One statistically meaningful technique to obtain a positive estimator for $E_\text{true}$ is the maximum-likelihood estimation procedure\index{maximum-likelihood (ML)} \cite{qstateest}. This can be applied to yield a unique estimator\index{unique estimator} $\hat E_\text{ML}$ as long as the measurement data obtained form a set of $\ID^2\OD^2$ linearly independent constraints. We say that this set of measurement data is informationally complete. However, the number of linearly independent parameters increases rapidly with the dimensions and a complete characterization of $E_\text{true}$ becomes unfeasible for complex processes. This is especially true when the quantum process acts on an infinite-dimensional Hilbert space \cite{csqpt}. The well-known method of \emph{Direct Characterization of Quantum Dynamics} (DCQD) \cite{qpt1} was introduced to reduce the amount of measurement resources\index{measurement resources} (the \emph{total} number $LN$ of copies measured) that are used for quantum process tomography. However, this method requires entangled input states and post-processing strategies that can be expensive when dealing with more complex quantum processes.

A more straightforward and conceptually different approach is to resort to informationally incomplete QPT. With this approach, less measurement resources\index{measurement resources} are used to obtain an estimator for the unknown quantum process to a fair amount of accuracy. As a consequence, there exists a convex set\index{convex set} of infinitely many ML estimators which are consistent with the measurement data. To choose the estimator which is least-biased from the convex set, we invoke the maximum-entropy\index{maximum-entropy (ME)} principle \cite{jaynes1,jaynes2} and choose the estimator with the largest entropy. Such an incomplete QPT can also give useful information about the quantum channel. In a typical tomography experiment, with data from measuring a finite number of copies, the resulting quantum process estimator can never be exactly equal to $E_\text{true}$ since experimental fluctuations are inevitable. One can only obtain an estimator that is close to $E_\text{true}$ within a certain tomographic precision. Thus, MLME QPT is typically useful in providing a unique estimator\index{unique estimator} for an unknown quantum process within a suitable tomographic precision using fewer incomplete measurement resources\index{measurement resources}. As will be shown, this reduction in measurement resources\index{measurement resources} is more pronounced for unitary quantum channels. Since $E_\text{true}$ is unknown, one common practice is to gauge such a tomographic precision with another operator $E_\text{prior}$\label{symbol:eprior}\index{prior information (knowledge)} that is close to $E_\text{true}$, based on some prior information\index{prior information (knowledge)} one has about the constructed quantum channel. The availability of such a $E_\text{prior}$\index{prior information (knowledge)} for a given $E_\text{true}$ will become useful and important in subsequent discussions.

The estimators obtained using the aforementioned method are least-biased with respect to the set of incomplete measurement data in the sense of the \emph{entropy of the quantum process}. In Ref.~\cite{iqpt}, which is an analytical study of the conventional maximum-entropy method\index{maximum-entropy (ME)}, the entropy functional for the Choi-Jami{\'o}{\l}kowski operator\index{Choi-Jami{\'o}{\l}kowski!-- operator or matrix} $E$\label{symbol:e} describing a quantum channel was introduced as $S\left(E\right)=-\tr{(E/\ID)\log (E/\ID)}$\label{symbol:se} and this was shown to exhibit nice properties. In particular, this concave channel entropy functional\index{concave function or functional} has a unique maximum in $E$ and is zero only when the quantum channel is unitary since $E/\ID$ is then a rank-1 projector. However, the analytical results in \cite{iqpt} apply only to simple qubit channels and are difficult to extend to general quantum channels of greater complexity. We shall extend the strategy in \S\ref{subsec:new_perspective} and establish adaptive iterative algorithms \cite{amlme_qpt} to search for the MLME estimator $\hat E_\text{MLME}$ which maximizes both the likelihood and entropy functionals using the channel entropy functional in \cite{iqpt}.

We first give some preliminary ideas on quantum process estimation in \S\ref{sec:qpt_prelim}. Then, in \S\ref{sec:qpt_mlme_iteralgo}, we will present the iterative MLME algorithm using variational principles to derive a steepest-ascent scheme and apply it to numerical simulations of two-qubit and three-qubit quantum channels. In \S\ref{sec:adaptstrat}, we will establish adaptive strategies to apply the MLME algorithm with the aim of minimizing the amount of measurement resources\index{measurement resources} needed to perform incomplete QPT. 

\section{Preliminaries of quantum process estimation}
\label{sec:qpt_prelim}

The estimation of the completely-positive map $\mathcal{M}$ that describes an unknown quantum process, in the manner presented in Eq.~(\ref{cptp}), is isomorphic to the estimation of an unknown quantum state. This is a consequence of the well-known Choi-Jami{\'o}{\l}kowski isomorphism\index{Choi-Jami{\'o}{\l}kowski!-- isomorphism} \cite{choi,jam,qstateest}. Let us define a maximally-entangled pure state $\ket{\Psi_+}=\sum_j\ket{j}_\mathcal{H}\otimes\ket{j}_\mathcal{H'}/\sqrt{\ID}$ in terms of the computational basis kets $\ket{j}_\mathcal{H}\otimes\ket{j}_\mathcal{H'}$. Here, the dimensions of the Hilbert spaces $\mathcal{H}$ and $\mathcal{H'}$ are both equal to the dimension $\ID$ of the input Hilbert space. Using this basis, there exists a one-to-one correspondence between the map $\mathcal{M}$ and a unique positive operator $E$ defined as follows:
\begin{align}
E\,\equiv&\ID\left(\mathcal{I}_\mathcal{H}\otimes\mathcal{E}_\mathcal{H'}\right)\left(\ket{\Psi_+}\bra{\Psi_+}\right)\nonumber\\
\widehat{=}&\sum_{jk}\left(\ket{j}\bra{k}\right)\otimes\mathcal{M}\left(\ket{j}\bra{k}\right)\,,
\end{align}
with $\mathcal{I}_\mathcal{H}$ being the identity map. From Eq.~(\ref{cptp}), the alternative expression
\begin{equation}
E=\sum_m\ket{\psi_m}\bra{\psi_m}\,,
\end{equation}
with
\begin{equation}
\quad\ket{\psi_m}=(1_\mathcal{H}\otimes K_m)\ket{\Psi_+}\sqrt{\ID}\,,
\end{equation}
implies that the rank of $E$ is equal to the number of linearly independent $K_m$s. It follows that $E$ is rank-1 if the completely-positive map is described by a single unitary Kraus operator\index{Kraus operators}, and only then.

The output state can be expressed in terms of $E$ by means of
\begin{equation}
\rho_\text{o}=\mathrm{tr}_\mathcal{H}\left\{E\left(\rho_\text{i}^\textsc{t}\otimes 1_\mathcal{K}\right)\right\}\,,
\end{equation}
where the transposition is defined with respect to the computational basis. Hence, reconstructing the quantum process amounts to estimating the positive operator $E$. To do so, one requires a total of $\ID^2\OD^2$ real parameters to specify the corresponding matrix. In the subsequent analyses, we shall consider trace-preserving maps, that is $\tr{\rho_\text{i}}=\tr{\rho_\text{o}}$ for any $\rho_\text{i}$, in which case the number of independent parameters is reduced to $\ID^2(\OD^2-1)$, with the constraints compactly written as
\begin{equation}
\mathrm{tr}_\mathcal{K}\left\{E\right\}=1_\mathcal{H}\,.
\label{constraints}
\end{equation}\label{symbol:identityh}

To estimate $E$, typically a set of $L$ input states $\rho^{(l)}_\text{i}$, with $N$ copies each, are sent through the quantum channel, one state at a time. The output state $\rho^{(l)}_\text{o}$\label{symbol:rhooutl} that corresponds to $\rho^{(l)}_\text{i}$ is measured with a POM consisting of $M$ outcomes $\Pi_m\geq0$ such that $\sum_m\Pi_m=1_\mathcal{K}$\label{symbol:identityk}. The probability of getting outcome $\Pi_m$ for the input state $\rho^{(l)}_\text{i}$ is given by $p_{lm}=\tr{E\left(\rho^{(l)\,\textsc{t}}_\text{i}\otimes\Pi_m\right)}{\Big/}L$\label{symbol:plm}. Here, $p'_l\equiv\sum_mp_{lm}=1/L$\label{symbol:plmprime}.

If the $LM$ parameters comprise $\ID^2\OD^2$ linearly independent ones, the measurement data will be informationally complete. One can thus perform a complete quantum process estimation using the maximum-likelihood (ML)\index{maximum-likelihood (ML)} algorithm \cite{qstateest} and so obtain a unique positive estimator\index{unique estimator} $\hat E_\text{ML}$ by maximizing the likelihood functional
\begin{equation}
\mathcal{L}(\{n_{lm}\};E)=\prod^L_{l=1}\left(\prod^M_{m=1}p_{lm}^{n_{lm}}\right)\,,
\end{equation}\label{symbol:likelihoode}where the number of occurrences\index{number of occurrences} $n_{lm}$\label{symbol:nlm} for the outcome $\Pi_m$ obtained in an experiment with the input state $\rho^{(l)}_\text{i}$ are such that $n'_l\equiv\sum_{m}n_{lm}=N$\label{symbol:nlmprime}. 

\section{The iterative algorithm}
\label{sec:qpt_mlme_iteralgo}

We consider the optimization of the information functional\index{information functional}
\begin{equation}
\mathcal{I}(\lambda;E)=\lambda S(E)+\frac{1}{LN}\log\mathcal{L}(\{n_{lm}\};E)\,,
\label{newinfo}
\end{equation}\label{symbol:infofunctionale}
where $\lambda$ is a parameter which scales the entropy relative to the normalized log-likelihood and should be chosen with a very small value. When the measurement data are informationally complete, one sets $\lambda$ to zero and optimizing $\mathcal{I}(\lambda=0;E)$ amounts to the ML problem \cite{qstateest,mlqpt}. In the same spirit as in \S\ref{subsec:new_perspective}, both our knowledge from the measurement data (contained in $\log\left(\mathcal{L}(\{n_{lm}\};E)\right)/LN$ which measures the information gain) and our ignorance (reflected in $S(E)$ which measures the lack of information) about the operator $E$ are taken into account in such a way that our ignorance takes an infinitesimal weight.\index{maximum-likelihood-maximum-entropy (MLME)} This introduces a small and smooth convex hill over the set of positive ML estimators which selects the one with the largest entropy. As in \cite{mlme}, the value of $\lambda$ may be chosen such that both $\log\left(\mathcal{L}(\{n_{lm}\};E)\right)/LN$ and $S(E)$ remain almost constant with respect to $\lambda$.

To maximize $\mathcal{I}(\lambda;E)$ with respect to $E$, we define the variation $E+\updelta E=(1+\mathcal{Z}^\dagger)E(1+\mathcal{Z})$, where $\mathcal{Z}$ is a small arbitrary operator such that Eq.~(\ref{constraints}) is satisfied, that is: $\tr{\updelta E}=0$. Thus the most general expression for $\mathcal{Z}$ is
\begin{equation}
1+\mathcal{Z}=\left(1+\updelta \mathcal{A}\right)\left[\sqrt{\mathrm{tr}_\mathcal{K}\left\{\left(1+\updelta \mathcal{A}^\dagger\right)E\left(1+\updelta \mathcal{A}\right)\right\}}\otimes 1_\mathcal{K}\right]^{-1}\,,
\label{variations}
\end{equation}
with an unrestricted infinitesimal $\updelta \mathcal{A}$. On the other hand, the variation of $\mathcal{I}(\lambda;E)$ with respect to $E$ gives $\tr{\updelta E\,W}$, where
\begin{equation}
W=\frac{1}{L}\sum_{lm}\frac{f_{lm}}{p_{lm}}\rho^{(l)\,\textsc{t}}_\text{i}\otimes\Pi_m-\frac{\lambda}{\ID}\left[1+\log\left(\frac{E}{\ID}\right)\right]\,
\label{auxop}
\end{equation}
and $f_{lm}=n_{lm}/LN$.
Since $\mathcal{Z}$ is small, the operator $1+\mathcal{Z}$ can be expressed as
\begin{equation}
1+\mathcal{Z}\approx\updelta \mathcal{A} + 1-\frac{1}{2}\,\mathrm{tr}_\mathcal{K}\left\{\updelta \mathcal{A}^\dagger E+E\updelta \mathcal{A}\right\}\otimes 1_\mathcal{K}\,
\end{equation}
in terms of the first-order variations $\updelta \mathcal{A}$ and $\updelta \mathcal{A}^\dagger$. In deriving the expression above, the approximation
\begin{equation}
\left(1+\phi\right)^{-\frac{1}{2}}\approx 1-\frac{1}{2}\,\phi
\end{equation}
for a small operator $\phi$ is used. The variation $\updelta E$ is thus evaluated as
\begin{align}
\updelta E=&\,(1+\mathcal{Z}^\dagger)E(1+\mathcal{Z})-E\nonumber\\
=&\,\left(\updelta \mathcal{A}^\dagger -\frac{1}{2}\,\mathrm{tr}_\mathcal{K}\left\{\updelta \mathcal{A}^\dagger E+E\updelta \mathcal{A}\right\}\otimes 1_\mathcal{K}\right)E\nonumber\\
&\,+E\left(\updelta \mathcal{A}-\frac{1}{2}\,\mathrm{tr}_\mathcal{K}\left\{\updelta \mathcal{A}^\dagger E+E\updelta \mathcal{A}\right\}\otimes 1_\mathcal{K}\right)\,.
\end{align}

Hence
\begin{align}
&\updelta \mathcal{I}(\lambda;E)\nonumber\\
=&\,\tr{\updelta E\,W}\nonumber\\
=&\,\text{tr}\Big\{\updelta \mathcal{A}^\dagger EW-\frac{1}{2}\,\left(\mathrm{tr}_\mathcal{K}\left\{\updelta \mathcal{A}^\dagger E\right\}\otimes 1_\mathcal{K}\right)EW\nonumber\\
&\,-\frac{1}{2}\,E\left(\mathrm{tr}_\mathcal{K}\left\{\updelta \mathcal{A}^\dagger E\right\}\otimes 1_\mathcal{K}\right)W+\text{h.c.}\Big\}\nonumber\\
=&\,\tr{\updelta \mathcal{A}^\dagger E\left(W-\frac{1}{2}\,\mathrm{tr}_\mathcal{K}\left\{WE+EW\right\}\otimes 1_\mathcal{K}\right)}+\text{c.c.}\,.
\end{align}
By imposing $\updelta \mathcal{I}(\lambda;E)>0$, the method of steepest ascent\index{steepest-ascent method} leads us to
\begin{equation}
\updelta \mathcal{A}=\updelta \mathcal{A}^\dagger=\frac{\epsilon}{2}\left(W-\frac{1}{2}\,\mathrm{tr}_\mathcal{K}\left\{WE+EW\right\}\otimes 1_\mathcal{K}\right)
\end{equation}
for some small $\epsilon>0$. Hence, to obtain the MLME estimator $\hat E_\text{MLME}$\label{symbol:emlme}, one simply fixes $\lambda\ll 1$ and iterates\index{algorithm} the equations
\begin{center}
\colorbox{light-gray}{\begin{minipage}[c]{12cm}
\uline{\textbf{MLME QPT iterative equations}}
\begin{align}
E_{n+1}=&\,(1+\mathcal{Z}^\dagger_n)E_n(1+\mathcal{Z}_n)\,,\nonumber\\
1+\mathcal{Z}_n=&\,\left(\updelta \mathcal{A}\right)_n + 1-\frac{1}{2}\,\mathrm{tr}_\mathcal{K}\left\{\left(\updelta \mathcal{A}\right)_n^\dagger E_n+E_n\left(\updelta \mathcal{A}\right)_n\right\}\otimes 1_\mathcal{K}\nonumber\\
\left(\updelta \mathcal{A}\right)_n=&\,\frac{\epsilon}{2}\left(W_n-\frac{1}{2}\,\mathrm{tr}_\mathcal{K}\left\{W_n E_n+E_n W_n\right\}\otimes 1_\mathcal{K}\right)\,,\label{iter1}
\end{align}
  \end{minipage}}
  \end{center}
where the expression for $\mathcal{Z}_n$ follows from Eq.~(\ref{variations}) and $W_n$ denotes the operator $W$ in Eq.~(\ref{auxop}) evaluated for $E_n$. One may do so by starting from a randomly chosen operator $E_0$ and continue until the extremal equation for $\hat E_\text{MLME}$ is satisfied with some pre-chosen numerical precision. To derive this extremal equation, we define the Lagrange functional\index{Lagrange functional} \cite{qstateest}
\begin{equation}
\mathcal{D}(E)=\mathcal{I}(\lambda;E)-\tr{\Lambda\,E}
\end{equation}
with the Lagrange operator\index{Lagrange operator} $\Lambda\equiv h\otimes 1_\mathcal{K}$\label{symbol:lagop} for the constraints in Eq.~(\ref{constraints}), where $h$ is a Hermitian operator. Setting the variation of $\mathcal{D}(E)$ to zero gives the extremal equation
\begin{equation}
\Lambda\hat E_\text{MLME}\Lambda=W_\text{MLME}\hat E_\text{MLME}W_\text{MLME}
\end{equation}
with $\Lambda=\sqrt{\mathrm{tr}_\mathcal{K}\left\{W_\text{MLME}\hat E_\text{MLME}W_\text{MLME}\right\}}\otimes 1_\mathcal{K}$.

Thus far, we have been assuming that the measurement outcomes $\Pi_m$ give perfect detection of quantum systems. The iterative equations in Eq.~(\ref{iter1}) can be generalized to the case of imperfect detection. As always, if each of the $M$ measurement outcomes $\Pi_m$ is assigned a detection efficiency\index{detection efficiency} $\eta_m\leq1$, one can define a new set of $M$ measurement outcomes $\tilde\Pi_m\equiv\eta_m\Pi_m$ such that $G\equiv\sum_m\tilde\Pi_m\neq1_\mathcal{K}$. It follows that the probabilities $p_{lm}=\tr{E\left(\rho^{(l)\,\textsc{t}}_\text{i}\otimes\tilde\Pi_m\right)}{\Big/}L$ do not sum to unity.

The likelihood functional, in this case, turns out to be
\begin{equation}
\mathcal{L}'(\{n_{lm}\};E)=\frac{\left(\sum_l\tilde{n}_l\right)!}{\left(\prod_{lm}n_{lm}!\right)\prod_{l'}\left(\tilde{n}_{l'}-n'_{l'}\right)}\left(\prod_{lm}p_{lm}^{n_{lm}}\right)\prod_{l'}\left(\frac{1}{L}-p'_{l'}\right)^{\tilde{n}_{l'}-n'_{l'}}\,,
\end{equation}\label{symbol:likelihoodeimp}
where $\sum_l\tilde{n}_l=N_\text{true}$\label{symbol:ntildel} is the unknown total number of copies and the primed quantities are defined as in \S\ref{sec:qpt_prelim}. Stirling's formula then gives
\begin{align}
\log\mathcal{L}'(\{n_{lm}\};E)\approx &\, N_\text{true}\left(\log N_\text{true}-1\right)-\sum_l\left(\tilde{n}_l-n'_l\right)\left(\log \left(\tilde{n}_l-n'_l\right)-1\right)\nonumber\\
&+ \sum_l\left(\tilde{n}_l-n'_l\right)\log \left(\frac{1}{L}-p'_l\right)+\sum_{lm}n_{lm}\log p_{lm}+\text{ const.}\,.
\end{align}
The corresponding derivative
\begin{equation}
\frac{\updelta\log\mathcal{L}'(\{n_{lm}\};E)}{\updelta \tilde{n}_{l}}=\log\left(\frac{N_\text{true}\left(\frac{1}{L}-p'_l\right)}{\tilde{n}_l-n'_l}\right)
\end{equation}
is zero for the most-likely $\tilde{n}_l$, which is given by
\begin{equation}
\tilde{n}_l=n'_l+N\frac{\frac{1}{L}-p'_l}{\sum_{l'}p'_{l'}}\,.
\end{equation}
Hence,
\begin{align}
\frac{\updelta\log\mathcal{L}'(\{n_{lm}\};E)}{\updelta E}=\sum_{lm}\left(\frac{n_{lm}}{p_{lm}}-\frac{N}{\sum_{l'}p'_{l'}}\right)\left(\frac{\rho^{(l)\,\textsc{t}}_\text{i}\otimes\tilde\Pi_m}{L}\right)\,.
\end{align}
In short, the iteration procedure of Eq.~(\ref{iter1}) can still be used with the new set of POM outcomes $\tilde\Pi_m$ provided that the operator $W$ in Eq.~(\ref{auxop}) is replaced by $W-W_0$, where
\begin{equation}
W_0=\frac{1}{L\sum_{l'}p'_{l'}}\sum_l\rho^{(l)\,\textsc{t}}_\text{i}\otimes G\,
\end{equation}
accounts for the copies that escape detection.

As an example, we apply the algorithm to numerical simulations on two-qubit channels, the \textsc{cnot}\index{cnot@\textsc{cnot}} gate described by the unitary operator
\begin{equation}
U_\textsc{cnot}\,\widehat{=}\begin{pmatrix}
\,\,1\,\, & \,\,0\,\, & \,\,0\,\, & \,\,0\,\,\\
\,\,0\,\, & \,\,1\,\, & \,\,0\,\, & \,\,0\,\,\\
\,\,0\,\, & \,\,0\,\, & \,\,0\,\, & \,\,1\,\,\\
\,\,0\,\, & \,\,0\,\, & \,\,1\,\, & \,\,0\,\,
\end{pmatrix}
\end{equation}
and a randomly generated non-unitary quantum channel described by a full-rank Choi-Jami{\'o}{\l}kowski matrix\index{Choi-Jami{\'o}{\l}kowski!-- operator or matrix}, as well as the three-qubit Toffoli gate described by the unitary operator
\begin{equation}
U_\text{Toffoli}\,\widehat{=}\begin{pmatrix}
\,\,1\,\, & \,\,0\,\, & \,\,0\,\, & \,\,0\,\, & \,\,0\,\, & \,\,0\,\, & \,\,0\,\, &\,\,0\,\,\\
\,\,0\,\, & \,\,1\,\, & \,\,0\,\, & \,\,0\,\, & \,\,0\,\, & \,\,0\,\, & \,\,0\,\, &\,\,0\,\,\\
\,\,0\,\, & \,\,0\,\, & \,\,1\,\, & \,\,0\,\, & \,\,0\,\, & \,\,0\,\, & \,\,0\,\, &\,\,0\,\,\\
\,\,0\,\, & \,\,0\,\, & \,\,0\,\, & \,\,1\,\, & \,\,0\,\, & \,\,0\,\, & \,\,0\,\, &\,\,0\,\,\\
\,\,0\,\, & \,\,0\,\, & \,\,0\,\, & \,\,0\,\, & \,\,1\,\, & \,\,0\,\, & \,\,0\,\, &\,\,0\,\,\\
\,\,0\,\, & \,\,0\,\, & \,\,0\,\, & \,\,0\,\, & \,\,0\,\, & \,\,1\,\, & \,\,0\,\, &\,\,0\,\,\\
\,\,0\,\, & \,\,0\,\, & \,\,0\,\, & \,\,0\,\, & \,\,0\,\, & \,\,0\,\, & \,\,0\,\, &\,\,1\,\,\\
\,\,0\,\, & \,\,0\,\, & \,\,0\,\, & \,\,0\,\, & \,\,0\,\, & \,\,0\,\, & \,\,1\,\, &\,\,0\,\,\\
\end{pmatrix}\,.
\end{equation}
To quantify the discrepancy between an MLME estimator and the true Choi-Jami{\'o}{\l}kowski operator\index{Choi-Jami{\'o}{\l}kowski!-- operator or matrix} $E_\text{true}$, we use the trace-class distance
\begin{equation}
\mathcal{D}_\text{tr}\left(\hat E_\text{MLME},E_\text{true}\right)=\frac{1}{2\ID}\tr{\big|\hat E_\text{MLME}-E_\text{true}\big|}\,.
\end{equation}
In these simulations, we take the $\ID$-dimensional projectors of a SIC POM as the input states.
\begin{figure}[htp]
  \centering
  \includegraphics[width=0.6\textwidth]{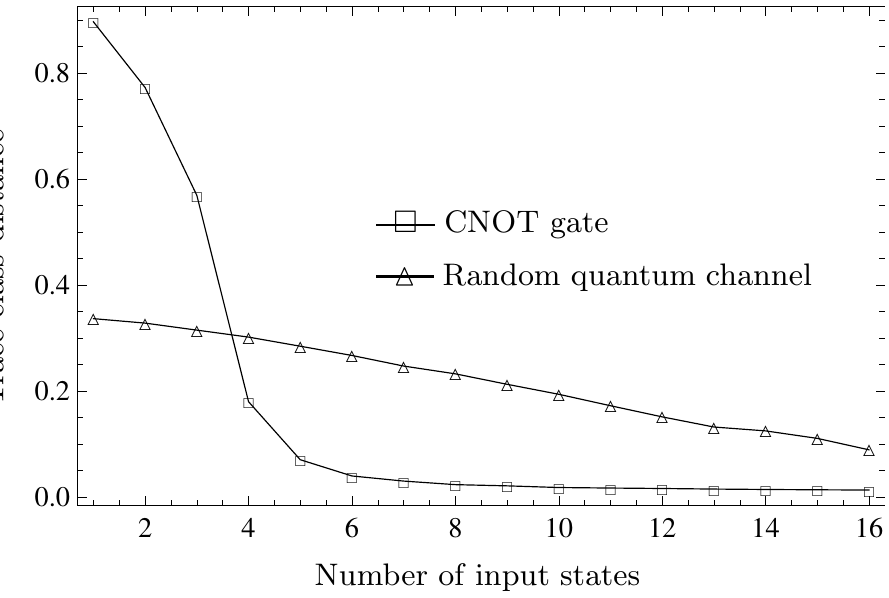}\\
  \includegraphics[width=0.6\textwidth]{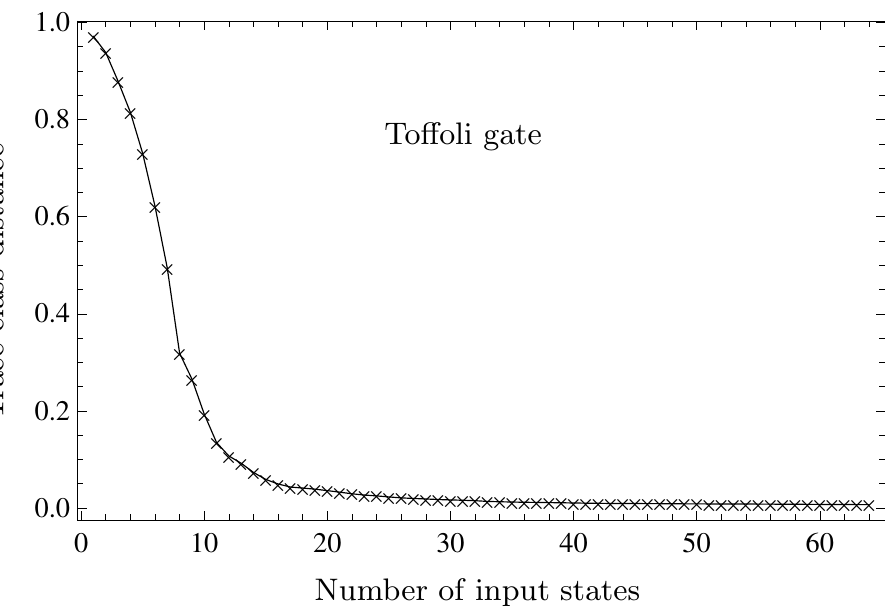}
  \caption{Numerical simulations on the two-qubit ($d=2^2$) and three-qubit ($d=2^3$) quantum channels where $\ID=\OD=d$. The projectors of symmetric informationally complete POMs (SIC POMs) are chosen as the linearly independent input states for all the simulations ($L=d^2$). For the measurements, informationally complete POMs consisting of tensor products of qubit SIC POMs are used ($M=d^2$). Each qubit SIC POM consists of a set of pure states whose Bloch vectors form the ``legs of a tetrahedron'' in the Bloch sphere. For the two-qubit channels, $N=10^4$ and an average over 50 experiments is taken to compute the trace-class distances. For the three-qubit channel, the measurement data are generated without statistical noise. For unitary channels, one can see that the MLME algorithm can still give fairly accurate estimations with a smaller number of input states than that of a linearly independent set. Numerical simulations of arbitrary two-qubit and three-qubit unitary channels suggest that the number is approximately $d^2/2$ for SIC POM input states, above which there is insignificant tomographic improvement.}
  \label{fig:channels}
\end{figure}
As shown in Fig.~\ref{fig:channels}, using the MLME algorithm for QPT can give fast convergence in terms of tomographic efficiency with a reduced number of input states as quantum resources. This reduction is especially significant for unitary processes, where the Choi-Jami{\'o}{\l}kowski operators\index{Choi-Jami{\'o}{\l}kowski!-- operator or matrix} are rank-1. For nonunitary quantum processes described by matrices of larger rank, the tomographic efficiency will be lower as shown in the first plot of Fig.~\ref{fig:channels}. This is expected in analogy with quantum state tomography where it is more difficult to reliably estimate highly-mixed states than nearly-pure ones. 

\section{Adaptive strategies}
\label{sec:adaptstrat}

An interesting question to ask with regard to incomplete QPT is whether one can perform it in an optimal way given the available resources by means of adaptive strategies. Here optimality refers to the minimization of the amount of resources (input states or measurements) used to perform incomplete QPT such that the distance between $\hat E_\text{MLME}$ and $E_\text{true}$ reaches a certain desired value. Very frequently, despite the fact that $E_\text{true}$ is always unknown, one has a rough idea of an operator $E_\text{prior}$ which may be close to $E_\text{true}$ based on some prior information\index{prior information (knowledge)} about the unknown $E_\text{true}$. This scenario is reasonable and typical when one designs a quantum channel experimentally which performs an expected quantum operation, with errors arising from imperfections of the components that make up the channel. We shall establish adaptive strategies which make use of such an operator in order to select, with the help of the MLME algorithm, resources for incomplete QPT in an optimal way. We refer to such tomography schemes as the \emph{adaptive MLME quantum process tomography} (AMLME QPT) schemes.

We will focus on adaptive strategies to choose the input states optimally. This can be reviewed in two separate cases: The case in which a fixed set of linearly independent input states is used (\S\ref{subsec:fixed_input}) and that in which arbitrary input states can be generated for incomplete QPT (\S\ref{subsec:all_input}). Adaptive strategies to choose the POM are relatively harder to formulate and this task is put aside for future studies.

\subsection{Optimization over a fixed set of linearly independent input states}\index{adaptive estimation scheme!-- adaptive MLME QPT}
\label{subsec:fixed_input}
In the previous section, we considered the projectors of the SIC POMs, which are known to have optimal tomographic efficiencies, as input states in the numerical simulations. Since these POMs are symmetric in the sense of Eq.~(\ref{sicpom}), any ordering of the input states in a given set gives the same plots in Fig.~\ref{fig:channels}. In practice, however, such entangled states are difficult to produce and one typically has access to a set of separable states \cite{iontrap} for measurements instead. In this case, there no longer exists such a symmetry and the tomographic performance depends on the order of the input states chosen, possibly strongly so. We propose to optimize the tomographic performance by choosing the input states adaptively based on the measurement data collected from the previously chosen input states, thereby using the prior $E_\text{prior}$.

To describe the adaptive strategy, let us consider a set of $L\geq D^2_\text{i}$ input states in which $D^2_\text{i}$ of them are linearly independent. Suppose that $N$, which is a fixed integer for all input states, copies of a randomly chosen input state $\rho_\text{i}^{(1)}$ are sent through the quantum channel and the first set of measurement data $\{\nu_{11},\ldots,\nu_{1M}\}$, $\sum_m\nu_{1m}=1$, is collected. With these data $f_{1m}\equiv\nu_{1m}$, one obtains the first MLME estimator $\hat E^{(1)}_\text{MLME}$. To select the next input state out of the remaining $L-1$ states, we take $E_\text{prior}$ as a gauge for $E_\text{true}$ to generate $L-1$ sets of probabilities respectively from the $L-1$ states. Each set of probabilities is then treated as the set of frequencies $\{\nu^{(k)}_{21},\ldots,\nu^{(k)}_{2M}\}$, for the corresponding input state $k$. Hence, one has $L-1$ sets of measurement data, each set being the combined data $\{\nu_{11},\ldots,\nu_{1M},\nu^{(k)}_{21},\ldots,\nu^{(k)}_{2M}\}/2$ with the normalized frequencies $f_{1m}\equiv\nu_{1m}/2$ and $f^{(k)}_{2m}\equiv\nu^{(k)}_{2m}/2$ such that $\sum_m(f_{1m}+f^{(k)}_{2m})=1$ for each $k$, and the corresponding $L-1$ \emph{projected} MLME estimators $\hat E^{(2)}_{\text{MLME},k}$.

The value of $k$ is selected such that a chosen figure of merit which quantifies the distance between $\hat E^{(2)}_{\text{MLME},k}$ and $\hat E^{(1)}_\text{MLME}$ is the largest, so that there is a high chance for the next MLME estimator to be closer to $E_\text{true}$. As an example, the figure of merit is taken to be the trace-class distance $\mathcal{D}_\text{tr}\left(\hat E^{(2)}_{\text{MLME},k},\hat E^{(1)}_\text{MLME}\right)$. With this input state, the second estimator $\hat E^{(2)}_\text{MLME}$ is then obtained with MLME QPT. One repeats this procedure for subsequent input states until the distance $\mathcal{D}_\text{tr}\left(\hat E^{(l+1)}_\text{MLME},\hat E^{(l)}_\text{MLME}\right)$ is below some preset threshold. An alternative to this would be to \emph{minimize} the trace-class distance $\mathcal{D}_\text{tr}\left(\hat E^{(l+1)}_{\text{MLME},k},E_\text{prior}\right)$.

It is important to understand that in this strategy, the prior information\index{prior information (knowledge)} $E_\text{prior}$ is \emph{not} used to reconstruct the unknown quantum process in any way. It serves only as a means to optimally select the input states from the given set so as to maximize the tomographic convergence. This adaptive strategy also relies partially on the measurement data obtained in the experiment. We have thus introduced an operational method of using the prior information\index{prior information (knowledge)} to minimize the amount of resources needed to perform reliable MLME QPT without introducing any artifacts coming from the prior information\index{prior information (knowledge)} into the reconstruction procedure. To summarize, the adaptive MLME strategy\index{algorithm} is as follows:

\begin{center}
\colorbox{light-gray}{\begin{minipage}[c]{12cm}
  \textbf{\uline{Adaptive MLME algorithm (Fixed set of input states)}}
\begin{enumerate}
  \item Randomly choose $\rho_\text{i}^{(1)}$ from the set of $L$ input states and set $l=1$.
  \begin{enumerate}
    \item Perform QPT using $\rho_\text{i}^{(l)}$ and obtain the set of frequencies $\{\nu_{l1},\ldots,\nu_{lM}\}$, $\sum_m\nu_{lm}=1$.
    \item Set $\nu=\bigcup^l_{j=1}\{\nu_{j1},\ldots,\nu_{jM}\}/\,l$.
    \item Invoke the MLME algorithm with $\nu$ and obtain $\hat E^{(l)}_\text{MLME}$. Use $E_\text{prior}$ to compute the frequencies $\{\nu^{(k)}_{l+1\,1},\ldots,\nu^{(k)}_{l+1\,M}\}$, $\sum_m\nu^{(k)}_{l+1\,m}=1$, from the remaining input states, with $k$ labeling the remaining $L-l$ states.
    \item Define $L-l$ sets of accumulated frequencies \mbox{$(\nu\cup\{\nu^{(k)}_{l+1\,1},\ldots,\nu^{(k)}_{l+1\,M}\})/(l+1)$} and calculate the $L-l$ projected MLME estimators $\hat E^{(l+1)}_{\text{MLME},k}$.
    \item Set $\rho_\text{i}^{(l+1)}$ as the input state corresponding to $k$ such that $\mathcal{D}_\text{tr}\left(\hat E^{(l+1)}_{\text{MLME},k},\hat E^{(l)}_\text{MLME}\right)$ is largest.
  \end{enumerate}
  \item Set $l=l+1$ and repeat Steps 1(a)--1(e).\\
\end{enumerate}
  \end{minipage}}
  \end{center}

\subsection{Optimization over the Hilbert space}\index{adaptive estimation scheme!-- adaptive MPL-MLME QPT}
\label{subsec:all_input}

More generally, the adaptive strategy may be extended to the case in which one has access to the entire Hilbert space of states. In other words, the task is to search for the next optimal input state $\rho^{(L+1)}_\text{i}$ from the $\ID$-dimensional Hilbert space based on the measurement data $\nu_{lm}$ obtained in the experiment from $L$ previously chosen input states, where $\sum_m\nu_{lm}=1$ for all $l$, and the prior information\index{prior information (knowledge)} $E_\text{prior}$ about the unknown quantum process.

To this end, we define the normalized \emph{projected} log-likelihood functional\index{projected log-likelihood functional}
\begin{equation}
\log\tilde{\mathcal{L}}(\{\nu_{lm}\};E,\rho)=\sum_{lm}\frac{\nu_{lm}}{L+1}\log\left(\tilde p_{lm}\right)+\sum_m\frac{\tilde \nu_m}{L+1}\log\left(\tilde p_m\right)\,,
\label{crossent}
\end{equation}\label{symbol:likelihoodeproj}
where\setlength{\parskip}{0pt}
\begin{center}
\begin{equation}
\tilde p_{lm}=\tr{E\,\frac{\rho^{(l)\,T}_\text{i}\otimes\Pi_m}{L+1}}\,,
\end{equation}
\end{center}
\begin{equation}
\tilde \nu_m=\tr{E_\text{prior}\rho^T\otimes\Pi_m}\,\,\,\text{and}\,\,\,\tilde p_m=\tr{E\,\frac{\rho^T\otimes\Pi_m}{L+1}}\,
\end{equation}\label{symbol:projquantities}
with $l$ always running from $1$ to $L$ over all previously used input state labels.

This projected log-likelihood functional is a good approximation to the log-likelihood functional for the situation in which the state $\rho$ is chosen as the next input state for the experiment as long as $E_\text{prior}$ is not too far away from $E_\text{true}$. The projected frequencies $\tilde \nu_m$ estimate the actual frequencies one gets when measuring the input state $\rho$. An optimal input state $\rho^{(L+1)}_\text{i}$ and the corresponding Choi-Jami{\'o}{\l}kowski operator\index{Choi-Jami{\'o}{\l}kowski!-- operator or matrix} are chosen as the positive estimators that maximize this projected log-likelihood functional.

Coincidentally, this maximum projected log-likelihood\index{maximum projected log-likelihood (MPL)} (MPL) procedure is equivalent to minimizing the \emph{cross entropy}\index{entropy!-- cross} functional $\mathcal{C}(E,\rho)=-\log\tilde{\mathcal{L}}(\{\nu_{lm}\};E,\rho)$\label{symbol:crossent} \cite{mce1,mce2} over all positive operators subjected to the respective constraints for $\rho$ and $E$. Hence, another way of understanding this procedure is to first regard both the incomplete measurement data collected after using $L$ input states and $E_\text{prior}$ as the full prior knowledge\index{prior information (knowledge)} one has about the unknown $E_\text{true}$. The statistical motivation for MPL or minimizing $\mathcal{C}(E,\rho)$ is, loosely speaking, to obtain estimators which are as compatible with this prior knowledge\index{prior information (knowledge)} as possible by minimizing the entropy of the prior knowledge\index{prior information (knowledge)} $\mathcal{C}(E,\rho)$. We will provide some more arguments related to this optimization technique in the later part of this section.

To carry out the optimization, we consider the response of $\log\tilde{\mathcal{L}}(\{\nu_{lm}\};E,\rho)$ to small variations of both $\rho$ and $E$. After some similar calculations as in \S\ref{sec:qpt_mlme_iteralgo}, we obtain the MPL iterative equations
\begin{align}
E_{n+1}=&\,(1+\mathcal{Z}^\dagger_n)E_n(1+\mathcal{Z}_n)\,,\nonumber\\
\rho_{n+1}=&\,\frac{(1+\epsilon_2\Xi_n)\rho_n(1+\epsilon_2\Xi_n)}{\text{tr}_\mathcal{H}\{(1+\epsilon_2\Xi_n)\rho_n(1+\epsilon_2\Xi_n)\}}\,,
\label{iter2}
\end{align}
where $\mathcal{Z}_n$ is defined by Eq.~(\ref{variations}) with
\begin{align}
\left(\updelta A\right)_n=&\,\frac{\epsilon_1}{2}\left(\mathcal{X}_n-\frac{1}{2}\mathrm{tr}_\mathcal{K}\left\{\mathcal{X}_n E_n+E_n \mathcal{X}_n\right\}\otimes 1_\mathcal{K}\right)\,,\nonumber\\
\mathcal{X}_n=&\,\sum_{lm}\frac{\nu_{lm}}{\tilde p_{lm}}\frac{\rho^{(l)\,T}_\text{i}\otimes\Pi_m}{(L+1)^2}+\sum_m\frac{\tilde \nu_m}{\tilde p_m}\frac{\rho^T\otimes\Pi_m}{(L+1)^2}\,,
\label{xop}
\end{align}
and
\begin{align}
\Xi_n\equiv &\, \mathcal{Y}_n-\text{tr}_\mathcal{H}\{\mathcal{Y}_n\rho_n\}\,,\nonumber\\
\mathcal{Y}_n=&\,\text{tr}_\mathcal{K}\Bigg\{\Bigg[\sum_m\frac{1_\mathcal{H}\otimes\Pi_m}{L+1}\nonumber\\
&\,\times\left(\log\left(\tilde p_m\right)E_\text{prior}+\frac{\tilde \nu_m}{(L+1)\tilde p_m}E\right)\Bigg]^T\Bigg\}\,.
\label{yop}
\end{align}
The MPL estimators satisfy the extremal equations
\begin{align}
\tilde\Lambda\hat E_\text{MPL}\tilde\Lambda&=\mathcal{X}_\text{MPL}\hat E_\text{MPL}\mathcal{X}_\text{MPL}\,,\nonumber\\
\hat\rho_\text{MPL}\mathcal{Y}_\text{MPL}&=\mathcal{Y}_\text{MPL}\hat\rho_\text{MPL}=\text{tr}_\mathcal{H}\{\mathcal{Y}_\text{MPL}\hat\rho_\text{MPL}\}\hat\rho_\text{MPL}\,,
\label{MPLextremal}
\end{align}\label{symbol:mplestimators}
where
\begin{equation}
\tilde\Lambda=\sqrt{\mathrm{tr}_\mathcal{K}\left\{\mathcal{X}_\text{MPL}\hat E_\text{MPL}\mathcal{X}_\text{MPL}\right\}}\otimes 1_\mathcal{K}\,.
\label{MPLextremal2}
\end{equation}
The small parameters $\epsilon_1$ and $\epsilon_2$ are positive numbers. Thus, to carry out the MPL procedure, one iterates Eqs.~(\ref{iter2}) until Eqs.~(\ref{MPLextremal}) are satisfied with a preset numerical precision.

There is one important feature of this optimization scheme. From Eq.~(\ref{crossent}), we note that $\log\tilde{\mathcal{L}}(\{\nu_{lm}\};E,\rho)$ is neither convex nor concave in $\rho$ and hence can have multiple local maxima. Thus, the MPL optimization is nonconvex.

To generate these local-maxima estimators, one can start from multiple randomly chosen starting points and perform the iterations. Thereafter, the state estimator $\hat\rho_\text{MPL}$ to be chosen as the next input state $\rho^{(L+1)}_\text{i}$ is such that its corresponding $\hat E_\text{MPL}$ gives the largest trace-class distance away from the previous MLME estimator $\hat E^{(L)}_\text{MLME}$, which is obtained from the data of the previously chosen $L$ input states, over all generated pairs of MPL estimators $(\hat\rho_\text{MPL},\,\hat E_\text{MPL})$. Again, one may also minimize the distance between $\hat E_\text{MPL}$ and $E_\text{prior}$.

Let us summarize the adaptive MPL-MLME strategy with the following scheme\index{algorithm}:
\begin{center}
\colorbox{light-gray}{\begin{minipage}[c]{12cm}
  \textbf{\uline{Adaptive MPL-MLME algorithm}}
\begin{enumerate}
  \item Randomly choose $\rho_\text{i}^{(1)}$ as the first input state and set $l=1$.
  \begin{enumerate}
    \item Perform QPT using $\rho_\text{i}^{(l)}$ and obtain the set of frequencies $\{\nu_{l1},\ldots,\nu_{lM}\}$, $\sum_m\nu_{lm}=1$.
    \item Set $\nu=\bigcup^l_{j=1}\{\nu_{j1},\ldots,\nu_{jM}\}/\,l$.
    \item Invoke the MLME algorithm with $\nu$ and obtain $\hat E^{(l)}_\text{MLME}$.
    \item Using $E_\text{prior}$, generate a set of pairs of MPL estimators ($\hat\rho_\text{MPL}$,\,$\hat E_\text{MPL}$), where the states $\hat\rho_\text{MPL}$ were not part of the $l$ input states previously used, by iterating Eqs.~(\ref{MPLextremal}) from different, randomly chosen starting points.
    \item Set $\rho_\text{i}^{(l+1)}$ as the input state corresponding to the state estimator $\hat\rho_\text{MPL}$ such that $\mathcal{D}_\text{tr}\left(\hat E_{\text{MPL}},\hat E^{(l)}_\text{MLME}\right)$ is the largest.
  \end{enumerate}
  \item Set $l=l+1$ and repeat Steps 1(a)--1(e).\\
\end{enumerate}
  \end{minipage}}
  \end{center}

\begin{figure}[htp]
  \centering
  \includegraphics[width=0.6\textwidth]{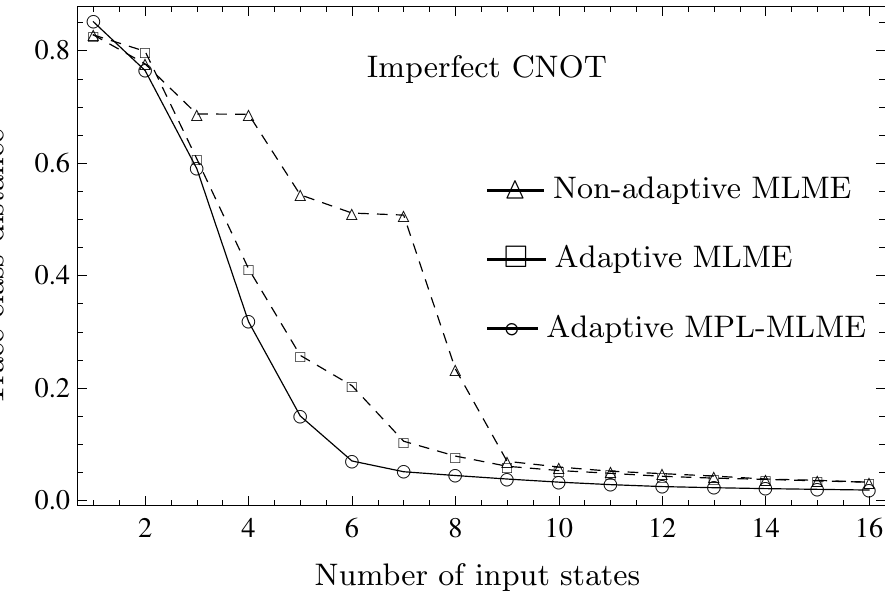}\\
  \includegraphics[width=0.6\textwidth]{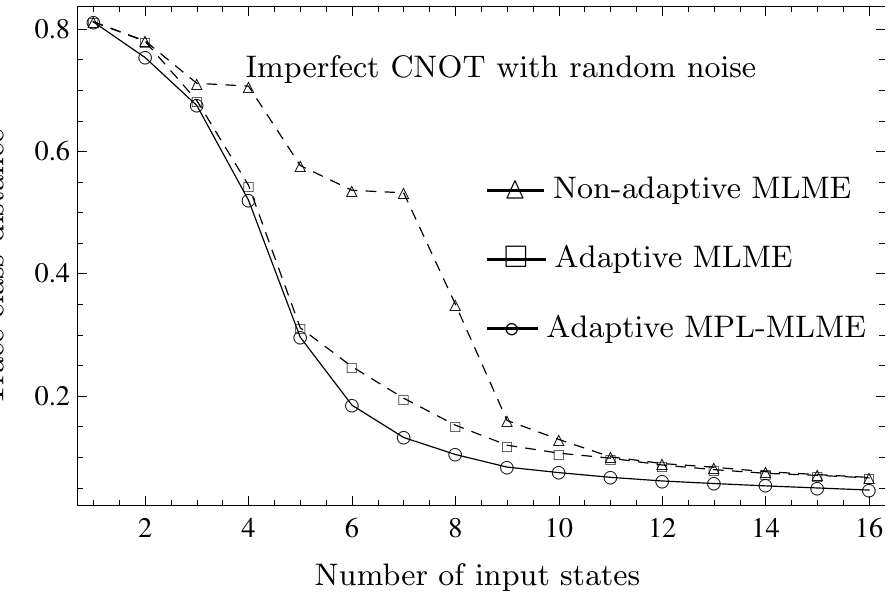}
  \caption{A comparison of three incomplete QPT schemes: the non-adaptive MLME scheme, the adaptive MLME scheme and the adaptive MPL-MLME scheme. Monte Carlo simulations are carried out on two different types of imperfect \textsc{cnot} gates described in the text. Here, $N=10^4$ and an average over 50 experiments is taken to compute the trace-class distances. For both the non-adaptive as well as the adaptive MLME schemes, the 16 linearly independent input states are chosen to be tensor products of projectors of the kets $\ket{0}$, $\ket{1}$, $(\ket{0}+\ket{1})/\sqrt{2}$ and $(\ket{0}+\ket{1}\I)/\sqrt{2}$. For all schemes, the POM outcomes are chosen to be the tensor products of qubit SIC POMs. The tomographic performance of the adaptive MPL-MLME scheme is the best out of the three. The plots show that the tomographic efficiency can be further improved by optimizing the input states over the Hilbert space instead of restricting to a fixed set of linearly independent input states, albeit the small difference in tomographic performance between the two adaptive schemes for some quantum processes.}
  \label{fig:comp}
\end{figure}

With this, let us first compare the performances of the three proposed schemes, namely the non-adaptive MLME scheme in \S\ref{sec:qpt_mlme_iteralgo}, the adaptive MLME scheme in \S\ref{subsec:fixed_input} and the adaptive MPL-MLME scheme. For this purpose, we consider two quantum processes, the first being an imperfect \textsc{cnot}\index{cnot@\textsc{cnot}} gate whose action is described by the Kraus\index{Kraus operators} operators
\begin{equation}
K_1=\sqrt{1-\epsilon}\,U_\textsc{cnot}\quad\text{and}\quad K_2=\sqrt{\epsilon}\,.
\label{cnot_id}
\end{equation}
This first channel is a \textsc{cnot}\index{cnot@\textsc{cnot}} gate with probability $1-\epsilon$ and does nothing to the input states with probability $\epsilon$, an imperfect \textsc{cnot}\index{cnot@\textsc{cnot}} gate represented by a rank-2 Choi-Jami{\'o}{\l}kowski operator\index{Choi-Jami{\'o}{\l}kowski!-- operator or matrix}. The second process is described by the Kraus\index{Kraus operators} operators
\begin{equation}
K_1=\sqrt{1-\epsilon}\,U_\textsc{cnot}\quad\text{and}\quad\left\{K_j=\sqrt{\epsilon}B_j\right\}^{16}_{j=2}\,,
\label{cnot_rand}
\end{equation}
where the 15 operators $B_j$ are randomly generated and satisfy the equation $\sum_jB^\dagger_jB_j=1_\mathcal{K}$. This second channel, which is represented by a full-rank matrix, is a \textsc{cnot}\index{cnot@\textsc{cnot}} gate with probability $1-\epsilon$ and randomly perturbs the input states with probability $\epsilon$ with additional noise. As an example, we set $\epsilon=0.1$. Figure~\ref{fig:comp}\footnote{The set of input states used in Fig.~\ref{fig:comp}, taken from Ref.~\cite{iontrap}, is just one of the many possible choices one can use in quantum process tomography. It is important to understand that this set is by \emph{no means} sanctioned to be the ``standard'' set of input states. Rather, these are four states of the six projectors of the standard six-outcome POM, but \textbf{any} four of the six states will serve the purpose equally well.} shows the numerical results.

Next, to understand how this adaptive MPL-MLME strategy can lead to an optimization in tomographic performance, we need to know how increasing the number of input states used in AMLME QPT can affect the corresponding MLME estimators. Since we are considering only a subset of the full linearly independent input states in general, there exists a convex set\index{convex set} of estimators $\hat E_\text{ML}$\label{symbol:eml} maximizing the likelihood functional for a given set of informationally incomplete measurement data. This means that the likelihood functional possesses a plateau\index{plateau} hovering over this convex set\index{convex set} of estimators. As the number of input states $L$ used increases, the likelihood plateau\index{plateau} will either remain unchanged (if no additional information about $E_\text{true}$ is gained after performing QPT with new input states) or decrease in size (if new independent information is obtained). Thus in general, the plateau\index{plateau} will continue to shrink to a point when a full set of linearly independent input states is used.

We conjecture that the adaptive MPL-MLME strategy optimizes the rate of decrease in the size of the likelihood plateau\index{plateau} by maximizing the normalized projected log-likelihood functional with respect to the next input state. A point of view to justify this conjecture is to interpret the maximum of the normalized log-likelihood functional $\log\left(\mathcal{L}(\{n_{lm}\};E)\right)/LN$ as the maximum information gain from the measurement data. When the number of copies $N$ is infinite, the data are noiseless and the resulting maximum information gain is $\sum_{lm}f_{lm}\log(f_{lm})$, which is the negative of the Shannon entropy of the measurement data. For finite $N$, the maximum information gain over the space of statistical operators will typically be lower than the true maximum due to the positivity constraint, especially when $E_\text{true}$ is highly rank-deficient. In this language, the MPL-MLME strategy attempts to maximize this maximum information gain as much as possible via the optimization of future input states over the entire Hilbert space of statistical operators, using the normalized projected log-likelihood functional as an estimate for the actual normalized log-likelihood functional describing future measurements. This is a possible explanation for the optimal decrease in the likelihood plateau\index{plateau} size since one has maximal knowledge about the unknown $E_\text{true}$ gained with the optimized input states and so the ambiguity in the estimators is minimized.

\begin{figure}[htp]
  \centering
  \includegraphics[width=0.6\textwidth]{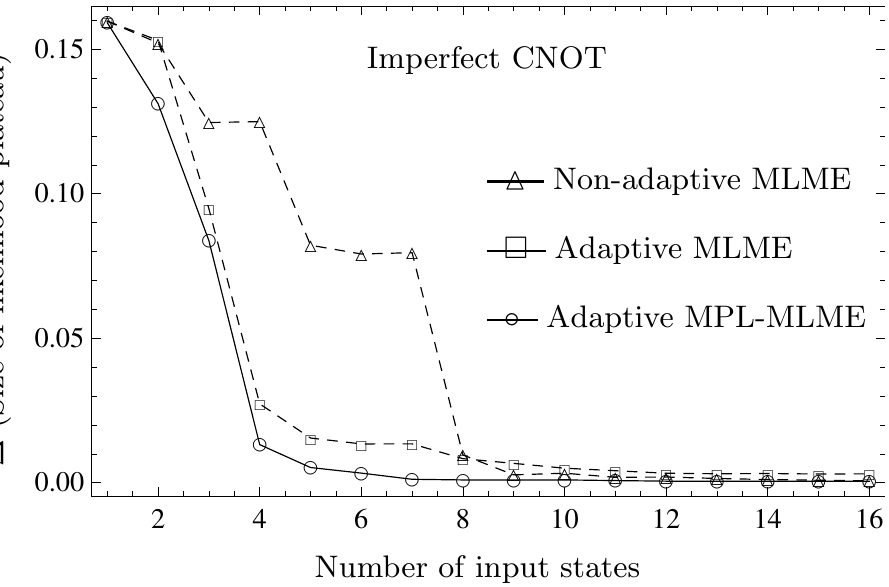}\\
  \includegraphics[width=0.6\textwidth]{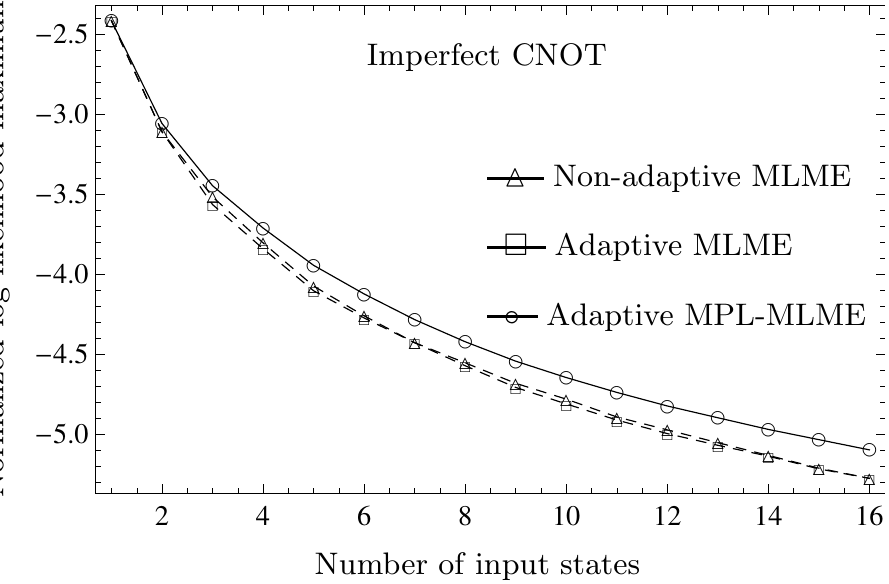}
  \caption{The dependence of the size of the likelihood plateau\index{plateau} ($\Delta$) and the normalized log-likelihood maximum on the number of input states. The respective performances of the non-adaptive MLME scheme, the adaptive MLME scheme and the adaptive MPL-MLME scheme are computed based on noiseless measurement data for an imperfect \textsc{cnot} gate with $\epsilon=0.1$. For both the non-adaptive MLME scheme and the adaptive MLME scheme, the 16 linearly independent input states are chosen to be tensor products of projectors of the kets $\ket{0}$, $\ket{1}$, $(\ket{0}+\ket{1})/\sqrt{2}$ and $(\ket{0}+\ket{1}\I)/\sqrt{2}$. For all schemes, the POM outcomes are chosen to be the tensor products of qubit SIC POMs. From the plot, the rate of decrease of $\Delta$ is the greatest with the adaptive MPL-MLME scheme. The increase in the normalized log-likelihood maxima with the adaptive MPL-MLME scheme may also be interpreted as greater maximum information gain after measurements using the optimal input states as compared to the other schemes.}
  \label{fig:plateau_height}
\end{figure}

We illustrate this point by considering the imperfect \textsc{cnot}\index{cnot@\textsc{cnot}} gate with $\epsilon=0.1$ described by Eq.~(\ref{cnot_id}). Since the boundary of the likelihood plateau\index{plateau} is complicated, we shall estimate its size numerically by first generating $N_0=10^3$ ML estimators $\hat E^{(j)}_\text{ML}$ labeled with the index $j$ for a given set of measurement data. Next, in the same spirit as in numerical sampling, we can define the operator centroid\index{centroid (operator)}
\begin{equation}
\overline{E}_\text{ML}=\frac{1}{N_0}\sum^{N_0}_{j=1}\hat E^{(j)}_\text{ML}
\end{equation}\label{symbol:emlbar}for this generated set of estimators and the normalized Hilbert-Schmidt standard deviation
\begin{equation}
\Delta=\frac{1}{\ID}\sqrt{\frac{\sum^{N_0}_{j=1}{\tr{\left(\hat E^{(j)}_\text{ML}-\bar E_\text{ML}\right)^2}}}{2N_0}}\,
\end{equation}\label{symbol:delta}away from the centroid. Thus, $0\leq\Delta\leq 1$. For sufficiently large $N_0$, the size of the plateau\index{plateau} may be well approximated by the spread $\Delta$. Figure~\ref{fig:plateau_height} compares the respective performances of the the three proposed schemes by analyzing the size of the likelihood plateau\index{plateau} and the maximum of the normalized log-likelihood functional. From Fig.~\ref{fig:plateau_height}, it is crucial to understand that $\Delta$ does not, strictly speaking, decrease monotonically with increasing height of the normalized log-likelihood functional. A counterexample is shown in the figure, that is a significant decrease in $\Delta$ for the adaptive MLME scheme as compared to the non-adaptive one with the corresponding slight decrease in its normalized log-likelihood maxima. We emphasize that what the adaptive MPL-MLME strategy exploits is the possible \emph{trend} of this behavior.

To end this part of the section, we comment that the aforementioned idea can be applied to adaptively choose the next set of POM outcomes $\Pi_j$ based on the collected measurement data. However, to perform the optimization successfully requires the solutions to more technical problems which include ensuring that the POM outcomes are linearly independent after the optimization. This project is left for future studies.

\subsection{A combination of both adaptive strategies}\index{adaptive estimation scheme!-- hybrid MLME QPT}
\label{subsec:combination}

Let us begin this final part of the section by reviewing the nonconvex feature of the MPL-MLME strategy discussed in \S\ref{subsec:all_input}. The presence of multiple local-maxima estimators which are linearly independent is an important element of the MPL-MLME strategy as it provides linearly independent input states which are optimal for measurement based on the data obtained from the experiments. In general, because of the nonlinearity of Eq.~(\ref{MPLextremal}), it is difficult to determine the number of such linearly independent extremal solutions for a given set of measurement data by analytical means. One can only search for as many linearly independent local-maxima estimators $\hat\rho_\text{MPL}$ as possible via numerical optimizations from different starting points within a reasonable time period.

\begin{figure}[htp]
  \centering
  \includegraphics[width=0.6\textwidth]{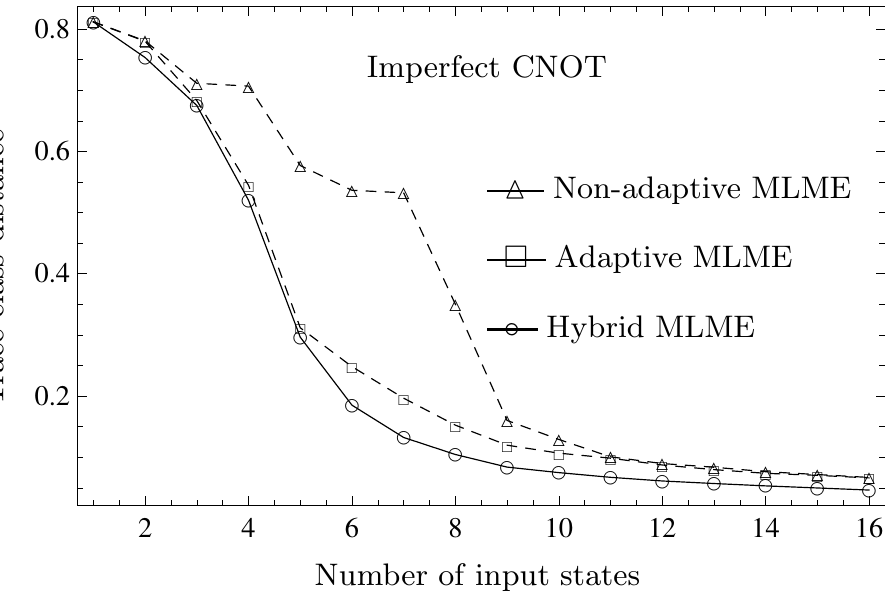}
  \caption{A comparison of three incomplete QPT schemes: the non-adaptive MLME scheme, the adaptive MLME scheme and a combination of the adaptive MPL-MLME scheme and the adaptive MLME scheme (hybrid scheme). Monte Carlo simulations are carried out on the imperfect \textsc{cnot} gate with $\epsilon=0.1$. Here, $N=10^4$ and an average over 50 experiments is taken to compute the trace-class distances. For both the non-adaptive as well as the adaptive MLME schemes, the default set of 16 linearly independent input states are chosen to be tensor products of projectors of the kets $\ket{0}$, $\ket{1}$, $(\ket{0}+\ket{1})/\sqrt{2}$ and $(\ket{0}+\ket{1}\I)/\sqrt{2}$. For all schemes, a set of 16 randomly generated positive operators, which are all linearly independent of one another, are used to form the POM. For this POM, the average repetition frequency of the adaptive MPL-MLME scheme is very high after four input states are used. The first input state for all schemes is chosen to be the same separable state $\rho^{(1)}_\text{i}=\ket{00}\bra{00}$. For the third scheme, the second to the fourth input states (shaded region) are optimized using the adaptive MPL-MLME strategy and the subsequent input states are chosen via the adaptive MLME strategy using the default set of input states which excludes $\ket{00}\bra{00}$. The plot shows that the overall performance of the combined strategy is better than the adaptive MLME strategy alone.}
  \label{fig:combined}
\end{figure}

Another technical subtlety is that these local-maxima estimators tend to repeat themselves during the optimization. Hence, a local-maxima estimator which was chosen as one of the input states earlier may reappear in later optimizations. The repetition frequency strongly depends on the POM chosen to measure the output states. The examples given thus far make use of the product tetrahedron measurements as the POM and the resulting MPL optimizations give linearly independent estimators with few repetitions. This may not be the case for other types of POM. In view of this, another way of doing AMLME QPT is to use both adaptive strategies in \S\ref{subsec:fixed_input} and \S\ref{subsec:all_input} interchangeably, the \emph{hybrid} MLME strategy. For example, one can start with the adaptive MPL-MLME strategy for tomography and when the repetition rate increases as more input states have been used, one may switch to the first adaptive MLME strategy. Figure~\ref{fig:combined} suggests that such a hybrid MLME strategy can further improve the tomographic performance as compared with the adaptive MLME strategy alone.

\subsection{Fixed measurement resources}
\begin{figure}[htp]
  \centering
  \includegraphics[width=0.6\textwidth]{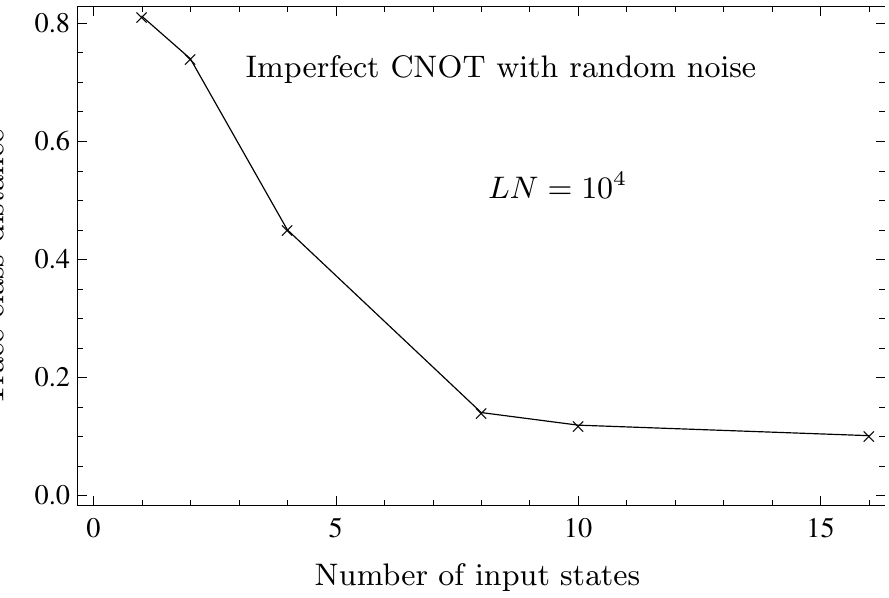}
  \caption{Numerical simulation on the imperfect two-qubit \textsc{cnot} gate with random noise for fixed $LN=10^4$. An average over 50 experiments is taken to compute the trace-class distances. The adaptive MPL-MLME strategy is used when the number of input states $L$ is less than 16.}
  \label{fig:fixed_LN}
\end{figure}

\index{measurement resources}
Finally, we try to answer, with numerics, the following question: For a fixed value of $LN$, is it more beneficial, in terms of tomographic performance, to measure more input states with fewer copies per input state or to measure fewer input states with more copies per input state? In quantum state estimation, it is well known that for a fixed number of measurement copies, it is better to measure more POM outcomes, an overcomplete set if possible \cite{overcomplete}. To see if there exists an analogous benefit to measure more input states in QPT, we performed a simulation with a fixed value of $LN$ and show the results in Fig.~\ref{fig:fixed_LN}.

It turns out that the average trace-class distance is a monotonically decreasing function of $L$, with the maximal $L=L_\text{ML}=16$. Hence, for a fixed amount of measurement resources\index{measurement resources}, the advantage of increasing the different types of measurements carries over to quantum process estimation. However, it is important to note that this does not contradict the fact that for a fixed average trace-class distance, one can use MLME to reduce the total number of measurement resources/settings\index{measurement resources} by simply reducing the number of input states necessary to achieve this distance. This is because, as discussed previously in \S\ref{sec:qpt_mlme_iteralgo} and also shown in Fig.~\ref{fig:fixed_LN}, the improvement gained by increasing the number of input states $L$ decreases rapidly with $L$, especially when the input states are chosen optimally. Put differently, it is not worth the trouble to increase $L$ after some point, beyond which there is very little tomographic improvement. This point, which is the essence of AMLME QPT, cannot be overemphasized. Experimentally, this means that one need not perform full tomography to obtain a quantum process estimator within a certain preset error margin since other confounding variables contribute to the total experimental error anyway. 

\section{Chapter summary}
\label{sec:qpe_conc}

We have established adaptive numerical strategies to perform incomplete quantum process tomography. One may choose whichever strategy is convenient to carry out tomography depending on the available types of measurement resources at hand. Each of these strategies combines the simplicity of incomplete quantum process tomography using quantum state estimation with good tomographic performances using optimization techniques. It can never be overemphasized that, although some prior information is necessary for each adaptive strategy, such information is \emph{never} used in the estimation of the unknown quantum process. Rather, the prior information is utilized to adaptively select future input states, the input states in our context, based on the current measurement data, to optimize the tomographic performance. The discussions presented in this chapter, therefore, provide a means of obtaining estimators for the unknown quantum process using incomplete resources which are typically within reasonably good experimental precisions. These estimators are statistically meaningful in that they are least-biased with respect to a set of informationally incomplete measurement data and are hence suitable for partial characterization of quantum processes. This is in contrast with the standard quantum process tomography which generally requires a huge amount of informationally complete measurement resources. 

\cleardoublepage
\chapter*{Conclusion}
\addcontentsline{toc}{chapter}{Conclusion}
\markboth{\hfill \,}{\, \hfill}
\label{chap:main_conc}

The frequentist's notion of quantum estimation serves as a very useful methodology for estimating the identity of a given source of quantum systems or a quantum channel. In this dissertation, we have touched on several aspects of this theory. They involve the two main statistical principles of maximum-likelihood and maximum-entropy, both of which are celebrated approaches in the subject of classical parameter estimation. Numerical techniques were developed to reconstruct quantum states and processes from the measurement data obtained. One important experimental application of these techniques, namely entanglement detection, was discussed in detail. Another important direction from the materials discussed in this dissertation would be to develop numerical methods for the construction of error bars that go with the reconstructed statistical or process operators. In view of this, we briefly mention that a methodology to construct what is called the \emph{region estimator} for a given set of measurement data was discussed in a recent Workshop on Quantum Tomography (WQT@CQT 28 November -- 02 December 2011). This estimator is a region of statistical operators that encloses the true state/process with a high probability, based on a pre-chosen likelihood ratio. Further improvements of this methodology with incomplete measurement data is a subject of future work. 

\appendix
\numberwithin{equation}{chapter}


\cleardoublepage
\chapter{Dual Superkets of the SIC POM}\index{dual}
\label{chap:dual_sic}
The superkets\index{superket} $\sket{\Pi_j}$ of a $D$-dimensional SIC POM follows the trace relation
\begin{equation}
\superinner{\Pi_j}{\Pi_k}=\frac{D\delta_{jk}+1}{D^2(D+1)}\,.
\end{equation}
These $D^2$ superkets\index{superket} are therefore not orthonormal\index{orthonormal} to one another. To facilitate the subsequent calculations, it is convenient to construct a set of $D^2$ orthonormal\index{orthonormal} superkets, denoted by $\sket{\Pi^\perp_j}$, out of the $\sket{\Pi_j}$s. To do this, we use the following ansatz:
\begin{equation}
\sket{\Pi^\perp_j}=\left(\sket{\Pi_j}\alpha+\sket{1}\beta\right)\left(\frac{D}{\sqrt{\alpha^2+\beta^2D^3+2\alpha\beta D}}\right)\,,
\end{equation}
where $\superinner{\Pi^\perp_j}{\Pi^\perp_j}=1$. The inner product
\begin{equation}
\superinner{\Pi^\perp_j}{\Pi^\perp_k}=\frac{D\alpha^2\delta_{jk}+\alpha^2+\beta^2D^3(D+1)+2\alpha\beta D(D+1)}{(D+1)(\alpha^2+\beta^2D^3+2\alpha\beta D)}
\end{equation}
suggests that $\alpha^2+\beta^2D^3(D+1)+2\alpha\beta D(D+1)=0$ for $\superinner{\Pi^\perp_j}{\Pi^\perp_k}=\delta_{jk}$. This equation allows for a free variable $\alpha$ or $\beta$. Choosing $\beta=1/D$, we find that $\alpha=\sqrt{D+1}-D-1$. Hence a good choice of orthonormal\index{orthonormal} superkets\index{superket} are
\begin{equation}
\sket{\Pi^\perp_j}=\frac{\sket{\Pi_j}\sqrt{D}\left(\sqrt{D+1}-D-1\right)+\sket{1}\frac{1}{\sqrt{D}}}{\sqrt{D+2-2\sqrt{D+1}}}\,.
\end{equation}
Using these orthonormal\index{orthonormal} superkets\index{superket} and after a tedious simplification, we obtain the matrix elements of $\mathcal{F}=\sum_l\sket{\Pi_l}\sbra{\Pi_l}$ to be
\begin{equation}
\mathcal{F}_{jk}=\sbra{\Pi^\perp_j}\mathcal{F}\sket{\Pi^\perp_k}=\frac{1}{D(D+1)}\delta_{jk}+\frac{1}{D^2(D+1)}\,.
\end{equation}
This means that
\begin{equation}
\mathcal{F}=\frac{1}{D(D+1)}\mathcal{I}+\sum_l\sum_{l'}\sket{\Pi^\perp_l}\frac{1}{D^2(D+1)}\sbra{\Pi^\perp_{l'}}\,.
\end{equation}
The form of the superoperator\index{superoperator} $\mathcal{F}$ is that of $\mathcal{F}=a\mathcal{I}+b\kappa\mathcal{P}$, where $\mathcal{P}=\mathcal{P}^2$ is a rank-1 projector. To invert this superoperator\index{superoperator}, we note that
\begin{align}
\mathcal{F}&=a\mathcal{I}+b\kappa\mathcal{P}\nonumber\\
&=a(\mathcal{I}-\mathcal{P})+(a+b\kappa)\mathcal{P}\,.
\end{align}
Since $\mathcal{I}-\mathcal{P}$ and $\mathcal{P}$ are orthogonal projectors, the inverse of $\mathcal{F}$ is given by
\begin{align}
\mathcal{F}^{-1}&=\frac{1}{a}(\mathcal{I}-\mathcal{P})+\frac{1}{a+b\kappa}\mathcal{P}\nonumber\\
&=\frac{1}{a}\mathcal{I}-\frac{b\kappa}{a^2+ab\kappa}\mathcal{P}\,.
\end{align}
Using the parameters $a=1/D(D+1)$, $b=1/D^2(D+1)$, $\kappa=D^2$ and
\begin{equation}
\mathcal{P}=\sum_l\sum_{l'}\sket{\Pi^\perp_l}\frac{1}{D^2}\sbra{\Pi^\perp_{l'}}\,,
\end{equation}
we obtain
\begin{equation}
\mathcal{F}^{-1}=D(D+1)\mathcal{I}-\sum_l\sum_{l'}\sket{\Pi^\perp_l}\sbra{\Pi^\perp_{l'}}\,.
\end{equation}
Thus,
\begin{equation}
\sket{\Theta_j}=\mathcal{F}^{-1}\sket{\Pi_j}=\sket{\Pi_j}D(D+1)-\sket{1}\,,
\end{equation}
where
\begin{align}
\superinner{\Theta_j}{\Pi_k}&=\Big[D(D+1)\sbra{\Pi_j}-\sbra{1}\Big]\sket{\Pi_k}\nonumber\\
&=D(D+1)\frac{D\delta_{jk}+1}{D^2(D+1)}-\frac{1}{D}=\delta_{jk}
\end{align}
as it should be. 

\cleardoublepage
\chapter{Wigner Functional in Fock Representation}\index{Wigner functional}\index{Fock states or representation}
\label{chap:wigner}
With the help of the relation between the Fock-state wave functions\index{Fock states or representation} $\langle x|n\rangle$ and the Hermite polynomials\index{Hermite polynomials} $H_n(x)$\label{symbol:hermite} given by
\begin{equation}
\langle x|n\rangle=\frac{1}{\pi^{-1/4}\sqrt{2^n\,n!}}\,\E^{-x^2/2}\,H_n(x)\,,
\end{equation}
the one-dimensional Wigner functional, defined as
\begin{equation}
\mathcal{W}(x,p)=2\int\D y\,\E^{2\I py}\bra{x-y}\rho\ket{x+y}
\end{equation}
for the dimensionless values $x$ and $p$, for a given statistical operator $\rho$ can be represented in the Fock basis as
\begin{align*}
\mathcal{W}(x,p)&=2\sum^{\infty}_{m=0}\sum^{\infty}_{n=0}\int\D y\,\E^{2\I py}\langle x-y|m\rangle\underbrace{\bra{m}\rho\ket{n}}_{=\,\rho_{mn}}\langle n|x+y\rangle\\
&=2\,\frac{\E^{-x^2-p^2}}{\sqrt{\pi}}\sum^{\infty}_{m=0}\sum^{\infty}_{n=0}\frac{\rho_{mn}}{\sqrt{2^{m+n}\,m!\,n!}}\int\D y\,\E^{-(y-\I p)^2}\,H_m(x-y)\,H_n(x+y)\\
&=2\,\frac{\E^{-x^2-p^2}}{\sqrt{\pi}}\sum^{\infty}_{m=0}\sum^{\infty}_{n=0}\frac{(-1)^m\rho_{mn}}{\sqrt{2^{m+n}\,m!\,n!}}\int\D y'\,\E^{-y'^2}\,H_m(y'-a)\,H_n(y'+b)\,,
\end{align*}
where $a=x-\I p$ and $b=x+\I p$.
\newpage
In obtaining the final equation, a new variable $y'=y-\I p$ is introduced and the property $H_n(-x)=(-1)^nH_n(x)$ is used. The job is thus to evaluate the integral of the general form
\begin{equation}
\mathcal{I}(a,b;m,n)=\int\D y'\,\E^{-y'^2}\,H_m(y'-a)\,H_n(y'+b)\,.
\label{prod_herm_int}
\end{equation}

We shall first consider the case where $m\geq n$. To proceed, it is useful for us to understand the response of $H_n(x)$ when the argument is shifted by a constant $a$. We begin with the generating function
\begin{equation}
\E^{2xt-t^2}=\sum^{\infty}_{n=0}H_n(x)\frac{t^n}{n!}\,
\end{equation}
for the Hermite polynomials.

It follows that
\begin{align}
\sum^{\infty}_{n=0}H_n(x+a)\frac{t^n}{n!}&=\sum^{\infty}_{k=0}H_k(x)\,\E^{2at}\,\frac{t^k}{k!}\nonumber\\
&=\sum^{\infty}_{j=0}\sum^{\infty}_{k=0}H_k(x)\,\frac{(2a)^{j}\,t^{j+k}}{k!\,j!}\nonumber\\
(n\equiv j+k)\Rightarrow&=\underbrace{\sum^{\infty}_{k=0}\sum^{\infty}_{n=k}}_{\displaystyle =\sum^{\infty}_{n=0}\sum^{n}_{k=0}}H_k(x)\,\frac{(2a)^{n-k}\,t^{n}}{k!\,(n-k)!}\nonumber\\
&=\sum^{\infty}_{n=0}\frac{t^{n}}{n!}\sum^{n}_{k=0}\binom{n}{k} H_k(x)\,(2a)^{n-k}\,,
\end{align}
so that
\begin{equation}
H_n(x+a)=\sum^{n}_{k=0}\binom{n}{k} H_k(x)\,(2a)^{n-k}\,.
\label{herm_trans}
\end{equation}
Using Eq.~(\ref{herm_trans}), the integral in Eq.~(\ref{prod_herm_int}) turns into
\begin{align}
\mathcal{I}(a,b;m\geq n)&=\sum^{m}_{j=0}\sum^{n}_{k=0}\binom{m}{j}\binom{n}{k}(-2a)^{m-j}(2b)^{n-k}\nonumber\\
&\quad\quad\quad\quad\times\underbrace{\int\D y'\,\E^{-y'^2}\,H_j(y')\,H_k(y')}_{=\,\sqrt{\pi}\,2^k\,k!\,\delta_{jk}\,\text{(Orthogonality relation)}}\nonumber\\
&=\sqrt{\pi}\,2^{m}\,(-a)^{m-n}\sum^{n}_{k=0}\binom{m}{k}\binom{n}{k}\,k!\,(-2ab)^{n-k}\nonumber\\
&=\sqrt{\pi}\,2^{m}\,n!\,(-a)^{m-n}\sum^{n}_{k=0}\binom{m}{k}\,\frac{(-2ab)^{n-k}}{(n-k)!}\nonumber\\
(j\equiv n-k)\Rightarrow&=\sqrt{\pi}\,2^{m}\,n!\,(-a)^{m-n}\sum^{n}_{j=0}\binom{n+(m-n)}{n-j}\,\frac{(-2ab)^{j}}{j!}\,.
\end{align}
From the definition
\begin{equation}
L_n^{(\nu)}(y)\equiv\frac{y^{-\nu}\E^y}{n!}\left(\frac{\D}{\D y}\right)^n\left(y^{n+\nu}\E^{-y}\right)=\sum^{n}_{j=0}\binom{n+\nu}{n-j}\,\frac{(-y)^{j}}{j!}\,,
\end{equation}
where $L_n^{(\nu)}(y)$ are the associated Laguerre polynomials\index{Laguerre polynomials (associated)} in $y$ of degree $n$ and order $\nu$, we have
\begin{equation}
\mathcal{I}(a,b;m\geq n)=\sqrt{\pi}\,2^{m}\,n!\,(-a)^{m-n}\,L_n^{(m-n)}(2ab)\,.
\label{mgreatn}
\end{equation}
The corresponding expression for $m< n$ requires the roles of $m$ and $n$, as well as those of $-a$ and $b$, to be interchanged. Thus
\begin{equation}
\mathcal{I}(a,b;m< n)=\sqrt{\pi}\,2^{n}\,m!\,b^{n-m}\,L_m^{(n-m)}(2ab)\,.
\label{mlessn}
\end{equation}
With Eqs.~(\ref{mgreatn}) and (\ref{mlessn}), we can write
\begin{equation}
\mathcal{W}(x,p)=2\,\E^{-x^2-p^2}\sum^{\infty}_{m=0}\sum^{\infty}_{n=0}\rho_{mn}\,\mathcal{M}_{mn}(x,p)\,,
\end{equation}
where
\begin{equation}
\mathcal{M}_{mn}(x,p)=
\begin{cases}
   (-1)^n\sqrt{\frac{2^m\,n!}{2^{n}\,m!}}(x-\I p)^{m-n}\,L_n^{(m-n)}(2x^2+2p^2) & \text{if } m \geq n \\
   (-1)^m\sqrt{\frac{2^n\,m!}{2^{m}\,n!}}(x+\I p)^{n-m}\,L_m^{(n-m)}(2x^2+2p^2) & \text{if } m < n
\end{cases}\,
\end{equation}
or, with $n_<=\min\{m,n\}$ and $n_>=\max\{m,n\}$,
\begin{equation}
\mathcal{M}_{mn}(x,p)=
(-1)^{n_<}\sqrt{\frac{2^{n_>}\,n_<!}{2^{n_<}\,n_>!}}(x+\I^{\,\text{sgn}(n-m)} \,p)^{n_>-n_<}\,L_{n_<}^{(n_>-n_<)}(2x^2+2p^2)\,.
\end{equation} 

\cleardoublepage
\chapter{Formula for Computing the Non-classicality Depth}\index{non-classicality!-- depth}
\label{chap:nonclass_depth}
From the definitions of the function
\begin{equation}
\mathcal{R}(\alpha;\tau)=\frac{1}{\pi\tau}\int (\D w)\,\text{exp}\left(-\frac{|\alpha/\sqrt{2}-w|^2}{\tau}\right)P(w)
\end{equation}
and the Glauber-Sudarshan $P$ function\index{Glauber-Sudarshan $P$ function} \cite{clmehta}
\begin{equation}
P(w)=\frac{\E^{|w|^2}}{\pi}\int(\D u)\,\langle -u^*|\rho|u\rangle\E^{|u|^2}\E^{wu^*-w^*u}\,
\end{equation}
with the overcomplete set of coherent states\index{coherent states} $\ket{u}$, similar manipulation in Appendix \ref{chap:wigner} gives
\begin{align}
R(\alpha;\tau)=&\,\frac{2}{\pi\,(1-\tau)}\,\E^{\frac{|\alpha|^2}{2(1-\tau)}}\sum^\infty_{m=0}\sum^\infty_{n=0}\frac{(-1)^m\rho_{mn}}{\sqrt{m!\,n!}}\nonumber\\
&\,\times\int(\D u)\,u^{*\,m}u^n\E^{-\frac{1}{1-\tau}\left[|u|^2\tau+\left(\alpha^*u-\alpha u^*\right)/\sqrt{2}\right]}\,.
\label{r_prelim1}
\end{align}
By defining $z=|z|\,\begin{small}\text{exp}\end{small}(\I\theta)\equiv\alpha/\sqrt{2}$, the necessary integral for subsequent calculations from Eq.~(\ref{r_prelim1}) is given by
\begin{equation}
\mathcal{I}(z,\tau;m,n)=\int(\D u)\,u^{*\,m}u^n\E^{-\frac{|u|^2\tau}{1-\tau}}\E^{-\frac{1}{1-\tau}(z^*u-zu^*)}\,.
\end{equation}
By introducing the polar coordinates $(\D u)=s\,\D s\,\D\varphi$, we have
\begin{align}
\mathcal{I}(z,\tau;m,n)&=\int^{\infty}_0\D s\,s^{m+n+1}\E^{-\frac{s^2\tau}{1-\tau}}\underbrace{\int^{2\pi}_0\D\varphi\,\E^{-\I\varphi(m-n)}\E^{-2\I |z|s\sin(\varphi-\theta)/(1-\tau)}}_{=2\pi\E^{-\I\varphi(m-n)} J_{m-n}\left(-\frac{2|z|s}{1-\tau}\right)}\nonumber\\
&=2\pi\E^{-\I\theta(m-n)}\int^{\infty}_0\D s\,s^{m+n+1}\E^{-\frac{s^2\tau}{1-\tau}}J_{m-n}\left(-\frac{2|z|s}{1-\tau}\right)\,,
\label{first_int}
\end{align}
where the second equality is obtained via the integral definition
\begin{equation}
J_\mu\left(y\right)=\frac{1}{2\pi}\int^{2\pi}_0\D\varphi\,\E^{-\I(\mu\varphi-y\sin\varphi)}\,
\label{besselintdef}
\end{equation}
of the Bessel function\index{Bessel functions (first kind)} (Friedrich Wilhelm Bessel\index{Friedrich Wilhelm Bessel}) of the first kind $J_\mu(y)$\label{symbol:jmuy} of integer order $\mu$. Using a new set of variables $t=s^2\tau/(1-\tau)$ and supposing that $m\geq n$,
\begin{align}
\mathcal{I}(z,\tau;m\geq n)=&\,(-1)^{m-n}\pi\E^{-\I\theta(m-n)}\left(\frac{1-\tau}{\tau}\right)^{m+1}\left(\frac{|z|}{1-\tau}\right)^{m-n}\nonumber\\
&\times\int^{\infty}_0\D t\,t^{n+(m-n)}\E^{-t}\frac{J_{m-n}\left(2\sqrt{\frac{|z|^2}{\tau(1-\tau)}t}\right)}{\left(\sqrt{\frac{|z|^2}{\tau(1-\tau)}t}\right)^{m-n}}\,.
\label{r_prelim2}
\end{align}
In deriving the identity above, we make use of the fact that $J_\mu(-y)=(-1)^\mu\,J_\mu(y)$, which follows immediately from the generating function
\begin{equation}
\E^{\frac{y}{2}\left(t-\frac{1}{t}\right)}=\sum_{\mu=-\infty}^\infty\,J_\mu(y)\,t^\mu\,,
\end{equation}
where $t$ is complex. To evaluate the integral in Eq.~(\ref{r_prelim2}), we need a few identities for $J_\mu(y)$. Let us start by establishing the power series expansion for $J_\mu(y)$ with an integer order $\mu$. For this, we need the expression for the $k$th derivative of $J_\mu(y)$ with respect to $y$. From Eq.~(\ref{besselintdef}),
\begin{align}
\left(\frac{\D}{\D y}\right)^kJ_\mu(y)\bigg|_{y=0}&=\frac{1}{2\pi}\int^{2\pi}_0\D\varphi\,\left(\I\sin\varphi\right)^k\E^{-\I(\mu\varphi-y\sin\varphi)}\bigg|_{y=0}\nonumber\\
&=\frac{\I^k}{2\pi}\int^{2\pi}_0\D\varphi\,\left(\sin\varphi\right)^k\E^{-\I\mu\varphi}\,.
\end{align}
Using the parametrization $q=\E^{\I\varphi}$,
\begin{align}
\left(\frac{\D}{\D y}\right)^kJ_\mu(y)\bigg|_{y=0}&=\frac{\I^k}{2\pi}\int^{2\pi}_0\D\varphi\,\left(\frac{\E^{\I\varphi}-\E^{-\I\varphi}}{2\I}\right)^k\E^{-\I\mu\varphi}\nonumber\\
&=\frac{1}{2^k}\frac{1}{2\pi}\oint_{\substack{\text{unit}\\\text{circle}}}\frac{\D q}{\I q}\,\frac{\left(q-q^{-1}\right)^k}{q^{\mu}}\nonumber\\
&=\frac{1}{2^k}\frac{1}{2\pi\I}\oint_{\substack{\text{unit}\\\text{circle}}}\D q\,\frac{(q^2-1)^k}{q^{k+\mu+1}}\,.
\end{align}The resulting contour integral\index{contour integral} can be evaluated using the Cauchy's Residue Theorem\index{Cauchy's Residue Theorem}\index{Residue Theorem|see{Cauchy's Residue Theorem}} (Baron Augustin-Louis Cauchy\index{Augustin-Louis Cauchy}), from which we have
\begin{equation}
\frac{1}{2\pi\I}\oint_{\substack{\text{unit}\\\text{circle}}}\D q\,\frac{(q^2-1)^k}{q^{k+\mu+1}}=\text{Res}\left(\frac{(q^2-1)^k}{q^{k+\mu+1}},0\right)\,.
\end{equation}
Since the pole\index{pole (complex analysis)} $q=0$ of the complex function in the argument is of order $k+\mu+1$, the corresponding residue\index{residue (complex analysis)} can be calculated using the formula
\begin{equation}
\text{Res}\left(\frac{(q^2-1)^k}{q^{k+\mu+1}},0\right)=\frac{1}{(k+\mu)!}\left(\frac{\D}{\D q}\right)^{k+\mu}(q^2-1)^k\bigg|_{q=0}\,.
\end{equation}

Since
\begin{align}
&\,\left(\frac{\D}{\D q}\right)^{k+\mu}(q^2-1)^k\bigg|_{q=0}\nonumber\\
=&\,\sum^k_{l=0}\binom{k}{l}(-1)^{k-l}\left(\frac{\D}{\D q}\right)^{k+\mu}q^{2l}\bigg|_{q=0}\nonumber\\
=&\,(-1)^k\sum^k_{l=0}\binom{k}{l}(-1)^{-l}\frac{(2l)!}{(2l-k-\mu)!}\,q^{2l-k-\mu}\bigg|_{q=0}\nonumber\\
=&\,(-1)^{\frac{k-\mu}{2}}(k+\mu)!\,\frac{k!}{\left(\frac{k-\mu}{2}\right)!\,\left(\frac{k+\mu}{2}\right)!}\,,
\end{align}
we have
\begin{equation}
\frac{1}{k!}\,\left(\frac{\D}{\D y}\right)^kJ_\mu(y)\bigg|_{y=0}=\frac{1}{2^k}\frac{(-1)^{\frac{k-\mu}{2}}}{\left(\frac{k-\mu}{2}\right)!\,\left(\frac{k+\mu}{2}\right)!}
\end{equation}
Thus, the Maclaurin series\index{Maclaurin series} (Colin Maclaurin\index{Colin Maclaurin}) of $J_\mu(y)$ is given by
\begin{align}
J_\mu(y)&=\sum^\infty_{k=0}\left(\frac{\D}{\D y}\right)^kJ_\mu(y)\bigg|_{y=0}\,\frac{y^k}{k!}\nonumber\\
&=\sum^\infty_{k=\mu}\frac{(-1)^{\frac{k-\mu}{2}}}{\left(\frac{k-\mu}{2}\right)!\,\left(\frac{k+\mu}{2}\right)!}\left(\frac{y}{2}\right)^k\,,
\end{align}
where we note that $\left[(k-\mu)/2\right]!=\infty$ when $k<\mu$. After a change of variable $k\rightarrow\,2k+\mu$, we finally obtain the power series expansion
\begin{equation}
J_\mu(y)=\sum^\infty_{k=0}\frac{(-1)^k}{k!\,(k+\mu)!}\,\left(\frac{y}{2}\right)^{2k+\mu}\,,
\end{equation}
which is very useful to obtain the necessary identities to proceed. We note that this formula is valid for \emph{any} real number $\mu$, although we have derived it from Eq.~(\ref{besselintdef}) for integer $\mu$. Since we are still considering the case where $m\geq n$, $\mu=m-n\geq0$.

The first identity
\begin{align}
\frac{\D}{\D y}\left(y^\mu J_\mu(y)\right)&=\sum^\infty_{k=0}\frac{(-1)^k}{k!(k+\mu)!}\left(\frac{1}{2}\right)^{2k+\mu}\frac{\D}{\D y}y^{2(k+\mu)}\nonumber\\
&=\sum^\infty_{k=0}\frac{(-1)^k}{k!(k+\mu)!}\frac{2(k+\mu)}{2^{2k+\mu}}y^{2(k+\mu)-1}\nonumber\\
&=y^\mu\sum^\infty_{k=0}\frac{(-1)^k}{k!(k+\mu-1)!}\left(\frac{y}{2}\right)^{2k+\mu-1}\nonumber\\
&=y^\mu J_{\mu-1}(y)\,,
\end{align}
relates the $y$-derivative of $y^\mu J_\mu(y)$ to another Bessel function\index{Bessel functions (first kind)} $J_{\mu-1}(y)$ that is one order lower. Next,
\begin{align}
&\int^\infty_0\D t\,\E^{-t}\left(\sqrt{y\,t}\right)^{n+\nu}J_{n+\nu}\left(2\sqrt{y\,t}\right)\nonumber\\
=&\,y^{\frac{n+\nu}{2}}\sum^\infty_{k=0}\frac{(-1)^k}{k!(k+n+\nu)!}y^{k+\frac{n+\nu}{2}}\underbrace{\int^\infty_0\D t\,\E^{-t}t^{k+n+\nu}}_{=(k+n+\nu)!}\nonumber\\
=&\,y^{n+\nu}\sum^\infty_{k=0}\frac{(-y)^k}{k!}\nonumber\\
=&\,y^{n+\nu}\E^{-y}\,.
\end{align}

With these two identities, and the definition of the associated Laguerre polynomials\index{Laguerre polynomials (associated)} in Appendix \ref{chap:wigner}, we have
\begin{align}
L_n^{(\nu)}(y)&=\frac{y^{-\nu}\E^y}{n!}\left(\frac{\D}{\D y}\right)^n\left(y^{n+\nu}\E^{-y}\right)\nonumber\\
&=\frac{y^{-\nu}\E^y}{n!}\int^\infty_0\D t\,\E^{-t}\underbrace{\left(\frac{\D}{\D y}\right)^n\left[\left(\sqrt{y\,t}\right)^{n+\nu}J_{n+\nu}\left(2\sqrt{y\,t}\right)\right]}_{=y^{\nu/2}\,t^{(n+\nu)/2}J_\nu\left(2\sqrt{y\,t}\right)}\nonumber\\
&=\frac{1}{n!}\int^\infty_0\D t\,\E^{y-t}t^{n+\nu}\frac{J_\nu\left(2\sqrt{y\,t}\right)}{\left(\sqrt{y\,t}\right)^\nu}\,.
\label{int_lag}
\end{align}

Thus, using the integral representation in Eq.~(\ref{int_lag}) for $\nu\equiv m-n\geq0$, we have
\begin{align}
\mathcal{I}(z,\tau;m\geq n)=&\,(-1)^{m-n}\,\pi\,n!\,\left(z^*\right)^{m-n}\E^{-\frac{|z|^2}{\tau(1-\tau)}}\nonumber\\
&\times\left(\frac{1-\tau}{\tau}\right)^{m+1}\left(\frac{1}{1-\tau}\right)^{m-n}L^{(m-n)}_n\left(\frac{|z|^2}{\tau(1-\tau)}\right)\,.
\label{i1}
\end{align}
For the case where $m<n$, we make use of the property
\begin{equation}
J_{-\mu}(y)=(-1)^\mu J_{\mu}(y)
\end{equation}
and evaluate the integral in Eq.~(\ref{first_int}) using, again, Eq.~(\ref{int_lag}), from which we obtain
\begin{align}
\mathcal{I}(z,\tau;m<n)=&\,(-1)^{m-n}\,\pi\,m!\,z^{n-m}\E^{-\frac{|z|^2}{\tau(1-\tau)}}\nonumber\\
&\times\left(\frac{1-\tau}{\tau}\right)^{n+1}\left(\frac{1}{1-\tau}\right)^{n-m}L^{(n-m)}_m\left(\frac{|z|^2}{\tau(1-\tau)}\right)\,.
\label{i2}
\end{align}
Finally, using the results in Eq.~(\ref{i1}) and (\ref{i2}), we have
\begin{align}
&\mathcal{R}(x,p;\tau)=\frac{\E^{-\frac{|\alpha|^2}{2\tau}}}{\tau}\sum^{\infty}_{m=0}\sum^{\infty}_{n=0}\bra{m}\rho\ket{n}\nonumber\\
\times \Bigg[&\,(-1)^{n_<}\sqrt{\frac{n_<!}{n_>!}}\left(\frac{1-\tau}{\tau}\right)^{n_>}\left(\frac{x+\I^{\,\text{sgn}(n-m)} p}{\sqrt{2}(1-\tau)}\right)^{n_>-n_<}L_{n_<}^{(n_>-n_<)}\left(\frac{|\alpha|^2}{2\tau(1-\tau)}\right)\Bigg]\,,
\end{align}
where $n_<=\min\{m,n\}$ and $n_>=\max\{m,n\}$. 

\cleardoublepage
\chapter{Uniqueness of the Hedged Likelihood Estimator}\index{unique estimator}
\label{chap:hedged_unique}
We suppose that there exist two estimators $\hat{\rho}_1$ and $\hat{\rho}_2$ that maximize the hedged likelihood functional $\mathcal{L}_\text{H}(\{n_j\};\rho)$. The concavity\index{concavity} of $\mathcal{L}_\text{H}(\{n_j\};\rho)$, which is equivalently expressed by the inequality
\begin{equation}
\mathcal{L}_\text{H}(\{n_j\};\lambda\hat{\rho}_1+(1-\lambda)\hat{\rho}_2)\geq \lambda\mathcal{L}_\text{H}(\{n_j\};\hat{\rho}_1)+ (1-\lambda)\mathcal{L}_\text{H}(\{n_j\};\hat{\rho}_2)\,
\end{equation}
for $0\leq \lambda \leq 1$, implies that the convex sum $\hat{\rho}'=\lambda\hat{\rho}_1+(1-\lambda)\hat{\rho}_2$ also maximizes $\mathcal{L}_\text{H}(\{n_j\};\rho)$. In other words, if we vary the parameter $\lambda$ along the direction from $\hat{\rho}_1$ to $\hat{\rho}_2$ and vice versa, the gradient of $\mathcal{L}_\text{H}(\{n_j\};\rho)$ will always be zero. Hence, making use of Eq.~(\ref{deltaf}), we obtain two simultaneous equations inasmuch as
\begin{align}
\tr{\big(\beta\hat{\rho}_1^{-1}+N\hat{R}_1\big)\hat{\rho}_2}=\beta D+N\,,\nonumber\\
\tr{\big(\beta\hat{\rho}_2^{-1}+N\hat{R}_2\big)\hat{\rho}_1}=\beta D+N\,.
\end{align}

Adding the two equations and dividing the sum by 2, we have
\begin{equation}
\frac{\beta}{2}\tr{\hat{\rho}_1\hat{\rho}_2^{-1}+\hat{\rho}_2\hat{\rho}_1^{-1}}+\frac{N}{2}\tr{\hat{R}_1\hat{\rho}_2+\hat{R}_2\hat{\rho}_1}=\beta D+N\,.
\label{defeqn}
\end{equation}
Since both $\hat{\rho}_1$ and $\hat{\rho}_2$ and their corresponding inverses are full-rank, each product of operators in the first trace term is also full-rank. Defining $M\equiv\hat{\rho}_1\hat{\rho}_2^{-1}$, we can express the first term in the eigenvalues $\lambda_k$ of the full-rank operator $M$, i.e.
\begin{align}
&\frac{\beta}{2}\tr{\hat{\rho}_1\hat{\rho}_2^{-1}+\hat{\rho}_2\hat{\rho}_1^{-1}}\nonumber\\
\equiv &\frac{\beta}{2}\tr{M+M^{-1}}=\beta\sum_k{\underbrace{\Bigg(\frac{\lambda^2_k+1}{2\lambda_k}\Bigg)}_{\geq 1}}\nonumber\\
\geq &\beta\sum_k=\beta D\,.
\end{align}
For the second term, denoting $\hat{p}_{k,j}\equiv\tr{\hat{\rho}_k\Pi_j}$, a similar argument follows \cite{rehacek3}, namely
\begin{align}
\frac{N}{2}\tr{\hat{R}_1\hat{\rho}_2+\hat{R}_2\hat{\rho}_1} = & \frac{N}{2}\sum_k{\Bigg(\frac{f_k}{\hat{p}_{1,k}}\hat{p}_{2,k}+\frac{f_k}{\hat{p}_{2,k}}\hat{p}_{1,k}\Bigg)}\nonumber\\
= & N\sum_k{f_k\underbrace{\Bigg(\frac{\hat{p}^2_{1,k}+\hat{p}^2_{2,k}}{2 \hat{p}_{1,k}\hat{p}_{2,k}}\Bigg)}_{\geq 1}}\nonumber\\
\geq  & N\sum_k{f_k}=N\,.
\end{align}
Therefore the left-hand side of Eq.~(\ref{defeqn}) is always larger than the right-hand side unless of course $\lambda_k=1$ in the first term, which leads to $\hat{p}_{1,j}=\hat{p}_{2,j}$ needed for the equality in the second term. It follows that the operator $M$ is the identity operator. This means that $\hat{\rho}_1\hat{\rho}_2^{-1}=\hat{\rho}_2\hat{\rho}_1^{-1}=1$ and so $\hat{\rho}_1=\hat{\rho}_2$, which concludes the proof. 

\cleardoublepage
\bibliographystyle{alpha}
\bibliography{ref}
\addcontentsline{toc}{chapter}{Bibliography}

\cleardoublepage
\chapter*{List of Publications}
\addcontentsline{toc}{chapter}{List of Publications}
\markboth{\hfill \,}{\, \hfill}
\label{chap:pub}
\begin{enumerate}
\setcounter{enumi}{6} 
\sitem Y. S. Teo, B. Stoklasa, B.-G. Englert, J. {\v R}eh{\'a}{\v c}ek, and Z. Hradil, \emph{Incomplete quantum state estimation: A comprehensive study}, Phys. Rev. A\,\textbf{85}, 042317 (2012).

\ssitem Y. S. Teo, B.-G. Englert, J. {\v R}eh{\'a}{\v c}ek, and Z. Hradil, \emph{Adaptive
schemes for incomplete quantum process tomography}, Phys. Rev. A\,\textbf{84}, 062125 (2011).

\ssitem Y. S. Teo, H. Zhu, B.-G. Englert, J. {\v R}eh{\'a}{\v c}ek, and Z. Hradil, \emph{Quantum-state reconstruction by maximizing likelihood and entropy}, Phys. Rev. Lett.\,\textbf{107}, 020404 (2011).

\ssitem H. Zhu, Y.S. Teo and B.G. Englert, \emph{Two-qubit symmetric informationally complete positive-operator-valued measures}, Phys. Rev. A\,\textbf{82}, 042308 (2010).

\ssitem H. Zhu, Y. S. Teo, and B.-G. Englert, \emph{Minimal tomography with entanglement witnesses}, Phys. Rev. A\,\textbf{81}, 052339 (2010).

\ssitem Y. S. Teo, H. Zhu, and B.-G. Englert, \emph{Product measurements and fully symmetric measurements in qubit-pair tomography: A numerical study}, Opt. Commun. \textbf{283}, 724 (2010).

\end{enumerate} 

\cleardoublepage
\addcontentsline{toc}{chapter}{Index}
\singlespacing
\phantomsection
\printindex
\end{document}